\documentclass[
12pt, 
a4paper, 
onecolumn,
oneside, 
hidelinks,
table]{book}

\usepackage[columnsep=2cm,
 left=1.575in,
 right=0.78in,
 top=1.18in,
 bottom=.88in]{geometry} 

\usepackage[T1]{fontenc}
\usepackage[utf8]{inputenc}

\usepackage{titlesec}
\titleformat{\chapter}[display]
{\normalfont\huge\bfseries}{\chaptertitlename\ \thechapter}{0pt}{\Huge}

\titlespacing*{\section}{0pt}{1em}{0.3em}

\usepackage{sectsty}
\chapterfont{\setstretch{1.5}}

\usepackage[titles]{tocloft} 

\usepackage{textcomp}
\usepackage{gensymb}
\usepackage{enumitem}
\usepackage{graphicx}
\usepackage[colorinlistoftodos]{todonotes}
\usepackage{fancyhdr}
\usepackage{setspace}
\usepackage{url}
\usepackage{verbatim}
\usepackage[bottom]{footmisc}
\usepackage[english]{babel}
\usepackage{multirow}

\usepackage{pifont}
\usepackage[nottoc, numbib]{tocbibind}

\usepackage{amsmath}
\usepackage{amsfonts}
\usepackage{amsthm}
\usepackage{amssymb}
\usepackage{longtable,tabularx}
\usepackage{csquotes}
\usepackage{siunitx}
\usepackage{booktabs}

\usepackage{moreverb}
\usepackage{bm}
\usepackage{algpseudocode}
\usepackage{algorithm}
\usepackage{subfig}
\usepackage{minitoc}
\usepackage{longtable}
\usepackage{multicol}
\usepackage{hyperref}
\usepackage{cleveref}
\usepackage{bookmark}
\usepackage{xcolor}
\usepackage{pdfpages}
\usepackage{tikz}
\usepackage[titletoc]{appendix}
\usepackage{mathtools}
\usepackage{etoolbox}
\usepackage[htt]{hyphenat}
\usepackage{soul}
\usepackage{braket}

\usepackage[acronym, nogroupskip, nopostdot]{glossaries} 
\newglossary[slg]{symbol}{sld}{sdn}{Nomenclature}
\makeglossaries
\glstoctrue
\usepackage{afterpage}

\usepackage{emptypage}

\usepackage{bbm}

\usepackage{tcolorbox}

\let\Algorithm\algorithm
\renewcommand\algorithm[1][]{\Algorithm[#1]\setstretch{2}}

\usepackage{lscape}
\usepackage{adjustbox}
\usepackage{array}
\usepackage{multirow}
\usepackage{rotating}

\usepackage{balance} 
\usepackage{tabularx} 
\usepackage{xcolor,colortbl}
\usepackage{bbding}
\usepackage{dblfloatfix}
\usepackage{color}
\usepackage{pifont}

\usepackage{listings}

\PassOptionsToPackage{table,xcdraw}{xcolor}

\usepackage{makecell}

\renewcommand{\texttt}[1]{%
	\begingroup
	\ttfamily
	\begingroup\lccode`~=`/\lowercase{\endgroup\def~}{/\discretionary{}{}{}}%
	\begingroup\lccode`~=`[\lowercase{\endgroup\def~}{[\discretionary{}{}{}}%
	\begingroup\lccode`~=`.\lowercase{\endgroup\def~}{.\discretionary{}{}{}}%
	\catcode`/=\active\catcode`[=\active\catcode`.=\active
	\scantokens{#1\noexpand}%
	\endgroup
}

\newcolumntype{L}[1]{>{\raggedright\let\newline\\\arraybackslash\hspace{0pt}}m{#1}}
\newcolumntype{C}[1]{>{\centering\let\newline\\\arraybackslash\hspace{0pt}}m{#1}}
\newcolumntype{R}[1]{>{\raggedleft\let\newline\\\arraybackslash\hspace{0pt}}m{#1}}

\makeatletter
\newcommand*\fsize{\dimexpr\f@size pt\relax}
\makeatother

\usepackage[backend=biber,
 giveninits=true,
 isbn=true,
 url=true,
 doi=true,
 style=ieee,
 citestyle=numeric-comp,
 dashed=false,
 maxnames = 5]{biblatex}
\newcommand{\RNum}[1]{\uppercase\expandafter{\romannumeral #1\relax}}
{%
  \fancyhf{}
  \fancyhead[R]{\nouppercase{\leftmark}}
  \fancyfoot[C]{\thepage} 
  \renewcommand{\headrulewidth}{0pt}
  \renewcommand{\footrulewidth}{0pt}
}

\usepackage{hhline}
\definecolor{GrayCell}{HTML}{DDDDDD}
\definecolor{LightGrayCell}{HTML}{EEEEEE}
\definecolor{RedCell}{HTML}{FF0000}
\definecolor{GreenCell}{HTML}{32CB00}
\definecolor{YellowCell}{HTML}{F8FF00}
\definecolor{WhiteCell}{HTML}{FFFFFF}
\newcolumntype{L}[1]{>{\raggedright\let\newline\\\arraybackslash\hspace{0pt}}m{#1}}
\newcolumntype{C}[1]{>{\centering\let\newline\\\arraybackslash\hspace{0pt}}m{#1}}
\newcolumntype{R}[1]{>{\raggedleft\let\newline\\\arraybackslash\hspace{0pt}}m{#1}}

\newcolumntype{M}[1]{%
	>{\begin{turn}{90}\begin{minipage}{#1}%
				\raggedright\hspace{0pt}}l%
			<{\end{minipage}\end{turn}}}

\newcommand{\myverb}{\fontsize{10}{48}\usefont{OT1}{lmtt}{b}{n}\noindent }

\newcommand{\ie}{{\em i.e., }}
\newcommand{\eg}{{\em e.g., }}

\newcommand{\hlc}[2][yellow]{{%
    \colorlet{foo}{#1}%
    \sethlcolor{foo}\hl{#2}}%
}

\newacronym{iot}{IoT}{Internet of Things}
\newacronym{ai}{AI}{Artificial Intelligence}
\newacronym{mae}{MAE}{Mean Absolute Error}
\newacronym{mape}{MAPE}{Mean Absolute Percentage Error}
\newacronym{ui}{UI}{User Interface}
\newacronym{csv}{CSV}{Comma-Separated Values}
\newacronym{mlr}{MLR}{Multiple linear regression}
\newacronym{rf}{RF}{Random forests}
\newacronym{svr}{SVR}{Support Vector Regression}
\newacronym{svm}{SVM}{Support Vector Machine}
\newacronym{mse}{MSE}{mean square error}
\newacronym{rmse}{RMSE}{Root Mean Squared Error}
\newacronym{wmae}{WMAE}{Weighted Mean Absolute Error}
\newacronym{cp}{CP}{Constraint Programming}
\newacronym{lpr}{LPR}{License Plate Recognition}
\newacronym{anpr}{ANPR}{Automatic Number Plate Recognition}
\newacronym{ocr}{OCR}{Optical Character Recognition}
\newacronym{sd}{SD}{Secure Digital}
\newacronym{ftp}{FTP}{File Transfer Protocol}
\newacronym{vdc}{VDC}{Volts Direct Current}
\newacronym{db}{DB}{Database}
\newacronym{osl}{OSL}{Ordinary Least Squares}
\newacronym{fte}{FTE}{Full-Time Equivalent}
\newacronym{milp}{MILP}{Mixed Integer Linear Programming}
\newacronym{pdu}{PDU}{People Detector Unit}
\newacronym{lorawan}{LoRaWAN}{Low Power Wide Area Network}
\newacronym{pv}{PV}{Private Vehicle}
\newacronym{pgis}{PGIS}{Parking Guidance Information System}
\newacronym{sv}{SV}{Shared Vehicle}
\newacronym{aes}{AES}{Advanced Encrption Standard}
\newacronym{bw}{BW}{Bandwidth}
\newacronym{dr}{DR}{Data Rate}
\newacronym{pdr}{PDR}{Packet Delivery Ratio}
\newacronym{txp}{TXP}{Transmission Power}
\newacronym{sp}{SP}{Spreading Factor}
\newacronym{snr}{SNR}{Signal to Noise Ratio}
\newacronym{gtfs}{GTFS}{General Transit Feed Specification}
\newacronym{ga}{GA}{Genetic Algorithm}
\newacronym{cdf}{CDF}{Cumulative Frequency Function}
\newacronym{rfid}{RFID}{Radio Frequency Identification}
\newacronym{lstm}{LSTM}{Long Short Term Memory}
\newacronym{ble}{BLE}{Bluetooth Low Energy}
\newacronym{arima}{ARIMA}{Auto-Regressive Integrated Moving Average}
\newacronym{rnn}{RNN}{Recurrent Neural Network}
\newacronym{mlp}{MLP}{Multi Layer Perceptron}
\newacronym{nn}{NN}{Neural Network}
\newacronym{pir}{PIR}{Passive Infrared}
\newacronym{rssi}{RSSI}{Received Signal Strength Indicator}
\newacronym{ap}{AP}{Access Point}
\newacronym{pso}{PSO}{Particle Swarm Optimization}

\DeclareUnicodeCharacter{0301}{*************************************}

\begin{document}
\frontmatter
\thispagestyle{empty}
\pagenumbering{alph}
\begin{titlepage}
\thispagestyle{empty}
\begin{center}



\Huge Modelling and Optimisation of Resource Usage in an IoT Enabled Smart Campus \\
\vspace{2cm}
\huge Thanchanok Sutjarittham\\
\vspace{3.5cm}
\large A Thesis submitted in fulfillment of the requirements for the Degree of\\
\vspace{0.5cm}
\large Doctor of Philosophy\\
\vspace{2cm}
\includegraphics[width=0.4\columnwidth]{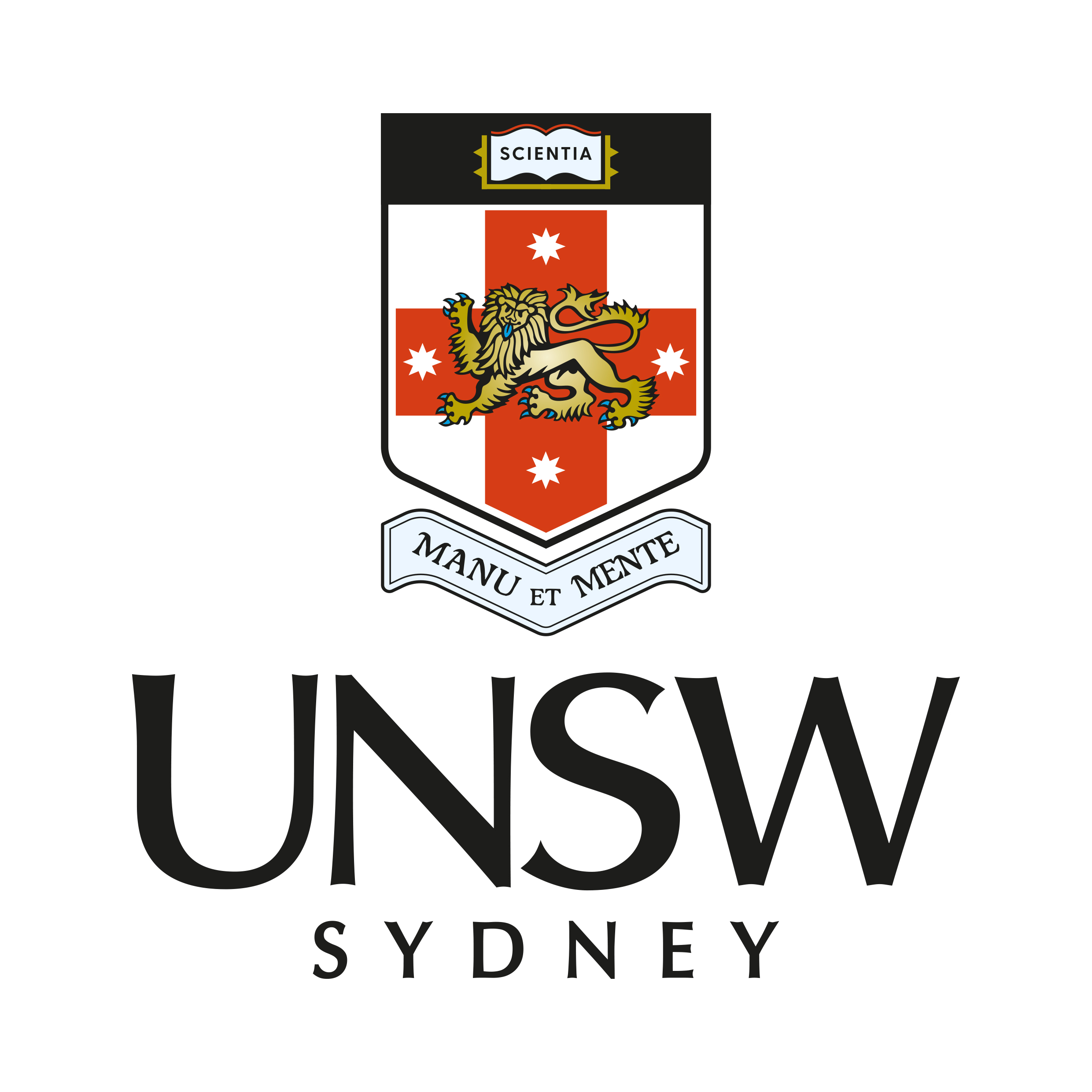}\\
\vspace{0.5cm}
\large School of Electrical Engineering and Telecommunications\\
\vspace{0.5cm}
\large Faculty of Engineering\\
\vspace{2cm}
\large January 2021
\end{center}

\end{titlepage}

\pagenumbering{roman}
\setstretch{1.18}

\chapter*{Abstract}
\bookmarksetupnext{level=section}
\addcontentsline{toc}{chapter}{Abstract}
\onehalfspace
\bookmarksetupnext{level=section}
\thispagestyle{plain}


University campuses are essentially a microcosm of a city. They comprise diverse facilities such as residences, sport centres, lecture theatres, parking spaces, and public transport stops. Universities are under constant pressure to improve efficiencies while offering a better experience to various stakeholders including students, staff, and visitors. Nonetheless, anecdotal evidence indicates that campus assets are not being utilised efficiently, often due to the lack of data collection and analysis, thereby limiting the ability to make informed decisions on the allocation and management of resources. Advances in the Internet of Things (IoT) technologies that can sense and communicate data from the physical world, coupled with data analytics and Artificial intelligence (AI) that can predict usage patterns, have opened up new opportunities for organisations to lower cost and improve user experience. This thesis explores this opportunity via theory and experimentation using UNSW Sydney as a living laboratory.

The building blocks of this thesis consist of three pillars of execution, namely, IoT deployment, predictive modelling, and optimisation. Together, these components create an end-to-end framework that provides informed decisions to estate manager in regards to the optimal allocation of campus resources. The main contributions of this thesis are three application domains, which lies on top of the execution pillars, defining campus resources as classrooms, car parks, and transit buses. Specifically, our contributions are:
i) We evaluate several IoT occupancy sensing technologies and instrument 9 lecture halls of varying capacities with the most appropriate sensing solution. The collected data provides us with insights into attendance patterns, such as cancelled lectures and class tests, of over 250 courses. We then develop predictive models using machine learning algorithms and quantile regression technique to predict future attendance patterns. Finally, we propose an intelligent optimisation model that allows allocations of classes to rooms based on the dynamics of predicted attendance as opposed to static enrolment number. We show that the data-driven assignment of classroom resources can achieve a potential saving in room cost of over 10\% over the course of a semester, while incurring a very low risk of disrupting student experience due to classroom overflow;
ii) We instrument a car park with IoT sensors for real-time monitoring of parking demand and comprehensively analyse the usage data spanning over 15 months. We then develop machine learning models to forecast future parking demand at multiple forecast horizons ranging from 1 day to 10 weeks, our models achieve a mean absolute error (MAE) of 4.58 cars per hour. Finally, we propose a novel optimal allocation framework that allows campus manager to re-dimension the car park to accommodate new paradigms of car use while minimising the risk of rejecting users and maintaining a certain level of revenue from the parking infrastructure;
iii) We develop sensing technology for measuring an outdoor orderly queue using ultrasonic sensor and LoRaWAN, and deploy the solution at an on campus bus stop. Our solution yields a reasonable accuracy with MAE of 10.7 people for detecting a queue length of up to 100 people. We then develop an optimisation model to reschedule bus dispatch times based on the actual dynamics of passenger demand. The result suggests that a potential wait time reduction of 42.93\% can be achieved with demand-driven bus scheduling. 
Taken together, our contributions demonstrates that there are significant resource efficiency gains to be realised in a smart-campus that employs IoT sensing coupled with predictive modelling and dynamic optimisation algorithms.

\chapter*{List of Publications} %
\bookmarksetupnext{level=section}
\addcontentsline{toc}{chapter}{List of Publications}
\bookmarksetupnext{level=section}
\thispagestyle{plain}

\vspace{-5mm}
During the course of this thesis project, a number of publications have been made based on the work presented here and are listed below for reference.

\vspace{5mm}
\noindent{\underline{\large Journal Publications}}
\begin{enumerate}

\item \textbf{T. Sutjarittham}, H. Habibi Gharakheili, S. S. Kanhere and V. Sivaraman, ``Optimizing Bus Scheduling by Intelligent Queue Estimation using LoRaWAN Sensors,'' (Under review at IEEE Internet of Things Journal)

\item \textbf{T. Sutjarittham}, H. Habibi Gharakheili, S. S. Kanhere and V. Sivaraman, ``Monetizing Parking IoT Data via Demand Prediction and Optimal Space Sharing,'' in IEEE Internet of Things Journal, doi: 10.1109/JIOT.2020.3044900.

\item \textbf{T. Sutjarittham}, H. Habibi Gharakheili, S. S. Kanhere and V. Sivaraman, ``Experiences With IoT and AI in a Smart Campus for Optimizing Classroom Usage,'' in IEEE Internet of Things Journal, vol. 6, no. 5, pp. 7595-7607, Oct. 2019, doi: 10.1109/JIOT.2019.2902410.

\end{enumerate}

\vspace{5mm}
\noindent{\underline{\large Conference Publications}}
\begin{enumerate}
\setcounter{enumi}{5}

\item \textbf{T. Sutjarittham}, G. Chen, H. Habibi Gharakheili, V. Sivaraman, and S. S. Kanhere, ``Measuring and Modeling Car Park Usage: Lessons Learned from a Campus Field-Trial," in IEEE International Symposium on a World of Wireless, Mobile and Multimedia Networks (WoWMoM), Washington DC, United States, 2019.

\item I. Pasquel Mohottige, \textbf{T. Sutjarittham}, H. Habibi Gharakheili, and V. Sivaraman, ``Role of Campus WiFi Infrastructure for Occupancy Monitoring in a Large University," in International Conference on Information and Automation for Sustainability (ICIAfS), Colombo, Sri Lanka, 2018.

\item \textbf{T. Sutjarittham}, H. Habibi Gharakheili, S. S. Kanhere, and V. Sivaraman, ``Realizing a Smart University Campus: Vision, Architecture, and Implementation," in IEEE International Conference on Advanced Networks and Telecommunications Systems (ANTS), Indore, India, 2018.

\item \textbf{T. Sutjarittham}, H. Habibi Gharakheili, S. S. Kanhere, and V. Sivaraman, ``Data-driven monitoring and optimization of classroom usage in a smart campus,” in ACM/IEEE International Conference on Information Processing in Sensor Networks (IPSN), Porto, Portugal, 2018, pp. 224-229.

\item \textbf{T. Sutjarittham}, H. Habibi Gharakheili, S. S. Kanhere, and V. Sivaraman, ``Demo Abstract: A Tool to Access and Visualize Classroom Attendance Data from a Smart Campus," in ACM/IEEE International Conference on Information Processing in Sensor Networks (IPSN), Porto, Portugal, 2018, pp. 140-141.

\end{enumerate}

\chapter*{Acknowledgment}
	\bookmarksetupnext{level=section}
	\addcontentsline{toc}{chapter}{Acknowledgment}
	\onehalfspace
	\bookmarksetupnext{level=section}

First and foremost, I would like to express my gratitude to my primary supervisor, Prof. Vijay Sivaraman. My Ph.D. journey would not have happened with you, thank you for your unwavering enthusiasm for research, your invaluable advice, and your constant support throughout my research journey. It has truly been a great privilege and honour to work under your guidance. I am equally thankful to my co-supervisors, Dr. Hassan Habibi Gharakheili and Prof. Salil Kanhere. Dr. Hassan, I will forever be grateful for your advice, insightful discussion, and your constructive criticism, that have helped me overcome any blockades I had during my research. I also profoundly thank you for countless of hours you've spent reviewing my manuscripts and your immense support in shaping my research work. Prof. Salil, I am very grateful for your constant support and insights throughout my journey.

Research is a team effort and this thesis is not the result of my own hard work alone but also that of my co-authors and collaborators. In addition to my supervisors, I want to extend my thank to Iresha Pasquel Mohottige, Gary Chen, Nixon Raju, William John Hales, and Akiran Pather. I graciously acknowledge the substantial contributions made by you to uplift the quality of the work presented in this thesis. 

I should not forget to thank my colleagues in my research team, Iresha, Ayoob, Arunan, Sharat, Jawad, Minzhao, Iresha, Chinaei, and Arman. Thank you all for the interesting conversations and support given to me during my time at UNSW.  Also a huge shout out to a great team of undergraduate and postgraduate students who have assisted my research work throughout my Ph.D. journey, Hales, Sean, Liam, Rachel, Jessica, Exia, Gary, Akiran, Nixon, Josh, and Tim. 

I avail this opportunity to thank all my closed friends for making my time as a Ph.D. candidate an enjoyable one. Their delightful friendship and hospitality have always mitigated my stress during this hard time. Special thank goes to Amarin Siripanich, for his love and support.

Words cannot express my gratitude to my family, especially my parents and siblings, whose unconditional love and supports have given me strength to pursue my dream. Thank you.

\onehalfspace

\vspace{-8mm}

\onehalfspace
\cleardoublepage
\dominitoc
\setcounter{tocdepth}{2}
\clearpage
\tableofcontents
\adjustmtc
\cleardoublepage

\clearpage
\listoffigures
\adjustmtc
\cleardoublepage

\clearpage
\listoftables
\adjustmtc
\doublespacing
\adjustmtc
\glsaddall
\printglossary[style=super, type=\acronymtype, nonumberlist]


\onehalfspace
\raggedbottom


\mainmatter
\doublespacing
\pagenumbering{arabic}
\pagestyle{fancy}{%
    \fancyhf{}
    \fancyhead[L,C]{}
    \fancyhead[R]{\nouppercase{\leftmark}}
    \fancyfoot[L]{}
    \fancyfoot[C]{\thepage}
    \fancyfoot[R]{}
    \renewcommand{\headrulewidth}{0pt}
    \renewcommand{\footrulewidth}{0pt}
}

\thispagestyle{fancy}
\setstretch{1.5}
\chapter{Introduction}

	\adjustmtc[2]
	\mtcsetfeature{minitoc}{open}{\vspace{1em}}
	\minitoc





University campuses are essentially a microcosm of a city which can encompass vast area of lands. They include a diverse range of facilities such as  residences, sport centres, lecture theatres, parking spaces and public transit stops. Hence, Universities are under constant pressure to improve efficiencies while offering better services to various stakeholders including students, staff, and visitors. 
Nonetheless, a number of campuses are not efficiently managing their resources. As evidenced in existing literature, real educational institutions can often have their teaching space utilisation as low as 20-40\% ~\cite{beyrouthy2009towards} and  up to 45\% of their parking spaces empty during school period, despite the increase in parking demand \cite{filipovitch2016:excessparking}.
These problems stem from the lack of monitoring system to provide the current usage of assets, which hinders informed decision in regards to the allocation and management of these resources. As a result, universities' assets are not optimally utilised.

With higher education institutes continuing to experience steady growth in enrolment number \cite{BritishCouncil}, real-estate will become a factor that prevent Universities from fulfilling this demand as enrolment to a course is capped by the capacity of the classroom to which it is allocated to. On the other hand, there is ample anecdotal evidence that classroom attendance is often well below its expected enrolment in response to digitally uplifted courses and greater accessibility to online contents. This discrepancy leads to an underutilisation of classroom, where the wasted seats could have been utilised to serve the increasing student demand. Despite the fact that course timetabling and classroom allocation have long been studied extensively in the past in order make optimal use of the available spaces \cite{lewis2008survey}, the formulated problems were based upon provided enrolment number of students rather than expected actual attendance. The lack of information on student attendance pattern has prevented educational institutions from making an informed decision with regards to their classroom assignments that could have yielded the optimum utilisation of their resources.
 
This surge in student enrolments does not only put pressure on classroom space usage, but also contributes to an increase in demand for on campus parking. Since many universities are large and spread out, despite the increase use of private vehicles, as many as 10-45\% of available parking spaces are not utilised \cite{filipovitch2016:excessparking}. This problem has also been observed at our campus in UNSW Sydney, where one of the multi-storey parking lots fills up by 10 am while the other often has availability. 
For many university campuses, there is no visibility into how car parking spaces are being utilised. Lack of information on real-time availability of spots not only leads to poor user-experience, but also worsen road traffic congestion due to prolonged periods for which cars stay on the road. Furthermore, a lack of data and analytics on how parking spaces are being used at an aggregated level (e.g. by time-of-day, by day-of-week, and by session) can prevent Estate Management from making informed decisions such as pricing policy and shared-transport space renting.

Furthermore, public transport stop at universities are notorious for overcrowding during peak hours, often times commuters need to wait for multiple services to arrive before there is sufficient room to board. Prolonged waiting is highly undesirable on multiple fronts including loss of productivity, fatigue, and discouragement from using public transport. In an effort to assuage this problem, many public transport operators have deployed vehicle tracking services in order to inform real-time arrival time of transit services to passengers \cite{ferris2010onebusaway}. Nonetheless, travel demand may outstrip the available capacity of the next arriving service, causing a number of passengers to wait for longer. Understanding passenger demand pattern as well as real-time transit schedule allows for a greater impact beyond enhancement of transit experience, in particular, the data can inform campus manager and public transport operator in transport scheduling decision that is more responsive to the actual transit demand. 

The emergence of Internet of Things Technology, where the world wide number of connected devices is forecasted to increase to 43 billion by 2023 \cite{gartner:iotforecast}, is transforming companies and organisations across industries. Accompanied by Intelligent data analytics, it creates myriad opportunities for new products, services, and business models. Undoubtedly, IoT is going to revolutionise the ways educational institutions around the world manage their valuable resources, allowing them to create a smarter campus that utilise resources efficiently, reduce waste, promote sustainable environment, and improve student and staff campus experience. 

Despite the potential benefits IoT can bring to improve operations and resource management of educational institutions, the majority of existing literature in this area only focuses on either creating sensing solution (\eg occupancy monitoring solutions \cite{ahmad2020occupancy}), or conventional optimisation of resources without considering the dynamics in demand. Yet there is a limited study that explore the full promise of IoT by combining such emerging technology with Artificial Intelligence and advanced data analytics in order to create well informed decisions on management and allocation of resources in the Smart Campus context. This research therefore aims to bridge this gap by proposing an end-to-end platform that combines practical deployment of IoT devices and data intelligence in order to solve current problems educational institutions are facing in regards to the inefficient use of their resources. By exploring such applications via theory and experimentation at a real university campus, the principles developed in our work for a smart campus can be applied at larger scale to future smart cities.

\section{Thesis contributions}

\begin{figure}[t!]	
	\centering
	\includegraphics[width=\linewidth]{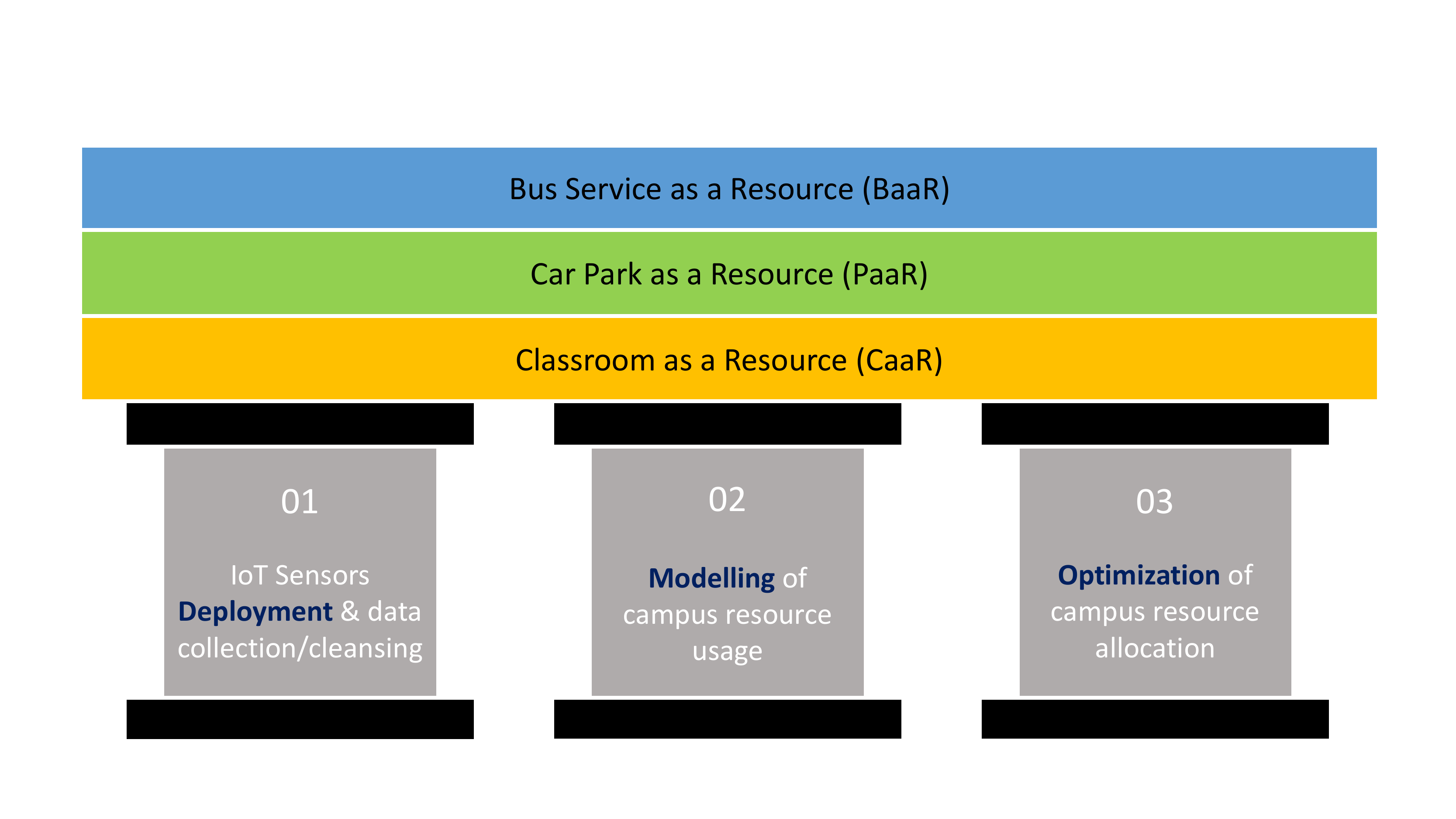}
	\vspace{-2mm}
	\caption{Building blocks of Smart Campus Framework}
	\label{fig:building_blocks}
\end{figure}

In this research, we propose an end-to-end framework that leverage Internet of Things technology, Artificial Intelligence, and advanced analytics, to optimise resource usage within a Smart Campus. 
We present the building blocks of our framework in Fig.~\ref{fig:building_blocks}, which consists three main pillars of execution: 1) The first pillar involves the deployment of IoT devices to collect relevant resource usage data in real-time; 2) The second pillar is the modelling of campus resource usage, this uses the historical collected IoT data to develop predictive models that allows us to understand the dynamics of demand and forecast the future usage of campus resources; 3) The last pillar is the optimisation of campus resources which can be achieved through new resource allocation schemes, this takes into account minimising wastage or maximising utilisation while providing exceptional services to the campus community.

By using UNSW Sydney as an experimental platform, we showcase how this framework can be employed to solve the existing problems related to the inefficient use of campus resources, namely classroom space utilisation, parking space utilisation, and allocation of public transit resources. Therefore, we define campus resources stacking on top of the Smart Campus pillars as classroom, car parks, and transit buses. Our specific contributions of this thesis are:

\begin{enumerate}

\item We showcase the usage of IoT and AI in optimising space usage of nine real lecture theatres at UNSW campus. Firstly, we undertake a systematic evaluation of several IoT sensing  for measuring classroom occupancy by benchmarking them in terms of cost, accuracy, privacy, and ease of deployment and operation. The most suitable solution is then selected and deployed across nine lecture theatres of varying sizes, allowing us to collect live occupancy data over a period of 18 weeks that can be used to infer attendance patterns of over 250 courses. Secondly, we develop predictive models using Machine Learning (ML) algorithms and quantile regression technique to predict future attendance of classes based on collected historical data from the previous semesters. Our models achieve a root mean square error (RMSE) of less than 16\% for attendance ranged between 0 and 100\%.  Lastly, we formulate an optimisation model for the allocation of classes to rooms based on the predicted attendance. We show that dynamic assignment of classrooms based on the predicted demand can save the usage of classroom resources by over 10\% while incurring a very low risk of disrupting student experience (\ie room overflows due to allocating classes to rooms that cannot accommodate actual attendance). Note that the current prediction models do not capture the effects  of the disruption to student demand by possible weekly changes to class timetable. This problem should be considered in the future work where timetabling optimisation and attendance prediction are of concern. (Chapter 3).

\item We showcase the usage of IoT in creating a novel dynamic parking space allocation framework that allows campus estate manager to re-dimension the car park to accommodate new paradigms of car use. We first instrument an on campus car park for a real-time monitoring of parking demand. We comprehensively analyse the usage data that spans over a period of 15 months, covering both teaching and non-teaching periods, and highlight interesting insights into car arrival and departure patterns. Secondly, we forecast future car park demand based on the historical data using ML models and compare predictive performance of various learning algorithms across multiple forecast horizons, ranging from a day up to 10 weeks. Our model achieves a mean absolute error (MAE) of 4.58 cars per hour for a 5-day prediction horizon. We then develop a continuous-time non-homogeneous Markov model using predicted arrival and departure rate of cars as an input in order to simulate a dynamic partitioning scheme for allocation of parking spaces to private and shared vehicles. Finally, we formulate an optimisation problem that decides the optimal allocation (\ie fraction of spaces to be allocated to car sharing companies) that improves space utilisation while minimising the probability of rejecting users due to a lack of parking spots. (Chapter 4).

\item We showcase the usage of IoT in the domain of optimising bus scheduling. Firstly, we design and implement people detecting system using ultrasonic sensors and LoRaWAN communications module. We then deploy our implemented devices at one of the busiest bus stops on campus and apply an algorithm to infer passenger queue length from the collected data. Our solution yields a reasonable accuracy with MAE of 10.7 people for a queue length of up to 100 people. Based on this actual transit demand, we develop an optimisation model to reschedule bus dispatch times in order to minimise total wait time of passengers at the stop. Our result suggests that total wait time of passengers can be reduced by up to 42.93\% with demand-driven bus scheduling. (Chapter 5).

\end{enumerate}

\section{Thesis Organisation}

The rest of this thesis is organised as follows:

\begin{description}

\item \textbf{Chapter 2} surveys Smart Campus initiatives around the world and existing IoT technologies and their applications in resource management, particularly in the domains of classroom resources, car park resources, and transit bus resources.

\item \textbf{Chapter 3} presents a novel dynamic allocation of classrooms based on predicted class attendance. The approach exploits IoT for occupancy data collection and AI for predictive modelling \cite{sutjarittham2019experiences,sutjarittham:classroomusage}.

\item \textbf{Chapter 4} presents a novel dynamic parking space allocation framework that allows campus estate manager to re-dimension a car park to support the prevalence of new transportation paradigms such as car-sharing \cite{sutjarittham2020monetizing,sutjarittham:carparkusage}.

\item \textbf{Chapter 5} presents our method of optimising bus scheduling through Intelligent passenger queue estimation using LoRaWAN sensors.

\item \textbf{Chapter 6} concludes the thesis with pointers to directions for future work.

\end{description}

\clearpage

\chapter{Literature Review}

\minitoc

The growing number of smart devices, with the world wide number forecasted to increase to 43 billion by 2023 \cite{gartner:iotforecast}, has envisioned organisations and institutions to revolutionise their operational and resource management strategies by exploiting data collected from such technologies. In this chapter, we first explore use cases of Internet of Things (IoT) technologies in university settings (generally defined by Smart Campus Initiatives). We then survey existing IoT technologies and their applications for resource management in the three focused domains namely classrooms, car parks, and transit resources management.

\section{Smart Campus Applications}

A smarter campus, equipped with Internet of Things technologies, offers myriads of potential benefits including better student learning outcomes, reducing energy, higher efficient resource management, and better informed decisions to improve student and staff experience. This section explores Smart Campus initiatives around the world and their various applications. In the following, the literature has been categorised into three main types of applications namely space utilisation, campus experience, and energy, water, and waste management.

\subsection{Space Utilisation}

One of the main challenges in the field of corporate real estate management arises due to the fact that real estate is static with a 50-year life span while being very cost intensive, whereas demand is dynamic. This problem has also been faced by universities with ageing real-estate portfolios who have to deal with an increase in student numbers and a higher expectation to provide better services. This implies the need for universities to better manage their assets using the limited resources. Despite the promise of IoT to provide real-time information on space usage and solve the aforementioned problem, the use of these technologies for space resource management is still a sparsely researched area \cite{valks2019smart}.

A few universities attempt to exploit benefits of IoT technology to monitor and utilise their learning spaces though a collaboration with industrial partners. Curtin University in Western Australia \cite{curtin:smartcampus} has collaborated with Hitachi to optimise the usage of leaning spaces at the Bentley Campus. They co-create a system using facial recognition technology to understand who is in a classroom and deploy a pupil counting solution to study course attendance patterns as well as how spaces are being used. Facial recognition is beneficial in identifying occupants inside a room, however privacy issue is still a main concern and the solution can be very expensive, making it challenging for the university to scale this up to all rooms across the campus. Although this engagement between Curtin University and Hitachi Data System has started since January 2016, no outcome of the research is available online at the time of writing this thesis.

University of Glasgow is one of the most leading universities in driving its smart campus project \cite{gasglow:smartcampus}. The university has successfully installed sensor network at in its library to report on environmental and occupancy, and developed optimised timetabling using predictive analytics based on student's course choice. Other technologies such as smart parking and footfall sensors are currently being explored. Furthermore, University of Glasgow has built and deployed its initial super sensor prototype using Raspberry Pi devices\cite{hentschel2016supersensors}, in order to monitor temperature and light at several staff offices. Other use cases of the sensors including room occupancy census monitoring, meeting rooms availability, and robotic support infrastructure, have yet to be implemented. Their proposed methods of measuring room occupancy census are to use WiFi Data to count number of unique WiFi MAC address and to utilise motion sensors. 

At a smaller scale, a study has been carried to investigate the usefulness of a novel WiFi-based indoor location system in providing indoor occupancy patterns and user behaviour, using a library of the University of Reading, UK, as an experimental platform \cite{wang2018understanding}. The work has demonstrated the effectiveness in providing occupancy and user behaviour information, which can be used to inform various space policies including library opening hours, redesign or relocation of spaces, and recommendation to improve health and well being implications for users. \textcolor{red}{}


In addition to learning spaces like classrooms and libraries, parking spots are also considered scarce resources of the university campus that should be managed efficiently. 
A number of smart campus applications have been implemented to provide campus services for smart parking. West Texas A\& M University has established a strategic initiative to develop a smart campus, where part of the effort has been put into a pilot project for smart connected parking using video surveillance cameras \cite{webb2018campus}. OpenCV, an open source library for computer vision, is used to track parking space availability. The data is then made available through a mobile applications to provide real-time usage of spaces.
A similar mobile application system has been proposed by the University of Peradeniya \cite{bandara2016smart} to report the status of parking areas within the campus, in this work magnetic sensor and distance sensor are used for vehicle detection. Soochow University has also planned to construct an intelligent transportation system on their campus by installing sensors on each parking area and collecting real-time information in order to provide campus community with parking availability data \cite{hengliang2016construction}. 

Some technologies used to monitor the utilisation of space, particularly facial recognition and location-based system, can pose significant ethical concerns due to the collection and use of individuals' personal data. These concerns are most likely related to unauthorised surveillance, unauthorised use of data, and transmission security \cite{caron2016internet2}.

\subsection{Campus Experience}

Another important aspect of IoT applications is the enhancement of user experience on campus. For instance, the technology will enable campuses to personalise the student experience, increase comfort within an indoor environment, and improve safety of the campus community.


In relation to indoor comfort of occupants, existing WiFi infrastructure for indoor positioning system has also been exploited for overcrowding detection in indoor events in one study \cite{lopez2017overcrowding}. Wireless routers are set along the monitored building to capture wireless frames from WiFi-enabled devices, the data is then sent to a processing server for position estimation. The real-time crowding information is aimed to support event organisers with regards to access control of attendees and prevention of overcrowding spots. The system is reported to locate 73.96\% of the attendees present in the building.

University of Melbourne has partnered with Cisco to leverage location-based analytics using data from Cisco’s wireless infrastructure (CMX) to examine foot traffic patterns from multiple granularities such as day, time, and location \cite{melborne:smartcampus}. They claimed that the solution is a better alternative to video-based traffic monitoring where \$15000 per day would have had been spent. The analytics also allows the university to compare predicted and actual occupancy level throughout the day which helps them in decision making for building modernisation projects; gives insights into footfall dwell time to reduce congestion and provide better campus experience for occupants; and helps increase revenue by informing the best pinpointing locations for advertisement and social media during on-campus events. Deakin University has also partnered with Cisco and Dimension data to provide information and services through mobile application and digital signage \cite{deakin:smartcampus}. It informs students which parts of the library are most crowded and allows them to request assistance and check out books.

IoT technology can also be used in learning applications such as attendance checking system. Traditionally, taking attendance can be a very time consuming process especially for classes with large number of students. To address this issue, RFID technology can be utilised to automatically track attendance of students and staff in order to save time and prevent human errors. Studies have shown that such technology can effectively assist in student management by eliminating the need for manual attendance checking \cite{nainan2013rfid}. In addition to RFID, near field communication (NFC) for attendance system has also been studied \cite{chew2015sensors}. NFC based attendance checking requires student to tap an NFC tag on a NFC smart phone in order for the attendance to be automatically stored in the server. 

Apart from attendance recording system, RFID technology has been employed to create smart laboratories \cite{wang2013toward}. By having students wearing RFID tags as they enter the lab, a computer can be assigned to them automatically. And since the system can track the usage of the lab stations, a new lab can be opened when the current capacity is full. Furthermore, the system has a mechanism to shut down idle computers automatically to save energy.

Another aspect of campus experience is safety of the campus community. In an event of emergency such as natural disasters and terrorist attacks, data on building occupancy and location of occupants is critical to accelerate search and rescue operation in order to ensure safety of the occupants. A work in \cite{nyarko2013cloud} utilises data from bidirectional passive infrared sensors and WiFi access points in order to determine occupancy levels of various zones within a building structure for attack and disaster response. Historical WiFi trace data has also been utilised in one study to evaluate human behaviour in planned and unplanned evacuation events in a university campus setting \cite{mohottige2020evaluating}. The study analyses data spanning a period of 180 days across 14 buildings and reports evacuation behaviour in terms of speed and occupancy count.

\subsection{Energy, Waste, and Water Management}

Smart buildings are buildings which integrate and account for intelligence, enterprise, control, and materials and construction as an entire building system, with adaptability at its core. The main drivers for such building system include reducing energy, maximising comfort and satisfaction of those within the building, and maintaining value of the buildings over a long period of time under changing use and external conditions \cite{buckman2014smart}.

In one study, Birmingham City University has implemented intelligent buildings as part of its new campus development. The development includes a deployment of extensive range of sensors throughout the buildings to turn-off services when no presence is detected in spaces. Carbon-dioxide detectors have also been affixed in lecture theatres for capacity detection in order to improve comfort of occupants while optimising the use of services and energy. As a result, the campus reported substantial energy savings of £140k per year and a reduction of CO2 emissions by 40\% \cite{hipwell2014developing}. 
University of Brescia has also aimed to achieve energy efficiency through smart grid management, control system and automation \cite{de2015brescia}. The project focuses on evaluating effectiveness of different renovation strategies using business intelligence and big data analysis. By using a classroom building as a pilot site, the campus reported the maximum energy reduction of 37.3\%  by enhancing the thermal properties of the envelope and using an effective ventilation. 

The campus of the Technical University of Crete has implemented a web based building energy management system which manages the campus buildings and public spaces use in an energy efficient way by monitoring the energy load, performing energy analysis per building, and allowing interactions between building energy simulation models \cite{kolokotsa2016development}. In addition to reducing the campus energy consumption, the system also aims to improve indoor environmental quality of residents by controlling  air quality, thermal, and visual comfort. The energy reduction within a range of 15\% to 30\% is reported as a result. 
With an increasing interest in real-time monitoring of environmental data aiming to reduce energy consumption while maintaining comfortable conditions, the work in \cite{habibi2016smart} presents a prototype of a smart system to monitor and control indoor environment data in order to obtain an optimal indoor conditions. This work concludes that users can take advantage of natural environment and passive design strategies by monitoring and controlling both outdoor and indoor environmental parameters in real-time in order to provide an efficient real-time energy use feed back.

One of the essential yet expensive services offered by a university campus is waste and water management. A number of studies published in the recent years have focused on the deployment of IoT technologies to improve efficiency of such services by equipping sensing devices at rubbish bins and waste trucks for data collection and analysis \cite{anagnostopoulos2015robust,folianto2015smartbin,anagnostopoulos2015top}. The analytic results are then used to suggest optimal cleaning schedule and route for waste contractors. 
A study in \cite{khan2020iot} developed a prototype of IoT-based intelligent bin to monitor waste content status in real-time for a university campus using ultrasonic sensors. The aim is to prevent the bins from overflowing and ultimately provide a clean and healthy environment to students. 

IoT technologies have also been demonstrated to improve effectiveness in the domain of water management, aiming at increasing productivity, improving efficiency, and expanding of new and existing business models \cite{robles2014internet}. 
Several works focus on developing smart water metering for monitoring water consumption in order to increase users' awareness and improve efficiency of the infrastructure management \cite{mudumbe2015smart,gabrielli2014smart}.
An experiment of water balance monitoring system deployment using ultrasonic sensors has been conducted in a medium-sized campus \cite{verma2015towards}. Based on the preliminary collected data, the study could report several issues with the campus's water distribution system including a low level leak and an improper management of the campus reservoirs.

\section{Classroom Space Monitoring and Management}

This section survey sensing technologies available for occupancy monitoring and existing studies on management of classroom resources and allocations.

\subsection{Occupancy Sensing technologies}

\begin{figure}[t!]	
	\centering
	\includegraphics[width=0.6\linewidth]{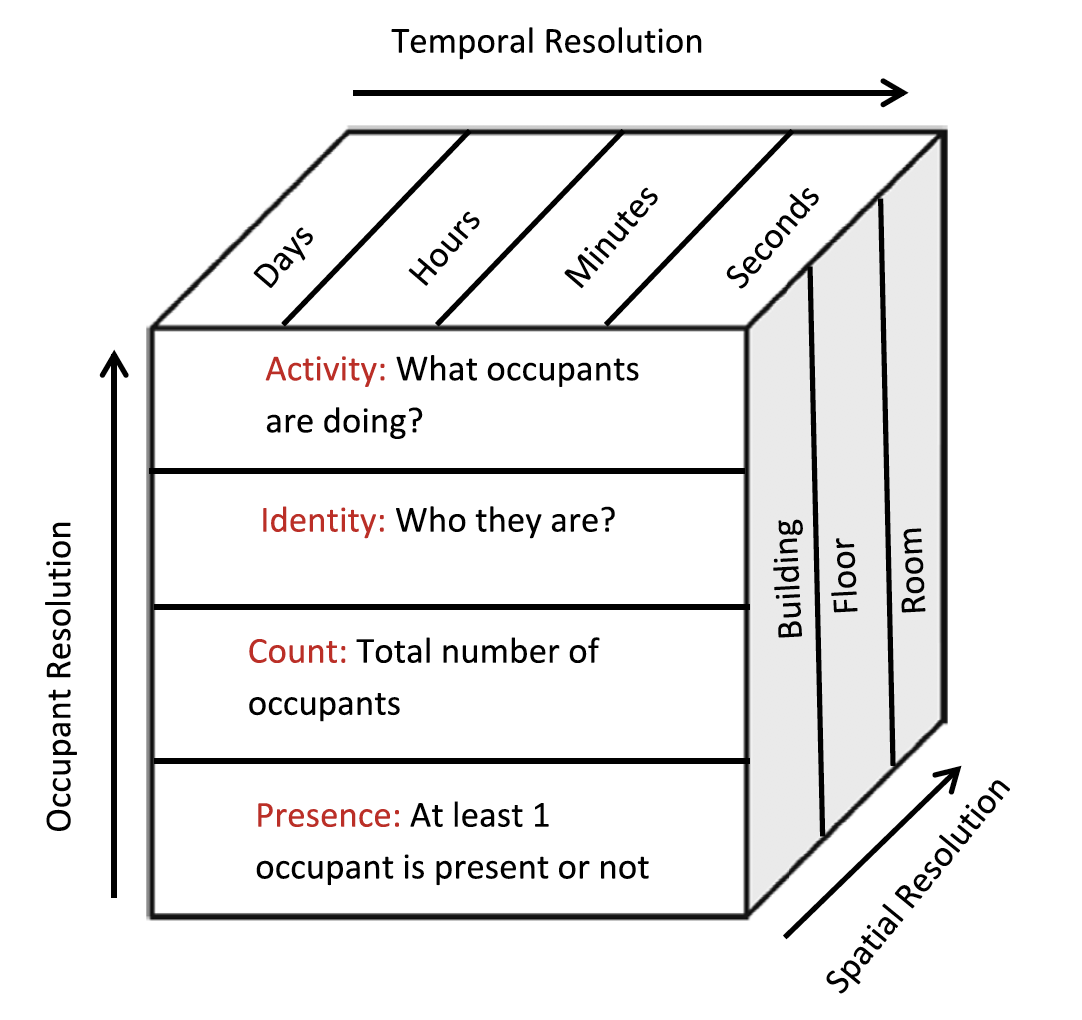}
	\vspace{-2mm}
	\caption{Concept of presence across three dimensions \cite{melfi2011measuring}}
	\label{fig:occupancy_concept}
\end{figure}

The concept of monitoring presence can be divided into three dimensions namely occupancy, spatial and temporal resolutions \cite{melfi2011measuring}. This concept can be visualised in Figure~\ref{fig:occupancy_concept}. The occupancy resolution indicates the level of detailed knowledge about occupancy and occupant's behaviour. Higher resolution of occupancy information also constitute a violation of privacy, especially when consent of occupants were not obtained prior to the monitoring process. In our study, we are only interested in count resolution where total number of occupants are determined.

There are various approaches to occupancy monitoring available in today's literature. The methods can be broader divided into two categories, namely explicit and implicit, based on the extent of additional hardware required. Explicit occupancy sensing require installations of supplementary sensors, predominantly with the sole purpose of occupancy detection, such as passive infrared sensors (PIR) and Radio Frequency Identification (RFID) sensors. On the other hand, implicit occupancy sensing exploits data collected from the existing built infrastructure to infer occupancy information. The most common examples are WiFi-based and smart meters based occupancy detection.

\subsubsection*{PIR Sensors}
Naturally, all objects including humans emit thermal energy in the form of infrared radiation, which is electromagnetic radiation with a frequency lower than visible light. PIR sensors can detect changes in the amount of infrared radiation emitted by these objects, which varies depending on the temperature and surface characteristics of the object.
In the literature, PIR sensors are commonly used to detect the presence or absence of occupants\cite{dodier2006building,duarte2013revealing,liu2015occupancy} or count the number of occupants within a space due to its ease of implementation and cost effectiveness. In the later use case, occupancy count can be obtained using a distributed PIR sensors \cite{wahl2012distributed} or a single PIR sensor \cite{raykov2016predicting}.
 
Despite being a low-cost sensing technique, occupancy detection based on PIR sensors can be unreliable due to its line-of-sight limitation, which prevent a detection of multiple people simultaneously passing through the line of sight of the sensor. Furthermore, the method is prone to error due to its inability to detect static occupants. Improvement to keep the sensor in constant motion is proposed to solve this issue \cite{movingpir:dynamicpricing}, nonetheless most of the commercial PIR sensors still operate in a conventional way and therefore do not offer an accurate outcome.

\subsubsection*{Ambient-based Sensors}

\newcommand\coo{\ensuremath{\mathrm{CO_2}}\ }

Since occupancy is one of the factors that change indoor environmental conditions such as temperature and \coo level, many studies utilise special-purpose sensors to infer occupancy count for a given space. 
A large number of works have employed \coo sensors for occupancy monitoring \cite{jiang2016indoor,zuraimi2017predicting,rahman2017occupancy} due to a strong relation between  \coo concentration and occupancy \cite{wang1999experimental}. However, key limitations of \coo sensing are its slow response time due to a gradual process of spread of gas and its vulnerability to external environment such as ventilation. 

To overcome significant errors that come with a single sensor detection, a multitude of studies have focused on a fusion of multiple sensors for occupancy detection and estimation.
A work in \cite{dong10:ITEST} used a network of sensors to obtain various environmental parameters such as \coo, carbon monoxide ($CO$), total volatile organic compounds (TVOC), acoustics, motion, temperature, and humidity, to derive occupancy count in an open office space using machine learning techniques. The method achieved an average accuracy of 73\%, however it has only been tested in a space that can accommodate less than 10 people at a time.
Works presented in \cite{deyr:Namatad, zimmermann:2017fusion, yang2012:multi} have also used indoor environmental sensors in combination with supervised learning methods to infer occupancy and achieved good accuracy results. Nevertheless, none of these studies have evaluated their methods in a larger room scenario where over 100 occupants can be accommodated. This is important for our study as typical lecture rooms can have capacities within a range of 100 to 500.
Since human body emits energy in the form of heat to the surrounding environment, thermal sensor can be used to detect occupants within a space \cite{beltran2013thermosense}. Some technologies utilise thermal detection in order to count the number people passing through the door by placing the device on the ceiling of the entryways \cite{evolveplus:overhead}.

\subsubsection*{Camera-based Sensors}

By mounting a camera to monitor a space of interest, not only count of people but their locations can also be obtained. Several studies uses video camera and complex image processing algorithms for pedestrian counting in an outdoor or public scenario \cite{hou_pang_2011, li2011:robust}. These studies achieved good accuracy of results, but the method requires significant computational resources. A high-resolution camera network has also been used to detect number of occupants inside a room in order to optimise energy processes \cite{stancil2008active,trivedi2000intelligent}. For a classroom room scenario, a study has successfully used image processing and Support Vector Machine (SVM) to measure classroom occupancy in large rooms with more than 100 occupants \cite{paci2014:classroomimage}. However, their method only produces accurate results when movements of students within a classroom are minimal.

Since camera sensors have shown to yield highly accurate occupancy count, the approach is commonly used as a mean of obtaining ground truth occupancy information in several occupancy studies \cite{jiang2016indoor,masood2015real,chen2016fusion}. The main concern about this method of occupancy detection is privacy of people under surveillance, especially if images or videos are taken without their explicit consent.

\subsubsection*{RFID based Sensors}
Radio Frequency Identification Tags (RFID) based system consists of three components including a transceiver, a transponder, and an antenna. In general, an RFID reader is made up of the scanning antenna and transceiver while the transponder is located in the RFID tag. There are two different methods in RFID based localisation and occupancy detection, active and passive mode. In an active mode \cite{want1992active}, an antenna is installed at the target location and information about occupants and their locations is transmitted to the central location from RFID tags. Consequently, this method can be used to obtain count resolution of occupancy like occupant density and profiling. 
The active mode can also be used for localisation by utilising Received Signal Strength Indicator (RSSI) information of occupants, which varies depending on the proximity of RFID tags and the reader. The challenge coming with this method of localisation is the fact that RSSI is easily affected by factors such as diffraction, reflection and transmission of radio waves. A study has attempted to overcome these shortcomings by proposing to install multiple readers in order to reduce the probability of error \cite{zhen2008indoor}, the method achieved 93\% of accuracy in localisation.  

RFID based localisation in a passive mode is introduced in \cite{tesoriero2010improving}, where coordinate of an entity is determined within a sensing surfaces consisted of grids of passive RFID tags. In this case, an RFID reader is attached with the entity. Several other methods also leverage passive RFID for occupancy estimation \cite{li2012measuring,ni2003landmarc}. Nonetheless, this approach suffers some serious drawbacks including inaccuracy in occupancy measurements caused by larger proximity from the target. 

\subsubsection*{WiFi and Bluetooth based Sensors}

Another approach to deduce occupancy count is to leverage information from existing WiFi access points (AP) infrastructure where network connection parameters, such as connections count and received signal strength indicator (RSSI), are used to infer occupancy \cite{fierro2012:wifi, balaji2013:wifi, ghai2012:wifi}. 
The advantage of this approach over the others is the fact that no additional hardware is required. However, there is a number of factors that may impact accuracy of the count. For instance, people who do not carry WiFi enabled devices would not get counted, individuals with multiple devices (e.g., laptop and phone) would get counted twice, and people outside the room may be connected to the AP inside the room and get included in the room occupancy count.
Furthermore, obtaining WiFi connectivity data may also constitute a violation of privacy if the identities of connected users can be deduced.

Some studies install Bluetooth Low Energy (BLE) beacons in buildings to estimate the number of occupants by counting the number of Bluetooth activated mobile devices \cite{conte2014bluesentinel,yang2016using}. BLE is more energy efficient than classical Bluetooth or WiFi, however major drawbacks of this approach in occupancy monitoring is the expenses associated with beacons installation and maintenance. The method also suffers a similar issue with WiFi-based detection where the accuracy of the measurement highly depends on the requirement of users carrying Bluetooth activated devices.

\subsection{Occupancy Forecasting}

Majority of people counting techniques exist to collect real-time occupancy information in order to support demand-driven control systems such as Demand Controlled Ventilation (DCV) and Heating Ventilation and Air Conditioning (HVAC)  \cite{ahmad2020occupancy}. The goal of these systems is to achieve reduction in energy consumption while maintaining comfortable indoor conditions for occupants. Despite the ample studies on real-time occupancy detection, studies on forecasting of occupancy using historical data is still limited in the literature. Having an informed knowledge about future occupancy demand, from 1 week to 1 academic term in advance, is important in the decision making process of managing of assets such as classrooms.

Some studies have attempted to perform future forecasting of occupancy from historical data. The authors in \cite{sangogboye2017:performance} compared two approaches of occupancy estimation based on indoor climate parameters and 3D stereo-vision camera. Both approaches were tested in two rooms, a classroom and a study zone. As a result, the camera-based approach was shown to outperform the indoor climate-based approach.
Historical occupancy estimation with additional contextual features including day name, day type (weekday or week end), season, and holiday (binary) were used to perform future count prediction using decision tree and random forest. The prediction achieved the best accuracy of 3\% error, yet it has to be noted that the models have only been trained using 20 days data with 1 day test data (using sliding window method). Moreover, for classroom occupancy, the dynamic variation of class attendance, which is likely to be influenced by factors such as courses and weeks of semester, are are captured by the models.

A long term collection of indoor occupancy data can be found in a few studies \cite{wren2007merl}. In one study, depth sensor (Kinetic for XBox One) is used to count the number of people entering and exiting through doors, covering a common office space \cite{liu2017:kinect}. The data is collected for a duration of 9 months and occupancy forecasting is performed using historical occupancy pattern. A simple solution of leveraging occupancy count from the previous day for the prediction suffers a very high error of 30.1\% while leveraging occupancy count of more than two previous days is not shown to improve the prediction significantly.

A study in \cite{yuan2020leveraging} develops multiple predictive models (time-inhomogeneous Markov chain, Linear Regression, and Sequential and contextual Neural Network) to forecast future occupancy in a commercial space. An adaptive model predictive HVAC controller, aiming to reduce energy consumption and improve thermal comfort of occupants, is then developed using future occupancy prediction. Based on a simulation using real-world occupancy data, the study shows that the proposed solution can reduce energy consumption by 29.5\%.


\subsection{Classroom Management and Allocations}


Several studies have revealed that many educational institutions do not use their teaching space resources efficiently, with the utilisation can often be as low as 20-40\% \cite{mccollum2004cornerstone}. For this reason, space managers are under pressure to improve this situation in order to reduce costs or allocate spaces for other uses while improving services. According to \cite{beyrouthy2009towards}, operation concerning the usage of spaces can be divided into two broad areas, space management of the usage of existing resources and space planning for long-term decisions regarding the provision of space resources. In this section, we survey existing studies related to the teaching space management in real-world universities. 
Since space allocation problem is highly correlated with timetabling procedure \cite{reis2000language}, various timetabling methods and techniques are discussed here.

Teaching space allocation in academic institutions, defined as an assignment of a set of lectures to a set of rooms and timeslots, has been studied extensively in the past. One study has been conducted to define and discover different dimensions and requirements of space allocations at 96 universities in the United Kingdom through questionnaire surveys \cite{burke1997space}. Based on the analysis of this survey, space allocation can be applied in four different ways including fitting all resources into a limited number of rooms, minimising the number of rooms required for a set of resources, adding or removing rooms or resources from a current allocation, and reorganising the current allocations. 

These types of space allocation can be subjected to various constraints relating to timing requirement of events or availability of resources. Some examples of hard constraints that applied to the majority of studies are students are only allowed to attend one event at any timeslot, events have to be assigned to rooms with adequate number of seats, and lecturers must not be assigned to more than one class in any timeslot \cite{badoni2015hybrid,el2015genetic,socha2003ant}. Additional soft constraints adopted by some studies include the avoidance of scheduling events in the last day slots, minimising the probability of students having more than one class per day \cite{socha2003ant}, and respecting lecturers' and students' timing preference \cite{sabar2012honey}.
Several meta-heuristic approaches have been employed to solve teaching space allocation problems. Some of the most popular ones are genetic algorithm \cite{abdelhalim2016:utilization}, simulated annealing \cite{kostuch2004:university}, and tabu search \cite{burke2003:time}.

In the existing literature of teaching space allocation in higher educational institutions, the expected number of students per class used in the optimisation is based on the enrolment number, which has been shown to be well higher than actual attendance. Consequently, even though class schedule has been optimised, under utilisation of classroom spaces is still under way in the real-world scenarios. 
We believe that our work is the first that combines occupancy monitoring system based on IoT technology to solve class allocation problem where courses can be allocated to classrooms based on their predicted attendance rather than the traditional enrolment information.

\section{Parking Space Monitoring and Management}

This section presents related work on car park monitoring technologies and methodologies for predicting parking demand.

\subsection{Car park Sensing Technologies}

Car park occupancy information is an important enabler for many smart parking systems existing today. There is an abundance of technology choices that can be deployed at parking lots to provide this information which we will discuss in the following.

\subsubsection*{Monitoring individual parking spot}

At a fine-grain level, monitoring car park occupancy is done by collecting information on the status of each individual parking spots. This can be achieved by installing a sensor at each parking bay, usually on the ceiling for an indoor parking structure and in the ground for an outdoor parking space. This approach of monitoring car occupancy does not only provide data on the number of vacant spots, but also their positions. 

In the existing literature, both passive and active infrared sensors have been used for vehicle detection. Passive infrared sensor (PIR) identifies the status of a parking spot by detecting changes in the energy emitted  by the vehicle. Since the sensor is extremely sensitive to environment stimuli and is likely to create false triggers, PIR is often used in combine with other types to create a detection scheme \cite{larisis2012u,manni2010smart}.
On the other hand, active infrared sensor actively emits infrared energy and detects the amount of reflected energy, allowing it to measure the distance to any object in front in order to detect a vehicle \cite{mimbela2007summary,mouskos2007technical}. Because infrared sensor is sensitive to environmental conditions, sensing accuracy remains a problem. 

Similar to active infrared sensor, ultrasonic sensor can be used to detect the presence of a vehicle by transmitting sound of frequency between 25kHz and 50kHz, and using the reflected energy to determine the status of a parking spot \cite{kianpisheh2012smart,lee2008intelligent,vishnubhotla2010zigbee}. Despite the fact that ultrasonic sensor offers a low cost and easy to install solution, the method is sensitive to temperature changes and extreme air turbulence. 
There are many studies that propose a hybrid approach for an intelligent parking system using a combinations of sensors such as ultrasonic sensor, magnetic sensor, and optical sensor, to detect individual parking slot \cite{tang2006intelligent,yang2012smart}. This approach usually creates a network of low cost sensors that communicate wirelessly to a central server and can be used by the upper layer management system for reporting and analytics purposes.

The most common sensor for vehicle detection that has been adopted by municipal deployments for detecting on-street parking \cite{laexpresspark,sfparkoverview2014,melbournedata} is magnetometer. It detects a vehicle by measuring the current magnetic field in order to identify large metal objects presence. This approach provides an advantage of not being sensitive to external weather condition with a trade off of a higher cost compared to the previous options.

The approaches to monitor each spot individually is expensive, especially for a large parking lot with hundreds of parking bays, hence such solutions are typically employed only in commercial parking areas such as shopping centres and airports.

\subsubsection*{Monitoring overall parking space}

Another approach to monitor car park status is to exploit existing network of video surveillance cameras. Several studies obtain videos of a car park view from the deployed CCTV cameras, and perform image processing \cite{Bong2008:cctv} or video analytics \cite{sevillano2014towards,bin2009design,albiol2011detection} to detect the status of parking bays. This approach offer a much more cost effective solution than installing sensor on each parking lot. However, continuously recording images (or videos) of users' vehicles may raise privacy concerns, especially when images are collected without users consent.


\subsubsection*{Crowdsensing and Crowdsourcing based detection}

The most economical way to obtain parking availability is via mobile crowd sourcing and crowd sensing, where the information is obtained from the drivers themselves through a mobile application. The system can collect parking information status from explicit feedback by the users or sensor based estimation \cite{rinne2014mobile,yang2013ipark}. Various types of data can be collected from users to infer parking availability including movements \cite{koster2014recognition,nandugudi2014pocketparker}, Bluetooth connections \cite{stenneth2012phonepark}, WiFi signatures \cite{nawaz2013parksense}, and the combinations of all data \cite{ma2014updetector}. 
The biggest issue with this approach is privacy concern, especially when the collected data is uploaded to a public database. Furthermore, crowd sourcing based approach requires a certain number of participants to contribute to the system, otherwise this can lead to poor accuracy of the results.

\subsubsection*{Monitoring entering and leaving vehicles of the parking lot}


Radio frequency identification (RFID) technology can also be employed to identifying vehicles for parking management. The system involves installing an RFID reader at an entrance and exit of the car park in order to detect RFID tags of arriving and departing cars \cite{pala2007:RFID,bagula2015design,abdullah2013integrating}. The drawback of this method is that car park users need to have a tag attached to their vehicles. Casual visitors who have not installed the tags would be excluded.
Another approach to detect vehicles at the entrance and exit of the car park is to through License Plate Recognition (LPR) technology \cite{rashid2012automatic,tian2014design}. LPR cameras can be deployed to automatically record license plate number and provide access control in restricted parking lot. This solution of monitoring usage of parking spaces is much more cost effective then the sensor network solution where a sensor is required at each parking bay.



\subsection{Parking Prediction and Modelling}

Parking prediction is one of the essential processes in smart parking services. In particular, real-time prediction of parking availability allows drivers to make informed decisions about their trips and therefore overall congestion and inefficient cruising can be reduced. In the literature, various prediction models have been proposed to forecast parking demands, ranging from stochastic models to provide the probability of availability to data-driven models where historical data is used to predict future parking availabilities. Long term estimation does not need to be done as prediction based on current occupancy. It could also be a seasonal model based on past history.

A number of studies in the field of vehicular ad-hoc networks have employed stochastic models to predict real-time parking availability by collecting parking information such as arrival rate, departure rate, and current occupancy of the car park from the network. These studies assume arrival and departure behaviour to follow certain distributions and model parking lot as a queue using Markov Chain. Among several queue models, $M/M/m/m$ queue, where arrival and departure flows are assumed to follow Poisson distributions, is commonly adopted \cite{caliskan2007:MMCC, klappenecker2014finding}. The computed parking lot occupancy can then be provided to cruising vehicles.
Other queue models that have been found in the literature include $M/G/c/c$ queue \cite{lu2010:MGCC} where stay duration of cars is assumed constant and Wilkinson queuing model \cite{pack2011using}. In these systems, blocking probability is calculated and disseminated to users via vehicular communications in order to prevent more users from approaching the parking lot when it is fully occupied. 
In addition to parking prediction, stochastic models have also been applied in other use case. For instance, one study has modelled parking process as a birth-death stochastic process for revenue prediction \cite{gongjun2011smartparking}. The numerical results are then obtained through simulation with fixed arrival and departure rate.

The aforementioned studies have obtained numerical results through simulation, often by using fixed arrival and departure rate. This limits the efficacy of the method as real parking demand displays seasonal behaviour, where parking demand is notably impacted by factors such as time-of-day and day-of-week. To account for these seasonal patterns, a study in \cite{rajabioun2013:intelligent} proposes a prediction approach based on a probabilistic model that uses historical data capturing temporal dynamics of car park users. The author models arrival and departure rate as non-homogeneous Poisson distributions. Consequently, by knowing the current car park occupancy, the conditional expected value of available parking spots at multiple horizons in the future are calculated.

With the emergence of IoT devices to collect real-time and historical parking information, many works have adopted data-driven approach for their occupancy estimation. An extensive range of methods that have been explored inculde classical time-series models like ARIMA \cite{burns1992econometric}; non-parametric models such as regression trees and support vector regression \cite{zheng2015parking}; and spatiotemporal models \cite{rajabioun2015street} when data of multiple parking locations are available.
Furthermore, new techniques involving deep learning have been gaining popularity in the last decade, in additional to classical multi-layer perceptron (MLPs) for predicting future parking occupancy \cite{vlahogianni2016real}, several variants of recurrent neural networks (RNNs) have been proposed including LSTM \cite{shao2018LSTM} and RNN using evolutionary algorithms \cite{camero2018evolutionary}.

All aforementioned studies predict parking occupancy in the near future in order to provide real-time or up to 1-hour in advance parking information to the users. Despite the fact that accurate short-term parking prediction can undoubtedly enhance informed parking decisions for users, longer term estimation is required for future space management such as space allocation problem that our work aims to address.

\subsection{Parking Management}

According to Litman \cite{litman2016parking} parking problem is typically perceived as inadequate parking supply that has led planners to provide more supplies in order to fulfil the growth of demand. Consequently, abundant parking facilities have resulted in urban sprawl, causing parking demand and supply to get further inflated. New parking management paradigms, on the other hand, aim to use parking facilities more efficiently, instead of expanding the physical infrastructure. This not only reduces development costs but also supports more strategic objectives such as reduction of motor vehicle use by encouraging the adoption of alternative modes of transport, thereby reducing traffic congestion. 

Parking management strategies can range from parking regulations (\ie controlling who, when and how long vehicles can park), to parking pricing (\eg performance-based pricing \cite{shoup1997high}), to mobility management which aims to change travel behaviour in terms of travel frequency, mode, destination, or time. Advancement in technology has brought numerous applications in the field of parking management in the recent decades. There are various parking systems that have been implemented at a car park-level including parking guidance information system (PGIS) where users are informed on availability and location of parking spaces; transit based information system which combines PGIS with public transit information to allow users to plan for their trip in advance; smart payment system where traditional payment method via cash is replaced by parking meter, mobile devices or RFID technology \cite{pala2007smart}; E-parking which allows users to enquire the availability of a parking facility and reserve a parking space \cite{wang2011reservation}; and automated parking which allows user to drive up to the bay where machine automatically locates the vehicle to the allocated space.

Since parking is a city-level problem, various large-scale deployment of smart parking technologies can be seen in the present day. SFpark \cite{sfparkoverview2014} is one of the most well-known example program by the Municipal Transport Agency in San Francisco, where parking availability information is collected and distributed in real-time. The data is used to periodically adjust meter and garage pricing in order to achieve the ideal level of parking availability. This demand responsive pricing scheme helps in demand distribution across both spatial and temporal space by encouraging car users to park in underused areas, reducing demand in overused areas. LA ExpressPark \cite{laexpresspark} is another pilot program that adopts demand-based pricing as a parking management strategy. It aims to achieve 10-30\% of available parking spots throughout the day.

Beijing has also launched a smart parking pilot program \cite{Chinaparking}, where 35\% of on-street parking is monitored. The collected real-time information is used by the government to regulate traffic. Other large scale-deployment of intelligent systems include an integrated parking system in Spain \cite{quinones2015design} where each parking is equipped with sensors and occupancy information is available through an application for drivers; the AREA system in Barcelona \cite{AREASpain} which is a road-side parking management system allowing users to enter their vehicle number into a smart meter and the updated information is sent to Barcelona Municipals Services; and a smart city initiatives at the City of Melbourne where data from 4300 in-ground sensors in the on street parking bays is made available through an open data platform \cite{melbournedata}.

Based on our survey on smart parking management systems, tangible benefits of real-time parking data has been demonstrated in various applications. Understanding the long-term parking dynamics allows for a greater impact beyond operations and management, by giving cities and institutions a more granular approach in regards to the definition of spatial use overall.


\section{Bus Demand Monitoring and Management}

\subsection{Queue sensing technologies}
Many existing works have proposed methods to estimate queue length \cite{wang2014:tracking,aubert1999:passengers}, waiting-time \cite{okoshi2015:queuevadis,shu2016:queuing}, or crowd density in general \cite{schauer2014:estimating,elhamshary2018:crowdmeter}. This section surveys the most commonly used approaches to deduce crowding information.

\subsubsection*{WiFi/Bluetooth Signals}

Due to the prevalence of mobile devices that support wireless communication such as WiFi and Bluetooth, many researchers have attempted to leverage these signal traces in estimating human crowd information. For instance, Wang et al. \cite{wang2014:tracking} uses a single WiFi monitor to track human queue in retail environment by analysing RSS trace from mobile devices, Shu et al. \cite{shu2016:queuing} uses WiFi positioning data to estimate queueing time at an airport, and several works \cite{schauer2014:estimating,tang2018:indoor} estimate crowd density within an indoor area by exploiting WiFi and Bluetooth captures. These methods, require WiFi access points or at least a signal monitoring device installed at certain locations, which may not be suitable for an outdoor scenario.


\subsubsection*{Crowd-sensing or Crowd-sourcing}
Crowd sensing through the use of information obtained from smart phones has been explored to estimate people crowding information. 
Several studies \cite{okoshi2015:queuevadis,li2014:queuesense} use crowd sourced data from sensors embedded in smart phones, such as accelerometer and magnetic compass, to estimate queue waiting time in a retail environment. The proposed methods are able to distinguish queuers from non-queuers with good accuracy, yet high number of participants (who are willing to install and have the application running on their mobile phones) are required to achieve such good results.
Elhamshary et al. \cite{elhamshary2018:crowdmeter} also leverages data sensed from users' smart phones in order to estimate crowding level in railway stations by exploiting a wider range of sensed data including individual's trajectory from entrance and ambient sound from smart phone's microphone. 
However, the method is not suitable for queue-specific detection, but instead can be used for monitoring of human crowd-level in general.
An even more user-centric approach is using crowd-sourcing kiosks that asks users about their estimated waiting time through an interactive interface \cite{goncalves2016:crowdsourcing}. This method poses several drawbacks including a high likelihood of insufficient contributions from queuers and deployment difficulty.

\subsubsection*{Camera-based queue detection}
A more traditional approach for crowd detection is to use camera-based systems \cite{segen1996:camera, masoud2001:novel, aubert1999:passengers} and employ video or image analysis. However, this approach is expensive, endangers privacy, and its accuracy can largely depend on the camera's field of view. 
The camera-based approach also poses deployment difficulties, especially in an outdoor environment where power and communications are not easily available.

Most of existing methods for measuring crowding and queue information are only applicable to indoor settings such as retail environments or airports. Works that have deployed a monitoring system in outdoor settings either aim to estimate overall crowd density \cite{weppner2013:bluetooth} or focus on tracking individuals \cite{leibe2005:pedestrian,musa2012:tracking}, as opposed to measuring queue length. To the best of our knowledge, none of these works attempt to estimate the length of an orderly queue and waiting time in an outdoor scenario.

\subsection{Passenger demand prediction}

Real-time monitoring and prediction of passenger demand is an essential component to improve the quality of bus service. The information can provide informed decisions to bus operators with regards to the allocation of bus resources, therefore services can be provisioned to passengers in a proactive manner as opposed to a reactive manner. Accurate forecast of public transport demand is foreseen to minimise operation cost while improving bus service quality to customers \cite{tirachini2013crowding}. In the literature, techniques used to predict passenger flows can be categorised into parametric and non-parametric models.

In the parametric approach, models such as auto-regressive integrated moving average (ARIMA), exponential smoothing, and historical averaging, have been employed to predict transport demand \cite{smith1997traffic,faraway1998time}. ARIMA model, in particular, has been applied in various applications including short-term traffic data prediction (\eg traffic flow, travel time, speed, and occupancy) \cite{suwardo2010arima}, international air passenger flow prediction, and passenger demand prediction in a metro station \cite{chen2016prediction}. Nonetheless, ARIMA model poses a limitation due to its assumption of linear relationships between the prediction variable and its lagged variables, hence non-linear relationships cannot be captured. 

In the non-parametric approach, machine learning models and neural networks (NN) are used to forecast transportation demands. In a study, Support Vector Machine (SVM) has been applied to predict passengers flow at urban rail transit stations using historical data, where the model is calibrated by particle swarm optimisation (PSO) \cite{zhou2014direct}. Another study applies NN based model to forecast short-term railway passenger demand by proposing multiple temporal units neural networks and parallel ensemble neural network to improve the accuracy of the prediction \cite{tsai2009neural}. The result has shown that the proposed network structures outperform the conventional multilayer perceptron (MLP) by up to 10.5\%. Additionally, Kalman filter \cite{wang2007real} and Gaussian maximum likelihood \cite{tang2003comparison} have also been used in real-time estimation of traffic.

Some studies have adopted a hybrid approach in predicting passenger flow. In \cite{zhang2017short}, Phase Space Reconstruction-Long Short Term Memory (PSR-LSTM) model has been proposed, where the theory of PSR is first used to extract a regular trajectory, then LSTM is used to predict short term passenger flow in a rail way station.

\subsection{Optimisation of bus scheduling}

There is an abundance of studies on optimising bus scheduling in the existing literature. While the majority of early research in this area adopted deterministic models where passenger demand is assumed to be constant over time \cite{sun2016optimization,mokhtari2018integration}, recent studies have been focusing on dynamic scheduling. Several studies model the optimisation problem as a real-time dynamic events, for instance, real-time dead heading scheme to optimally skip stops in the event of disruptive incidents \cite{eberlein1998real}, dynamic bus holding scheme to reduce in-vehicle passenger delay \cite{daganzo2009headway}, and dynamic control scheme that introduces additional headways aiming to evenly distribute bus headways at different stops \cite{zhang2018two}. Although these strategies can improve bus service reliability and address immediate disruptive events, the decision is only made based on real-time data and long term passenger demand patterns are overlooked.

Several studies have proposed dynamic bus scheduling scheme aiming to improve passengers' transit experience. In one study, a dynamic bus dispatch model that can minimise passenger wait time has been proposed. The model has been tested under multiple cases of real-time passenger flow information and genetic algorithm (GA) is used to solve the optimisation problem. The study shows that the proposed dynamic scheduling strategy can effectively utilise the real-time information from IoT to improve the performance of the transit system \cite{luo2019new}. 
A study in \cite{huang2019novel} has also presented a novel bus dispatching model based on bus arrival time and predicted passenger demand. By considering historical origin-destination data, this study aims to minimise passenger wait time during transfer. 
Apart from optimising bus scheduling to improve passenger transit experience, other studies also aim to minimise costs of bus operations \cite{chen2020multiobjective}, improve on-time bus performance \cite{sun2017unsupervised}, and relieve unbalanced spatial and temporal distribution of buses in areas with less passenger demand \cite{pang2017scheduling}.

Since scheduling buses involves multiple stakeholders who often have conflicting objectives, some studies have adopted multi-objectives optimisation approach to solve the problem. A study in \cite{chen2020multiobjective} develop a model that aims to minimise bus operational cost while minimising passenger wait time and total bus overload by using predicted passenger arrival data for each time horizon. In this study, the problem has been solved using two popular evolutionary methods namely NSGA-II and MOED/D. This conflicting objectives between the minimisation of passenger wait time and operational efficiency have also been modelled in a Smart Campus environment \cite{feng2018design}. In this study, simulated annealing algorithm is employed to solve the problem.

\section{Conclusion}

In this chapter, we surveyed various applications of IoT in the context of smart campus, focusing on the domains of space utilisation, campus experience, and management of energy, waste and water. We showed that the emerging technology of IoT has presented a myriad of opportunities for campuses to become smarter, operate more efficiently, while providing enhanced services to its community. We then focused on the application domain of classroom space management where we explored numerous sensing technologies for occupancy detection, approaches for occupancy prediction, and existing methodologies of classroom allocation and management. We revealed that while a number of studies have focus on developing sensing approach for room occupancy, the method has never been applied in the field of classroom management where dynamics of attendance is considered during the allocation process. This thesis has deeply explored the use case of occupancy measurement in improving the utilisation of classrooms by incorporating predictive modelling of class attendance and combinatoric optimisation for optimum class scheduling.
Next, we focused on the parking space application domain where car park sensing technologies and existing parking management systems are explored. Despite a plethora of sensing technologies and smart parking applications available in the existing literature, there is still a limited number of research aiming at using collected data for long-term parking space planning. This thesis has shown that long-term prediction of parking demand can aid the space partitioning decision, allowing the campus to accommodate the new paradigm of future transportation. 
Finally, we focused our review on the transit bus domain where we surveyed various sensing technologies on human queue detection, passenger flow prediction, and approaches to optimise bus scheduling. We found that most of the queue detection technologies presented in the current literature only cater for indoor scenarios, which is not suitable for our on campus bus stop. The majority of existing studies on optimising bus scheduling based on dynamics of passenger demand also does not consider cases when passenger demand exceeds bus capacity and causes some passengers to wait for the next service. This thesis addressed this gap by capturing passenger demand at bus stop within the transit scheduling optimisation model.

The rest of this thesis investigates the potential of IoT technology in addressing the aforementioned research gaps, beginning with data-driven classroom management in an IoT enabled campus.

\chapter{Monitoring and Optimisation of Classroom as a Resource}
	\minitoc
	
	Increasing demand for university education is putting pressure on campuses to make better use of their real-estate resources. Evidence indicates that enrolments are rising, yet attendance is falling due to diverse demands on student time and easy access to online content. In the previous chapter, we explored various occupancy sensing technologies that are potential enablers for informed management of classroom resources. This chapter outlines our efforts to address classroom under-utilisation at UNSW, arising from the gap between enrolment and attendance. Our first contribution undertakes an evaluation of several IoT sensing approaches for measuring classroom occupancy, and comparing them in terms of cost, accuracy, privacy, and ease of deployment and operation. Our second contribution instruments 9 lecture halls of varying capacity across the campus, collects and cleans live occupancy data spanning about 250 courses over two sessions, and draws insights into attendance patterns, including identification of cancelled lectures and class tests, while also releasing our data openly to the public. In our third contribution, we use AI techniques for predicting classroom attendance, applying them to real data, and accurately predicting future attendance with an RMSE error as low as 0.16. Our final contribution is to develop an optimal allocation of classes to rooms based on predicting attendance rather than enrolment, resulting in over 10\% savings in room costs with very low risk of room overflows.

\vspace{7mm}
\section{Introduction}\label{sec:introduction}
Higher education institutes continue to experience steady growth in enrolment demand \cite{BritishCouncil}. A major factor limiting universities in fulfilling this demand is real-estate, since enrolment in a course is capped by the capacity of the classroom to which the course is allocated. However, with recent trends towards student lifestyles that mix study with work and other commitments, as well as greater access to online content, there is ample anecdotal evidence that classroom attendance is often well below the enrolment number. This presents an opportunity for education institutes to better optimise the usage of classroom space based on attendance rather than enrolments. Since class attendance can vary significantly between courses and across weeks of semester, visibility into actual class attendance and ability to predict future attendance based on historical data are needed to dynamically re-allocate courses to rooms while minimising risk of overcrowded lecture rooms where class attendance exceeds room capacity.

Several methods are available to count the number of people in an indoor space, such as WiFi-based approach \cite{wifiLCN2018}, camera image processing, thermal imaging, ultrasound imaging, and beam counters affixed to entryways \cite{sutjarittham:classroomusage}. Each method has its own pros and cons across various dimensions such as cost, power, communications, ease of deployment and operations, privacy, and accuracy. For example, using WiFi data and cameras endanger privacy, thermal and ultrasound imaging have low accuracy, and camera-based image processing is computationally expensive. Furthermore, a method that works well in a small room may not be as effective in a larger lecture theatre, and cost/accuracy may also be impacted by the layout of the room, the number/width of doorways, and the availability of power and wired/wireless network connections. 
Hence understanding both benefits and challenges of various approaches in order to adopt the most suitable methods for the nature of the room is important for the real deployment of classroom occupancy monitoring system.

This chapter describes our experiences in adopting IoT to measure and AI to predict the attendance of lectures in courses at our University campus, and to use these to optimise the usage of lecture rooms. The work presented in this chapter has been published in \cite{sutjarittham2019experiences,sutjarittham:classroomusage}. Our specific contributions are four-fold:

\begin{enumerate}
\item We begin by testing several sensing methods in a lab environment and characterising their trade-offs in aspects such as cost, ease of installation, method of data extraction, privacy, and accuracy. 
\item We then make appropriate sensor selections, build a full system, and deploy it across 9 lecture theatres of varying size across the university campus. We collect and clean the data to obtain visibility of occupancy across these rooms in real-time over a period of 18 weeks (\ie a full semester in 2017 and half a semester in 2018), integrate it with University timetabling data to infer attendance patterns of over 250 courses, and highlight interesting findings such as attendance trends, cancelled lectures, and class tests. We also make our occupancy data openly available to the research community. 
\item We develop machine-learning models to predict classroom attendance using three algorithms namely multiple regression, random forest, and support vector regression (SVR). We employ quantile regression technique, allowing asymmetric penalties for under-prediction and over-prediction of attendance. Our models are able to predict attendance in advance with a root-mean-square error (RMSE) of less than 0.16. We also make our attendance dataset openly available to the research community.
\item Finally, we develop an optimisation algorithm for allocating classes to rooms based on predicted attendance rather than static enrolments, and show potential saving of over 10\% in room costs.
\end{enumerate}

The rest of this paper is organised as follows: In \S\ref{sec:mon}, we present our lab evaluation of various sensing methods and their trade-offs, while \S\ref{sec:visual} describes our field deployment across campus and interesting insights obtained therein. We present our techniques for predicting classroom attendance in \S\ref{sec:pred}. Our optimisation formulation for dynamic classroom allocation is described in \S\ref{sec:optim}, and the chapter is concluded in \S\ref{sec:con}.

\section{Sensing Classroom Occupancy}\label{sec:mon}
In this section, we describe various sensing methods for counting people, 
outline their relative trade-offs with a view towards making
appropriate selections suitable for a larger-scale deployment across
the campus, and briefly explain our system architecture for collecting, cleansing, and visualising sensing data.

\subsection{People Counting Methods}

We investigated several commercial sensors and straight-away eliminated those that send data to the vendor's cloud servers, since we wanted to: (a) keep the data entirely on-premises and not risk it leaving our campus infrastructure; and (b) not be beholden to a vendor to access our own data, hence freeing us from ongoing service costs. In other words, we wanted a ``sale'' model of the device so we could have unfettered access to our data without any ongoing ``service'' fees. We were quite happy to buy spares of the units to cover for device failures; further, this model allows us to integrate data into a centralised repository to facilitate better analytics across the many data feeds we have on campus.

We narrowed our lab trials to four types of commercial sensors: EvolvePlus Wireless Beam Counter \cite{evolveplus:beamcounter}, EvolvePlus Overhead Camera \cite{evolveplus:overhead},  Steinel HPD Camera (pre-market release), and Steinel Presence Detector \cite{steinel:PIR}. In addition, the University IT department provided us with timestamped connections logs from two WiFi access points (one inside our lab and one just outside), so we could compare our approaches to those obtained from WiFi logs. We note that the WiFi logs gave us personal user information such as their device MAC address, user-ID, and connection durations; we therefore obtained ethics clearance (UNSW Human Research Ethics Advisory Panel approval number HC17140) for this experiment.

The {\bf Beam Counter} comprises a pair of infrared (IR) break-beam sensors mounted on the door frame, and counts the number of people passing through in each direction. It communicates the counts (for ``in'' and ``out'' directions) to a gateway every 30 seconds using a propriety wireless protocol, and the gateway then posts these readings via Ethernet to an SQL database (DB) server hosted on a VM in our on-premises cloud infrastructure. The {\bf Overhead Camera} is a thermal sensor mounted on the ceiling close to the entrance facing downwards, and counts the number of people passing below it. It also communicates the counts in each direction to the same gateway as the beam counter, which then forwards it on to the SQL DB. We wrote a script that pulls data from the SQL DB, stamps the data with the time and the unique UUID of the gateway, and posts as a JSON string to our master database (which holds data from many sources) via a REST API. 

The {\bf HPD Camera} (pre-market release) is a people counting sensor mounted in a corner with full view of the room. It uses built-in image processing to compute the number of people present within a configurable zone of interest. It is powered over Ethernet, and comes pre-configured with a server that be queried via a REST API. We wrote a ``broker'' script that polls the camera every 30 seconds to get the people count, and posts the time-stamped and sensor UUID-stamped data in JSON format to our master database. The {\bf Presence Detector} is a passive infrared (PIR) sensor mounted on the ceiling in the middle of the room, and detects motion. Though it does not count the number of people in a room, it gives a binary indication on whether the room is occupied or not -- this sensor can be used as a way to calibrate the other counting sensors which may accumulate errors with time. The PIR sensor sends its binary occupancy state every 60 seconds to its corresponding gateway via a propriety wireless interface, which then posts it to a broker script that again time- and sensor-UUID-stamps the data and posts to our master database.

Lastly, we receive a CSV file of daily WiFi connection logs for the two access points from our IT department every morning at 7am -- real-time feed of data was not possible due to technical limitations of the AP vendor. We wrote a script to parse the log file and compute the number of unique users connected to each AP every 30 seconds -- this was also posted to our master database.  

With possibility of sourcing data from various sensing devices, one may want to perform sensor fusion for an accurate occupancy measurement. At a very minimum, a combination of PIR sensor and passing people counters (\ie beam counter and overhead camera) seems reasonable. PIR sensors are fairly accurate in detecting whether a room is empty which can be useful for resetting the errors accumulated over time via the people counting sensors. It is important to note that detecting presence is not a trivial task for a large lecture theatre due to limited coverage of PIR sensors, and thus configuring non-overlapping zones for multiple units of PIR sensors can be quite challenging. For a second step of fusion, adding WiFi data or HPD camera would help infer an accurate occupancy since these methods measure occupants count instantaneously without keeping states (\ie not cumulative). But, as explained next, these sensors come with their own shortcomings. We note that deploying a collection of sensors at the scale of a university campus can significantly increase the cost. Therefore, our primary focus in this paper is to select and deploy one sensor type for each classroom, and demonstrate its value in optimal allocation of rooms to courses.

\subsection{Sensor Evaluation and Selection}
Our lab trial helped us compare the various counting methods in terms of their ease of installation, calibration, power and communications requirements, accuracy, cost, and privacy, as summarised in Table~\ref{table:occupancy_sensors_compare}. 


\begin{table}[t!]
 \centering
	\begin{adjustbox}{width=1\textwidth}
		\vspace{-3mm}
		
		\begin{tabular}{l c c c c c c c}
			\toprule 
			& \textbf{Installation} & \textbf{Calibration} & \textbf{Power} & \textbf{Communications} & \textbf{Accuracy} & \textbf{Cost} & \textbf{Privacy}\\
			\midrule
			Beam counter &  easy & easy & battery & wireless & high & medium &  high     \\
			Thermal sensor &  hard & medium & AC  & wireless & low & high &  high   \\
			HPD Camera & medium & hard & PoE & Ethernet & medium & medium &  medium      \\
			PIR Sensor & hard & easy & AC & wireless & binary & medium & high \\
			WiFi Data & existing & existing & existing & existing & existing & existing & low \\
			\bottomrule
		\end{tabular}
		\end{adjustbox}
		\caption{Sensors Comparison}
		\label{table:occupancy_sensors_compare}
\end{table}

Our comparison across these measures is qualitative rather than quantitative. Even aspects such as accuracy, that can be quantified, depend on factors like room size and layout, mounting position, number of doors, and width of doorways, which can vary widely across deployment environments. We therefore resort to qualitative measures (low, medium, and high) in this table, derived from our experience across the rooms we instrumented, and we back these up with several data points presented later in the paper.

\textbf{Installation:} The thermal camera, HPD camera, and PIR sensor needed professional installation by certified tradesmen, since each needed special mounting brackets and extra wiring for mounting on (or near) the ceiling. We could install the beam counter sensor easily by ourselves using two-sided adhesive strips on the door frame at around waist-height.  

\textbf{Calibration and Positioning:} Sensor positioning is another key factor in our comparison. The thermal camera needs to be positioned at a certain height range (i.e. 2.2m - 4.4m) recommended by the manufacturer and close to the entrance allowing the best coverage to count everyone that passes underneath. This requirement makes it hard or impossible to use the thermal camera in very large lecture halls with high ceilings. Beam counters require to be mounted at around waist-height (too low causes each leg to get counted separately, and too high causes the swinging arms to get counted!). Once an appropriate height is chosen for the beam counters, doors of all classrooms need to be outfitted in the same way. The HPD camera needs prior configuration for zone of interest that can vary across rooms depending on the room size and the place at which the camera is mounted. The PIR sensor is positioned at the center of the room (on the ceiling) to have a symmetrical coverage over an area that can also vary across rooms depending on their seating arrangement.

\textbf{Power and Communications:} Provisioning power was challenging for the thermal camera and PIR sensor, since the campus has pre-built and fixed wiring only in certain locations in each classroom. Therefore our Facilities Management was required to supply new exterior wiring for these three sensors. The beam counters are battery powered (with stated battery life in excess of a year), and the HPD camera required a special PoE switch that provides Ethernet for both power and communications. The corresponding gateways for the beam counter, thermal camera, and PIR sensor were hidden inside a closet with available power and Ethernet.

\textbf{Accuracy:} We performed several spot measurements in our lab to extract ground truth on occupancy. We found that the beam counter is the most accurate among the four techniques. We note that the beam counter has very good accuracy when the the door is narrow, like in our lab. However, for a wider doorway its accuracy is worse, since it does not always capture individuals walking in/out side-by-side (this became more evident in our field-trial, described in the next section). We found the accuracy of the thermal camera to be very sensitive to mounting position and distance from the entrance. Moreover, since the door of our lab opens inwards, it was not very conducive for the overhead thermal camera (mounting it on the outside of the room was not an option as it was a busy corridor). The HPD camera tended to have a non-zero absolute count error, which made its relative error high when the number of people in a room is small (e.g. less than 10) and low when the number of people is high (e.g. more than 40). We could not test its accuracy scaling to larger counts as our lab can only accommodate around 40 people. Lastly, the people count derived from the WiFi access points was wildly inaccurate, because our lab is adjacent to a busy corridor and study space that is busy with students during regular hours, and we could not distinguish who was inside versus outside the room. 

\textbf{Cost:} The beam sensors and PIR sensors are priced in the range of a few hundreds of dollars, while the cameras are in excess of a thousand. The beam counter and thermal camera both need a gateway to send their readings to the back-end server, and each gateway is priced in at nearly a thousand dollars. Bear in mind that each gateway can connect up to 20 sensors (though our deployment described in the next section maps at most 4-5 sensors to a gateway in large lecture theatres). The beam counters therefore end up as a more cost-effective solution than the cameras for large-scale deployment across campus.  

\textbf{Privacy:} Among the four sensing techniques, WiFi clearly endangers students privacy as their IDs are visible (due to PEAP authentication their devices perform to connect with the campus WiFi network). The HPD camera does on-board processing and does not store or transmit any images of people (though it is possible to log in to it to view the current image), and can hence be deemed to preserve privacy. The beam counter and the thermal camera are truly privacy-preserving, since they can only sense the number of people passing through the doorways without sensing any private attributes of the individuals.

\textbf{Summary:} The trade-offs discussed above are summarised in Table~\ref{table:occupancy_sensors_compare}. WiFi is not an option as it compromises privacy and is inaccurate. The cameras are eliminated as being expensive, difficult to install/position, and poor in accuracy (though we are considering them for open spaces that do not have doorways). The PIR sensor has only binary output (\ie 0 for unoccupied and 1 for occupied), and is used for re-calibration rather than counting. We therefore decided on a larger-scale deployment of the beam counter, based on its relatively lower cost, easy deployment, high accuracy, and good protection of privacy. Our deployment in classrooms is described next.

\vspace{-5mm}
\subsection{System Architecture and Data Collection}

\begin{figure}[t]
	\centering
	\includegraphics[width=0.7\textwidth]{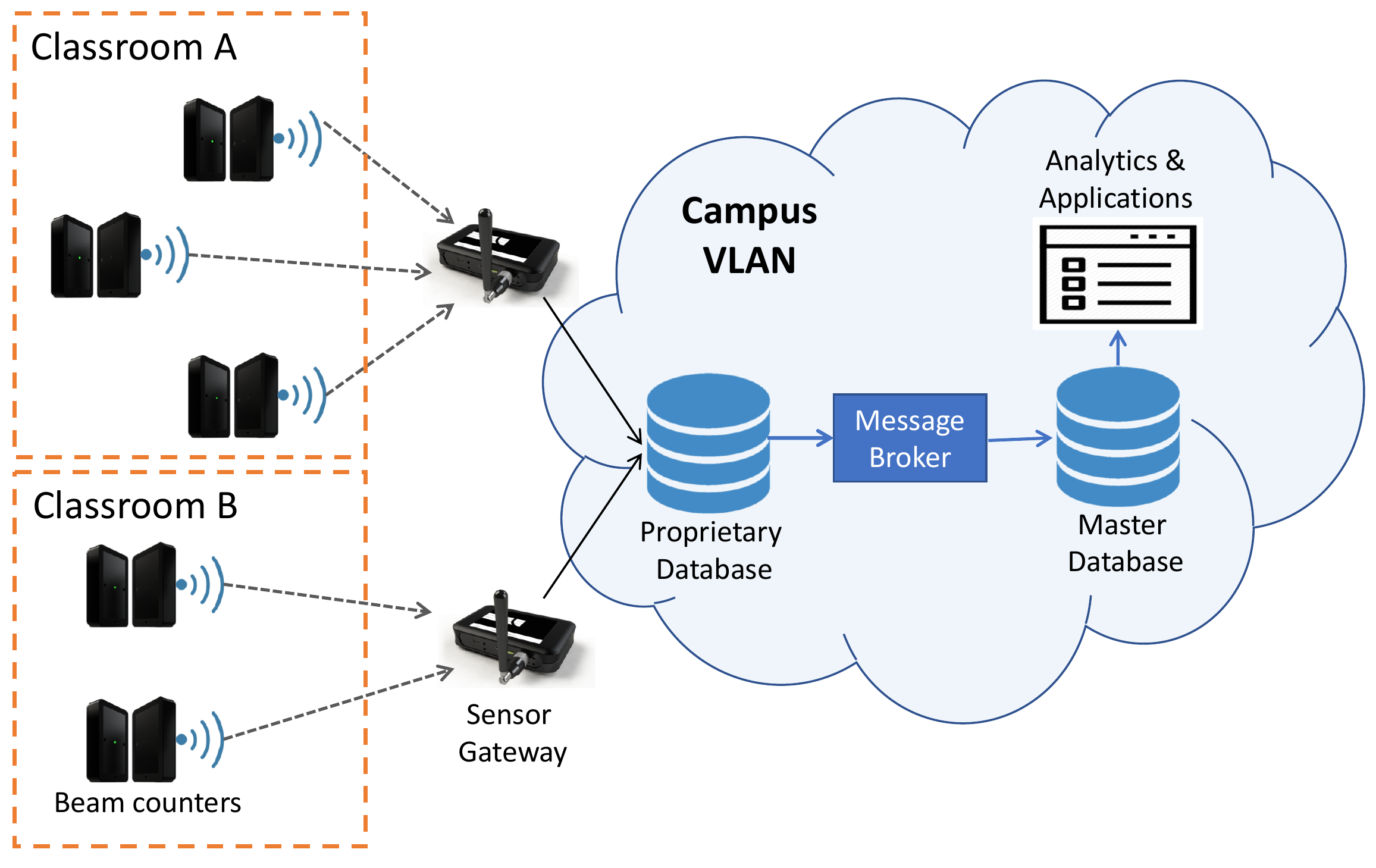}
	\vspace{-3mm}
	\caption{System architecture of classroom occupancy monitoring.}\label{fig:sys-arch}
	\vspace{-5mm}
\end{figure}

Figure \ref{fig:sys-arch} shows a high-level system architecture of the classroom occupancy monitoring system using beam counters. First the beam counters communicate their count data to sensor gateways that are installed in each room via a proprietary wireless protocol. The gateway is directly connected to an Ethernet port which has been provisioned to allow connection to our university private VLAN, where the data is being stored in a proprietary database. We then wrote a message broker script to unify the data format into a commonly agreed structure, where the sensor data is converted into a JSON string as well as being tagged with timestamp and sensor UUID, before getting posted to our master database via a REST API. This provides the feasibility for the platform to collect data from various sensors which generate heterogeneous data format.

Similarly, the retrieval of data for analysis or as an input to applications can be easily performed through a GET RESTful API. This raw beam counter data only contains timestamped count in and count out from each sensor installed at the doorway, hence data preprocessing is required to infer room occupancy and finally class attendance number.

\section{Data Processing and Visualisation}\label{sec:visual}

We worked with campus staff to identify appropriate classrooms for a field trial, and picked 9 rooms of varying size, as shown in the first column in Table~\ref{table:occupancy-error}.
Some of the doorways to the lecture-halls posed a challenge as they were very wide, increasing the likelihood that multiple students walking out side-by-side get counted as one. The data collected over the first few days was manually verified (volunteers were used to do head-counts) so as to obtain ground-truth and calibrate the errors. In what follows we describe our methods for data cleansing, linking with class-timetabling information, processing, and visualisation using a web-UI.

\begin{table}[t!]
	\begin{center}
		\caption{Measured error from ground-truth of occupancy.}
		\label{table:occupancy-error}
		\begin{tabular}{lllclcl c l c l }
			\toprule 
			&   &&\multicolumn{2}{c}{Error} \\
			\cmidrule{4-5}
			Room &  \#seats & \#doors & Room-based & Course-based \\
			\midrule
			BUS105 & 35 & 1  & $27.7$ \% & $13.0$ \%       \\
			BUS115 & 53  &  1  &  $34.2$ \%  & $17.3$ \%  \\
			CLB7 &  497  &  4 &  $89.5$\% & $4.6$ \%    \\
			CLB8 & 231  &  3 & $26.3$ \%  & $16.1$ \%   \\
			MAT A & 472  &  6 & $25.5$ \% & $8.0$ \%    \\
			MAT B & 246  &  3 & $6.3$ \%  & $9.1$ \%   \\
			MAT C & 110  &  2 & $14.6$ \% & $24.4$\%    \\
			MAT D & 110  &  2 & $16.9$ \% & $9.2$ \%    \\
			PhTh  &  369  & 4  & NA  & NA  \\
			\bottomrule
		\end{tabular}
		\vspace{-2mm}
	\end{center}
\end{table}

\subsection{Occupancy and Attendance Calculation}\label{sect:attd-processig}

We compare two methods of data processing to deduce the occupancy from the number of entries and exits at each door:

\textit{Method 1: Room-based}: Our first (naive) method for deriving occupancy is to set it to the cumulative number of entries minus the cumulative number of exits across all doorways of a classroom. However, errors arise when students walk in/out in groups; though we reset counts to zero at midnight each day, errors accumulating during the day can become significant. 

\textit{Method 2: Course-based}: To reduce the errors accumulating during the day, we enhance our method by computing course attendance independent of each other by linking our sensor data with course timetable databases obtained from our University. We assume that students may enter the room up to 10 minutes prior to start of the scheduled lecture time, and may leave up to 10 minutes after the scheduled lecture time. Attributing each entry and exit to a specific lecture therefore allows us to compute attendance per-course, and errors are not carried over from one lecture to the next even if they are adjacent in time to each other.

\begin{figure}
	\centering
	\includegraphics[width=.7\textwidth]{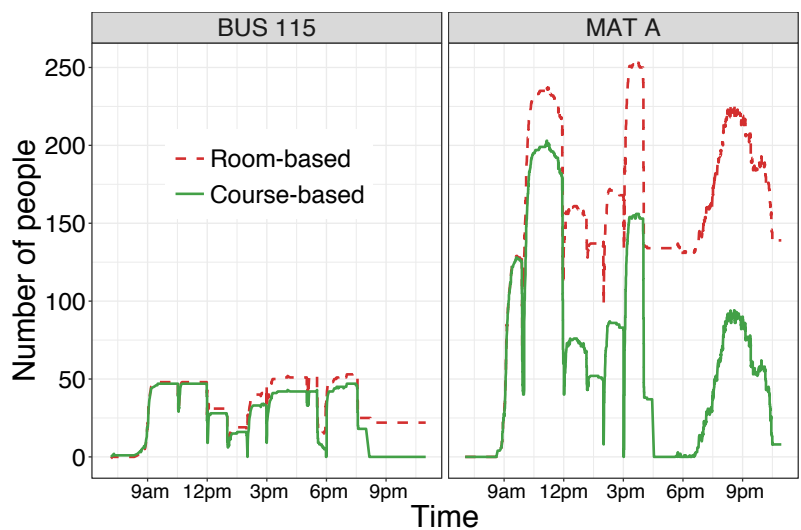}
	\vspace{-2mm}
	\caption{Error of room-based method is higher than course-based and increases with the size of classroom; `BUS 115': small room and `MAT A': large theater.}\label{fig:course_room_based}
	\vspace{-3mm}
\end{figure}

\textbf{Accuracy of Counting}: To evaluate the accuracy of our counting methods, we obtained ground-truth information by having volunteers physically count attendance during the lectures. We collected a total of 50 samples covering 31 lectures over 4 days. The ground-truth samples were collected from 8 out of 9 classrooms in which the sensors have been deployed. Table~\ref{table:occupancy-error} shows the average error of the computed occupancy using the two methods described above applied to the various rooms. As expected, course-based occupancy computation yields lower errors (average: $12.71$\%) compared to room-based occupancy computation (average error: $30.60$\%). This is because the room-based method gradually built-up errors over the course of a day, whereas the course-based method had a stable error irrespective of time-of-day (since the errors do not accumulate). However, it should be noted that the course-based method requires access to timetabling information, which may not be generalisable to other environments.


\begin{figure}
	\centering
	\includegraphics[width=.7\textwidth]{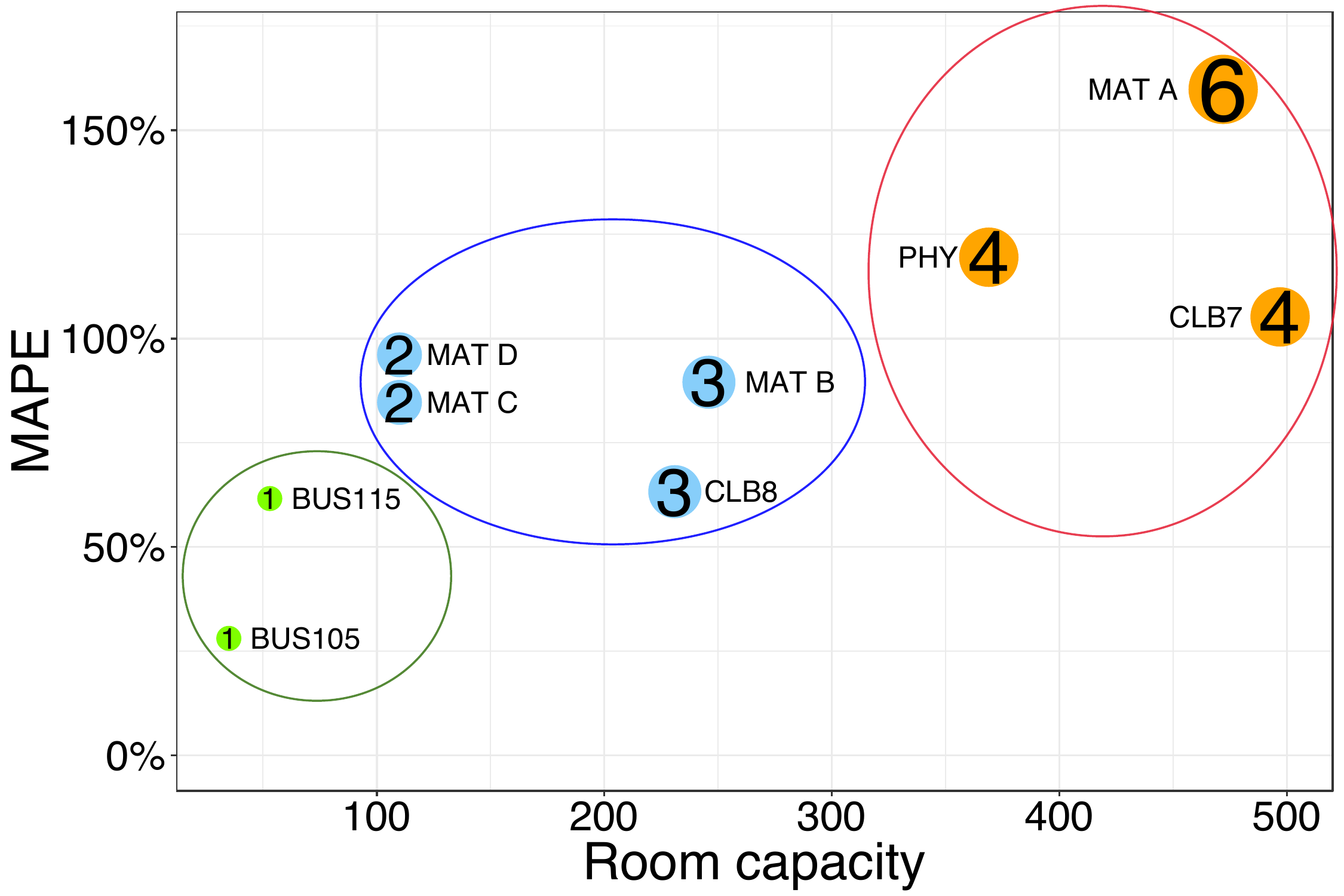}
	\vspace{-2mm}
	\caption{The mean absolute percentage error (MAPE) versus room size, highlighting three clusters.}\label{fig:sensor-MAPE}
	\vspace{-2mm}
\end{figure}


{\textbf{Typological Analysis of Error in Calculating Occupancy}: We now analyse the impact of room characteristic (\ie size and number of doorways)  on the accuracy of estimating occupancy. Fig.~\ref{fig:course_room_based} shows the occupancy computed by room-based and course-based methods, for two representative rooms (one small classroom `BUS 115' with 53 seats and one large lecture theatre `MAT A' with 472 seats) on 9-August-2018. Our first observation is that the room-based method, shown by dashed red lines, results in higher residuals (\ie accumulated error in calculated occupancy) than the course-based method, shown by solid green lines, at the end of the day. Additionally, it is seen that the gap between room-based and course-based is dilated in larger classrooms -- for example a significant gap of more than 100 people is observed in `MAT A' (on the right plot), compared to a relatively smaller gap in `BUS 115' (on the left plot).} 

\vspace{1mm}	
{For our analysis, we use deployed sensors data (without ground-truth) collected in entire Semester 2, 2018 to investigate the impact of the two key factors (\ie rooms size and number of doorways). We employ the mean absolute percentage error (MAPE) metric computed by the average of absolute difference between room-based and course-based divided by course-based occupancy (\ie ratio of the gap between green and red lines in Fig.~\ref{fig:course_room_based} to green lines). We illustrate in Fig.~\ref{fig:sensor-MAPE} the value of MAPE as a function of room characteristics. Note that x-axis is the room capacity and each circle on the plot shows the number of doorways for the corresponding classroom -- size of each circle is proportional to doors count. We observe that the MAPE is positively correlated with both the room capacity and the number of doorways, highlighted by three clusters namely small rooms with less than 100 seats and one door (green circles), medium-size rooms with 100 to 300 seats and 2 to 3 doors (blue circles), and large theatres with more than 300 seats and more than 4 doors (orange circles). We note that for larger classrooms, the chance of students walking in/out side-by-side is relatively higher (such instances are usually counted as a single individual by the beam counters) which results in a larger MAPE value.}


\textbf{Occupancy and Attendance data}: Our weekly occupancy dataset, computed using the Method-2 above, is openly available for download \cite{data2018occupancy}. Each row in a CSV file represents the real-time measurement from beam sensors comprising \textit{time-stamp}, \textit{week} of semester, room information including \textit{room name}, \textit{number of doorways}, and \textit{number of seats}, course information including \textit{course-id} (we have intentionally obfuscated the actual names of courses), \textit{course start-time}, and \textit{course end-time}, sensor measurements including \textit{count-in}, \textit{count-out}, and computed number of attendance (i.e. \textit{occupancy}). Note that count-in and count-out are available for the entire day (even during times with no lectures scheduled), whereas occupancy is available only when a course is scheduled.

Additionally, we release class attendance dataset \cite{tara:data2018} which will be used for prediction in \S\ref{sec:pred}. We derive class attendance by taking the maximum value of occupancy count in the room during the period when the class is operational.

\subsection{Data Visualisation}

\textbf{Tool:} To provide an intuitive user interface (UI) for real-time occupancy monitoring, we developed a web application using R Shiny -- our tool is available at \cite{webUI2018}. The tool allows the user to view the attendance pattern of a course (by choosing from the course dashboard tab), as well as the utilisation rate (number of attendees divided by the total number seats available for each classroom) for different time-slots. 

\begin{figure}[t]
	\centering
	\includegraphics[width=0.7\textwidth]{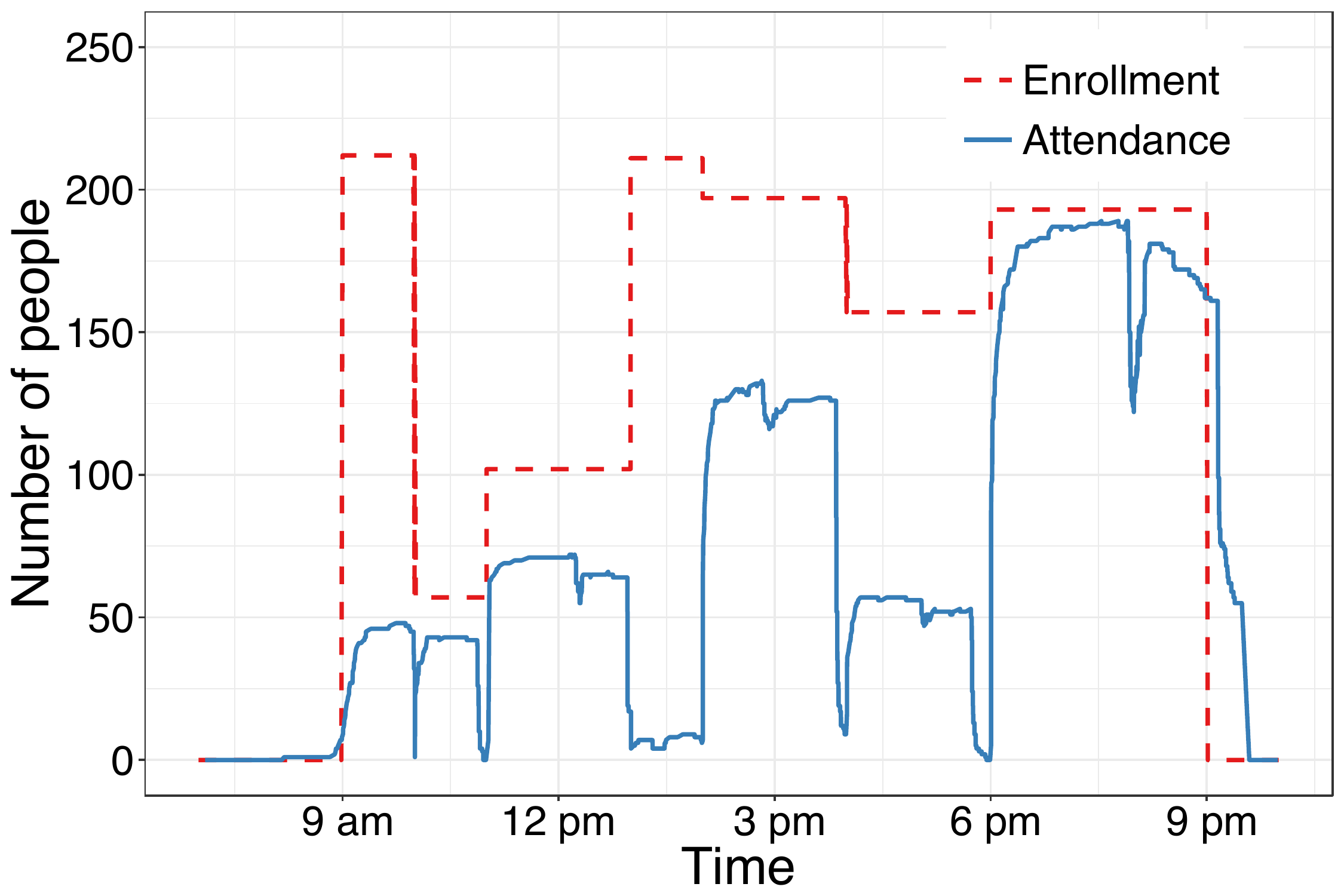}
	\caption{Occupancy pattern of a classroom on 16 August 2017.}\label{fig:room-occupancy}
\end{figure}

\textbf{Insights:} Our UI provides some interesting insights into attendance patterns.
Fig.~\ref{fig:room-occupancy} shows our UI output for an occupancy pattern of a selected room (CLB8) on a selected day, comparing the number of attendees (red line) and the associated enrolments (blue line) for 7 courses scheduled between 9am-9pm. From the plot, attendance is seen to vary widely across courses, in the range of 10\% to over 90\% of the enrolments. Interestingly, we observe that the lecture scheduled between 1pm-2pm has an enrolment of 211 but close to zero attendance; this indicates the cancellation of lecture which has led to a wastage of room spaces on the day. The visibility of room occupancy monitoring allows us to quantify space utilisation that is otherwise largely unknown to facility managers.

\begin{figure}[t]
	\centering
	\includegraphics[width=.7\textwidth]{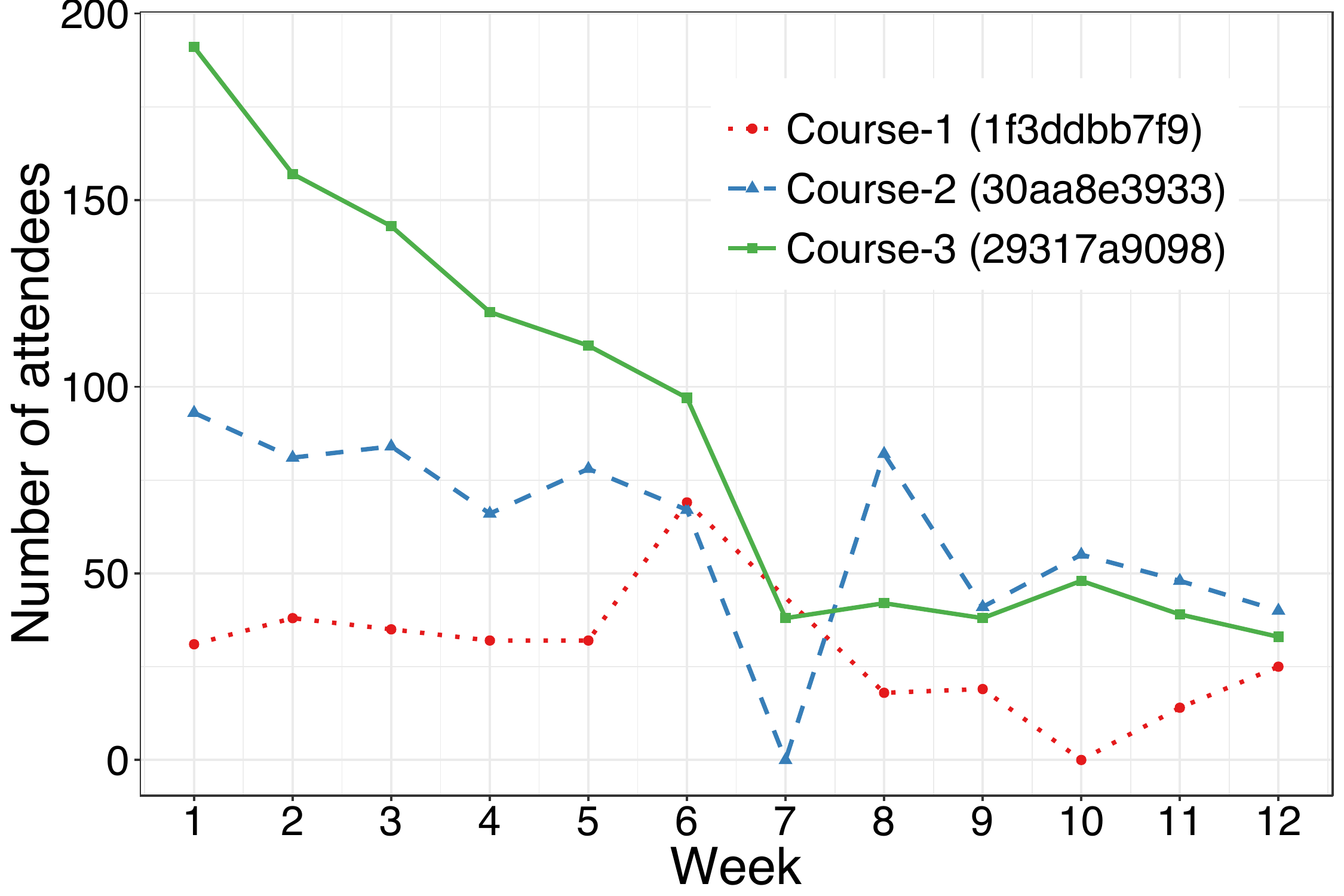}
	\caption{Attendance pattern of three courses across weeks.}\label{fig:course-pattern}
\end{figure}

Our visualisation tool also provides visibility into attendance pattern of all courses scheduled in the classrooms where sensors were installed. Figure \ref{fig:course-pattern} shows an example of attendance patterns for 3 selected courses (we have obfuscated course names) across the whole semester from week 1 to week 12. We can observe some interesting trends such as a general decline in attendance over week, represented in blue line; class cancellation on week 7, represented in green line; and mid session class examination during week 6 (red line), which has been verified by looking up the course web page. 

\begin{figure}[t]
	\centering
	\includegraphics[width=.8\textwidth]{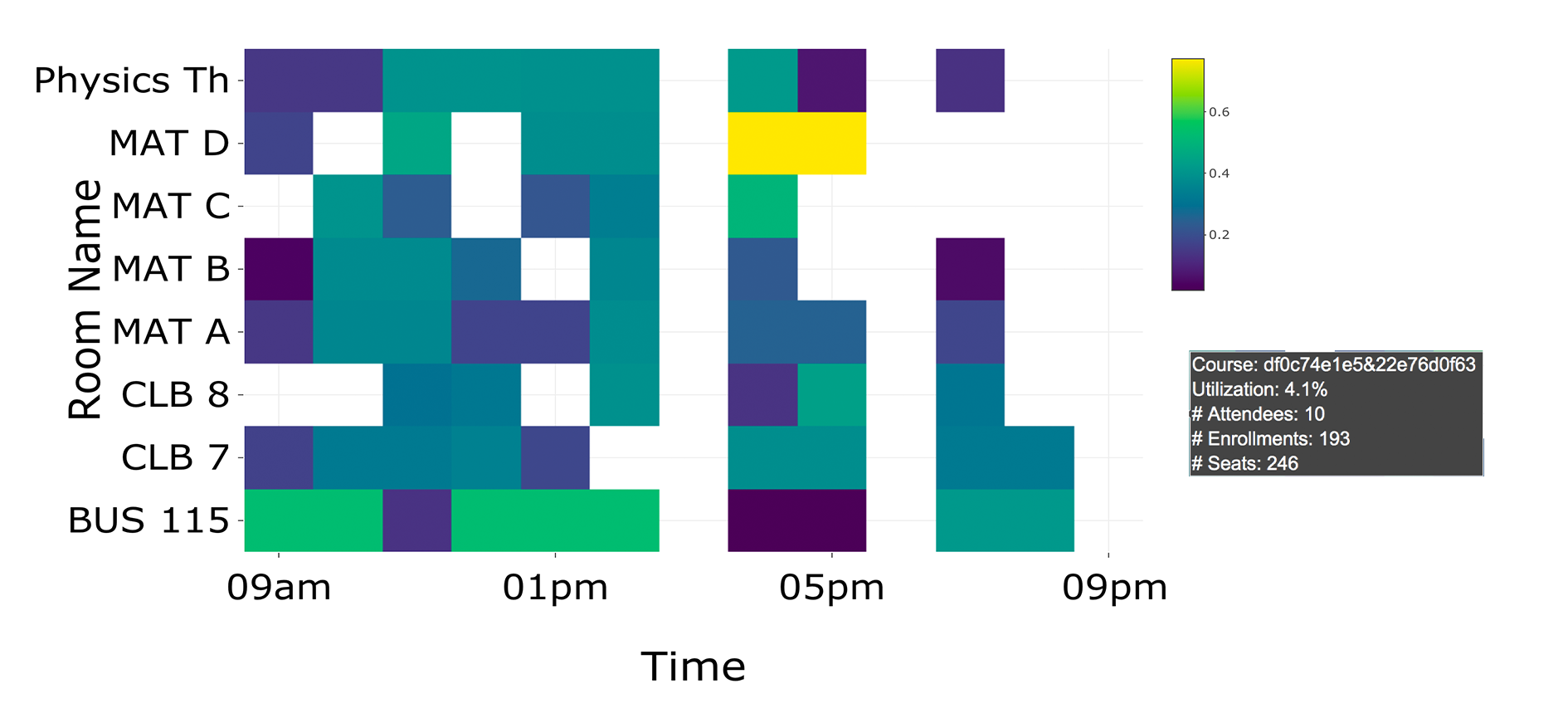}
	\caption{Heat-map of classrooms occupancy on a chosen day.}\label{fig:heat-map}
\end{figure}

Furthermore, the tool allows us to generate a room utilisation heat-map on a chosen day, as shown in Figure \ref{fig:heat-map}. Bright yellow cells represent time slots where rooms are being mostly occupied with utilisation rate approaching 1, whereas color scale towards dark blue represents time slots when rooms are under utilised. On the web interface, hovering over a cell shows further details on the usage of classroom and the scheduled course. For instance, we can see from the plot that course {\fontsize{10}{48}\usefont{OT1}{lmtt}{b}{n}\noindent df0c74e1e5} has an enrollment number of 193 but only 10 students attending the lecture, leading to a poor utilisation of $4.1$\%. The interface allows campus managers to track classroom utilisation, with a view towards more
optimal allocation, as described in \S\ref{sec:optim}.

\section{Prediction of Classroom Attendance }\label{sec:pred}

In previous section, we observed that the falling attendance pattern results in underutilisation of classrooms. In order to improve the overall utilisation, universities may want to dynamically re-allocate classrooms in advance based on attendance -- if attendance is much lower than the number of enrolments of a class then a smaller room may potentially be allocated to it, thus saving cost.
{We note that while our system is typically useful to obtain real-time data, room scheduling is based on the predicted attendance from historical data and should be decided several weeks prior to actual classes.} 
This entails a fairly accurate prediction of attendance for all classes operated on campus. It is important to note that underestimated prediction may lead to class overflow (\ie significant discomfort for students), and
overestimation would lead to wasted capacity and thus not achieving optimal cost reductions.


In this section, we compare three learning-based algorithms in predicting class attendance, each using two different functions of regression. We train models using historical labelled dataset from semester 2, 2017 and test the models with attendance dataset from semester 2, 2018. 
Note that it is infeasible to perform spot measurement at scale to collect ground-truth data needed for training of our prediction models. We, therefore, use course-based attendance count (which is deduced from data measured by sensors) to generate supervised learning models. The error in sensor measurement and thereafter in prediction, will be considered in \S\ref{sec:optim} when the predicted attendance is used for dynamic allocation of classrooms.



\subsection{Attributes Impacting Attendance}
There are several factors that can influence class attendance in universities, especially student motivations, quality of teaching, and characteristics of class lectures \cite{devadoss:attendance}. However it is infeasible to measure students motivation without conducting extensive surveys of a large population of students. Further, quality of teaching largely depends on the course lecturer, which can vary from semester to semester. We instead consider attributes related to individual course (\eg engineering faculty, undergraduate, tutorial) and temporal aspects (\eg week 3, Monday) which can be readily obtained.

In order to generalise our prediction model, we do not use specific course attributes such as course code or course instructor as inputs to our model. This allows us to perform prediction for courses for which no training data is available. 
Note that since our data collection started from semester 2 in 2017, there are insufficient data samples that span across multiple ``semester'' and ``year'' combinations. Thus, we do not include these attributes for our prediction model in this paper -- even though they may have impacts on class attendance. 

In summary, the following attributes are used in our prediction models that are specifically used for class attendance prediction:

\newcommand\litem[1]{\item{\itshape #1:}}
\begin{itemize}[label=\textbullet]
	\litem{\textbf{class type}} 
	type of the class, \eg lecture, lab, and tutorial.
	\litem{\textbf{faculty}} 
	faculty the course belongs to, \eg engineering, medicine, and science.
	\litem{\textbf{school}} 
	school the course belongs to (\eg Material Science).
	\litem{\textbf{enrollment}} 
	number of students enrolled in the course.
	\litem{\textbf{course duration}} 
	duration of the course (in hours).
	\litem{\textbf{degree}} 
	degree of study (\eg undergraduate, postgraduate). 
	\litem{\textbf{course status}} 
	enrollment status of the course (\eg open, full).
	\litem{\textbf{joint}} 
	a binary indicating if the course is combined with other courses.
	\litem{\textbf{week}} 
	week of semester (\eg week 3).
	\litem{\textbf{day}} 
	day of week (\eg Monday).
	\litem{\textbf{time-of-day}} 
	categorical value of time the class begins (\eg morning: 9am-12pm, afternoon: 12pm-3pm, evening: 3pm-6pm, and night: 6pm-9pm).
\end{itemize}

For the output of our prediction model,  we use normalised attendance  which is the ratio of maximum classroom occupancy (from course-based method in \S\ref{sect:attd-processig}) to the enrolment count. Hence the output varies between 0 (no student attended, or class cancellation) and 1 (all enrolled students attended).

Prior to generating and testing a model, we cleansed our data instances. This process involved removing classes that have no attendance (or cancelled class), removing classes with excessive overflow (\ie normalised attendance more than 1.5) due to probably over-counting, and capping the normalised attendance to 1.
Our attendance dataset with corresponding attributes for both training (\ie semester 2, 2017) and testing (\ie semester 2, 2018) is openly available for download \cite{tara:data2018}.
Each row in a CSV file represents a class comprising all attributes described above along with the actual attendance number obtained from the occupancy sensors data (\ie the attendance field is not normalised in our released dataset).

\subsection{Prediction Modelling}\label{section:modelling}
We choose supervised machine learning algorithms to perform attendance prediction given our labelled dataset from 2017. We considered three common regression learning algorithms including multiple linear regression, random forest, and support vector regression.
For each of these algorithms we apply two regression functions namely ordinary least square and quantile regression. The models are trained using caret package \cite{kuhn:caret} in R \cite{R}.

We first explain the algorithms used to build our prediction models: 



\begin{table}[!t]
	\begin{center}
		\caption{Summary of datasets for training and testing.}
		\label{table:summary_data}
		\vspace{-2mm}
		\begin{tabular}{ l  lll }
			\toprule 
			\textbf{Dataset} & \textbf{Description} & \textbf{Sample Size}\\
			\midrule
			DS1 & Sem2, 2017 - train set & 1497       \\
			DS2 & Sem2, 2017 - test set & 639  \\
			DS3 & Sem2, 2018 - test set & 940    \\
			\bottomrule
		\end{tabular}
		\vspace{-4mm}
	\end{center}
\end{table}

\textbf{Multiple Regression}:
Multiple linear regression (MLR) is one of the simplest prediction algorithms. It is the most common form of linear regression where the value of a variable is predicted based on the value of two or more attributes. The algorithm finds the best fit for the training data by minimising the sum of the squares of residues to obtain the resulting model. 


\textbf{Random Forest}:
Random forests (RF) is an ensemble learning method that can be used for both classification and regression problems. During training, multiple decisions trees are created from the training set, where a random selection of features is used to split each node of a tree. The final prediction result is obtained from majority voting, where mean is used for regression trees \cite{Breiman:rf}.
	


\textbf{Support Vector Regression}:
Support vector regression (SVR) applies the same principles as support vector machine (SVM) to the data but for a regression problem. The algorithm involves transforming the training data into a higher dimensional feature space, where linear regression is performed with a tolerance margin provided by the boundary lines \cite{drucker:SVR}. In our model, radial basis kernel is used to map data to higher dimensions.


Machine learning algorithms attempt to minimise a loss function on the training data as part of their modelling process. Commonly used loss functions for regression models, \ie mean square error (MSE) and mean absolute error (MAE), are symmetric. They treat error of under-prediction and over-prediction equally. This may not be desirable for class attendance estimation as under-prediction (\ie class allocated to a room with capacity lower than actual attendance) is more harmful than over-prediction since it leads to students' discomfort.

To address unequal treatment of errors (\ie differentiating under-prediction and over-prediction), we employ quantile loss function \cite{koenker2001:quantile} for generating prediction models. Quantile loss function is defined by:

\begin{equation} \label{eq:quantile_loss}
L(y, \hat{y} ) = \sum_{i=1}^{n} {(y_{i} - \hat{y_{i}}}) (\tau - \mathbbm{1}_{y_{i} - \hat{y_{i}} < 0})
\end{equation}

where $y$ is the actual attendance and $\hat{y}$  is the predicted attendance, $\mathbbm{1}$ is an indicator function that equals to 1 in case of over-prediction and 0 otherwise, quantile $\tau$ is a tuning parameter that takes a value from 0 to 1 allowing us to ascribe varying emphasis on over/under-prediction (\ie $\tau < 0.5$ causes over-prediction to impose higher cost than under-prediction while $\tau > 0.5$ increases the cost of under-prediction). Our choice of quantile $\tau$ will be explained next.


\subsection{Performance Evaluation of Prediction Models}\label{sect:classroom-pred-performance}
We build our prediction models using the three learning algorithms mentioned earlier and
consider both ordinary regression and quantile regression as loss functions for each, thus a total of 6 models are evaluated. We use 10-fold cross validation to select the best tuning parameters that yield the lowest error for each model. This subsection explains the dataset we used for model training and testing. We also compare the prediction results of the six models using a range of metrics. 


\begin{table}
	\begin{center}
		\caption{Summary of courses per faculty.}
		\label{table:faculty_count}
		\vspace{-2mm}		
		\begin{tabular}{l l c l }
			\toprule 
			\textbf{Faculty} &  \textbf{Count} \\
			\midrule
			Faculty of Science & 698       \\
			UNSW Business School & 662  \\
			Faculty of Engineering & 436    \\
			Faculty of Arts \& Social Science & 203    \\
			Faculty of Medicine& 97    \\
			Faculty of Built Environment & 22    \\
			Faculty of Engineering \& Science& 18    \\
			\bottomrule
		\end{tabular}
		\vspace{-2mm}
	\end{center}
\end{table}

\begin{figure}
	\centering
	\includegraphics[width=0.7\textwidth]{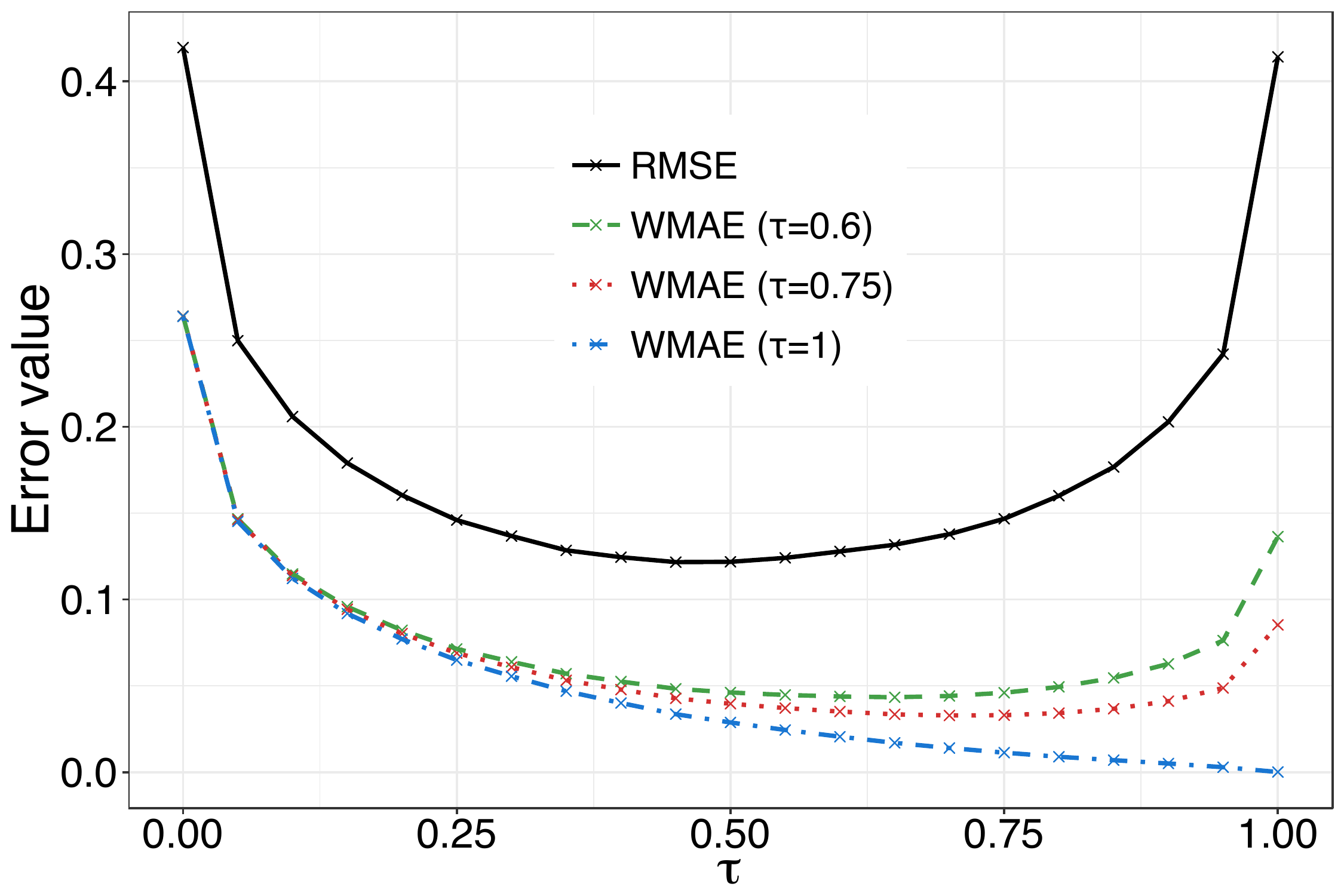}
		\caption{Impact of quantile parameter on prediction error.}
		\label{fig:compare-tau}
\end{figure}

We partitioned the attendance data from entire semester 2 of 2017 into a train set and test set at a ratio of 70\% (\ie DS1) and 30\% (\ie DS2) respectively. Additionally, we obtained attendance data for the first 6 weeks from semester 2 of 2018 (\ie DS3) and used it as another test set to evaluate the performance of the models in predicting future attendance.
A summary of our datasets along with their sample sizes is shown in Table \ref{table:summary_data}. We note that only data from semester 2  (and not semester 1) was used for the prediction, eliminating any biases that could arise from the effect of semester on our attendance prediction.
Additionally, a summary of courses per various faculties is shown in Table \ref{table:faculty_count}.
 It is also important to note that not all courses are run every semester and the number of enrolment for courses that run every semester may vary vastly (for example the course ``Computer Networks and Applications'' in our university may have around 300 students in semester 1, but over 600 students in semester 2).

Both mean absolute error (MAE) and root mean squared error (RMSE) measure the average magnitude of the errors in a set of predictions. RMSE is most useful when large errors are particularly undesirable since errors are squared, giving a relatively higher weight to larger errors, before being averaged. 
Additionally, we employ the average weighted absolute error (WMAE) \cite{haupt2011:cross} to evaluate the  performance of prediction when under-prediction is considered more costly than over-prediction. Our custom WMAE is defined as the mean of the quantile loss function and is written as:

\begin{equation} \label{eq:wmae}
\frac{1}{n} \sum_{i=1}^{n} {(y_{i} - \hat{y_{i}}}) (\tau - \mathbbm{1}_{y_{i} - \hat{y_{i}} < 0})
\end{equation}



As noted in \S\ref{section:modelling}, we consider a quantile loss function to differentiate under-prediction and over-prediction. We now build quantile regression models by varying the value of quantile parameter $\tau$ in order to quantify its impact on performance metrics. 
Fig.~\ref{fig:compare-tau} shows error values of RMSE and WMAE versus the quantile parameter for a random-forest-based model as an example. Note that for WMAE, the $\tau$ value is fixed to 0.60, 0.75, and 1.00 as shown by dashed green, dotted red, and dashed dotted blue lines respectively.
As expected, it is seen that RMSE values, shown by solid black lines, produce a symmetrical curve with a central dividing line at the second quantile (\ie $\tau = 0.50$), as this metric treats under-prediction and over-prediction equally. 
On the other hand, we observe asymmetric curves for WMAE that have the minimum error values at its corresponding $\tau$ (\eg WMAE with $\tau=0.75$, shown by dotted red lines,  gets minimum at 0.75 quantile). 

As mentioned earlier, the quantile parameter $\tau$ with a value greater than 0.50 will penalise under-prediction more than over-prediction. We note that for $\tau=1$, only under-estimation is penalised in the prediction process with no penalty for over-estimation. This may provide a very safe margin for campus managers during dynamic classroom allocation. But, this choice yields a high RMSE of over $0.4$ which makes it unattractive. On the other hand, $\tau=0.5$ gives the minimum RMSE but it weights over-prediction and under-prediction equally which does not match our requirement. Hence, we choose a quantile value of $0.75$, which still results in a reasonable value of RMSE (\ie below 0.2) while giving more weight to under-prediction. This value is used as the quantile parameter for both loss function and WMAE performance metric.

\textbf{Attendance Prediction within Semester:}
We first apply our prediction models to a testing set containing attendance data of classes from the same semester/year (\ie semester 2, 2017). For this, we use DS1 to train the model and DS2 for testing. The performance results of our six models are shown in Table \ref{table:pred_results}.
In terms of RMSE, we can see that RF and SVR algorithms achieve the best predictive performance in both cross-validation and testing, with values within a range of $0.120$ and $0.135$. Linear models (both MLR and quantile MLR), on the other hand, yield the highest errors of over $0.16$ RMSE on the validation set. This suggests that attendance and course attributes are not linearly correlated and such non-linear relationship is better modelled by RF and SVR.
By imposing higher cost to under-prediction, we can see that quantile regression methods give better performance in terms of WMAE when compared to their default regression counterparts. RF achieves the best performance with WMAE of $0.033$ during cross validation and $0.034$ during testing.

\begin{table*}[!t]
	\centering
		\caption{Performance of  prediction models.}
		\label{table:pred_results}
		\vspace{-2mm}
		
		\begin{adjustbox}{width=\textwidth}
		\begin{tabular}{l cl c l clclclclcl}
			\toprule 
			& \multicolumn{3}{c}{2017 train set DS1 (cross-validation)} && \multicolumn{3}{c}{2017 test set DS2 (testing)} && \multicolumn{3}{c}{2018 test set DS3 (testing)} \\
			\cmidrule{2-4} \cmidrule{6-8} \cmidrule{10-12}
			Models                     & RMSE        & MAE         & WMAE       & & RMSE        & MAE         & WMAE      & & RMSE        & MAE        &  WMAE   \\
			\midrule
			\textbf{Multiple Linear Regression (MLR)} & $0.163$       & $0.123$       & $0.060$       && $0.149$       & $0.118$       & $0.060$      && $0.193$  & $0.145$      & $0.063$      \\
			\textbf{Random Forest (RF)}            & $\color{blue}{0.120}$       & $0.086$       & $0.043$    && $\color{blue}{0.121}$       &  $0.089$       & $0.044$      && $0.157$    & $0.122$   & $0.051$      \\
			\textbf{Support Vector Regression (SVR)}  & $\color{blue}{0.135}$    & $0.094$       & $0.041$       && $\color{blue}{0.125}$    & $0.086$   & $0.042$      && $0.193$    & $0.148$    & $0.062$     \\

			\textbf{Quantile Linear Regression}  & $0.188$    & $0.142$       & $0.048$       && $0.179$    & $0.135$   & $0.045$      && $0.240$    & $0.183$    & $0.059$     \\ 
			
			\textbf{Quantile Regression Forests}  & $0.147$    & $0.102$       & ${\color{blue}0.033}$       && $0.154$    & $0.108$   & $\color{blue}{0.034}$      && $0.217$    & $0.167$    & $\textbf{\color{green!55!blue}0.047}$     \\ 
			\textbf{Quantile Regression using SVM}  & $0.141$    & $0.095$       & $0.035$       && $0.147$    & $0.100$   & $0.036$      && $0.216$    & $0.170$    & $0.061$     \\ 
			
			\bottomrule
		\end{tabular}
		\end{adjustbox}
\end{table*}



\begin{figure}[!t]
	\centering
	\includegraphics[width=0.7\textwidth]{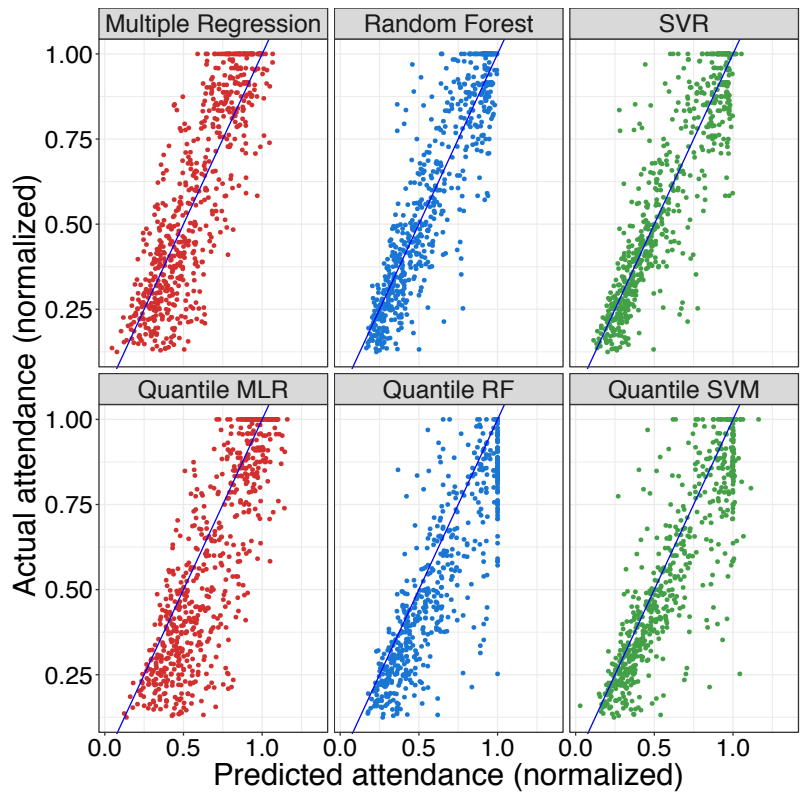}
	\caption{Predicted vs. actual attendance for test set in 2017 (\ie DS2).}\label{fig:pred-actual-test}
\end{figure}

To visualise the performance of the six models, we show in Fig.~\ref{fig:pred-actual-test} scatter plots of the predicted normalised attendance (x-axis) versus the actual attendance (y-axis) with a blue fitted line of $y=x$ (\ie predictions are expected to match the actual output) -- points on the left side of this line represent under-prediction, and right side points represent over-prediction. We also note that the predicted attendance can sometimes go beyond 1.  From the plot we can obviously see that points are more dispersed for linear regression models (red dots) compared to SVR (green dots) and RF (blue dots), which matches the results of RMSE metrics discussed earlier. Furthermore, we  observe that quantile regressions (shown by bottom plots in Fig.~\ref{fig:pred-actual-test}) result in more over-predictions than under-predictions, which is an expected output.

We note that the prediction result is fairly consistent in cross-validation and testing, indicating that a good prediction can be obtained if part of the data from one semester/year are used to predict the attendance for the remainder of the same semester. A more practical implementation that involves using historical data to predict future year/semester attendance is discussed next.

\begin{figure}[!t]
	\centering
	\includegraphics[width=0.7\textwidth]{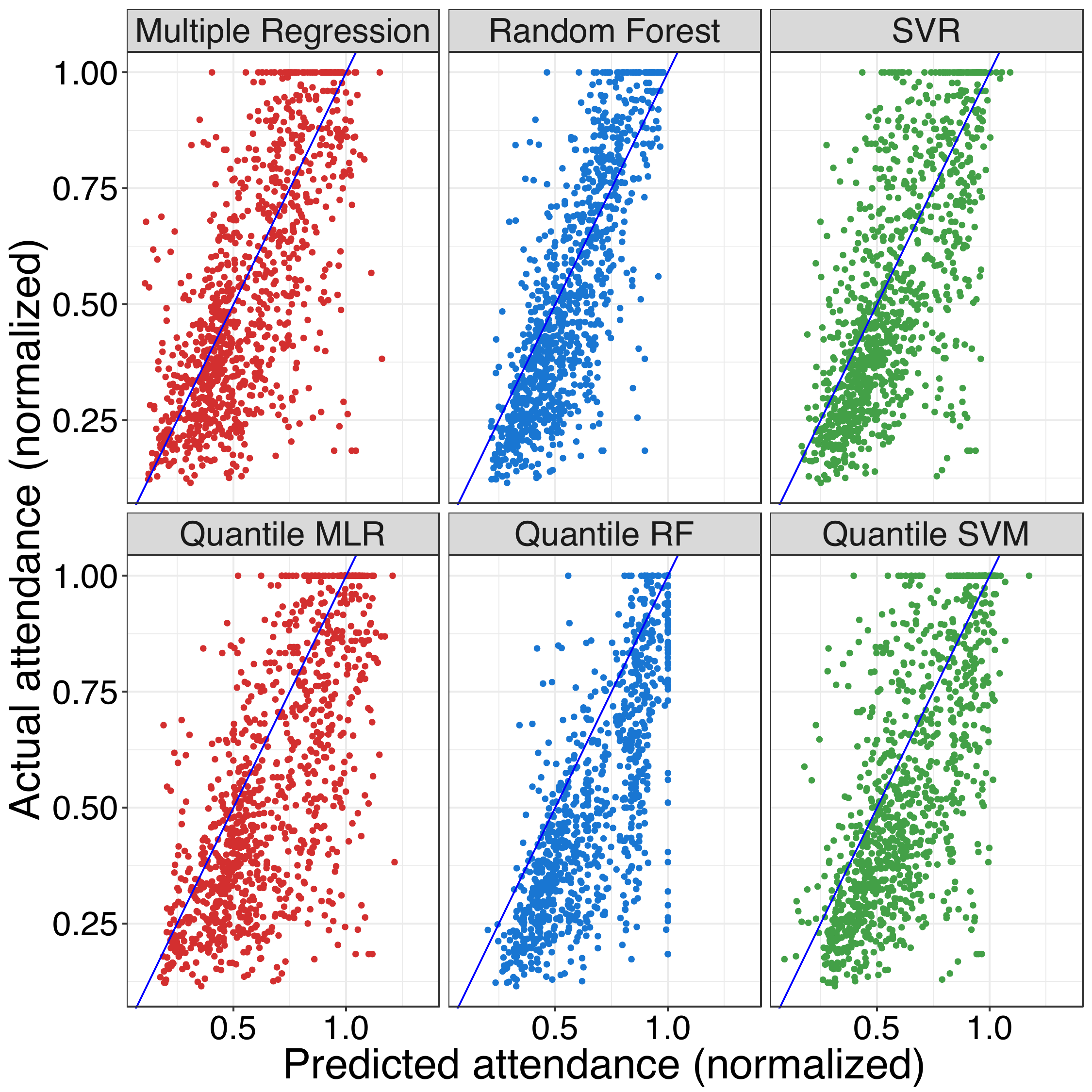}
	
		\caption{Predicted vs. actual attendance for test set in 2018 (\ie DS3).}
		\label{fig:pred-actual-2018}
\end{figure}

\textbf{Attendance Prediction for Future Semester:}
We now predict the attendance for semester 2, 2018 (\ie DS3) while the model was trained by the data from the same semester but in 2017 (\ie DS1). Results of all the models are shown by the last column in Table~\ref{table:pred_results}. 
In terms of RMSE, it is seen that RF achieves the best performance with a value of $0.157$ -- slightly higher than the result from testing DS2. Looking at WMAE, quantile RF give the most accurate result with the error of $0.047$ (highlighted in green).


From the predictions vs. actual attendance plot in Fig.~\ref{fig:pred-actual-2018}, as expected, we observe wider dispersions from the fitted line compared to the scatter plots in Fig.~\ref{fig:pred-actual-test} for the within-semester attendance prediction. RF and quantile RF (blue dots) show the lowest deviation from the fitted lines compared to predictions from linear regression models (red dots) and SVR models (green dots).

The lower performance in prediction of future-semester attendance is caused by several factors when changing year/semester -- for example, variations in enrolments from year to year due to popularity of courses, changes made to the programs  (\eg a certain course may no longer be considered a core course while another course may become a core), changes in lecturing staff, changes in students lifestyle (motivation for attending classes), and increases in availability (or better quality) of online lecture recordings. Despite all these factors, we believe that the future-semester prediction results obtained from quantile RF (with WMAE of $0.047$ and RMSE of $0.217$) is acceptable given the non-trivial variations mentioned above.


\section{Optimal Allocation of Classrooms}\label{sec:optim}
Our empirical results from Section \ref{sec:visual} highlight the significant variations in occupancy on a weekly basis which leads to under-utilisation of classrooms. This presents an opportunity for campus managers to employ a dynamic allocation scheme to save cost. In this section, we develop an optimisation formulation to determine the potential cost savings. Our formulation uses prospective attendance counts computed from our prediction model (discussed in \S\ref{sec:pred}). A practical implementation could develop a dynamic schedule for a course using our optimisation algorithm, while leaving some margin for error arising from the prediction algorithm. 

We first formulate a problem with an objective function and related constraints. We then apply the optimisation to our dataset and quantify the cost savings versus students discomfort (due to room overflows) when adopting dynamic classroom allocation in comparison to the traditional fixed enrolment-based allocation. 

\subsection{Problem Formulation}
First, let there be $L$ courses, each with its own start time, duration, and the attendance number. Each course is allocated to one of $R$ classrooms available on campus, and each room has a cost associated with it (proportional to its capacity). The cost of room $j$ is denoted by $C_{j}$ where $1 \leq j \leq R$.   
We consider our optimisation problem over a one-day window with a total of $S$ slots -- each slot accounts for an hour period.


We define our variables to be $x_{i, s}$ that indicates the room to which the course $i$ is allocated during timeslot $s$, where $1 \leq i \leq L$ and $1 \leq s \leq S$:

\begin{equation} \label{eq:input_varX}
x_{i,s} \in \{ 0, 1, 2,....., R \}
\end{equation}

Note that  $x_{i, s} = 0$ indicates that course $i$ is not allocated to any room during $s$.
From variables $x_{i, s}$, we derive another variable denoted by $y_{i, j}$ which is a binary value indicating whether or not course $i$ is allocated to room $j$ during any given slot:

\newcommand{\twopartdef}[4]
{
	\left\{
		\begin{array}{ll}
			#1 & \mbox{if\  } #2 \\
			#3 & \mbox{\  } #4
		\end{array}
	\right.
}

\begin{equation} \label{eq:input_varY}
y_{i,j} = \twopartdef { 1 } {\  course~{i}~allocated~to~room~{j} } { 0 } {\ otherwise }
\end{equation}

Furthermore, room-slot allocation variables, $z_{j,s}$ indicate the number of courses that are allocated to room $j$ at timeslot $s$:

\begin{equation} \label{eq:room-slot-var}
z_{j, s} \in \{ 0, 1, 2,....., L \}
\end{equation}

Therefore, the total cost of allocation for a given day is specified as:

\begin{equation} \label{eq:cost_func}
J= \sum_{j=1}^{R} \bigg\{{C_{j} \sum_{s=1}^{S} z_{j,s} }\bigg\}
\end{equation}

where $C_{j}$ is the capacity of room $j$. 

Our aim is to minimise the total cost $J$ in Eq. (\ref{eq:cost_func}). Note that allocation of a course to a room incurs a full cost of that room, and an unallocated room incurs no cost. Thus, the total cost of the allocation is the sum of capacities of rooms that are used across all timeslots of the day.


We have four sets of constraints in our optimisation problem listed as follows.
\begin{itemize}
\item \textit{Course Constraint}: Each course can only be allocated to one room during a timeslot. These $L$ constraints are captured by:

\begin{equation} \label{eq:course_constraint}
\sum_{j=1}^{R} {y_{i,j}} = 1\  \forall\ i
\end{equation}

\item \textit{Room Constraint}: A room cannot be occupied by more than one course at a time. Hence, Eq. (\ref{eq:room-slot-var}) can be expressed as:

\begin{equation} \label{eq:room_constraint}
z_{j, s} \in \{ 0, 1 \}
\end{equation}

\item \textit{Capacity Constraint}: Course $i$ with attendance number of $o_{i}$ needs to be assigned to room $j$ that has sufficient capacity $C_{j}$ accommodating students attended. These constraints are captured by:

\begin{equation} \label{eq:enrol_contraint}
o_{i}  \leq \sum_{j=1}^{R}{C_{j} y_{i,j}}\	\forall\	i
\end{equation}

\item \textit{Schedule Constraint}: Each course has a fixed schedule (\eg course 1 is scheduled from slot 3 to slot 5, equivalent to 11 am to 1pm), thus room allocation can not change over these consecutive slots (\ie slots 3 to 5  for this example).

\end{itemize}

\subsection{Optimisation Algorithm}\label{sect:opt-algo}

To solve the optimisation problem, we employed constraint programming (CP) algorithm using Google Optimisation Tools for Python \cite{google:ORTOOL} as a solver. CP is widely used for solving combinatorial problem, where search space with possible values of variables (domains) are defined and constraints are declaratively stated in order to limit all possible assignments to a set of feasible solutions. There are two phases involved in CP: (a) a propagation phase where infeasible regions are methodically removed from the search space, (b) a search phase where the browsing of the search space is performed using a complete search algorithm such as backtracking, or an incomplete search such as local search algorithm \cite{rossi:handbook}. 
CP can be used to solve for all feasible solutions, or an near optimal solution. Due to time complexity, we choose the latter option.

\begin{figure}[t]
	\centering
	\includegraphics[width=0.7\textwidth]{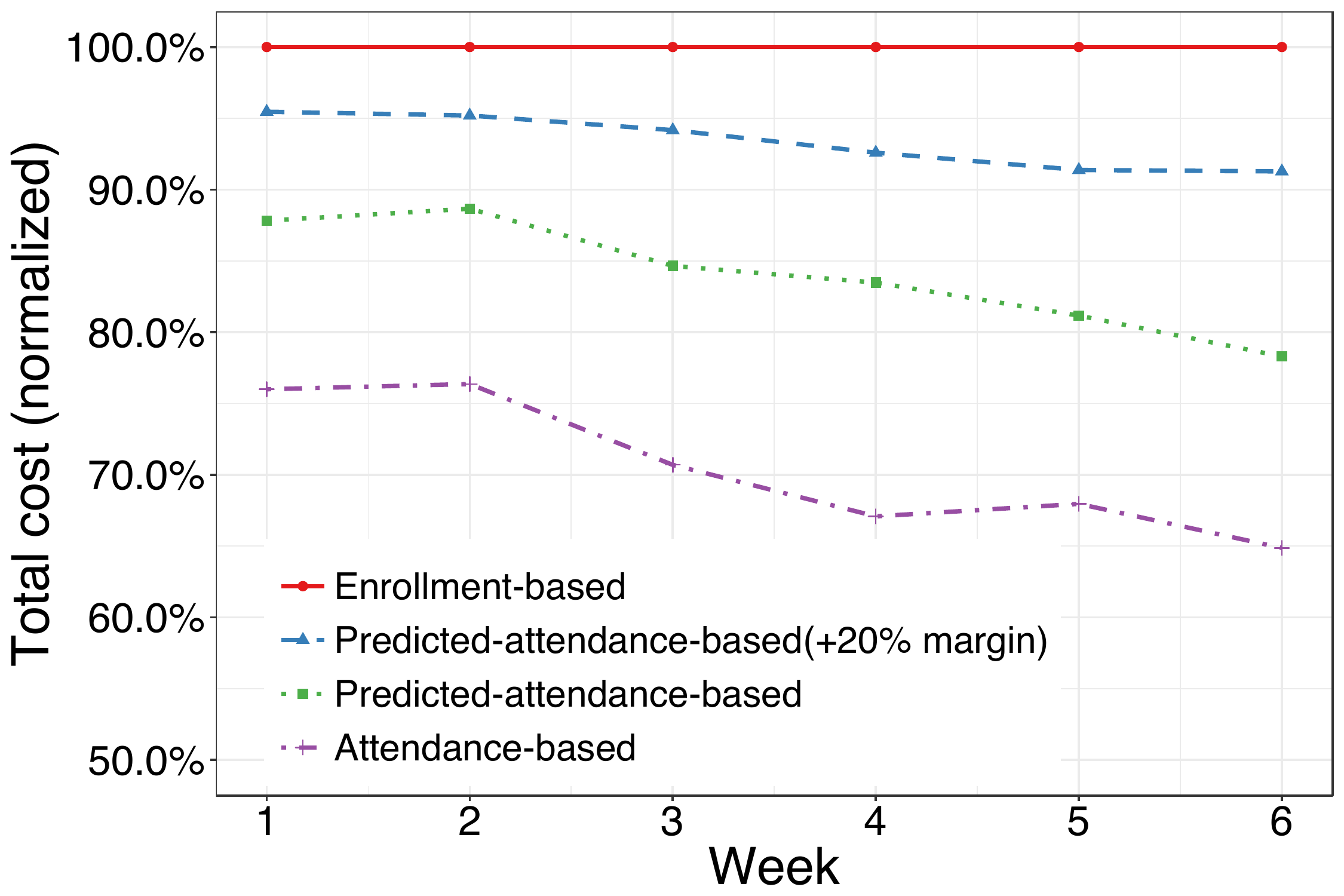}
	\vspace{-3mm}
	\caption{Allocation cost across weeks in 2018.}\label{fig:cost-by-week}
\end{figure}

\begin{figure}[!t]
	\centering
	\includegraphics[width=0.7\textwidth]{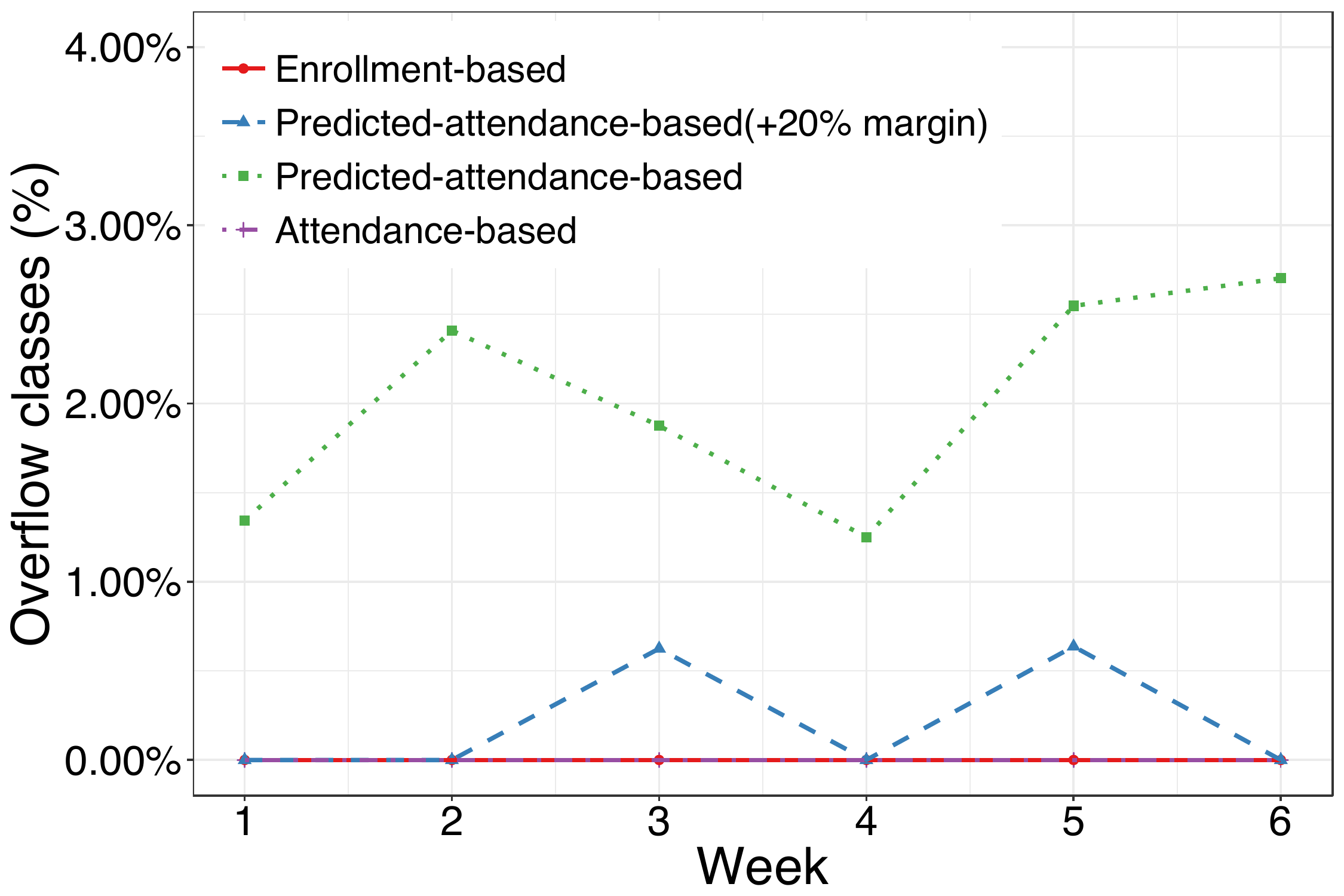}
	\vspace{-3mm}
	\caption{Students discomfort across weeks in 2018.}\label{fig:overflow-by-week}
\end{figure}


For our optimisation problem, we used  the real data obtained from semester 2, 2018 (\ie DS3) consisted of 748 courses (as mentioned in \S\ref{sect:classroom-pred-performance}) operating in the 9 classrooms. 
We assume that there are some spare rooms available for optimal allocation -- we use only one spare room of
100 seats capacity. There are 12 timeslots for each operation day, starting from 9 am (corresponding to timeslot 1) to 9 pm (corresponding to timeslot 12).
The optimisation was performed for a complete set of courses accounted for a period of 6 weeks.
We considered several scenarios, both enrolment-based and attendance-based. For the latter scenario, we use different attendance count including actual attendance (measured by our sensors), predicted attendance, and also predicted attendance with additional margin (\ie 5\%, 10\%, 15\%, 20\%, 25\%, and 30\%).  
For each round of our optimisation, we obtain the total cost of allocation. 
Note that for some cases when predicted attendance is used, the classroom may overflow due to under-prediction and cause students discomfort. We, therefore, measure the number of ``overflow classes'' in each round of allocation.

\subsection{Optimisation Results}
We run our optimisation algorithm and obtain the cost of allocation for each day. We then compute the weekly cost of allocation by adding daily costs across the week. 
We plot the weekly total cost in Fig.~\ref{fig:cost-by-week} -- total costs are normalised with respect to the enrolment-based scenario as a baseline (solid red line).
Unsurprisingly, the enrolment-based approach results a constant cost (\ie upper bound) as it tries to meet the fixed constraints every week. On the other hand, total cost obtained from various scenarios of attendance-based allocation  (dashed lines) falls gradually due to falling pattern of attendance for majority of the courses.
We note that even though allocation based on actual attendance yields the best result with lowest cost, it is not feasible in practice.
It is seen that the prediction-based allocation yields a cost slightly higher than when actual attendance is considered, but still it is beneficial to the campus by saving on average of $12$\% per week of operation (as shown by dotted green lines with no margin). Obviously, adding margin to the predicted attendance count would reduce the cost saving, for example 20\% margin gives a saving of about $5$\% per week.

\begin{figure}[!t]
	\centering
	\includegraphics[width=0.7\textwidth]{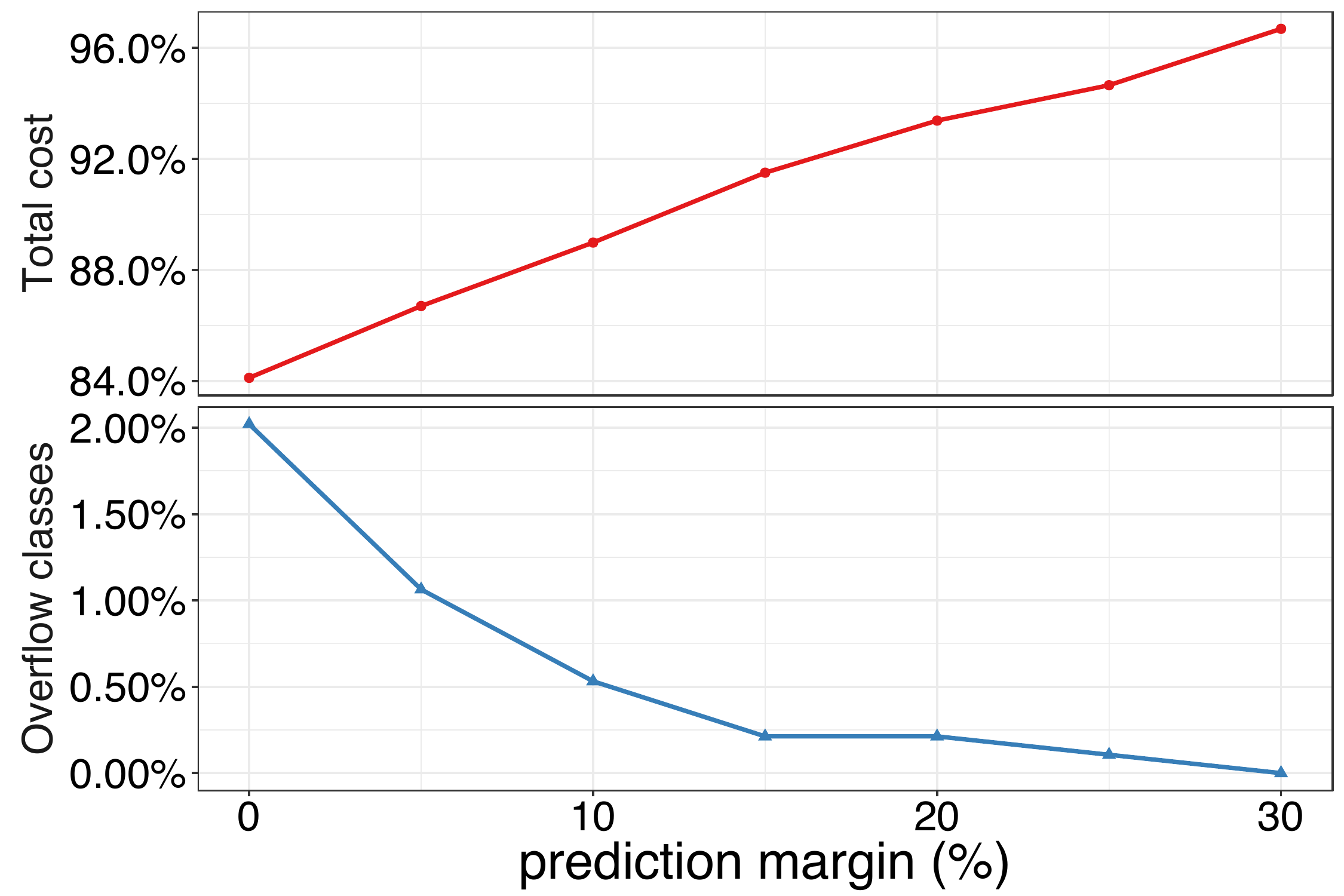}
		\caption{Allocation cost versus students discomfort.}\label{fig:cost-by-margin}
\end{figure}

In addition, we capture the number of overflow classes (\ie the room capacity is lower than the actual attendance of the class) as a proxy for the students experience (or discomfort) from dynamic allocation of classrooms. Fig.~\ref{fig:overflow-by-week} shows the fraction of overflow classes (out of all allocated classes) across weeks for various scenarios. As we expect, allocations based on enrolment and actual attendance count yield no overflow since rooms capacities are well provisioned in advance. Allocation based on predicted attendance, on the other hand, causes overflow between $1.25$\% to $2.70$\% of classes per week when no margin is considered (as shown by dotted green lines). The measure of overflow is reduced to less than $1$\% by applying a margin of 20\% (as shown by dashed blue lines).

Lastly, we focus on the impact of margin on our evaluation metrics. We plot in Fig.~\ref{fig:cost-by-margin} the total normalised cost (top) and the total overflow fraction of classes (bottom) for a duration of 6 weeks by varying the prediction margin from 0\% to 30\%. 
We can see that the normalised cost monotonically increases from $84$\% (with no margin) to $96$\% (with 30\% margin). Conversely, the overflow falls from $2$\% (accounted for 19 impacted classes) to $0$\% where no class is impacted.
This suggests that campus can benefit from at least $4$\% cost saving over 6 weeks of operation by employing a dynamic allocation of classrooms with no impact on student experience.
  

\section{Conclusion}\label{sec:con}
In this chapter we have outlined our efforts to address classroom under-utilisation in a real University campus arising from the gap between enrolment and attendance. We instrumented classrooms with IoT sensors to measure real-time usage, used AI to predict attendance, and performed optimal allocation of rooms to courses minimising space wastage. 
We undertook a lab evaluation of various commercial IoT sensors and compared them in terms of cost, ease of operation, and accuracy. We then deployed our sensors in 9 real classrooms of varying sizes across campus and collected data over two semesters covering over 250 courses, which we release to the public. Our data and visualisation reveal interesting insights into course attendance patterns and class utilisation measures. Based on this real data, we developed AI based methods to predict classroom attendance which fed into our optimisation algorithm for dynamic allocation of classes to classrooms based on predicted attendance rather than enrolments, and showed gains of 10\% in room costs with a very low risk of room overflows.


\chapter{Monitoring and Optimisation of Car park as a Resource}
	\minitoc

In the previous chapter we demonstrated the potential of IoT technologies in enhancing the efficiency of classrooms management at a university campus through applying the three pillars Smart Campus Framework. In this chapter, we apply the same framework to demonstrate IoT technology as an enabler for a more efficient space use in a parking domain. We show how IoT technology can provide an opportunity for many organisations with large on-premise parking spaces to rethink the use of their car parks by re-dimensioning spaces to accommodate new prevalence transportation paradigms such as car sharing, allowing them to better utilise this space, reduce energy footprint, and create new revenue stream.

Our first contribution describes experiences and challenges in measuring car park usage on the university campus and removing noise in the collected data. Our second contribution analyses data collected during 15 months and draws insights into usage patterns. Our third contribution employs machine learning algorithms to forecast future car park demand in terms of arrival and departure rates, with a mean absolute error of 4.58 cars per hour for a 5-day prediction horizon. Lastly, our fourth contribution develops an optimal method for partitioning car park space that aids campus managers in generating revenue from shared cars with minimal impact on private car users.



\section{Introduction}\label{sec:introduction}
\vspace{-3mm}

Universities worldwide are experiencing a surge in student enrolments \cite{BritishCouncil}, accompanied by an expansion in staff numbers, which together have contributed to an increase in demand for on-campus parking. 
Despite the increasing trend of private car usage to commute to campus \cite{dell2018:methodology}, as many as 10-45\% of available parking spaces are empty since they are distributed across a large campus area \cite{filipovitch2016:excessparking}.
This problem has also been observed at our campus in UNSW Sydney, where one of the multi-storey parking lots fill up by 10 am while the other often has availability.

Parking spaces can be thought of as perishable goods with sunk cost and an empty space at any time resembles an unsold item that can not be resold later \cite{pierce2015optimizing}. Hence, 
effective management is required to ensure their efficient use. The rapid advances in information technology in conjunction with data monetisation have improved the efficiency of parking management systems with the push towards adopting dynamic and data-driven policies. A good example is the adoption of a dynamic pricing scheme for public parking spaces by Municipal Transport Agency in San Francisco \cite{SFpark:dynamicpricing}, where parking rates are adjusted based on demand data, in order to maintain their target utilisation. Capturing fine-grained dynamics of parking usage over a long-term period is likely to offer far-reaching benefits. In particular, high-resolution data can be used to inform strategic decisions such as expansion of parking capacity, developing new parking facility, or partitioning spaces to accommodate new paradigms of car use \cite{litman2016parking}.



The medium-term future is likely to evolve around shared transport, leading to autonomous vehicles in the longer-term. This will not only enhance commuter experience, but also cut down the use of fuel, alleviate the number of cars on the road, and improve traffic congestion. Car sharing (offered by companies such as GoGet, ZipCar, Car2Go, etc.) is projected to grow at a rate of over 20\% between 2018 and 2024 \cite{gminsight:carsharing} and is becoming a more mainstream mode of transport. Accordingly, universities such as UNSW can leverage such trends to go green by encouraging car sharing schemes in order to reduce on-campus parking congestion and savings in infrastructure spending for new parking facilities to keep up with the growth of the campus population.

Many universities are moving towards this trend by partnering with car sharing companies to offer shared transport services to their campus community \cite{stasko2013carsharing}. Current schemes are predominantly based on static allocation where a fixed number of parking bays are reserved for car sharing vehicles, and only support round-trip transport services where vehicles are required to be returned to their dedicated based station. In recent years, there has been a rapid adoption of one-way car sharing model that provides a more flexible service to users by allowing them to leave shared vehicles at locations different from their pick-up point  \cite{shaheen2015one}. This recent trend will likely capture new shared transport users, hence motivating universities to adopt a more efficient dimensioning method for their parking infrastructure. Accordingly, existing static allocation of parking spaces can potentially be replaced with a dynamic scheme where spaces allocation changes dynamically based on predicted usage demand.




This chapter describes our experience in instrumenting an on-campus car park for real-time monitoring of space utilisation, and our developed novel framework for monetising the collected data by dynamically allocating parking spaces  to car-sharing and private car users. The work presented in this chapter has been published in \cite{sutjarittham2020monetizing,sutjarittham:carparkusage}. Our specific contributions are three-fold:

\begin{enumerate}
\item We first comprehensively analyse the car park usage data that spans over a period of 15 months, covering both teaching and non-teaching periods, and highlight interesting insights into car arrival and departure patterns.
\item We then apply machine learning models to forecast future car park usage demand and compare predictive performance of various learning algorithms across multiple forecast horizons, ranging from a day up to 10 weeks.
\item We develop a framework that assists estate manager in selecting an optimal allocation  of car spaces to car sharing vehicles. This involves development of a continuous-time non-homogeneous Markov model based on the predicted usage to simulate dynamic partitioning scheme and an optimisation model to assist campus managers in selecting an optimal allocation (\ie fraction of spaces allocated to car sharing vehicles). The framework aims to enhance parking space utilisation while avoiding situations where users are turned away due to lack of parking spots.
\end{enumerate}

The rest of this chapter is organised as follows: 
\S\ref{sec:datacollection} outlines our experiences in implementing an IoT car park monitoring system for real-time data collection while \S\ref{section:analytics} presents interesting insights obtained therein.  \S\ref{section:prediction} compares machine learning algorithms in forecasting multi-step ahead arrival and departure rate of cars. We show how the predictions are used to approximate the number of rejected users (given their respective allocation) using Markov modelling and an optimisation formulation that can aid car park dimensioning decisions in \S\ref{section:optimization}. The chapter is finally concluded in \S\ref{sec:con}.

\begin{figure*}[t!]
	\centering \includegraphics[width=0.9\textwidth]{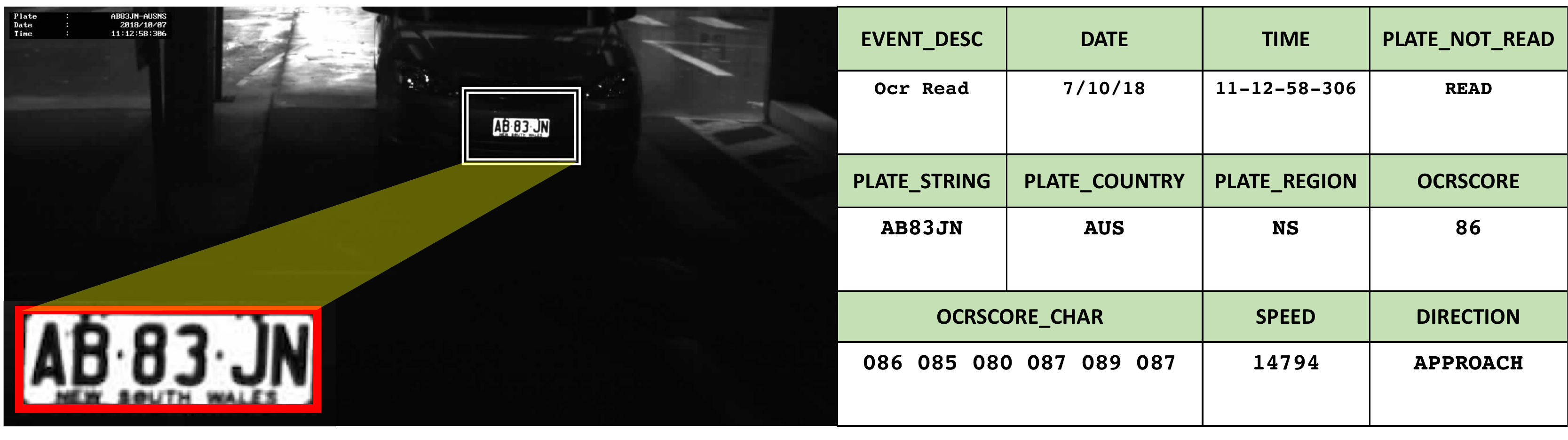}
	\caption{LPR camera outputs for author's car entering into the campus car park.}
	\label{fig:license_plate}
\end{figure*} 

\section{Data Collection and Cleansing}\label{sec:datacollection}

In this section, we first outline our experience with license plate recognition (LPR) technology to measure car park usage. 
Since license plate information is private and confidential, we obtained appropriate ethics clearances for this study (UNSW Human Research Ethics Advisory Panel approval number HC171044) prior to conducting the experiment. We then briefly explain our system architecture for collecting, storing, and analysing data collected by the LPR cameras. Lastly, we discuss measurement challenges, quantify errors and present our method for data cleansing.


\subsection{License-Plate-Recognition Camera}

We investigated several commercial sensors for our car park monitoring solution with two goals. First, we want to have a complete ownership of the data without risking it leaving our campus infrastructure. Second, we aim to access our own data without relying upon a vendor, hence freeing us from ongoing service costs. This would allow us to integrate the collected data into a central repository of our overarching Smart Campus project in order to facilitate better analytics across many data feeds we have on campus \cite{sutjarittham:ANTS}.


We selected Nedap's automatic number plate recognition (ANPR) camera \cite{nedap:camera} as our parking monitoring solution. To read license plates of vehicles, the camera uses LPR technology which involves two main stages of operation: 

(a) locating license plate in the captured image by isolating a rectangle area (of the license plate number), using physical characteristics such as the shape, symmetry, width to height ratio and alphanumeric characters;  and (b) separating and recognising characters inside the isolated image \cite{chang2004:LPR}.


The camera unit consists of several components including a high definition camera, infrared (IR) illumination, and ANPR, allowing the camera to read complex number plate at various lighting conditions (due to weather or different hours of the day). 
The recognition engine relies on an on-board library that supports license plates readings from specific countries, each of which uses its own characters, colors, and designs.
The camera also  provides a management console (a web-page accessible via its IP address) that allows users to configure various parameters such as shutter time, strobo time (activation time of IR illuminator), and gain. We configured the camera to use the default ``Autoiris'' mode, which is recommended by the manufacturer guidelines, in order to allow the camera to automatically adjust the parameters based on the current lighting condition.



The ANPR system is capable of real time optical character recognition (OCR) processing, which runs when a license plate number is presented within the camera's frame. For each detection, the algorithm outputs two types of data; a JPEG image (with adjustable quality value between 0 and 100, where 100 is the highest resolution 1080p) of the vehicle, and a data record containing parameters such as timestamp, license plate string, OCR score, speed, country, state/region, type of vehicle, and up to 50 more fields. Fig.~\ref{fig:license_plate} shows an example of a real license plate (for the private vehicle of an author of this thesis chapter) recognised by the camera. On the left is the JPEG captured image showing the isolated license plate, and on the right is a list of selected key data fields generated by the OCR algorithm. The definition of the fields are as follows:


\begin{itemize}
	\item {\myverb{EVENT\_DESC}}: The value ``OCR Read'' means that the algorithm was able to recognise individual characters in the isolated license plate.
	\item {\myverb{DATE}} and {\myverb{TIME}}: Timestamp of the record.
	\item {\myverb{PLATE\_NOT\_READ}}: This field indicates whether the license plate was successfully isolated or not -- the value could be either ``READ'' or ``NOTREAD''.
	\item {\myverb{PLATE\_STRING}}: The output string of the recognised license plate. Australian cars and motorcycles have 6 and 5 alphanumeric characters on their license plate respectively.
	\item {\myverb{PLATE\_COUNTRY}} and {\myverb{PLATE\_REGION}}: Country and state of the license plate -- AUS for Australia, and NS for New South Wales.
	\item {\myverb{OCRSCORE}}: Overall confidence value (between 0 and 100) given by the camera on how accurately the entire license plate number is recognised.
	\item {\myverb{OCRSCORE\_CHAR}}: OCR score for individual characters of the string -- in this example, character ``J'' has the highest score 89 and character ``8'' has the lowest score 80.
	\item {\myverb{SPEED}}: Speed of the vehicle in 100$\times$ actual speed (km/h) -- we found that this data field is unreliable since reported values ranged from 1000 to 10,000,000. 
	\item {\myverb{DIRECTION}}: Direction of the vehicle relative to the camera, \ie ``APPROACH'' for entry camera and ``GOAWAY'' for exit camera.
\end{itemize}

\begin{figure}[t!]
	\centering \includegraphics[width=0.7\textwidth]{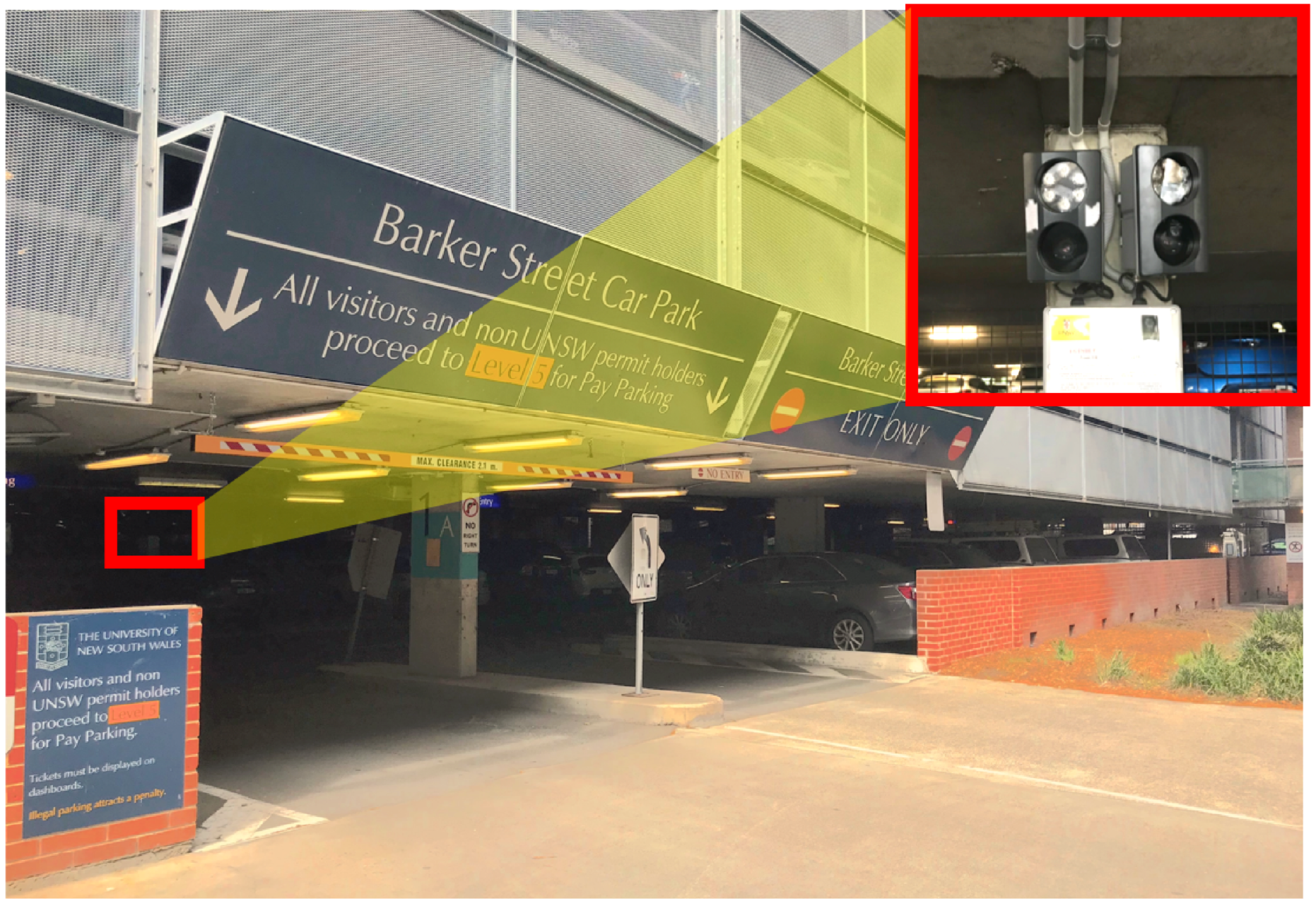}
	\caption{A real picture showing entrance and exit of our campus car park and a pair of LPR cameras installed side-by-side close to the ceiling of the car park ground floor.}

	\label{fig:carpark}
\end{figure}

During our field trial, we only captured and stored text output of the camera and not the JPEG image in order to maintain users' privacy and save space for data storage.
In order to tune various parameters of cameras and collect ground-truth records (for measuring accuracy), we temporarily recorded a number of image captures. The camera supports both local storage on an Secure Digital (SD) card and an external File Transfer Protocol (FTP) for data collection. The later option was selected for continuous data collection, where the camera creates and automatically updates a text file in Comma Separated Values (CSV) format containing information about every detected vehicle by the ANPR.

For the deployment, we chose an on-campus 5-storey car park that serves 895 parking spaces for students, staff, visitors, and contract workers. The first four levels of the facility are reserved for permit holders and the top level for hourly-based paid parking. The car park has a one-lane entrance and exit on the ground floor, which we installed two LPR cameras to capture the arriving and departing cars as shown in Fig.~\ref{fig:carpark}. For installation, the devices require 24 volts direct current (VDC) power supply and communicates via an Ethernet port.

Hence, we (with help from our campus Estate Management) supplied new power points and provisioned Ethernet ports for the cameras. We note that the flow rate of cars at our deployed car park is fairly moderate, with an average of 3 cars per minute during peak hours on busy days, and hence the workload on the OCR system is not a concern, especially when the response time of cameras (for generating output records) is in the order of tens of milliseconds.

\begin{table}[t!]
	\begin{center}
		\caption{Measurement accuracy of LPR camera from ground-truth of cars and their license-plates.}
		\label{table:LPR-acccuracy}
		\begin{tabular}{lllclcl }
			\toprule 
			&\multicolumn{2}{c}{\textbf{Accuracy}} \\
			\cmidrule{2-3}
			Camera & Capture rate & Read rate \\
			\midrule
			\textbf{Entry}  & $94$\% & $85$\%       \\
			\textbf{Exit}  &  $88$\%  & $42$\%  \\
			\bottomrule
		\end{tabular}
	\end{center}
\end{table}

\subsection{Measurement System Architecture}

Fig.~\ref{fig:sys-arch} shows the architecture of our car park monitoring system, comprising of three main layers: (a) ``sensing layer'' is where the LPR cameras record license plates information of  arriving and departing cars; (b) ``data layer'' is the core of our system, hosting an FTP server, a software engine for cleansing and anonymising data, message broker, and multiple databases for backup and load balancing. Once the data is cleaned, it is passed onto the message broker for unifying records into a JSON format. Each record is first tagged with time-stamp and sensor UUID, and then is posted via a RESTful API to our master database; (c) ``analytics layer'' includes health check monitoring (whether cameras are active and functioning, or not), data analysis, and visualisation modules -- this layer retrieves raw data from the master database (DB) and writes computed occupancy (real-time) and stay duration (per vehicle) into another DB that is used as a backend for visualisation.

The cameras are connected via a high-speed wired Ethernet cable to the campus network, sending real-time data records to backend servers on-premise. Given the reliable connection between sensing layer (cameras) and analytics layer (servers) we do not consider a stochastic process for the communication within our system.

\subsection{Measurement Accuracy} \label{sec:accuracy}
There are various factors that can impact the accuracy of the LPR cameras including  placement (\eg height and angle of installation), lighting conditions, speed of vehicles, angle of the license plate, and physical condition of the license plate.
We quantify the accuracy using two metrics: (a) ``capture rate'' which is a fraction of cars correctly detected, and (b) ``read rate'' which is a fraction of license plate numbers that are recognised correctly by the OCR algorithm running inside the camera. According to the ANPR's  manufacturer \cite{AnprGuidance}, it is expected to have capture rate and read rate of 98\% and 95\%, respectively.

\begin{figure}[t!]
	\centering \includegraphics[width=0.7\textwidth]{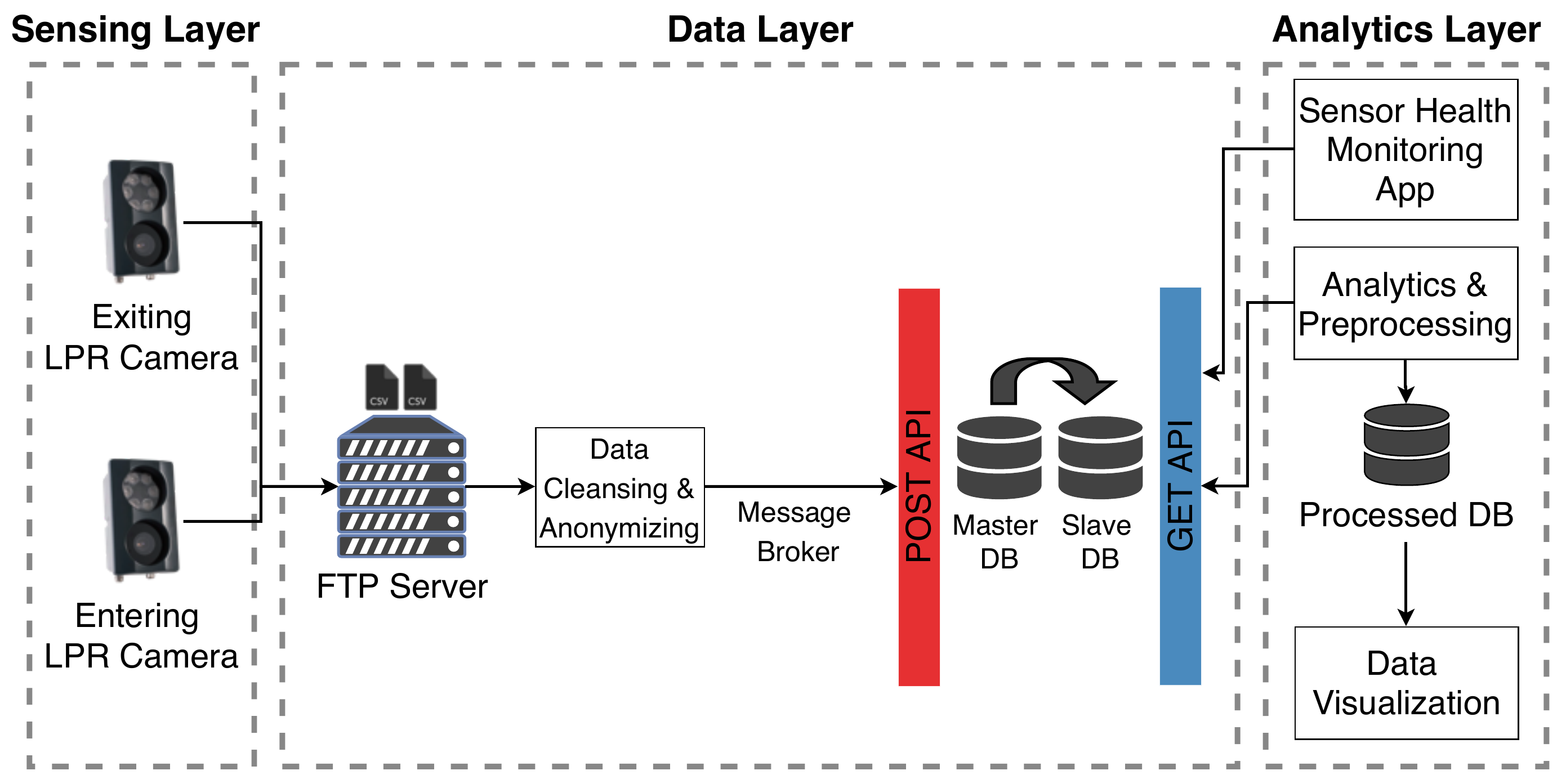}
	\caption{System architecture of collecting, analysing, and visualising data from LPR cameras.}

	\label{fig:sys-arch}
\end{figure}

To measure the capture rate, we performed several spot measurements over 5 days (Monday to Friday), each for a period of 5 hours (11am to 4pm) in order to collect ground-truth data of car park usage. We used a GoPro camera to record video logs and captured a total of 1400 vehicles entering and exiting the car park. For measuring read rate, we first enabled the JPEG recording on both cameras for a week and manually inspected 700 images for each of the LPR cameras (randomly selected) in order to obtain the ground-truth license plate labels.

Table~\ref{table:LPR-acccuracy} summarises the accuracy of both entry and exit cameras. It is seen that the exit camera under performs by both accuracy metrics, especially with a very low read rate of $42$\%. This means that more than half of the departing vehicles are not correctly recognised. The poor read rate of the exit camera was mainly due to its non-ideal placement which causes the detection to happen at a slight angle, and thus negatively affecting the performance of the OCR algorithm.
Re-positioning the camera was a nontrivial task due to difficulty (and cost) of provisioning Ethernet port and power outlet for the new position. 
With the majority of read rate errors resulted from misinterpretation of only one of the six characters, this can be used to our advantage in the cleansing process.

\begin{figure}[t!]
	\centering \includegraphics[width=0.7\textwidth]{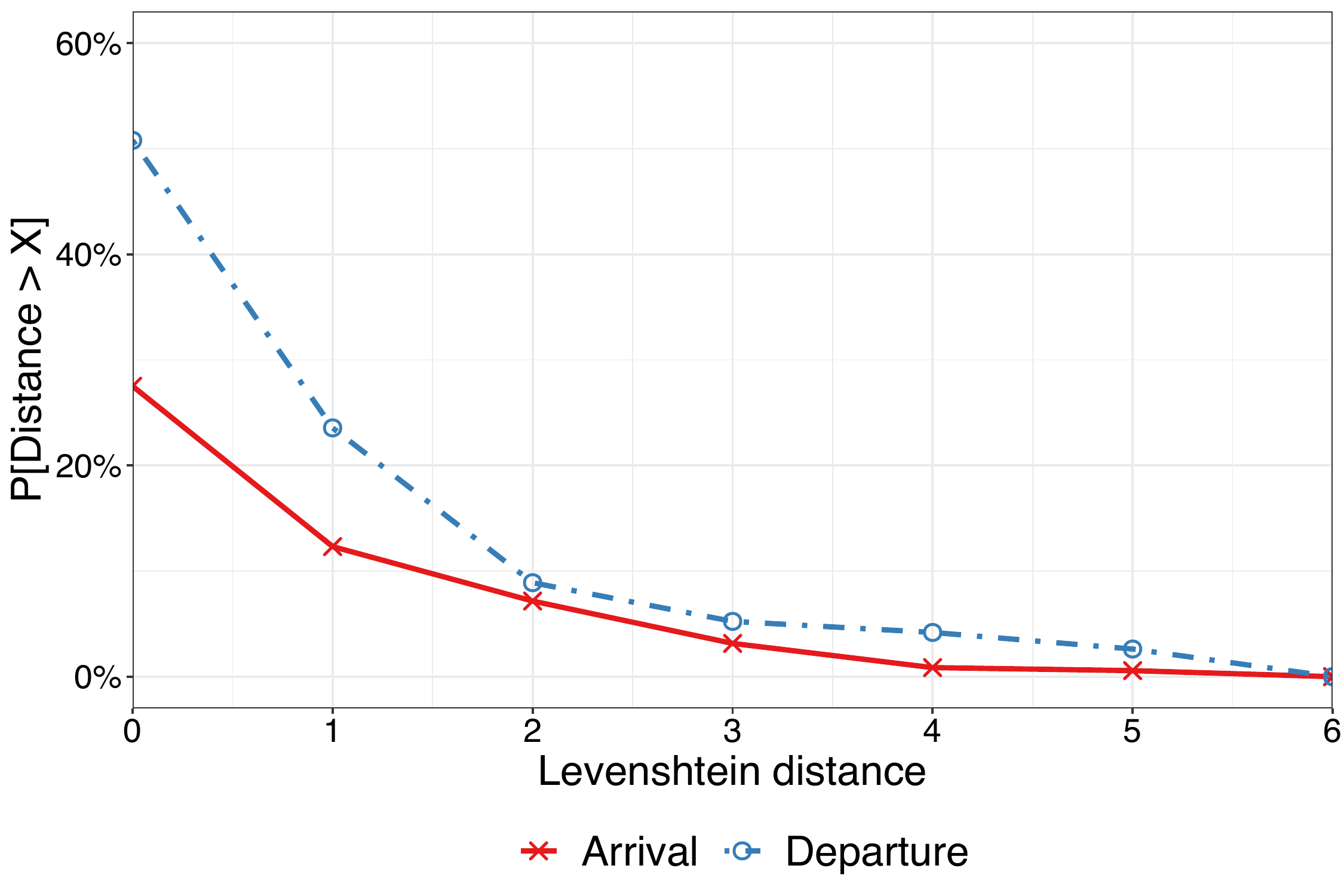}
	\vspace{-3mm}
	\caption{CCDF of Levenshtein distance between captured and ground-truth plate numbers.}
	\vspace{-3mm}
	\label{fig:levenshtein_distance}
\end{figure}

\subsection{Measurement Errors}\label{sec:errors}

The errors observed from the outputs of cameras can be categorised into three types:
(a) multiple recognitions of the same license plate, (b) incorrect recognition of license plate location, and (c) incorrect recognition of license plate characters.

\textbf{Multiple recognitions:} This error type occurs when the camera takes multiple images of a single vehicle (possibly because of its speed or moving pattern), and thus triggers the OCR algorithm multiple times. This generates multiple data records for the vehicle in the CSV file, resulting in over-counting of vehicles. These multiple records do not necessarily have the exact same license plate string -- it may output slightly different strings due to the angle of the moving car and its distance to the camera in a sequence frames captured.

\textbf{Incorrect locating:} 
This type of error occurs when the OCR algorithm incorrectly locates a license plate in the captured image and attempts to recognise characters. The output license plate strings from this error are almost always ``{\myverb{OCR NOT READ}}'' as the read plate does not match any of known formats of license plates available in its embedded library. There are rare cases where a non-license plate object gets partially recognised that their output appears as incomplete strings with low OCR scores of below 60.

\begin{figure}[t!]
	\centering \includegraphics[width=0.7\textwidth]{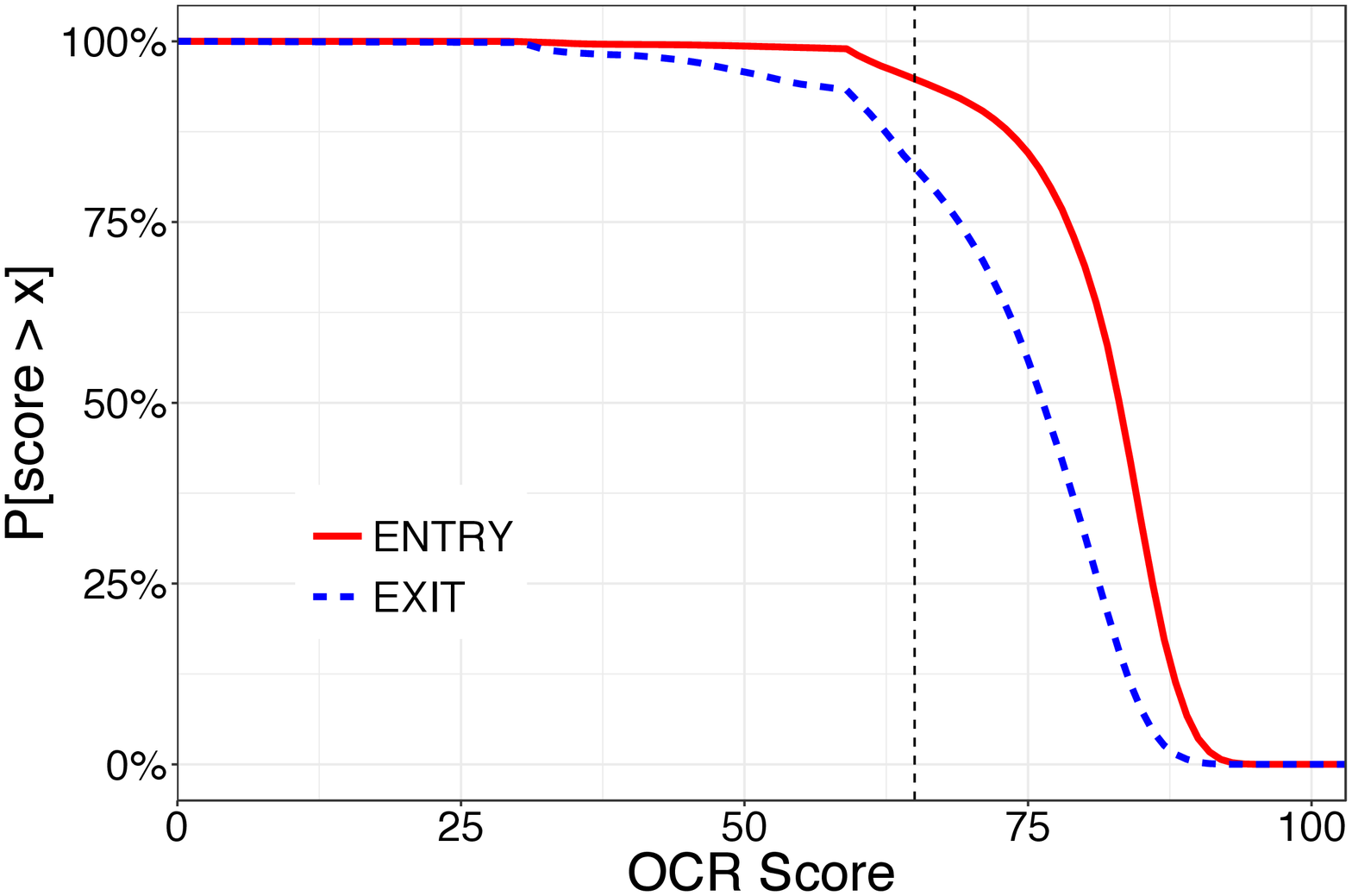}
	\vspace{-3mm}
	\caption{CCDF of OCR score for entry and exit cameras.}
	\vspace{-3mm}
	\label{fig:low ocr}
\end{figure}

\textbf{Incorrect recognition:} This type of error occurs when the camera successfully locates a license plate, but fails to recognise the characters correctly.
To quantify the severity of the read errors, we employ Levenshtein distance \cite{LevenshteinDist} to measure the difference between two string words, \ie the minimum amount of single character addition, substitution or deletion required to make two strings identical. For example, the Levenshtein distance between the string ``{\myverb{ABC123}}" and ``{\myverb{AC123}}" is 1, two strings will be identical by inserting a character ``{\myverb{B}}" into the string ``{\myverb{AC123}}". We compute the Levenshtein distance between ground-truth and recognised strings, the complementary cumulative distribution function (CCDF) plot of the distance is shown in Fig.~\ref{fig:levenshtein_distance}. It is seen that up to $28$\% of records from the entry camera and $52$\% from the exit camera have at least 1 character misrecognised (\ie distance of more than 0 character). We also observe that the majority of these  errors are caused by 1 misread character, accounting for $15$\% for entry and $30$\% for exit of the total observed records.
Note that incorrect recognition also results in lower OCR scores for the output record. 
Fig.~\ref{fig:low ocr} shows the distribution the OCR score of our deployed cameras. 
It can be seen that for the exit camera, $83$\% of the data records come with an OCR score more than 65. For the entry camera, on the other hand, records seem more reliable where $85$\%  of them have the score greater than 75.

\begin{figure}[t!]
	\centering \includegraphics[width=0.7\textwidth]{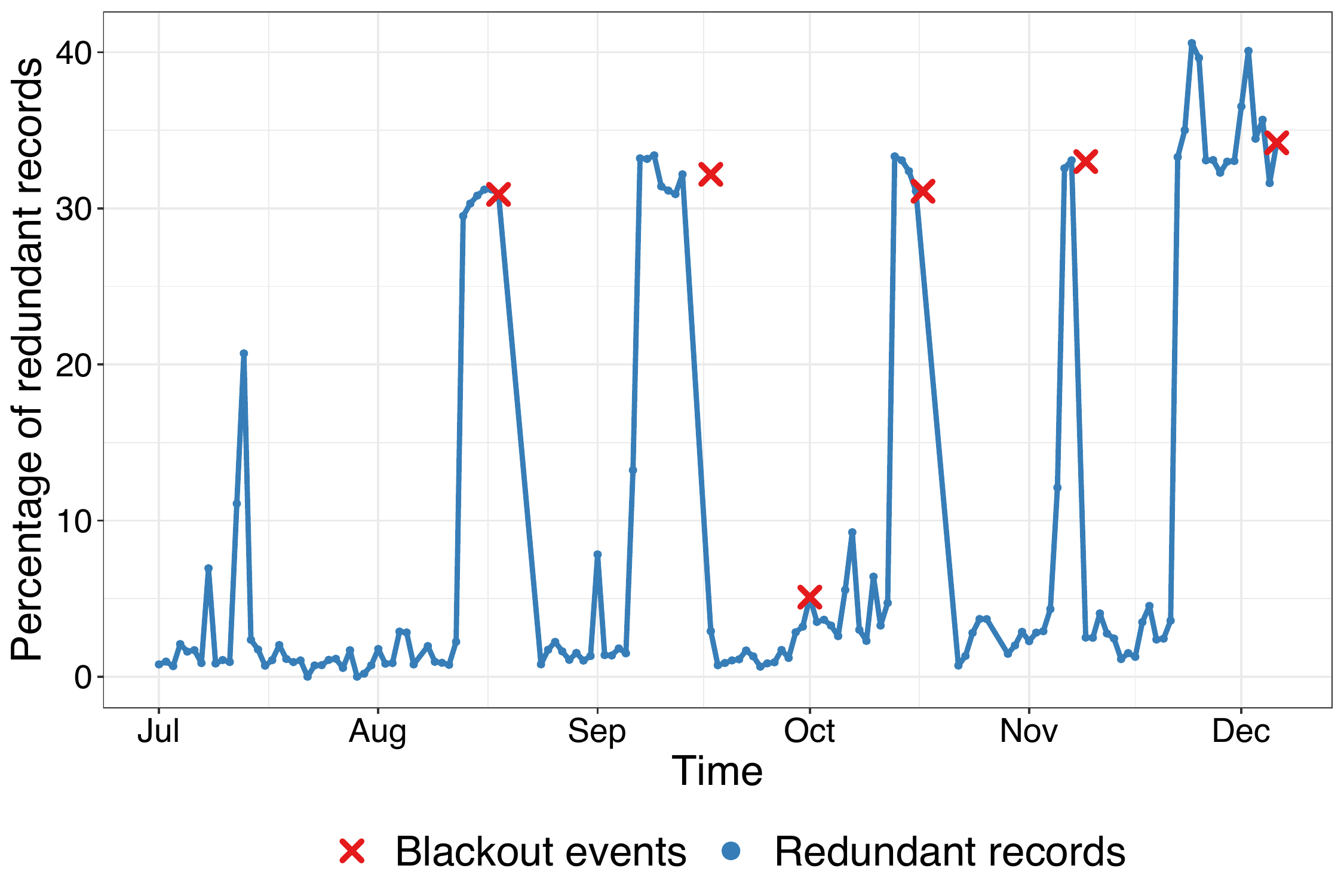}
	\vspace{-3mm}
	\caption{Rate of error due to redundant (multiple) records rises prior to camera black-outs.}
	\vspace{-3mm}
	\label{fig:dups-reboot}
\end{figure}

\begin{table}[t!]
	\begin{center}
		\caption{Distribution of error types for each camera $\break$(ground-truth dataset).}
		\label{table:error-stats}
		\begin{tabular}{lllclc l c l|}
			\toprule 
			&\multicolumn{3}{c}{Error types} \\
			\cmidrule{2-4}
			Camera & Multiple recognitions &  Incorrect locating & Incorrect recognition \\
			\midrule
			\textbf{Entry}  & $59.3$\% & $4.2$\% & $36.4$\% \\ 
			\textbf{Exit}  & $13.5$\% & $34.9$\% & $51.6$\% \\
			\bottomrule
		\end{tabular}
	\end{center}
\end{table}

We summarise the distribution of each error type for both cameras in Table \ref{table:error-stats}. 
As can be observed, the errors from entry records are largely stemmed from multiple-recognitions which accounted for $59.3$\% of the total errors. This is attributed to the recurrence of a peculiar problem where the camera's view unexpectedly blacked out, resulting in no data being recorded (\ie black-out events).
To understand this relationship, we plot in Fig.~\ref{fig:dups-reboot} the time-trace of daily error rate due to redundant records, overlaid by black-out events which required camera reboot. It is interesting to observe that the rate of redundant records steeply rises a few days prior to black-out events (evidenced by red cross markers in the middle of August to middle of December). 
The errors in data records of the exit camera, on the other hand, are mostly ($51.6$\%) from incorrect recognition. 
We believe this is primarily due to the non-ideal placement of the camera that causes it to capture license plates at an angle which is non-ideal, hence resulting in poor performance of the OCR algorithm.
Furthermore, we can see that $34.9$\% of the errors are due to incorrectly locating of license plates, while the same measure is only accounted for $4.2$\% of the total errors from the entry camera. 
By manually inspecting the collected images, we found that moving grass (close to the exit point of the car park, as visible in Fig.~\ref{fig:carpark}) gets occasionally detected as a moving object by the exit camera. This issue does not affect our entry camera, and thus it displays a much lower rate of incorrectly locating errors.

\subsection{Data Cleansing and Preprocessing}\label{sec:cleansing}

We cleanse raw data collected from the two cameras with the following objectives: 
(a) removing multiple records to obtain the correct count of arrivals/departures, (b) removing records of non-vehicle objects incorrectly captured by cameras, and (c) matching license plates captured by both cameras to deduce the distribution of stay-duration in our campus car park.

\begin{figure}[t!]
	\centering \includegraphics[width =0.8\textwidth]{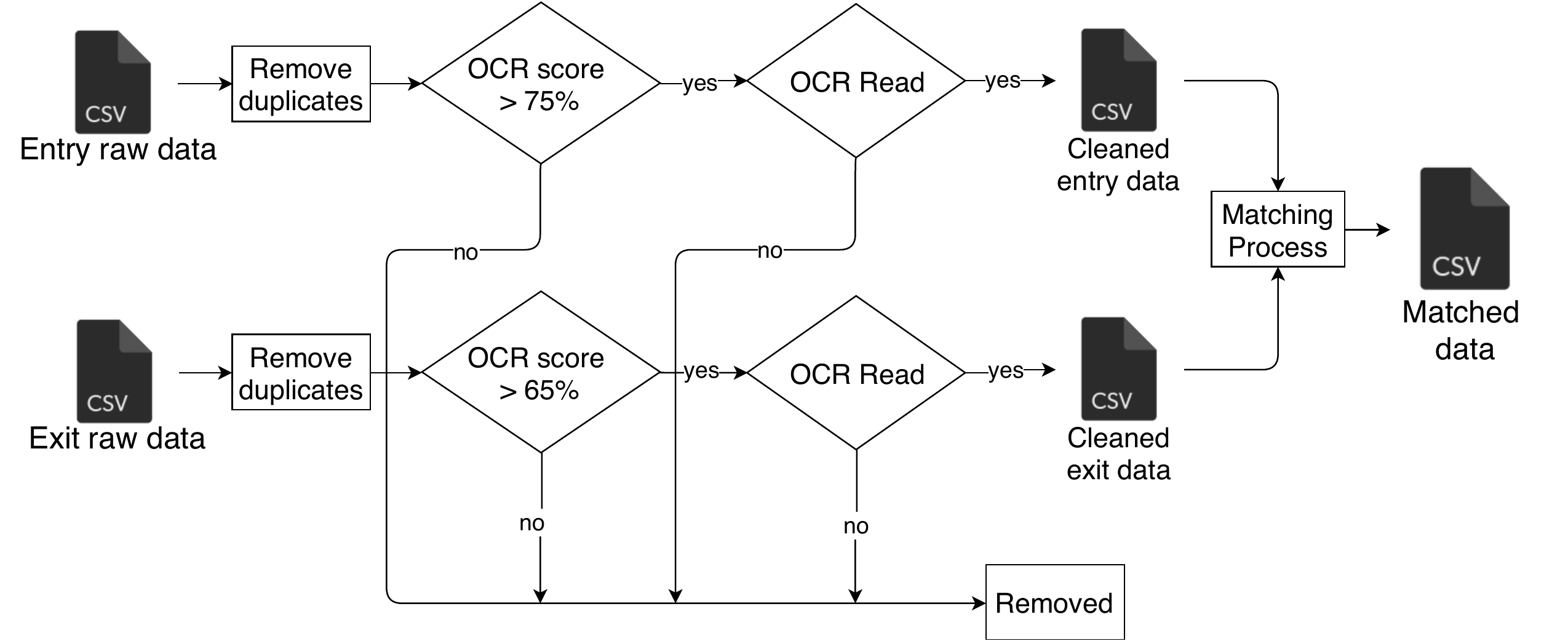}
	\vspace{-2mm}
	\caption{Data cleaning process.}
	\vspace{-3mm}
	\label{fig:cleansing_matching}
\end{figure}

Fig.~\ref{fig:cleansing_matching} shows our cleansing process with various stages involved.
We first remove duplicate records caused by multiple recognitions. A license plate is considered as ``redundant'' if it re-appears within the next five records after its first appearance with the Levenshtein distance between the plates of two or less. The distance threshold of two is selected based on our ground-truth analysis (Fig.~\ref{fig:levenshtein_distance}) -- applying this filter eliminates only  8\% and 10\% of entry and exit records, respectively. We found that increasing the threshold to three will only improve the coverage by less than 10\%, which is not substantial especially when the chance of plates getting mismatched increases (two different plates get incorrectly matched and treated as the same plate). For license plates identified as redundant, the one with the highest OCR score is kept and others are discarded.

We then remove records with low OCR scores. Most of these records relate to the erroneous captures (redundant records) of one vehicle.
As discussed earlier in Fig.~\ref{fig:low ocr}, we use filtering threshold of $75$ and $65$ for OCR scores for the entry and exit cameras respectively. 
Lastly, we remove all records with {\myverb{OCR NOT READ}} value -- those with incorrectly located license plate in the captured image. The pair of cleaned data will be used for arrival/departure counts.   
A summary of records removal due to each error type is shown in Table~\ref{table:error-removal-percentage}. 


As mentioned earlier, the goal of the last stage of our cleansing process is to match vehicles from the entry and  the exit datasets. The output of this stage will be used to compute stay duration of users. Similar to the multiple recognitions removal process, we use a Levenshtein distance of two or less for the matching. For one-to-many matched events, which rarely occur, we select the pair that yields the lowest Levenshtein distance and the highest OCR score. 
By running the matching process on daily data from July to the end of December, we were able to match $86$\% of the records on average. The remaining unmatched records correspond to: (a) overnight parking, (b) vehicles that are captured by one camera (mostly entry camera), but not the other.

We found from our spot measurements that only a small number of vehicles stayed overnight (\ie 23 cars per day on average for a sample size of 11 nights), and thus they would not have significant impact on the car park usage patterns. We therefore analyse our dataset on a per-day basis (\ie midnight-midnight).

\begin{figure}[t!]
	\centering
	\includegraphics[width=0.7\textwidth]{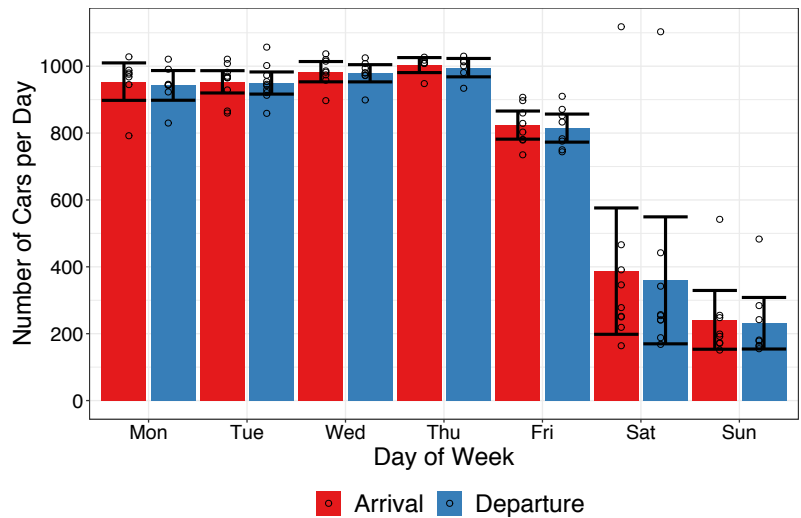}
	\caption{Number of arriving and departing cars per day during teaching period.}\label{fig:numcars_dayofweek}
\end{figure}

\begin{table}[t!]
	\begin{center}
		\caption{Percentage of records removed from the entire dataset at each stage of cleansing.}
		\label{table:error-removal-percentage}
		\vspace{-2mm}
		\begin{tabular}{lllclc l c l|}
			\toprule 
			&\multicolumn{3}{c}{Error types} \\
			\cmidrule{2-4}
			Camera & Multiples (redundant) & low OCR score & OCR-Not-Read \\
			\midrule
			\textbf{Entry}  & $9.7$\% & $12.9$\% & $2.5$\% \\ 
			\textbf{Exit}  & $7.9$\% & $8.5$\% & $8.2$\% \\
			\bottomrule
		\end{tabular}
		\vspace{-4mm}
	\end{center}
\end{table}

\section{Analysis and Insights into Usage Pattern}\label{section:analytics}

In this section, we analyse our cleansed data (obtained from \S\ref{sec:datacollection}) which spanned 15 months of  teaching and non-teaching periods in 2018 and 2019, to highlight the usage pattern of the campus car park across various temporal dimensions including time-of-day, day-of-week, week-of-semester, and semester break/exam periods. 


\subsection{Arrival and Departure Pattern}\label{sect:enter-exit-rate}

We begin with Fig.~\ref{fig:numcars_dayofweek} that depicts the average daily number of cars entering to (red bar) and exiting from (blue bar) the car park for each day-of-week during teaching periods.
The black points represent actual measurement values and the error bars represent 95\% confidence interval of data-points. On average, it is seen that there are about $950$ to  $1000$ cars using the car park on weekdays, except for Friday, where the number drops to about $830$ due to fewer number of classes running. During weekends, the number of cars occupying the car park are significantly lower, with an average of below $400$.
Furthermore, we observe narrower error bars on weekdays compared to weekends, suggesting predictable usage patterns for weekdays. In contrast, the car park usage varies significantly on weekends ranging from $200$ to $600$. Interestingly, it goes beyond $1000$ cars for one particular Saturday (\ie 1 September 2018). After checking the University event calendar, we found that this corresponds to the University Open Day which typically attracts a large number of high-school students and their families. The University provides free parking for all Open Day Visitors.


\begin{figure}[t!]
	\centering
	\includegraphics[width=0.7\textwidth]{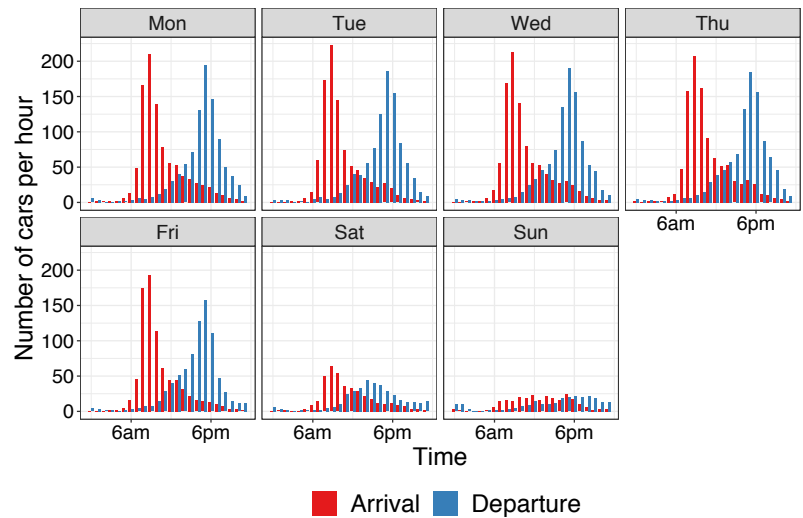}
	\vspace{-2mm}
	\caption{Average hourly arrival and departure rate of cars for each day-of-week.}\label{fig:numcar_tod_day}
	\vspace{-3mm}
\end{figure}

Fig.~\ref{fig:numcar_tod_day} illustrates an average hourly count of arriving cars (red bars) and departing cars (blue bars) by time-of-day, for each day of the week (separate graphs per day).
It is seen that during weekdays, the arrival rate starts rising steeply from 6am, peaks at 8am-10am, and falls slowly afterwards. The departure process displays a similar bell-shape pattern but shifted in time by about 8 hours, \ie rising in the afternoon, peaking at 5pm-6pm, and falling afterwards. 
During peak demand times, about $200$ cars enter and exit the parking lot per hour. This number is slightly lower for Friday (\ie  $194$ and $165$ cars per hour).
We also observe an irregular pattern for weekends where car park usage heavily depends on events hosted on campus. Our findings of weekday arrival/departure pattern corroborate with other studies \cite{rajabioun2013:intelligent}.

\begin{figure}
	\centering
	\includegraphics[width=0.7\textwidth]{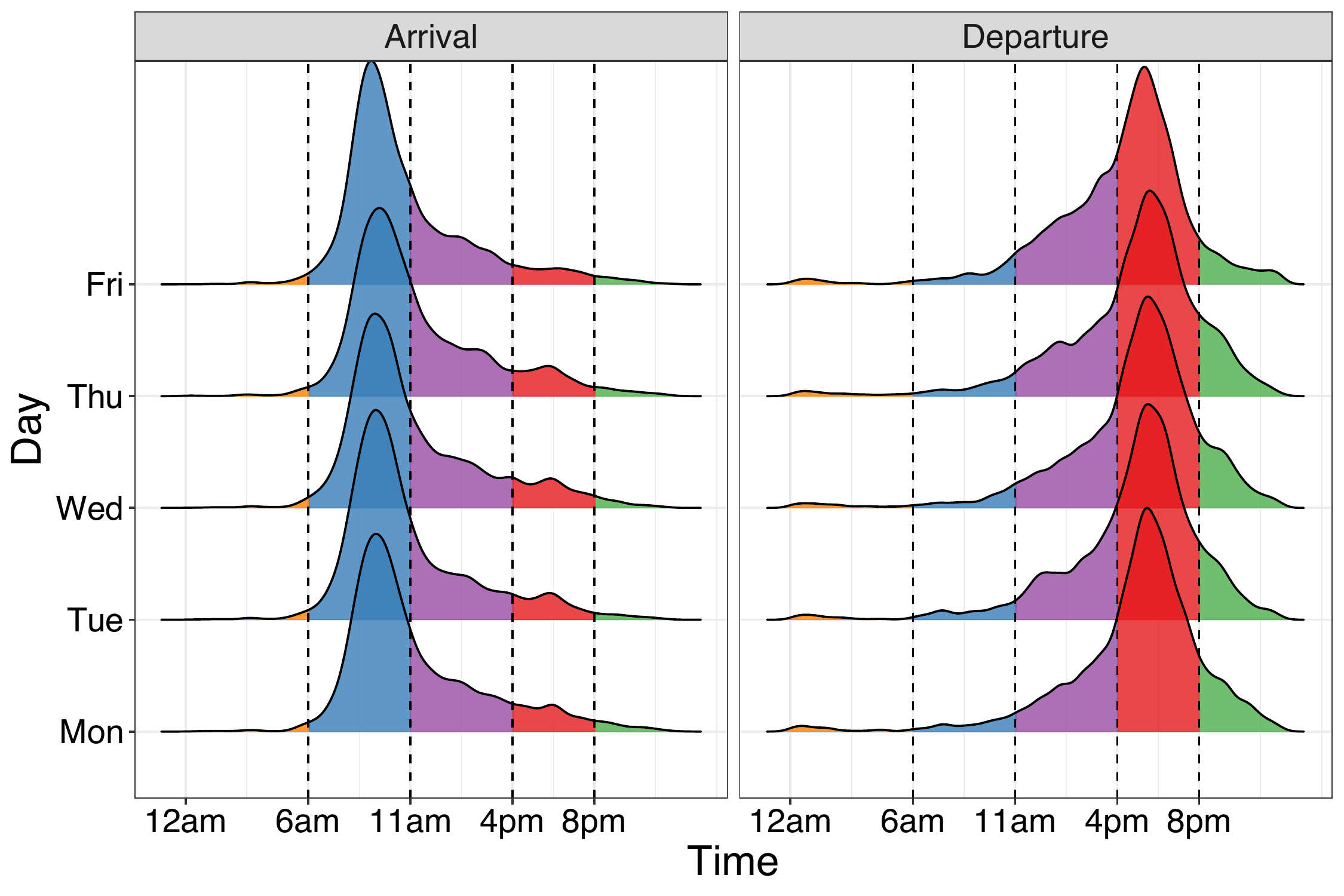}
	\vspace{-2mm}
	\caption{Distribution of arrival and departure time during teaching period, across each weekday}\label{fig:dist_byhourday}
	\vspace{-3mm}
\end{figure}

Fig.~\ref{fig:dist_byhourday} depicts the distribution of arrival and departure rate for the five weekdays (in a stacked representation with Friday on top and Monday at the bottom). 
To better illustrate the pattern, we color-coded five time intervals: orange for prior-sunrise (12am-6pm), blue for morning-peak (6am-11am), purple for afternoon-offpeak (11am-4pm), red for evening-peak (4pm-8pm), and green for night-time (8pm-12am). 
Note that the difference in absolute numbers is not seen for Friday due to normalisation.
Unsurprisingly the distribution plot shows that both arrival and departure patterns are consistent across the majority of time-slots on weekdays, with the strongest similarity observed during peak hours.
As before, we observe some different trends for Friday with the red part of the arrival curve (4-8pm) showing a less pronounced peak compared to other weekdays. This suggests fewer arrivals to campus on Friday evening, which can be attributed to the fact that there are very few evening lectures running on Friday. 
A closer examination of the exit curve for Friday reveals that the area in purple is larger than the corresponding regions of other weekdays. The green part of the curve is also considerably flatter than the other days. This suggests higher percentage of cars leaving during early afternoon time and lower percentage of cars  exiting the car park after 8 pm on Fridays compared to Mondays-Thursdays.

Additionally, we looked at the arrival and departure patterns during different periods of the academic calendar including orientation week (O-Week) which is largely geared for new entrants to get acquainted with the university, regular teaching weeks, a week long mid-semester break, a study-break week right before final exams, exam periods and a lull period before the end of year holiday shut down during which undergraduate students are away and campus attendance for everyone else progressively reduces. The general patterns are similar to Fig.~\ref{fig:numcar_tod_day}  for all periods. 
One main difference observed was the morning arrival peak occurred one hour earlier (\ie 8am-9am) during the O-Week and other non-teaching periods. The former can be attributed to the fact that a portion of students and administration staff arrive early during the O-week for setting up various activities.
The latter can be ascribed to the fact that students are less likely to be on campus during the non-teaching period and thus the morning peak time is largely determined by the arrival pattern of staff who typically arrive earlier (between 8-9 am) than students.
Furthermore, it can be observed that car park usage drops (\ie around 50 cars) during mid-semester break and other non-teaching periods, suggesting that the majority of car park users are staff members. This is not surprising due to high price of parking permits in our university. 

\subsection{Stay Duration Pattern}
Our dataset allows us to obtain insight into stay-duration of car park users given that the license plate numbers are captured at entry and exit points. We choose to analyse the distribution of these patterns across various user groups, instead of on an individual basis. 
For this analysis we use the cleansed dataset after the matching process mentioned in \S\ref{sec:cleansing}.

\begin{figure}
	\centering
	\includegraphics[width=0.7\textwidth]{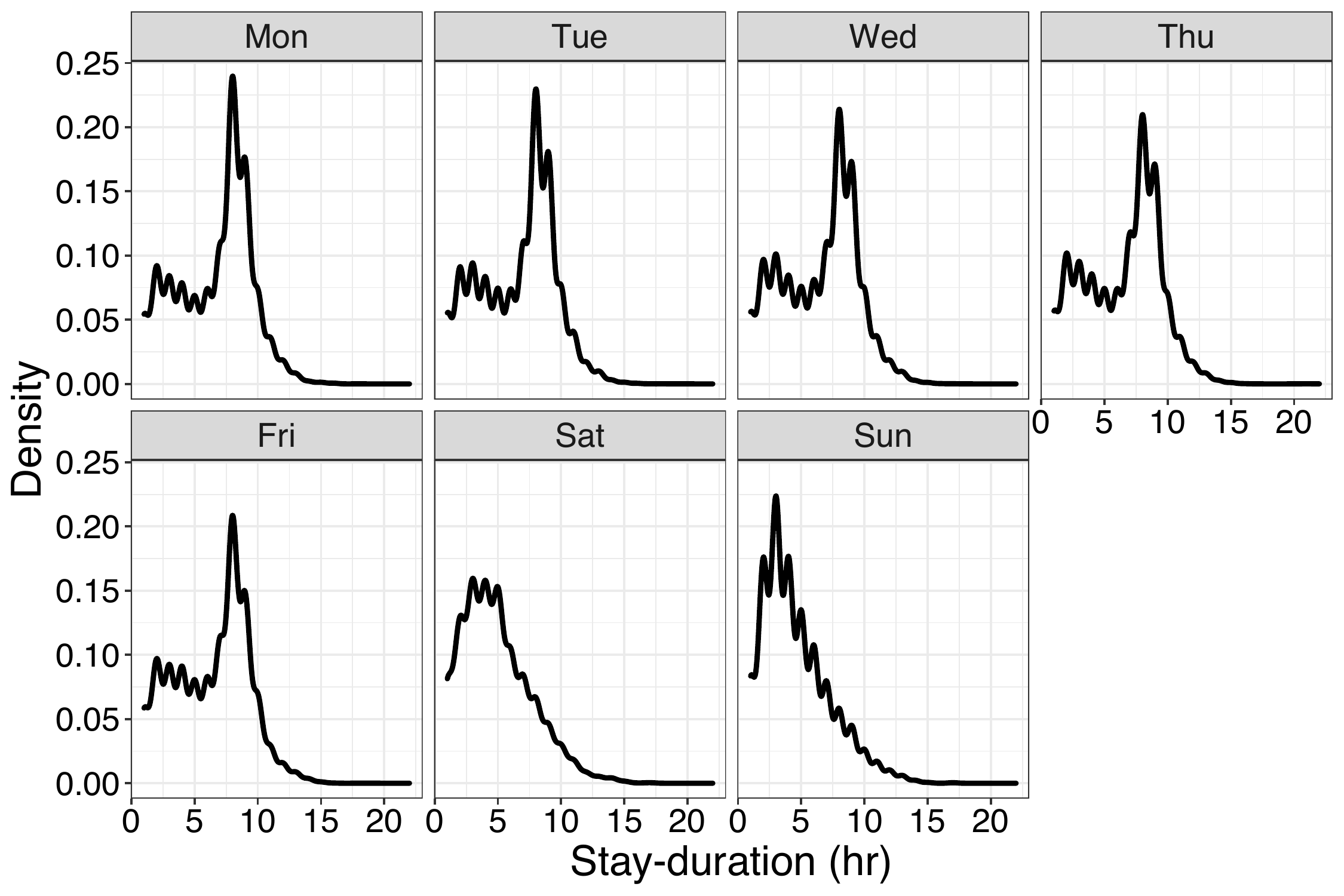}
	\vspace{-3mm}
	\caption{Density distribution of stay-duration (hour) for each day-of-week.}
	\label{fig:dist_stayduration_byday}
	\vspace{-3mm}
\end{figure}

Fig.~\ref{fig:dist_stayduration_byday} shows the distribution of stay duration (in hours) for each day-of-week.  It is clearly seen that the distribution is peak at around 8 hours for Mondays to Fridays -- this is consistent with the standard working hours in Australia, which is 7.6 hours a day \cite{fairwork:AUworkhour}. We also observe that users tend to use the car park for  about 2-4 hours during weekends - these users are likely to attend events hosted on or near-by the campus or Postgraduate students attending Saturday lectures.

\begin{figure}
	\centering
	\includegraphics[width=0.7\textwidth]{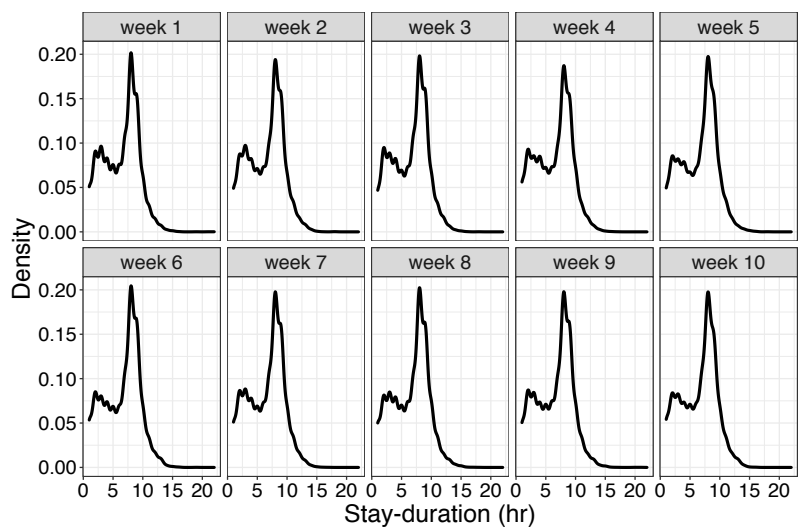}
	\caption{Density distribution of stay-duration (hour) for each week of term. }
	\label{fig:dist_stayduration_byweek}
\end{figure}

We further look at the stay duration pattern for each week of term in Fig.~\ref{fig:dist_stayduration_byweek}. 
A consistent pattern of bi-modal (double-peaked) distribution is observed for each week. The peak stay duration, as expected, centres at 8 hours, a typical full-time work day. The second peak centred between 2 to 3 hours, highlighting usage patterns for weekends as well as usage from students and visitors.
Note that we do not have access to the information about car park users due to privacy concern, therefore the above assumptions were made based on typical behaviour of different car park user types.


\subsection{Parking Behaviour Users}
%
%
%

\begin{figure}[t!]
	\begin{center}
		\mbox{
			\subfloat[Stay-duration versus arrival time $\break$ for each cluster.]{
				{\includegraphics[width=0.4\textwidth]{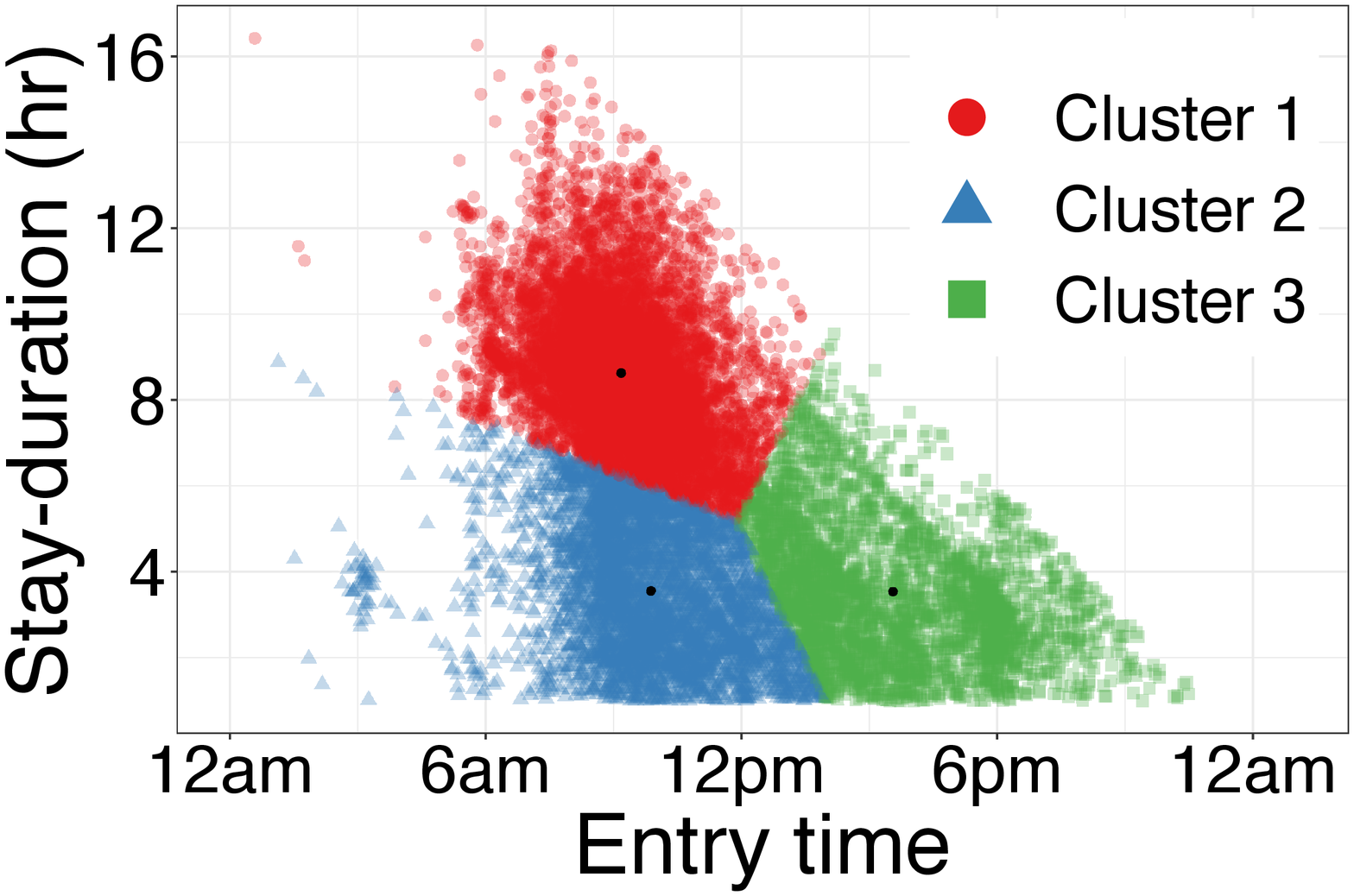}}\quad
				\label{fig:scatterClustering}
			}
			\subfloat[Distribution of arrival and $\break$ departure time for each cluster.]{
				{\includegraphics[width=0.4\textwidth]{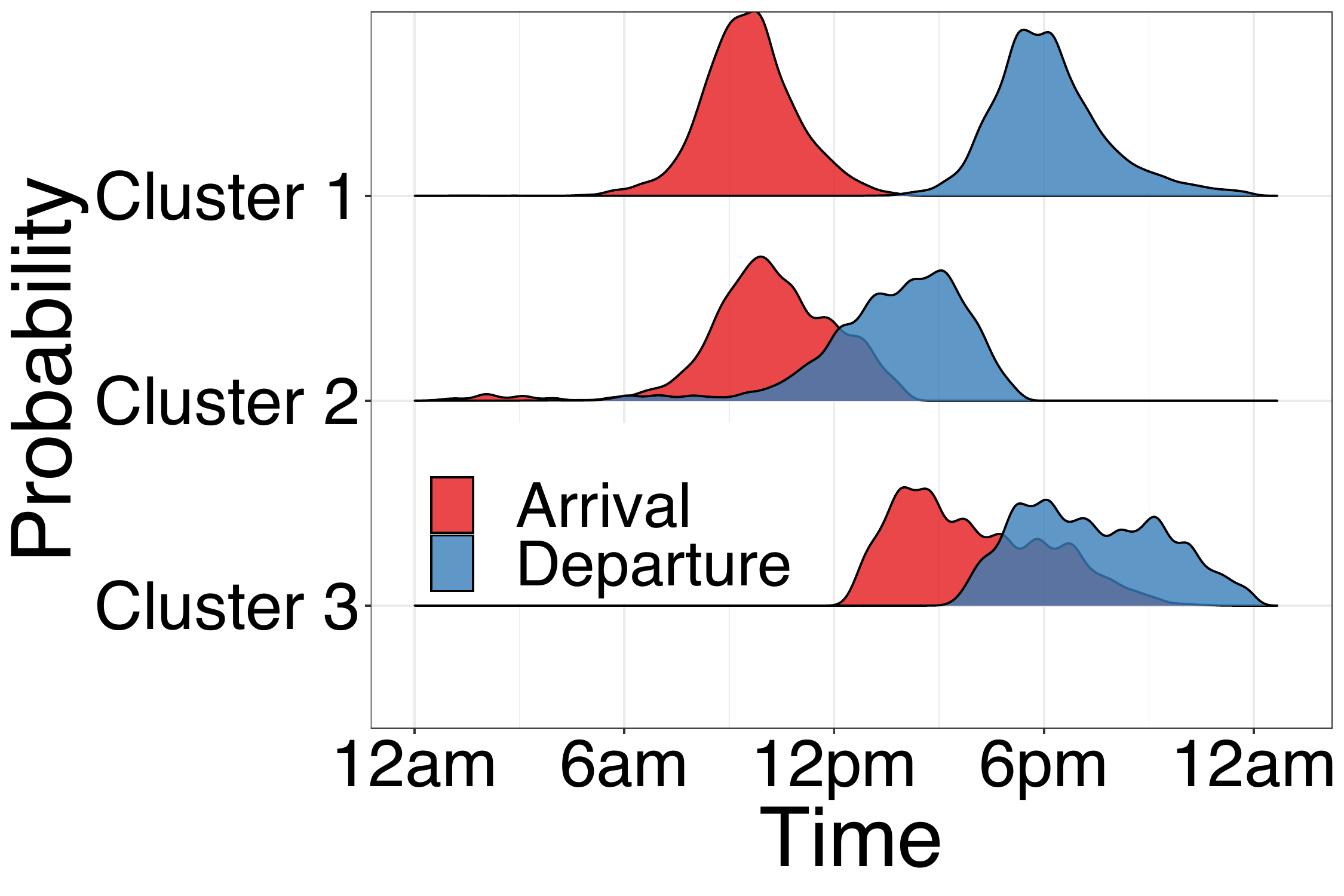}}\quad
				\label{fig:distClustering}
			}
		}
		\caption{Clustering of car park users using arrival time and stay duration.}
		\label{fig:clustering}
	\end{center}
\end{figure}

We now cluster car park users using k-means algorithm \cite{hartigan1979:kmean} to identify parking behavior of certain user groups on weekdays during teaching period.
For each car we extract three features including arrival time, departure time, and stay duration. Three number of clusters (optimal value is $k=3$) were selected based on the elbow method.
Fig.~\ref{fig:clustering} shows the result of our clustering.  We show in Fig.~\ref{fig:scatterClustering} the scatter plot of stay-duration versus entry time for individual vehicles in our dataset -- clusters are color-coded. It is seen that ``Cluster-1'' (shown by red region) corresponds to users who enter the car park early and stay more than 8 hours (center of cluster-1 is located at 9am entry time and 8.8 hours of stay duration) -- this cluster represents full-time staff. ``Cluster-2" users enter the car park at about the same time as Cluster-1 and stay shorter (\ie less than 5 hours) -- morning visitors and typical students. Lastly, ``Cluster-3" users are those who enter late and stay for a short period (with center at 3pm entry time and 3.5 hours of stay) -- this cluster is likely to denote afternoon visitors and postgraduate coursework students who attend evening classes.


In Fig.~\ref{fig:distClustering}, we show the distribution of arrival and departure time for each cluster. We observe that the distributions for Clusters-2 and -3 are relatively wider than of Cluster-1. Again, it is seen that full-time staff in Cluster-1 typically enter at about 9am and exit at about 5pm. Similar to the scatter plot, Cluster-2 users arrive early and leave early too. Lastly, as expected Cluster-3 users enter in the afternoon and exit in the evening. We also observe a significant overlap in the exit and entry distributions for these users, which suggests that they tend to stay on campus for a short period (about 3 hours). 

Lastly, to quantify the impact of choosing a larger number of clusters (\ie more than the optimal value 3), we experiment with $k=4$. In doing so, we observe that two of the original clusters, \ie Cluster-2 and Cluster-3 in Fig.~\ref{fig:clustering}, remain intact with the same arrival and departure characteristics. However,  the original Cluster-1 gets divided into two sub-clusters, both representing the behaviour of full-time staff. The first sub-cluster-1 displays a pattern fairly similar to Cluster-1; arrival time centred at 9am and departure time at 6pm. The other one, sub-cluster-2, which accounts for half of the size of the first one, presents a slightly different characteristic, with arrival and departure times shifted by an hour (centred at 10am and 7pm, respectively).

\section{Forecasting Parking Demand}\label{section:prediction}
We apply three machine learning models to forecast long-term parking usage, specifically, arrival and departure rate of cars per hour, and evaluate their performance across various forecast horizons.

\subsection{Demand Time-Series Forecasting}

Having an accurate forecast of user demand is essential in planning for future space allocation of the car park. In particular, long-term forecasting is useful for campus managers to implement car park partitioning schemes, where spaces are dynamically allocated to private and shared-car users, in order to improve the utilisation of parking facility. 
In this work, we adopt multi-step ahead time-series forecasting for predicting arrival and departure rates of cars. We perform the prediction for a range of time horizons, from one day to ten weeks, in order to determine an appropriate time horizon which provides sufficient demand forecast and yields an acceptable prediction accuracy.

The car park demand displays seasonal characteristics affected by several temporal factors like hour-of-day and day-of-week (as discussed in \S\ref{section:analytics}). Additionally, including past observations (lagged dependent variables) as part of features in a prediction model is a standard practice in time-series analysis which is known to produce robust estimation \cite{wilkins2018lag}.
Below we summarise features that we used in our forecasting models:

\begin{itemize}
   \item  Time-of-day (Fourier representation)
   \item  Day-of-week (Fourier representation)
   \item  Teaching/Non-teaching period
   \item  Lagged variables (daily seasonal lags of last 10 days, \eg to forecast arrival rate between 9am-10am, historical arrival rates of the same time slot from the last 10 days were used as features)
\end{itemize}

We use Fourier representation as features to capture seasonality (\eg time-of-day and day-of-week) in our time-series data. Given a seasonal period of $p$ hour, the Fourier terms with $K$ sine and cosine pairs is defined as:

\begin{equation}
\left[ 
	\sin\left( \frac{2\pi kt}{p} \right) , \cos\left( \frac{2\pi kt}{p} \right) 
\right]_{k=1}^{K}
\end{equation}

\begin{figure}
	\centering
	\includegraphics[width=0.7\textwidth]{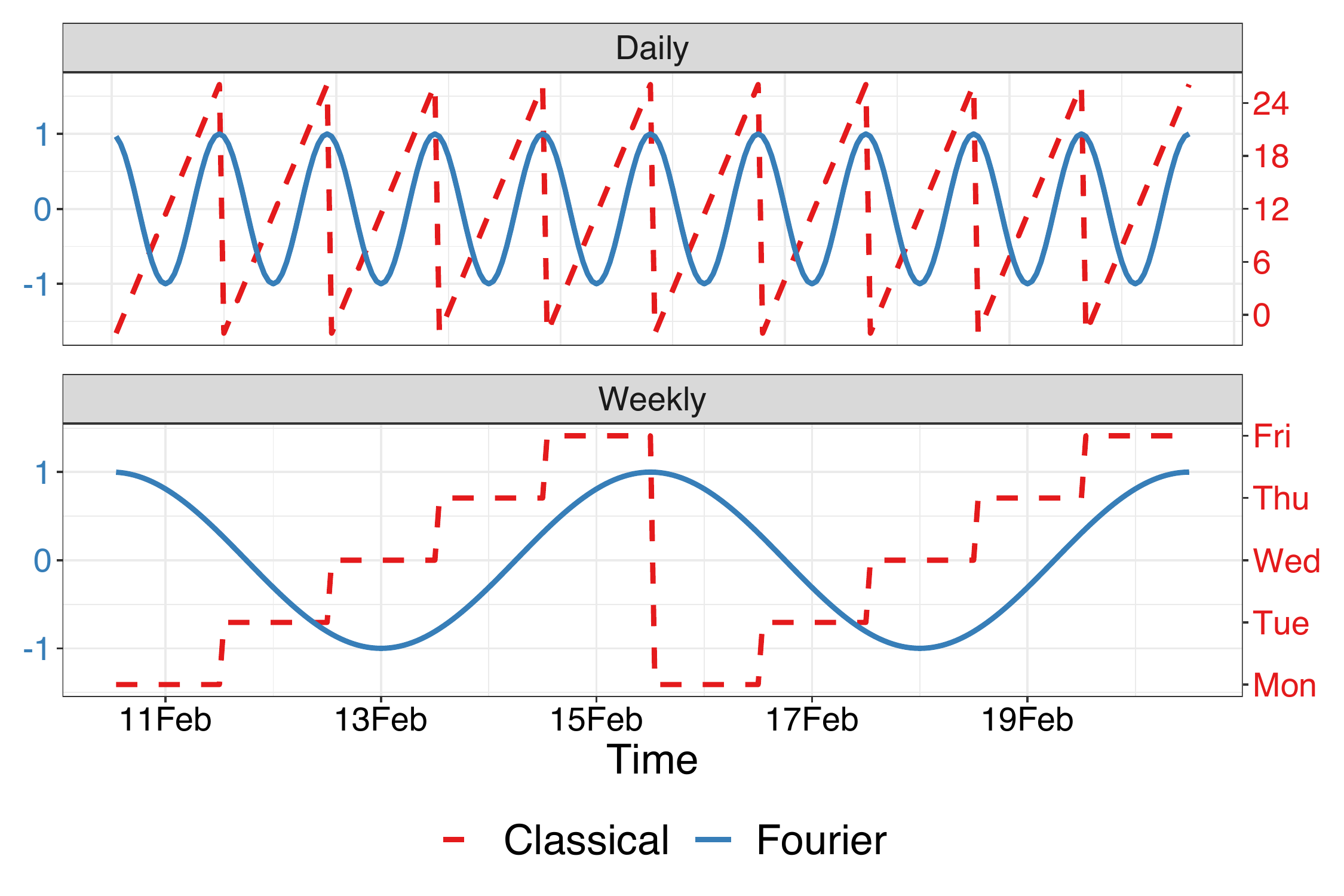}
	\vspace{-4mm}
	\caption{Comparison of classical representation and Fourier terms representation of seasonality.}\label{fig:fourier_terms}
	\vspace{-5mm}
\end{figure}

Fourier terms for time-of-day and day-of-week seasonality are obtained by setting $p=24$ and $p=120$ (5 working days) respectively. We show an example of the first sine terms for both daily and weekly seasonality, comparing with the typical representations of seasonal variables in Fig.~\ref{fig:fourier_terms}.
It can be seen that Fourier terms (given the smooth nature of sinusoidal waves) allow us to better model the relationship between the starting and ending of a day or week (\eg continuity of the rates profile  from 12am on a day to 1am on next day). We tested up to eight Fourier terms in our models and found that inclusion of two terms is sufficient for the prediction.



\subsection{Strategies for Long-Term Forecasting}



We predict future arrival and departure rates using machine learning models
\cite{bontempi2012machine}. In what follows we describe our strategy to predict long-term usage demands and the learning algorithms we employ.

\subsubsection{Multi-Step Ahead Prediction Strategies}

In general, long-term forecasting is performed to predict the future behavior of an observed time series over a long horizon.
Common strategies include: 
(a) recursive approach \cite{sorjamaa2007methodology}, where the one-step ahead prediction is fed back as an input to predict the next steps recursively (the predictor progressively takes estimated values as inputs, instead of actual observations); (b) direct approach \cite{chevillon2007direct}, where multiple forecasting models are built, each corresponding to a forecasting horizon (steps); (c) hybrid strategy, which combines the previous two strategies \cite{zhang2013iterated} (models specific to each forecast horizon with recursive inclusion of the prediction from all previous horizons); and (d) Multiple-Output strategy, where a single model predicts multiple steps of forecasting in one go (the predicted value is not a scalar quantity but a vector of future values  of the time series). This strategy, however is more complex and requires a longer time to train.

For our purposes, we adopted a direct approach as it does not suffer from the propagated errors like in recursive strategy and does not involve training complexity like in hybrid/multi-output strategy. However, our choice of approach requires multiple models for multiple forecast horizons.

\subsubsection{Forecasting Algorithms}
We employ three widely-used algorithms to build our forecasting models:



\textbf{Random Forest (RF) Regression} 
is an ensemble learning method where a collection of decision trees are combined to make the final prediction in order to give better accuracy while reducing the likelihood of overfitting. Each decision tree is generated by a random sampling of training observations. Also, random subsets of features are used for splitting nodes.

\textbf{Support Vector Regression (SVR)} applies the same principles as Support Vector Machine (SVM) by using the concept of maximal margin hyperplane. In SVR, a decision boundary with distance $\epsilon$ from the hyperplane can be tuned and the objective of the algorithm is to maximise the number of points (support vectors) within the boundary line. Additionally, hyper parameter $C$ is used to tune the tolerance of data points that fall outside the decision boundary. For a non-linear SVR, a kernel trick is employed to transform the training data into a higher dimensional feature space prior to performing linear regression. In our model, radial basis kernel is used for this mapping.

\textbf{Multiple Linear Regression (MLR)}
is one of the simplest prediction algorithms that involves multiple input variables (features) for the prediction. There are several techniques that can be used to train linear regression from the data. The most common one is ordinary least squares (OSL), where the objective of the algorithm is to minimise sum of squared errors when fitting the data.


\begin{figure}
	\centering
	\includegraphics[width=0.7\textwidth]{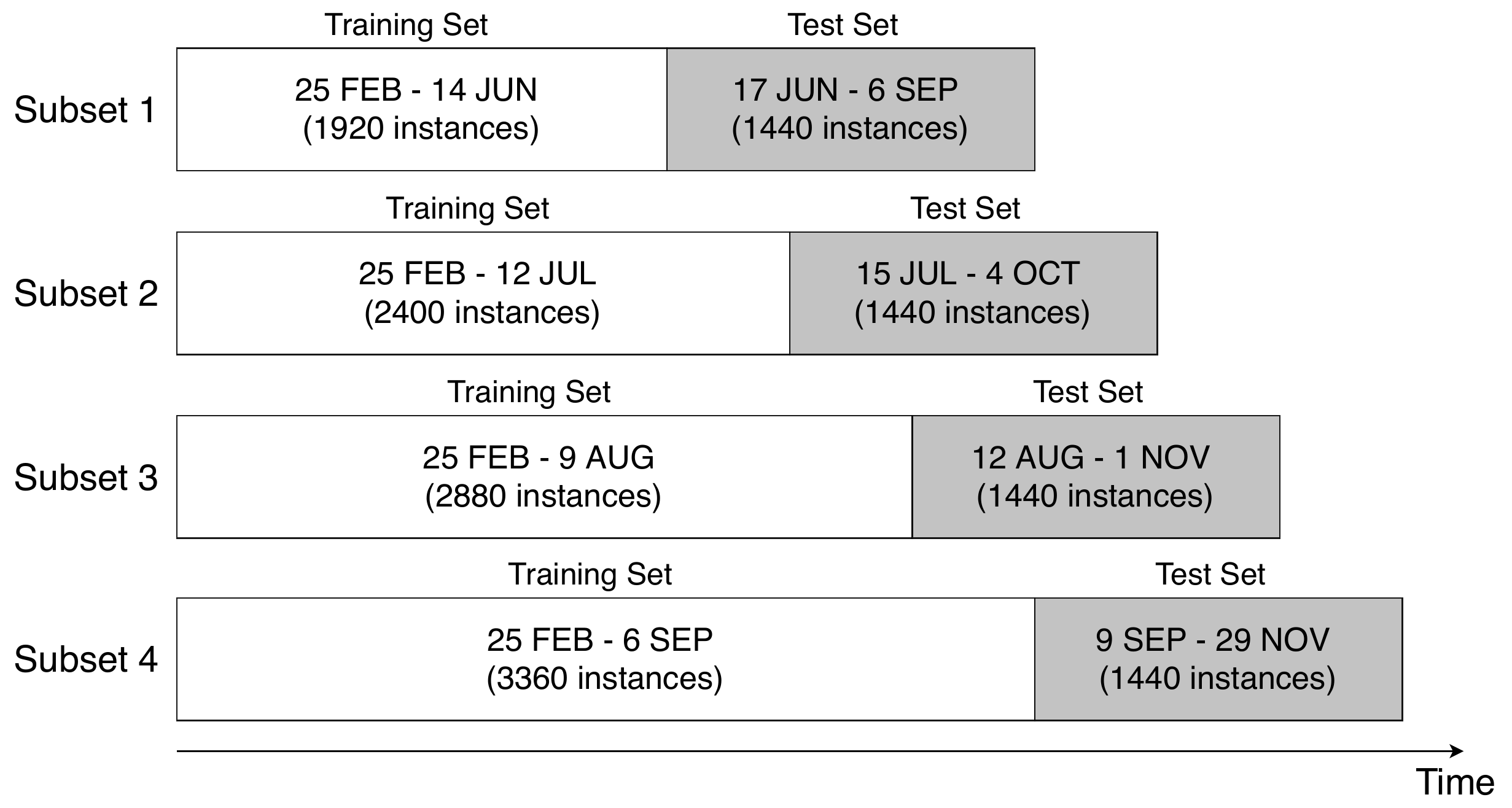}
	\vspace{-4mm}
	\caption{Summary of four train/test subsets. }\label{fig:train_test_data}
	\vspace{-3mm}
\end{figure}

\subsection{Performance Evaluation of Forecasting Models}

\subsubsection{Evaluation Methods and Dataset}
We use historical data from 2019 (from 25 Feb to 29 Nov) for training and testing our models. 
We employ nested cross validation procedure where the dataset is divided into multiple training and testing subsets, each training set is further divided into sub-cross-validation sets for tuning hyper parameters. Use of multiple train-test subsets allows for an unbiased and robust assessment of the performance of the models  in predicting unseen data. Note that we preserve the temporal order of observations during the train-test splitting and cross validation in order to prevent data leakage so that future data is not used to train the model that predicts past data. Hence, we emulate and demonstrate a real-world forecasting scenario.

Fig.~\ref{fig:train_test_data} shows a summary of how we divide our dataset into 4 subsets, each consisting of varying size data (from 16 weeks to 28 weeks worth of instances) used to train the model along with fixed-size data corresponding to test periods of 12 weeks starting right after their respective training period. 

We also use a simplistic baseline model in our comparisons to confirm that the more sophisticated approach that we have adopted is worth the effort. Our baseline model uses a mean average rate of each specific hour-slot (\eg 9am-10am) computed over historical data as a prediction of future values for the corresponding hour-slot. In other words, past observations are used to compute a daily profile of hourly arrival and departure rates which will be used for any day in the future.



%
%

\begin{figure}
	\centering
	\includegraphics[width=0.7\textwidth]{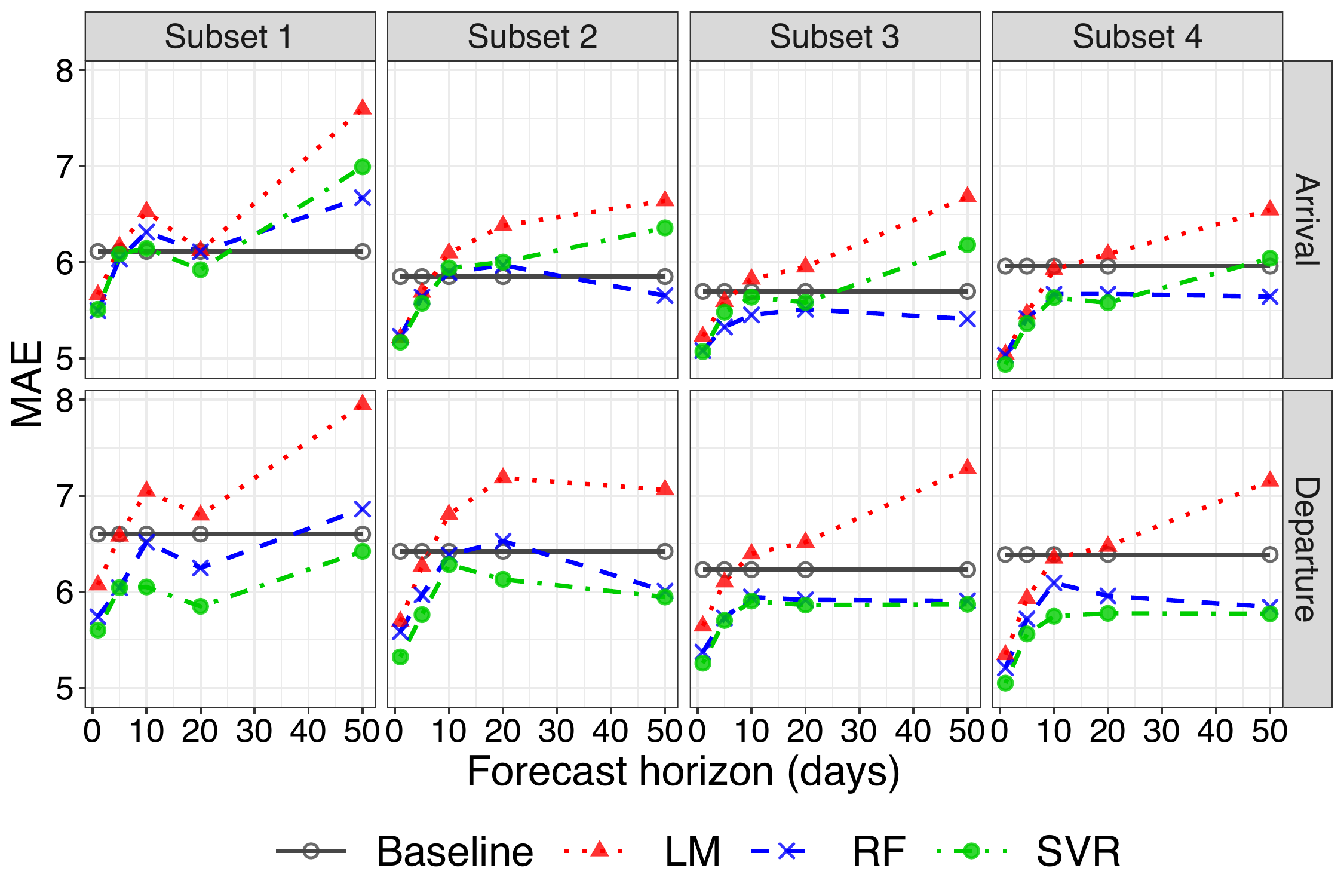}
	\vspace{-4mm}
	\caption{Performance of forecasting models per individual subsets during cross-validation phase.}\label{fig:validation_mae}
	\vspace{-5mm}
\end{figure}

\subsubsection{Evaluation Results}
For each train-test subset, we first perform cross-validation on the training set using time slicing in order to tune the hyper parameters of our machine learning models. We then train the final model of each algorithm using the parameters that minimise the error and apply the model to predict the test set. Note that we will use Mean Absolute Error (MAE) as our performance metric as we intend to penalize every error equally (\ie mispredicting 10 cars per hour is twice as costly as mispredicting 5 cars per hour).

\begin{table*}[t!]
		\caption{Mean absolute Error (MAE) of  prediction models.}
		\label{table:pred_results}
	\begin{adjustbox}{width=1\textwidth}

\begin{tabular}{lllllllllllllllllllll}
\toprule
\multicolumn{1}{c}{} & \multicolumn{10}{c}{\textbf{Arrival}} & \multicolumn{10}{c}{\textbf{Departure}} \\
\cmidrule(l{3pt}r{3pt}){2-11} \cmidrule(l{3pt}r{3pt}){12-21}
\multicolumn{1}{c}{} & \multicolumn{5}{c}{Cross-Validation Phase} & \multicolumn{5}{c}{Testing Phase} & \multicolumn{5}{c}{Cross-Validation Phase} & \multicolumn{5}{c}{Testing Phase} \\
\cmidrule(l{3pt}r{3pt}){2-6} \cmidrule(l{3pt}r{3pt}){7-11} \cmidrule(l{3pt}r{3pt}){12-16} \cmidrule(l{3pt}r{3pt}){17-21}
model & 1-day & 5-day & 10-day &20-day & 50-day &1-day & 5-day & 10-day &20-day & 50-day & 1-day & 5-day & 10-day &20-day & 50-day & 1-day & 5-day & 10-day &20-day & 50-day\\
\midrule
\addlinespace[0.3em]
\multicolumn{21}{l}{\textbf{Subset 1}}\\
\hspace{1em}Baseline & 6.11 & 6.11 & 6.11 & 6.11 & 6.11 & 5.65 & 5.65 & 5.65 & 5.65 & 5.65 & 6.60 & 6.60 & 6.60 & 6.60 & 6.60 & 6.14 & 6.14 & 6.14 & 6.14 & 6.14\\
\hspace{1em}LM & 5.66 & 6.16 & 6.53 & 6.12 & 7.60 & 4.51 & 4.84 & 5.30 & 5.79 & 6.45 & 6.07 & 6.58 & 7.05 & 6.80 & 7.95 & 4.80 & 5.32 & 5.83 & 6.07 & 6.80\\
\hspace{1em}RF & 5.50 & 6.04 & 6.32 & 6.11 & 6.67 & 4.62 & 4.85 & 4.98 & 5.08 & 5.22 & 5.74 & 6.04 & 6.52 & 6.25 & 6.86 & 4.83 & \hlc[green]{5.29} & 5.45 & 5.53 & 5.53\\
\hspace{1em}SVR & 5.51 & 6.09 & 6.15 & 5.92 & 6.99 & 4.43 & 4.98 & 5.05 & 5.42 & 5.74 & 5.60 & 6.04 & 6.05 & 5.85 & 6.42 & 4.73 & 5.15 & 5.34 & 5.50 & 5.81\\
\addlinespace[0.3em]\hline\hline
\multicolumn{21}{l}{\textbf{Subset 2}}\\
\hspace{1em}Baseline & 5.85 & 5.85 & 5.85 & 5.85 & 5.85 & 5.70 & 5.70 & 5.70 & 5.70 & 5.70 & 6.42 & 6.42 & 6.42 & 6.42 & 6.42 & 6.08 & 6.08 & 6.08 & 6.08 & 6.08\\
\hspace{1em}LM & 5.21 & 5.69 & 6.10 & 6.38 & 6.64 & 4.59 & 5.11 & 5.46 & 5.50 & 6.22 & 5.69 & 6.27 & 6.81 & 7.19 & 7.06 & 4.73 & 5.21 & 5.60 & 5.78 & 6.47\\
\hspace{1em}RF & 5.23 & 5.64 & 5.89 & 5.97 & 5.65 & 4.47 & 4.71 & 4.75 & 4.89 & 5.15 & 5.59 & 5.97 & 6.38 & 6.53 & 6.01 & 4.58 & 5.04 & 5.19 & 5.00 & 5.34\\
\hspace{1em}SVR & 5.17 & 5.57 & 5.94 & 6.00 & 6.36 & 4.74 & 4.98 & 5.21 & 4.98 & 5.83 & 5.32 & 5.76 & 6.28 & 6.13 & 5.95 & 4.66 & 4.93 & 5.03 & 5.08 & 5.50\\
\addlinespace[0.3em]\hline\hline
\multicolumn{21}{l}{\textbf{Subset 3}}\\
\hspace{1em}Baseline & 5.70 & 5.70 & 5.70 & 5.70 & 5.70 & 5.59 & 5.59 & 5.59 & 5.59 & 5.59 & 6.23 & 6.23 & 6.23 & 6.23 & 6.23 & 5.99 & 5.99 & 5.99 & 5.99 & 5.99\\
\hspace{1em}LM & 5.23 & 5.59 & 5.83 & 5.95 & 6.68 & 4.54 & 5.01 & 5.31 & 5.57 & 5.42 & 5.64 & 6.10 & 6.40 & 6.52 & 7.28 & 4.56 & 5.04 & 5.40 & 5.82 & 5.85\\
\hspace{1em}RF & 5.08 & 5.33 & 5.45 & 5.51 & 5.41 & 4.45 & 4.69 & 4.76 & 4.87 & 4.83 & 5.37 & 5.73 & 5.95 & 5.92 & 5.91 & 4.53 & 4.94 & 5.07 & 4.91 & 5.03\\
\hspace{1em}SVR & 5.07 & 5.48 & 5.64 & 5.58 & 6.18 & 4.41 & 4.76 & 5.00 & 4.75 & 4.89 & 5.26 & 5.70 & 5.90 & 5.86 & 5.87 & 4.59 & 4.92 & 4.92 & 5.00 & 5.02\\
\addlinespace[0.3em]\hline\hline
\multicolumn{21}{l}{\textbf{Subset 4}}\\
\hspace{1em}Baseline & 5.96 & 5.96 & 5.96 & 5.96 & 5.96 & 4.65 & 4.65 & 4.65 & 4.65 & 4.65 & 6.39 & 6.39 & 6.39 & 6.39 & 6.39 & 5.24 & 5.24 & 5.24 & 5.24 & 5.24\\
\hspace{1em}LM & 5.04 & 5.46 & 5.93 & 6.09 & 6.55 & 4.21 & 4.53 & 4.81 & 5.10 & 5.26 & 5.35 & 5.93 & 6.35 & 6.47 & 7.15 & 4.57 & 4.75 & 5.04 & 5.48 & 6.14\\
\hspace{1em}RF & 5.03 & 5.42 & 5.67 & 5.67 & 5.64 & {4.09} &  {4.06} &  {4.15} &  {4.18} &  {4.23} & 5.21 & 5.72 & 6.09 & 5.96 & 5.84 &  {4.32} &  \hlc[green]{4.56} &  {4.62} &  {4.60} &  {4.99}\\
\hspace{1em}SVR & 4.94 & 5.36 & 5.63 & 5.58 & 6.04 & 4.39 & 4.63 & 4.89 & 4.51 & 4.97 & 5.05 & 5.56 & 5.75 & 5.78 & 5.77 & 4.53 & 4.78 & 4.80 & 4.89 & 5.06\\
\bottomrule
\end{tabular}
		\end{adjustbox}
\end{table*}

Table~\ref{table:pred_results} shows the numerical results in both cross-validation and testing phases.
We observe that the MAE of prediction generally reduces from top (Subset1) to bottom (Subset4). For instance, RF model gives MAE of $5.29$ on Subset1 and $4.56$ on Subset4 for predicting the departure rate with 5-day horizon (highlighted in green). This improvement can be ascribed to the increase in size of training data, \ie 1920 and 3360 instances for Subset1 and Subset4, respectively. It also suggests that with more data, overfitting can be minimised and our models generalise better.

To visualise cross-validation results, we plot Fig.~\ref{fig:validation_mae} to illustrate the MAE for the three ML algorithms under consideration. For short term forecasting, we can see that all three algorithms exhibit similar performance, with SVR yielding slightly better results for 1-day ahead prediction (average MAE of $5.17$ and $5.31$ for arrival and departure rate respectively) and RF model performing marginally better over 5 days arrival rate (average MAE of $5.61$).  For longer term prediction (10+ days), where predictive performance is observed to deteriorate, RF (blue lines with cross markers) is seen to outperform the other two algorithms for predicting arrivals while SVR (green lines with circle markers) yields the best results in predicting departures. On the other hand, LM yields the worst predictive power, suggesting that the parking rate data can be better explained using tree-based (\ie RF) or non-linear (\ie SVR) models as opposed to a linear one.

\begin{figure}
	\centering
	\includegraphics[width=0.7\textwidth]{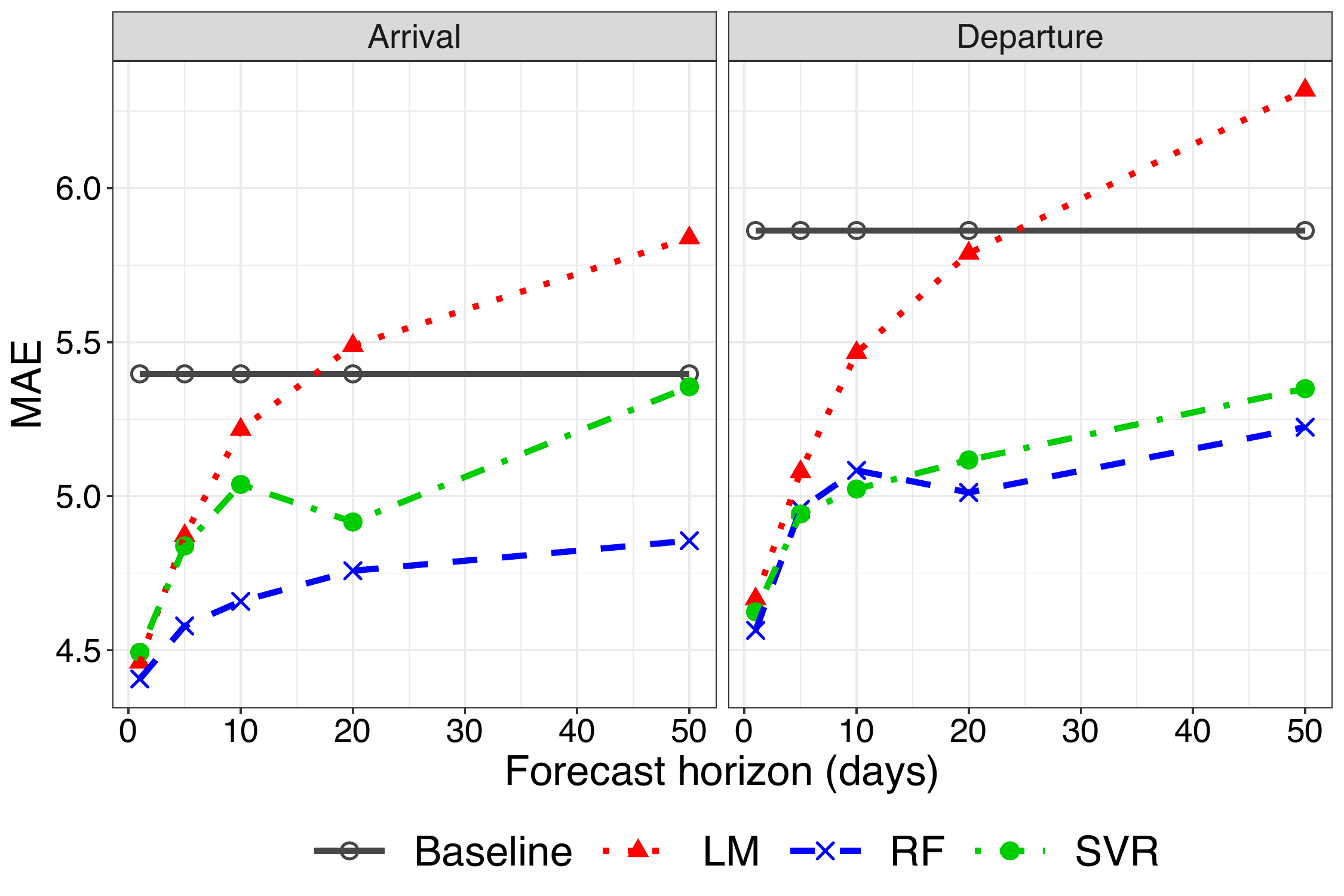}
	\caption{Performance of forecasting models (averaged over the four subsets) during testing phase.}\label{fig:prediction_mae}
\end{figure}

Next, we evaluate the performance of our models on the test set (unseen data). Fig.~\ref{fig:prediction_mae} shows the average MAE across all four subsets. In general, we can see that the forecast errors rise as the time horizon increases. For instance, the average MAE of departure prediction is as low as $4.67$ for LM model when predicting at 1 day forecast horizon, but increases to $6.32$ when predicting at 50 day horizon. This highlights the fact that our models learn the pattern better when more recent observations are available.
In predicting the arrival rates (left plot), RF is the clear winner with an MAE lower than $5.0$ across all the horizons. To predict the departures, however, both RF and SVR models perform fairly similarly.
Corroborating the validation results, LM yields the highest MAE for both arrival and departure among the three algorithms and even worse than the baseline for time horizons greater than 20 days.

From the results in Fig.~\ref{fig:prediction_mae}, we can see that RF and SVR models perform better in forecasting parking demand on our university campus, with an average prediction error of less than 5 cars per-hour for a medium-term horizon (5-day), and less than 5.5 cars per-hour for a longer-term horizon (10-week).

Note that the rate of car arrival and departure varies significantly (between 0 and more than 200 cars per hour) throughout the day. Therefore, we analyse the dynamics of MAE (for the RF model) during non-peak hours versus peak hours for a medium-term prediction horizon. During non-peak hours, with an average rate of 18 cars (entering/exiting) per hour, the model yields MAE of $3.60$ for both arrival and departure rates. We observe that more than half of the predictions come with an absolute error of  less than $2$. Also, it is found that about $20$\% of predictions give an error of more than $5$ -- these high errors predominantly correlate with a sudden change in parking usage due to irregular events/functions organised on campus during evening hours (between 4pm and 11pm), and hence can not be captured by our model. 

During peak hours (8am-10am for arrival and 4pm-7pm for departure), when an average of $180$ and $130$ cars per hour enter and exit the car park, the model yields MAE of $13$, translating to about 10\% of the total traffic. Note that large errors during peak hours are predominantly observed during non-teaching period when academics have flexible working arrangements, and hence a higher uncertainty in the parking usage is introduced.
We note that building RF and SVR models incurs an overhead of tuning their hyper-parameters, while MLR gives a closed-form solution with no hyper-parameters.

As previously mentioned, our existing models do not capture special events hosted on campus that could have a significant impact on parking demand. When such data is available publicly, this attribute should be considered as part of the model features in future work. Alternatively, the information could be used to flag the uncertainty of the predictions.

\section{Optimisation of Carpark Partitioning}\label{section:optimization}


As car sharing continues to grow in popularity, especially with new emergence of service models like one-way car sharing, university estate managers may want to set aside some of car park spaces to accommodate shared vehicles. Such a forward looking strategy is likely to increase the utilisation of campus parking facility, and hence generate new revenue. However, it is important to maintain the satisfaction of both private and shared vehicle users by minimising situations where they are turned away due to lack of parking spots (rejected users).

Since parking usage is highly seasonal depending on hour-of-day and day-of-week (\S\ref{section:analytics}), we envision a novel partitioning scheme that dynamically allocates a certain fraction of total spaces to car sharing vehicles, as opposed to the existing static allocations. This dynamic allocation can be adjusted using predicted demands (\S\ref{section:prediction}), so that rejected users are minimised while a better utilisation of the entire facility is achieved.
In order to highlight the value of dynamic partitioning (driven by ML-based forecast), we compare its efficacy with a baseline static approach. 

This section describes our methodology to decide an optimal space partitioning. We first develop a Markov model that  estimates the number of rejected cars for a given partitioning scheme (fraction of cars allocated to car sharing companies). We then formulate an optimisation problem to formally choose an optimal scheme based on minimising the total cost of rejections for both types of users, \ie existing private-vehicle (PV) users and prospective shared-vehicle (SV) users.  Finally, we evaluate the efficacy of our Markov modelling and optimisation framework via simulation of real usage behaviours recorded through our IoT parking system discussed in \S\ref{sec:datacollection}.

\subsection{Carpark usage as a Markov Process}\label{section:markov}

\begin{figure}
	\centering
	\includegraphics[width=0.7\textwidth]{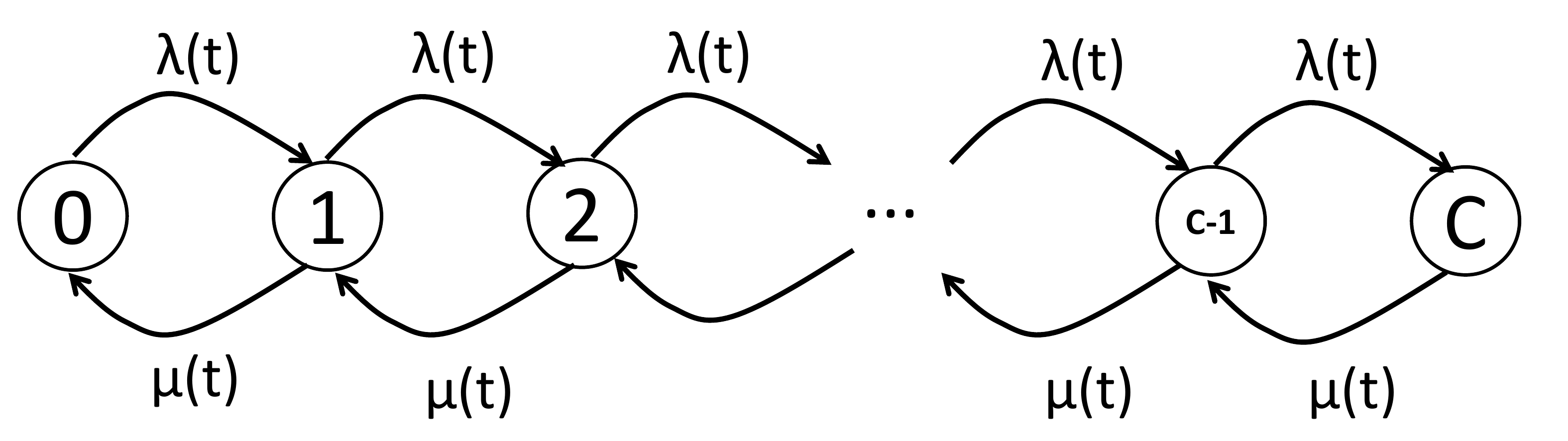}
	\caption{Usage dynamics of carpark with capacity $C$ -- cars arriving at rate $\lambda(t)$, and departing at rate $\mu(t)$.}\label{fig:markov-chain}
\end{figure}

\begin{figure}
	\centering
	\includegraphics[width=0.7\textwidth]{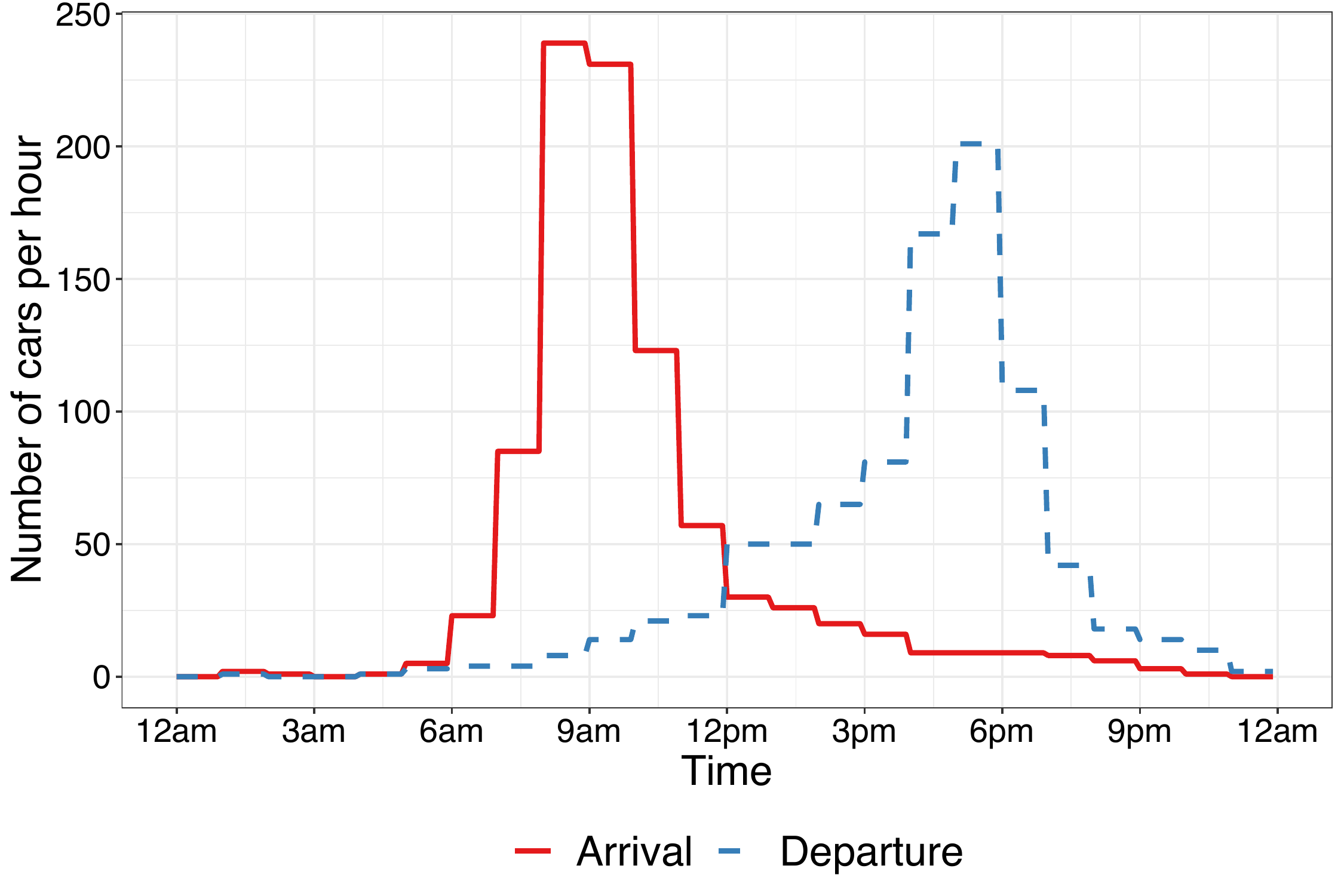}

	\caption{Hourly rate of arrival and departure on 11 Feb 2019.}
	\label{fig:lambda_mu}
\end{figure}

A car park can be abstracted as a queuing system, with an incoming and outgoing cars considered as arrival and service events respectively, and the number of parking spaces as the queue capacity. We therefore model the dynamics of car park usage as a M/M/1/C queue whose operation is visualised in Fig.~\ref{fig:markov-chain} by a continuous-time Markov chain. Each state of this chain represents the number of occupied spaces in the car park which has a total capacity $C$. An arriving car to the park at state $C$ (full) is rejected. 
An hourly arrival and departure profile on an example weekday (11 Feb 2019) is illustrated in Fig.~\ref{fig:lambda_mu}, where arrival rate peaks at 240 cars during 8am-9am and departure rate peaks at 200 cars during 5pm-6pm.


%
%
%

In continuous-time Markov process, transition rate matrix $Q$ can be used to calculate the transition probabilities at any time $t\geq0$. The $Q$-matrix, with dimension $(C+1)\times(C+1)$,  consists of elements $q_{i,j}$ denoting the transition rate from state $i$ to state $j$. 
For Poisson arrival and departure rate of $\lambda$ and $\mu$, the $Q$-matrix is represented by:

%

\begin{equation}
Q  = \begin{pmatrix}
     -\lambda 	 	&        \lambda			&       						&             	 	&              \\
        \mu    	  	& -(\lambda + \mu)	& \lambda 				& 			     	&          		\\
          	  			&		\ddots				 					& \ddots					&  \ddots   	&          \\
        					&     							&    	     \mu	 		& -(\lambda + \mu   )  & \lambda \\
          				&     							&    			 	& \mu  			& -\mu \\
    \end{pmatrix}
\end{equation}

Kolmogorov backward equations are used to characterize stochastic processes. In particular, they describe how the transition probability $P(t)$ that a stochastic process is in a certain state changes over time by differential equations \cite{anderson2012continuous}:



\begin{equation}
P'(t) = P(t)Q
\end{equation}

The solution of the differential equations is given by:  
\begin{equation}
P(t) = e^{Qt} = \sum_{n=0}^{\infty}\dfrac{(Qt)^{n}}{n!}
\end{equation}

In order to compute an estimate of the solution above (infinite sum of matrix powers), we use Krylov subspace projection method \cite{sidje1998:expokit}  which is capable of coping with sparse matrices of very large dimension.
The probability of the Markov process being at state $n$ can be obtained by $\pi(t) = \pi(0)P(t)$ where $\pi(0)$ is the initial state vector. 

Since a daily profile of transition rates is a step function (piece-wise constants) of hourly slots, as shown in Fig.~\ref{fig:lambda_mu}, our matrices Q and P become time-varying and hence are represented by $Q_{k}$ and $P_{k}$, with $k \in \mathbb{Z}:{0 \leq k \leq 23}$, each corresponding to an hour-slot.
Note that, we aim to track the dynamics of state probabilities over time during the forecast horizon. We first discretise time into fixed-size epochs, $\delta t$, each of 5 minute duration. 
We update fine-grained real-time state probabilities after every epoch $\delta t$ (probability matrix $P_{k}$ remains consistent across all epochs of each hour-slot) by:

\begin{equation}
\pi(t_{k} + n\delta t) = \pi(t_{k} + (n-1)\delta t)  P_{k}
\end{equation}

where $n \in \mathbb{Z}:{1 \leq n \leq 12}$ for 12 epochs (each 5-min) in an hour-slot, and $t_{k}$ is the timestamp at the beginning of hour-slot $k$.


To quantify car park user experience, we consider the number of rejected users, \ie those who cannot find a parking spot since their allocated partition is full at their time of arrival. In our Markov model, we compute the number of rejections after each epoch time by: 

\begin{equation}
r(t_{k} +n\delta t) = \pi_{C}(t_{k} + (n-1)\delta t) \times \lambda(k) \times \delta t
\end{equation}

where $ \pi_{C}(t_{k} + (n-1)\delta t)$ is the probability of the car park being full at time $t_{k} + (n-1)\delta t$ within the hour-slot $k$ and $\lambda(k)$ is the predicted arrival rate over that hour-slot. 
We accumulate rejected cars per each epoch across a day, and feed the daily count of rejected cars into  our optimisation formulation, explained next.

\subsection{Optimisation Formulation}\label{section:opt-formulation}

The goal of our optimisation is to select the best partitioning scheme (\ie what fraction of capacity to allocate to car sharing services) that can minimise the cost of rejected users, while maintaining a certain level of revenue from space leased to car sharing companies. We formulate an optimisation problem over a period of $D$ days.


Let there be $P$ capacity partitioning schemes and $D$ number of days within the scope of our optimisation. The number of rejected shared-vehicle and private-vehicle users for selecting scheme $j$ on day $i$ are respectively denoted by $r^{SV}_{ij}$ and $r^{PV}_{ij}$, where $1 \leq j \leq P$ and $1 \leq i \leq D$. Further, each scheme is identified  by a number of spaces allocated to car sharing users, this number is denoted by $s_{j}$.

Our decision variable is denoted by  $x_{ij}$, indicating if scheme $j$ is selected on day $i$. This can be represented by the following equation:

\begin{equation} \label{eq:input_var}
x_{ij} = \twopartdef { 1 } {\  scheme~{j}~is~selected~on~day~{i} } { 0 } {\ otherwise }
\end{equation}

We assign a constant cost for each rejected customer as $W^{SV}$ for shared vehicles and $W^{PV}$  for  private vehicles. Since our aim is to minimise the total cost of space allocation across $D$ days, the objective function can be written as:

\begin{equation} \label{eq:cost_func}
\begin{aligned}
\min \quad &  \sum_{i=1}^{D} \sum_{j=1}^{P}\bigg\{{x_{ij} \times ( W^{SV} r^{SV}_{ij}  +W^{PV} r^{PV}_{ij} ) }\bigg\}\\
\end{aligned}
\end{equation}

On a given day, one partitioning scheme will be selected, and hence the following constraint: 

\begin{equation} \label{eq:1PartPerDay}
\sum_{j=1}^{P} {x_{ij}} = 1\  \forall\ i
\end{equation}

Furthermore, campus managers would expect a minimum amount of revenue to be generated by leasing out parking spaces to car sharing companies in order to cover the investment made for implementing such an allocation scheme. 
Given a daily dollar price $M$ for each space leased out and the minimum revenue $R$ over $D$ days expected by the University, the revenue constraint is given by:

\begin{equation} \label{eq:Rev_constraint}
\sum_{i=1}^{D} \sum_{j=1}^{P} {M x_{ij}s_{j}} > R
\end{equation}


We set our optimisation period $D$ to 5 working days (1 week) as we can obtain a fairly accurate prediction of demand at this forecast horizon in \S\ref{section:prediction}.
We use a set of constant parameters in our optimisation that are summarised in Table~\ref{table:opt_constant}. The renting price (for car sharing companies) of each parking space  is estimated by converting the annual market price for car space lease in city of Sydney \cite{PBA:carsharingreport} to a daily rate. The cost of rejecting a shared vehicle user is considered to be equal to opportunity cost of losing the rent per car space, and the cost of rejecting a private user is equal to daily parking fee on our university campus \cite{UNSW:parkingfee}. Also, the revenue from parking space lease ($R$) is expected to at least cover the investment cost of operating a space sharing scheme. This includes (a) cost of planning \& administration: 0.5 full-time equivalent (FTE) staff at rate $\$100,000$ per annum, (b) cost of system installation and maintenance: $\$ 1,900$ per annum per car space, and (c)  5\% safety margin.


\begin{table}[t!]
	\begin{center}
		\caption{Constant parameters of our optimisation scheme.}
		\label{table:opt_constant}
		\vspace{-2mm}
		\begin{tabular}{ l p{3cm}  p{7cm}  }
			\toprule 
			Parameter      &  Value  & Description \\
			\midrule
			$M$   			&	\$15.8 /car /day		& Renting price per car space, computed from	annual market rate for city of Sydney \cite{PBA:carsharingreport}.\\ 
			\addlinespace[0.2cm]
			$W^{SV}$  	&	 \$15.8 /car	&	Opportunity cost of rejecting a shared-vehicle user is equal to the lease price per space.\\			
			\addlinespace[0.2cm]						 
			$W^{PV}$  	&	\$26 /car		&	Opportunity cost of rejecting a private-vehicle user is equal to day permit rate for parking on campus \cite{UNSW:parkingfee}.	\\ 
			\addlinespace[0.2cm]
			$R$  			&	\$36468.75 /week		&	Minimum revenue to cover the fixed cost of operating a Space Sharing Scheme, including planning \& administration (0.5 FTE staff at \$100,000 annual rate) +  system installation and maintenance (\$1,900/space/annum) + 5\% safety margin. \\ 
			\bottomrule
		\end{tabular}
		\vspace{-4mm}
	\end{center}
\end{table}

\subsection{Evaluation Results}
To simulate the dynamics of shared vehicles, we assume that 20\% of the existing users subscribe and use shared-vehicles (SV) service on a regular basis and 80\% of the current users continue using their private vehicles (PV) to commute to the University. 
Furthermore, car sharing have been found to serve other uses in addition to daily work commute, including shopping, business, and leisure \cite{becker2017comparing}. Hence, we further assume additional demand of SV users by considering a constant usage profile sampled from a binomial distribution with hourly $\lambda$ rate of $B(200, 0.5)$ and hourly $\mu$ rate of $B(200, 0.4)$. This corresponds to an assumption of 200 new subscribers with 0.5 and 0.4 probability of entering and exiting the car park in each 1-hour slot.

\begin{figure}
	\centering
	\includegraphics[width=0.7\textwidth]{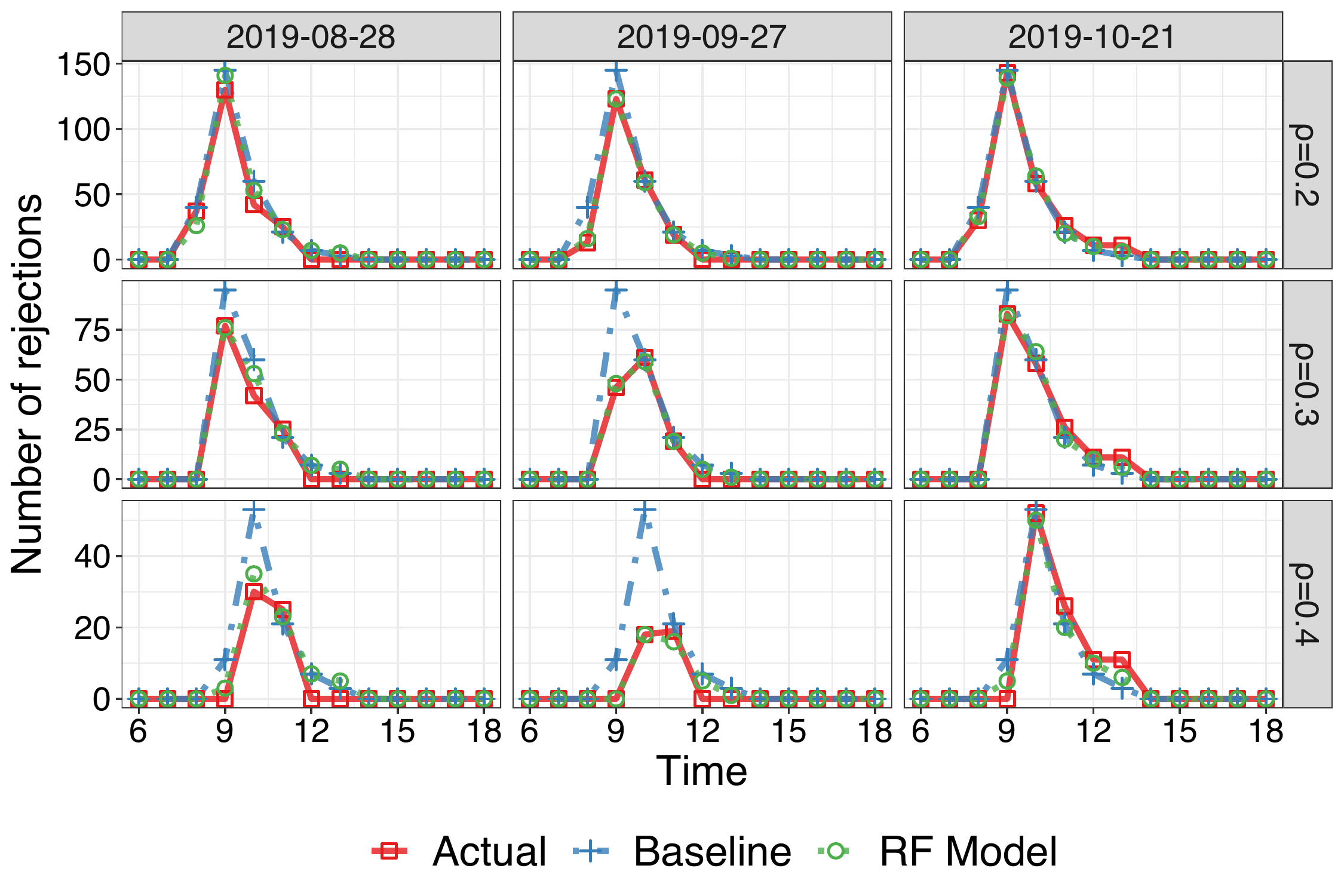}
	\vspace{-2mm}
	\caption{Time-trace of hourly rejections across three representative days for tighter allocation policies ($\rho=0.2$, $\rho=0.3$, and $\rho=0.4$) by which PV rejections may occur.}\label{fig:num_rejs_perday}
	\vspace{-3mm}
\end{figure}

Since we defined the scope of our allocation scheme as 1-week (equivalent to 5 working days), the 5-day ahead demand forecast obtained from the best performing model (RF) from Section~\ref{section:prediction}, are used in our optimisation formulation, where partitioning of spaces is done dynamically. In addition, the results are compared with static partitioning using the forecast demand from the baseline prediction model, where arrival and departure profile is constant for each day.

Using the aforementioned assumptions and data, we first evaluate the accuracy of our Markov model in estimating the number of rejection per each partitioning scheme. We then use the computed rejection number to find the optimal space partitioning scheme by applying the optimisation formulation from \S\ref{section:opt-formulation} and compare rejection cost incurred from both static and dynamic partitioning.

\subsubsection{Evaluation of Markov Model}
We evaluate the performance of our Markov model in estimating the total number of rejected PV users per day by comparing the output from Markov model to the actual rejected number calculated from the actual arrival and departure count (obtained from sensor data).

Fig.~\ref{fig:num_rejs_perday} compares hourly rejection number computed from baseline count data (blue line) and forecast count data from RF model (green line) with the actual rejections (red line). The results are shown across 3 different capacity partitioning schemes (row wise) where $\rho$ indicates the fraction of parking spaces allocated to PV users, for 3 example days. 
From observations, we can see that for $\rho=0.2$,  predictions from baseline and RF model can give a decent estimate of the rejections number with all the three lines aligning closely. On the other hand, for $\rho=0.4$, we can clearly see that rejection profile calculated using the baseline model stays constant across the 3 example days due to its static arrival and departure rate profile, while the actual rejection trend has changed across the days. This illustrates an example where the superior predictive power from RF model for usage demand can lead to a better approximation of rejection numbers.

\begin{figure}
	\centering
	\includegraphics[width=0.7\textwidth]{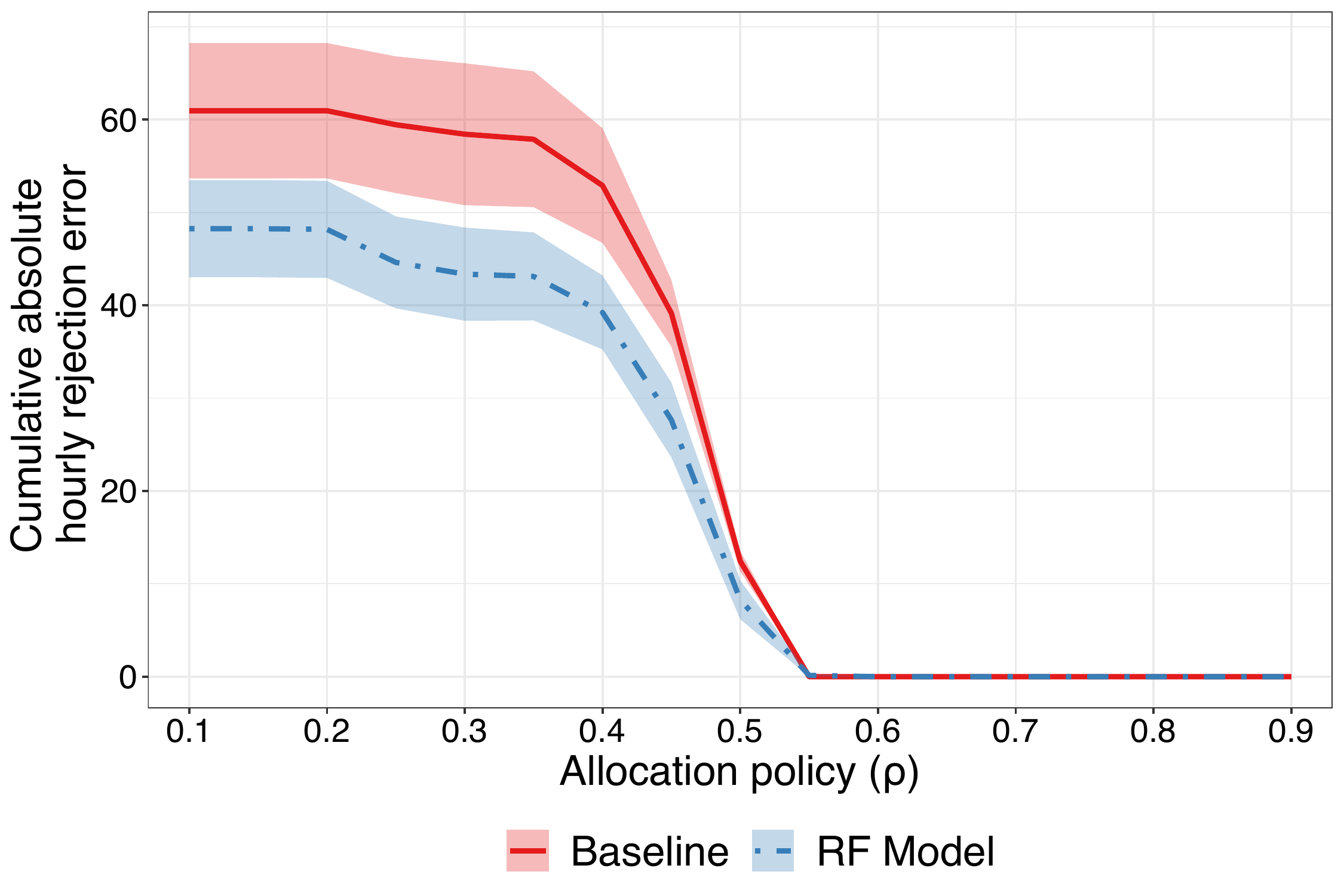}
	\vspace{-2mm}
	\caption{Sum of absolute hourly rejection errors per day (average) as a function of allocation policy $\rho$.}\label{fig:refs_error_perconfig}
	\vspace{-3mm}
\end{figure}

In order to quantity the outcome of the Markov modelling, we accumulate the absolute hourly error of rejections across each day, and obtain the daily average error. The average daily sum of errors (PV user case) for both baseline and RF model with 95\% confidence interval of mean (area bounding the line) are shown in Fig.~\ref{fig:refs_error_perconfig} for different capacity partitioning schemes ($\rho$).  Undoubtedly, the errors decrease as the fraction of spaces allocated to PV users increases due to the lower number of rejected users. From the graph, we can see that the error becomes zero when $\rho$ reaches 0.55 as there are sufficient spaces allocated to accommodate the PV users demand. By comparing the two lines, it is evident that the rejection error is lower for RF model (blue line) when compared with baseline model (red line) across all partitioning schemes. The highest average daily errors for RF model is 48.25 cars while baseline model 60.95 when $\rho=0.1$.


%

\subsubsection{Evaluation of Partitioning Scheme Optimisation}
We employed Mixed Integer Linear Programming (MILP) algorithm, which is conducive for a problem that has a linear objective function subjected to a set of linear constraints, to solve our optimisation problem described in Section~\ref{section:opt-formulation}. We calculated the total cost incurred from the optimum allocation scheme across each week for the two demand prediction: one using the predicted demand from the baseline mean model (static partitioning) and another using the 5-day ahead forecast from RF model (dynamic partitioning). In addition to the minimum cost obtained from the optimum allocation, we calculated the actual cost incurred if the optimum partitioning scheme was selected.

\begin{figure}
	\centering
	\includegraphics[width=0.7\textwidth]{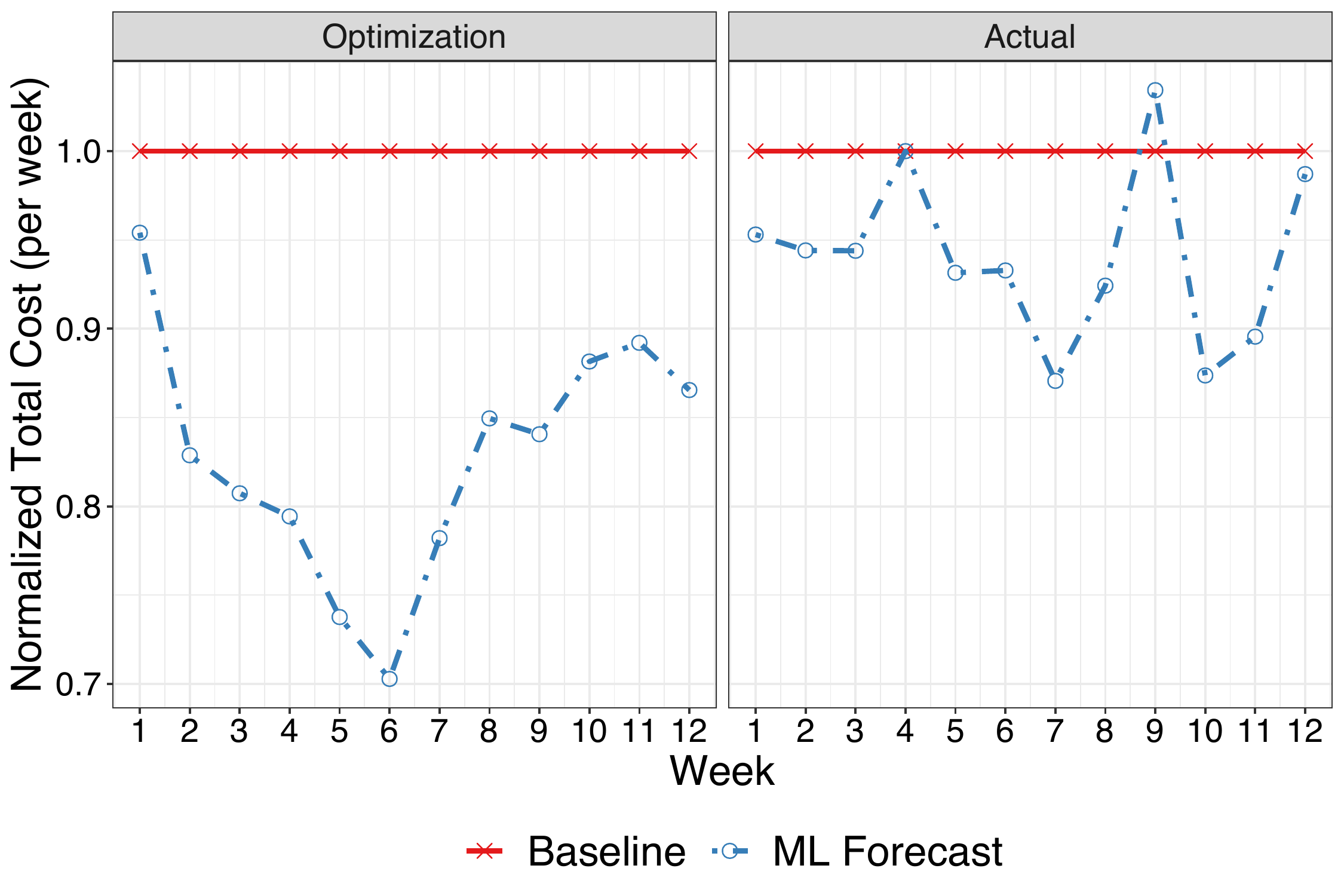}
	\vspace{-2mm}
	\caption{Weekly (normalised) cost:  dynamic (ML forecasting) versus static (baseline forecasting) partitioning. }\label{fig:compare_opt_cost}
	\vspace{-3mm}
\end{figure}

Fig.~\ref{fig:compare_opt_cost} compares weekly total costs for the baseline (red) and ML forecast (blue) approaches - the costs were normalised with respect to the baseline case. The left plot represents the expected cost from the optimum partitioning scheme, with the actual cost incurred shown on the right plot. 
By looking at the minimum cost obtained from our optimisation (left plot), it is apparent that the ML forecast approach yields lower cost compared to the baseline approach. This suggests that the variation in usage demand, which can be captured through ML forecasting, can provide flexibility in allocating the partitioning scheme and hence yield a lower optimum cost.


When examining the actual cost incurred from the selected partitioning scheme (right plot), we can see that there is a smaller gap between the ML and baseline approaches, indicating lower actual cost savings than expected. In particular during Week 9, the cost saving from dynamic partitioning using ML forecast is expected to be 16.74\% but the actual incurred cost is higher by 3.43\%. The reasons for this discrepancy is due to errors from predicting parking demand (arrival and departure rate) which can be attributed to the inaccuracy in approximating rejections. For instance, we have one particular day during week 9 where the ML predicted the rejections to be significantly greater than reality, leading the optimisation to select an alternate partitioning scheme that is not the optimum choice in reality and hence magnifying the actual objective function cost.


The results of our analysis show that dynamic dimensioning of car park spaces can provide a better user experience when compared to the static partitioning approach, with an average of $6.3$\% lower weekly customer rejections cost. 
Our proposed framework shows a potential benefit of dynamic allocation for future car park dimensioning which can be adopted by universities with an aim to optimise the usage of their parking facilities while keeping up with the evolution of shared mobility market.


\section{Conclusion}\label{sec:con}
Digital transformation in transport industry demands large organisations like universities to revisit the operation of their expensive on-campus parking facilities. With the rapid prevalence of IoT technologies, universities can benefit from the immense amount of data generated from smart devices, assisting them with planning and restructuring the usage of their parking facility. 
In this chapter we have outlined our experiences in designing and deploying a monitoring system for a real car park on our university campus. We collected data over 15 months (covering both teaching and non-teaching periods) and cleaned it for analysis. We then analysed the usage data and highlighted insights into car arrival and departure patterns as well as users parking behaviour. Finally, we developed a novel framework for the university to optimally partition their parking infrastructure through dynamically allocating a fraction of parking spaces to car sharing operators. We showed that the University can reduce the cost of customer rejections by 6.3\% per week when adopting dynamic allocation using predicted arrival and departure rates.



\chapter{Optimising Bus Scheduling by Passenger Demand Prediction}
	\minitoc

Anecdotal evidence has shown that bus stops around our University campus, specifically those serving express buses to city center, can get very crowded during certain times. This not only causes immense frustration to students who experience large variations in wait-time, but also creates challenges for the university and transport authority in knowing when to schedule extra buses. 
This chapter applies the three pillars Smart Campus Framework to enhance the efficiency of on-campus bus schedule and passenger experience. We begin by developing a LoRaWAN ultrasonic sensor for detecting people in the queue. The sensor emits an ultrasonic tone pulse every few seconds, and determines whether someone is in front of it based on the reflections received, if any. Ten sensor units are built, tested, and tuned in a lab environment to achieve optimum detection accuracy and data transmission rate. Next, we install these sensors at 6-meter spacing along the campus fence bordering the bus-stop. We develop an algorithm to infer number of passengers in the queue from sensor data and demonstrate that it achieves reasonable accuracy with mean absolute error of 10.7 people (for a queue size of up to 100 people). Finally, we develop an optimisation model to reschedule bus dispatching time, aiming to minimise total wait time of passengers. We show that a reduction of up to 50\% in passengers' wait time can be achieved by adopting demand-driven bus scheduling.

\section{Introduction}
\vspace{-2mm}
On a typical weekday, thousands of people would arrive and depart the campus at various times of day. A vast majority of these would avail of public transport services for their daily commute. 
Transit stops at universities are notorious for overcrowding during peak hours \cite{daniels2013:paradox}. It is not uncommon for commuters to wait for a second or even a third service to arrive before there is sufficient room to board.

Prolonged waiting is highly undesirable on multiple fronts including loss of productivity, fatigue, tiredness and discouragement for using public transport. In an effort to assuage this problem, many public transport providers have deployed real-time vehicle tracking services, whereby the arrival time of the next service is displayed in real-time on a display installed at transit stops or through mobile applications. During peak times, the passenger demand may outstrip the available capacity on the next arriving service, resulting in many passengers having to wait for longer. A system that can measure the length of the queue at transit stops in real-time would therefore be favourable for passengers, by allowing a better estimation of passenger boarding time and aiding their decisions on transit stop selection and trip start time. The ability to estimate passenger demand would also enable a pragmatic approach to transit scheduling, which would lead to efficient transit resource management and thus an enhancement in commuter transit experience.

However, monitoring passenger queue length and wait time at transit stops still remains a challenge, especially in an outdoor scenario where power and wired connectivity are usually inaccessible, thus hindering sensor deployment efforts. Many studies have proposed mobile detection and crowd-sensing based approaches \cite{okoshi2015:queuevadis,li2014:queuesense} to measuring queue size or dwelling time of people, yet the solutions have only been tested and deployed in an indoor environment. Furthermore, such systems would require a fair fraction of queue members to actively participate in order to achieve a good accuracy. Existing works related to measuring crowding in an outdoor scenario are mostly camera based \cite{weppner2013:bluetooth}, and if applied to queue measurement can raise privacy concerns.

In this chapter, we propose a novel end-to-end sensor-based system for measuring queue length in an outdoor setting and showcase how the collected data on passenger demand can enhance operational transit scheduling in order to make a more efficient use of bus resources and improve public transit experience for the campus community. Our specific contributions are three-fold:

\begin{enumerate}
\item we design and implement people detector devices using ultrasonic sensing. The solution is weather resistant, battery-operated, and communicates wirelessly, allowing outdoor deployment where access to power and communications ports are infeasible. The sensors uses LoRaWAN for data communication by utilising a public LoRaWAN base-station located on our campus. We test and tune the sensor and LoRaWAN-based parameters to achieve optimum detection sensitivity and superior data delivery ratio for our deployed environment while achieving low energy consumption.
\item We deploy ten people detector devices at one of the busiest bus stops on campus. The sensed data and empirical observations are used to develop an algorithm to infer queue length. Our method yields a reasonable accuracy with Mean Average Error (MAE) of 10.7 people for a queue size of up to 100 people.
\item We develop an optimisation model to reschedule bus dispatch times based on actual transit demand in order to minimise total wait time of passengers at the stop. Our result suggests that 42.93\% passenger wait time reduction can be achieved without requiring the scheduling of additional buses.
\end{enumerate}

To the best of our knowledge, we are the first to employ an ultrasonic sensor-based solution using LoRaWAN to address the challenges in measuring queue length in an outdoor scenario. We use bus stops as an illustrative example for public transit stops. Our solution could also be applied to other transportation modes, and for any outdoor scenario (\eg stadiums, airports) where an orderly queue is formed. 

The rest of this chapter is organised as follows: In \S\ref{sect:queuesensing}, we present our design and implementation of people detector units. Our developed intelligent queue inference algorithm and deployment of the sensors are described in \S\ref{sect:Qlength}. We then show the potential of collected passenger demand in optimising transit bus scheduling in \S\ref{sect:queue_opt}. Finally, this chapter is concluded in \S\ref{sect:bus_conclusion}.

\section{Sensing People in Queue: Design, Implementation, and Tuning}\label{sect:queuesensing}

In this section, we describe our solution, including our choices of technology (ultrasonic sensing for people detection and LoRaWAN for communications), implementation of the people detector unit (PDU), and performance analysis and tuning. Note that we obtained appropriate ethics clearance (UNSW Human Research Ethics Advisory Panel approval no. HC180359) prior to conducting this research work.

\subsection{Design Decisions}\label{sec:decision}

\begin{figure}[t!]
	\centering
	\includegraphics[width=0.7\textwidth]{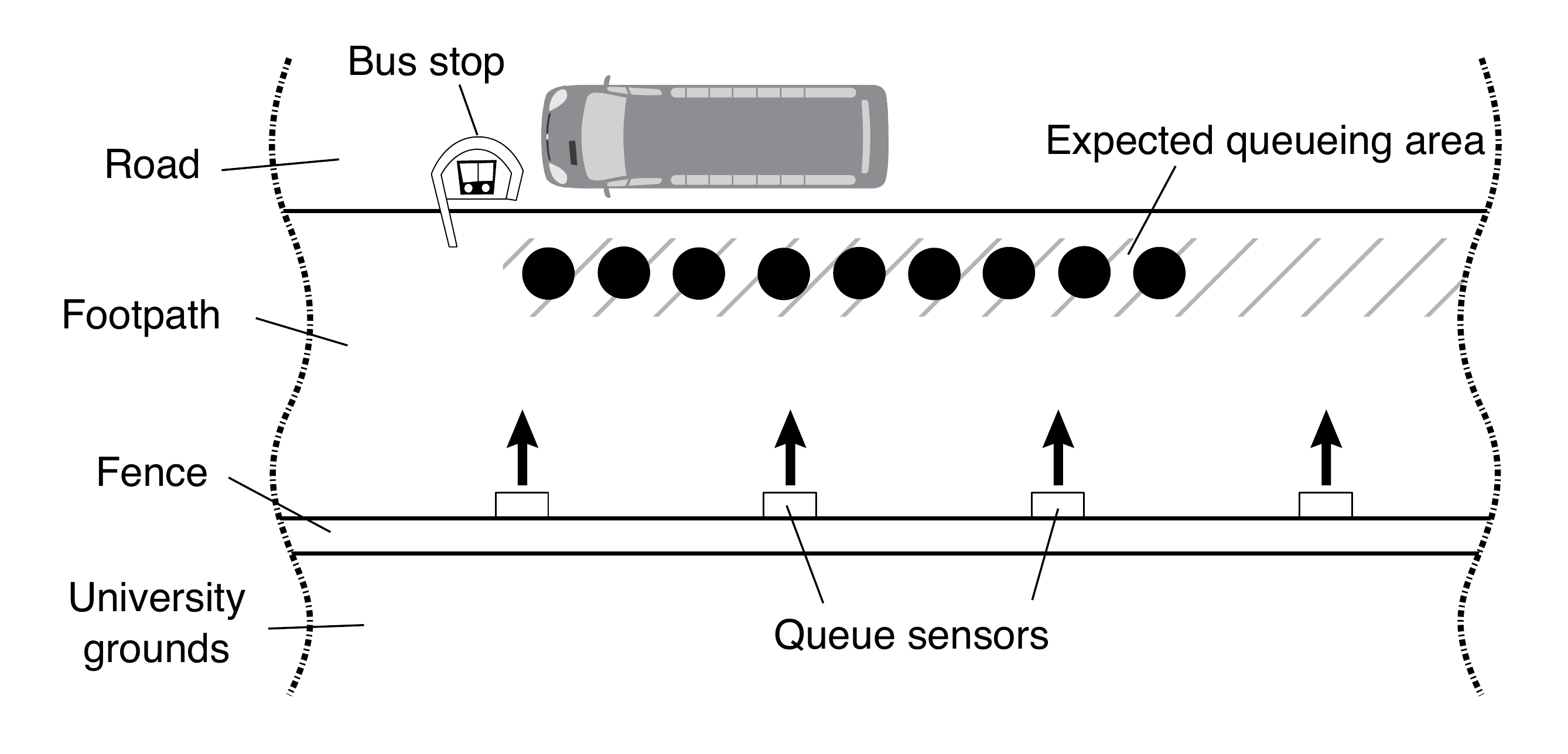}
	\caption{Intended installation locations of people detection devices at a bus stop.}
	\label{fig:busstop_topview}
\end{figure}

Our bus stop of interest is located outside our university campus. The formation of the queue is in a straight line, along the campus fence. Fig.~\ref{fig:busstop_topview} shows the presence of passenger queue relative to the road, bus stop, campus ground, and fence. The expected area where the queue forms is approximately at a distance of 2-3 meters from the fence and can on occasion extend beyond the edge of the fence line. 
Since power and communications points are not available within the proximity of the bus stop, we chose to use an array of small single spatial sensors rather than a single multi-spatial sensor (such as camera) due to portability and battery life requirements. Our choices of sensors and communications technologies are discussed next.

\subsubsection{Sensing Choices}\label{spaciality}

In what follows, we compare several sensing technologies for people detection as summarised in Table~\ref{tab:sensors_compare}. Criteria used for our comparison include whether the sensor is contact-based (contact based sensors require physical contact with queue members to take measurements, whilst non-contact sensors make their measurements from a distance, which is generally preferred to minimise risks of public interaction), whether the sensor is passive or active (active sensors do not require user participation to collect data as they actively sense by their own impetus, while passive sensors require conscious and deliberate effort from queue members to record readings), spaciality (whether or not the sensor returns a single reading from a single position in space, or a collection of readings spread over several areas in space - whilst more data is intrinsically more valuable for analysis, some multi-spacial sensing devices, such as cameras, collect so much data that privacy and transmission bandwidth become serious issues), 
privacy, cost, and power draw. 

\begin{table*}
	\begin{center}
		\caption{Summary comparison of sensor technologies for bus queue monitoring.}
		\label{tab:sensors_compare}
		\begin{adjustbox}{width=1\textwidth}
		\begin{tabular}{ l c c c c c c p{7cm} }
			\toprule
			\textbf{Technology} & \textbf{\makecell{Contact \\based}} & \textbf{\makecell{Passive/\\ Active}} & \textbf{Spacial}  & \textbf{\makecell{Privacy \\intrusive}}  & \textbf{\makecell{Unit cost \\(USD)}}  & \textbf{\makecell{Relative \\power draw}}  & \textbf{Other limitations} \\
			\midrule
			Scannable QR codes           & no  & passive & single & yes & 10\textsuperscript{-1} & None & Low participation, high abuse. \\
			Pressure pads                & yes & active  & either & no  & 10\textsuperscript{2}  & Low  & Council permission, tripping hazard. \\
			Passive infra-red            & no  & active  & single & no  & 10\textsuperscript{1}  & Low  & Only useful at night.\\
			Visible light (single-diode) & no  & active  & single & no  & 10\textsuperscript{0}  & Low  & Only useful at night.\\
			Visible light (camera)       & no  & active  & multi  & yes & 10\textsuperscript{2}  & High & Large data volumes. \\
			Microwave                    & no  & active  & single & no  & 10\textsuperscript{1}  & Low  & May require EMI approval. \\
			Laser time-of-flight         & no  & active  & single & no  & 10\textsuperscript{2}  & Low  & Expensive \& niche.\\
			Ultrasonic time-of-flight    & no  & active  & single & no  & 10\textsuperscript{0}  & Low  & \\
			WiFi session logs            & no  & active  & multi  & yes &  0                     & None & Unreliable, only covers certain WiFi users (not all queue members). \\ 
			\bottomrule
		\end{tabular}
		\end{adjustbox}
	\end{center}
\end{table*}

Based on the aforementioned selection criteria, we analyse various sensor technologies including QR code, pressure-pad, passive infrared, single-diode visible light, camera, microwave, existing WiFi infrastructure, and ultrasonic distance sensors.

\textbf{Scannable QR code} is a form of passive sensor that requires queue members to scan unique QR code posters placed along the length of a bus queue. This is a very cheap and flexible option as all the effort of data collection and communication is performed by participant mobile phones.  Unfortunately, the heavy reliance on sufficient and frequent participation from the users makes it difficult to obtain reliable data without an effective incentive mechanism and abuse prevention system.

\textbf{Pressure-pads} are flat devices that can be placed on the ground to detect the weight of users standing on them.  These devices are typically marketed in the aged care and burglar detection industries for installation under special floor mats, beds, and chairs \cite{arun_pressuremat}.  Unfortunately their installation on a footpath is complicated, requiring extensive approvals (council), mechanical protection (vandalism, weather, vehicles), and safety risk mitigation (tripping hazards).

\textbf{Passive infrared} sensors detect the infra-red light emitted by all warm entities, such as humans.  These sensors are typically placed on the walls of rooms to detect burglars.  Unfortunately these sensors cannot discriminate between people and other sources of heat (\eg vehicles,  sunlight), making them inappropriate for outdoor roadside installation and operation, especially during the day.

\textbf{Single-diode visible light} sensors only sense a single sample (or ``pixel'') of light.  Unfortunately there is no unique color or strength of light that is reflected by individuals standing in a queue to indicate their presence.  A transmissive-style sensor arrangement must instead be used, where a known light source is installed on one side of the queue and measured on the other side to see if it has been obscured.  This requires light sources to be installed on one side of the queue (\eg the road side) and sensors on the other (\eg the property side), requiring consideration of council approvals and safety (tripping hazards) for the general public.

\textbf{Camera visible light} sensors record significantly more data than single-diode sensors. A low-resolution of 640$\times$480 pixels camera already requires the transmission of 300,000 samples, which even with compression will still require significant time and energy. Additionally these devices collect personally identifiable information, which would require more complex data handling (privacy) and approval arrangements. 

\textbf{Microwave} sensors are similar to single-diode visible light sensors, however they have the advantage of being able to use a wavelength that is not naturally abundant.  This permits them to operate in a reflective style arrangement without suffering interference from other environment wave sources.  Unfortunately microwave sensors can be classified as ``intentionally radiating devices'' and many low-cost units do not come with emissions approvals from the relevant authorities, meaning they could be illegal to operate and/or interfere with the electronic devices of queue members. 

Using \textbf{existing WiFi infrastructure} session logs from existing WiFi infrastructure in the surroundings can potentially be used to determine queue occupancy without additional capital costs, however this method has severe accuracy and validity limitations.  Only users of the university WiFi systems will be measured. Moreover, it is difficult to determine exactly where users are relative to each access point (\eg in a bus queue or in a nearby cafe) and WiFi coverage can often be poor at bus stops as they are usually located at the edge of the campus.

\textbf{Ultrasonic distance sensors} and \textbf{laser time-of-flight} sensors take distance measurements between themselves and objects in front of them.  They operate by emitting a small amount of sound or light, and measure how long it takes for this wave to bounce off a nearby object and come back.  ``Ultrasonic'' waves are sound waves above human hearing, typically chosen to avoid irritating humans and to allow smaller physical sensor size.  As sound waves are many magnitudes slower than light waves, ultrasonic distance sensors are typically much simpler, cheaper, and available in more varieties than their laser counterparts.

\vspace{1mm}
\textbf{Summary}: We chose ultrasonic distance sensors because of their low power requirements, low cost, low implementation complexity and high likelihood of detecting the presence of individuals in a queue while not being adversely impacted by other objects in the environment.
Compared to light and infra-red solutions, they do not suffer interference problems (sunlight, car headlights, etc), and compared to solutions such as WiFi they are less likely to miss a fraction of queue members. Ultrasonic, whilst not a perfect sensing method, avoids most of the disadvantages of other sensing technologies.

\subsubsection{Communication Choices}

Data collected by sensor units must be communicated back to a central location for permanent storage and analysis. Real-time streaming of collected data permits immediate analysis and reporting.
Wired communication is infeasible due to its high infrastructure costs and its inflexibility for making positional adjustments. Thus we only consider wireless communications solutions for which we have a set of requirements including low power draw, simple to operate, long-range, low-cost, and multiple devices should be able to directly 
 communicate with a gateway. 
Note that high data-rate is not a requirement for our PDU, as the distance measurement data collected is at the scale of a few bytes per minute. In the following, we discuss several wireless communication options in detail.  A summary of our comparison is available in Table~\ref{table:comms_compare}.

\begin{table*}
	\begin{center}
		\caption{Summary comparison of communication technologies for bus queue monitoring.}
		\label{table:comms_compare}
		
		\begin{adjustbox}{width=1\textwidth}
		\begin{tabular}{c c c c c c c}
			\toprule
			\textbf{Technology name} & \textbf{Data-rates} & \textbf{Range} & \textbf{\makecell{Primary \\topology}} & \textbf{\makecell{Usage \\Complexity}} & \textbf{Power draw} & \textbf{Cost} \\
			\midrule
			WiFi               & Low to high & Medium & Star, P2P    & High & High & Low \\
			Bluetooth          & Low to med  & Short  & Master-slave & High & Low  & Low \\
			Wireless broadband   &   Low to high & Long   & Star    & High & Low  &    High   \\
			Zigbee, NRF24, etc & Low to med  & Short  & Various      & Low  & Low  & Low-Moderate \\
			LoRaWAN            & Low         & Long   & Star         & Low  & Low  & Moderate \\
			\bottomrule
		\end{tabular}
		\end{adjustbox}
	\end{center}
\end{table*}

\textbf{WiFi} (802.11) is a large collection of standards optimised for high data-rates and persistent sessions. Notable amount of time, along with exchanging of control messages is required to initiate wireless connections; hence demanding high software complexity and high power draw.  

\textbf{Bluetooth} is a collection of standards optimised for master-slave communications.  While this is optimal for operating simple devices such as microphones and speakers, it is not effective or simple for the operation of an arbitrary number of PDUs simultaneously.  Some methods allow for multiple client devices (``Bluetooth Piconet''). However, this requires further synchronisation complexity, and imposes a (low) total device limit of 7.

\textbf{Wireless broadband} systems such as 3G, 4G, and LTE can be optimised for low power signaling and sleeping after a connection is initially negotiated, however compliant radio modules are typically very expensive and complex to operate.  Many vendors require signature of non-disclosure agreements and external Linux-running microprocessors with proprietary drivers for these devices to function.

Proprietary low-power communications systems and modules such as \textbf{IEEE 802.15.4} (Zigbee) and \textbf{NRF24} provide low-cost, simple-interfaced and low-power solutions.  Unfortunately their range is very limited, and would require the installation of local (line-of-sight) reception towers near instrumented bus stops. 

\textbf{LoRaWAN} and \textbf{LoRa} are a set of low-power and long range communications standards. Radio modules are easily available with simple UART-style interfaces. The typical network topology is star, with an arbitrary number of low-power devices transmitting to a centralised (infrastructure) base-station for data collection. Base-stations themselves can either be shared (public) resources run by third parties or private infrastructure \cite{lorawan_spec}, in both cases the standards use an advanced encryption standard (AES) implementation to provide message security \cite{lorawan_security}.
Unfortunately LoRa is implemented on a wide variety and disparate set of frequency standards \cite{ttn_freqplans}, with many LoRaWAN radio modules on the market only supporting a single frequency band.
LoRaWAN radio modules that meet local regulatory requirements are also not necessarily cheap.

\textbf{Summary:} We chose LoRaWAN as the communications platform for this work because of its longer range, lower power draw, and simpler implementation complexity compared to other alternatives.

\begin{figure}[!t]
	\centering
	\includegraphics[width=0.7\textwidth]{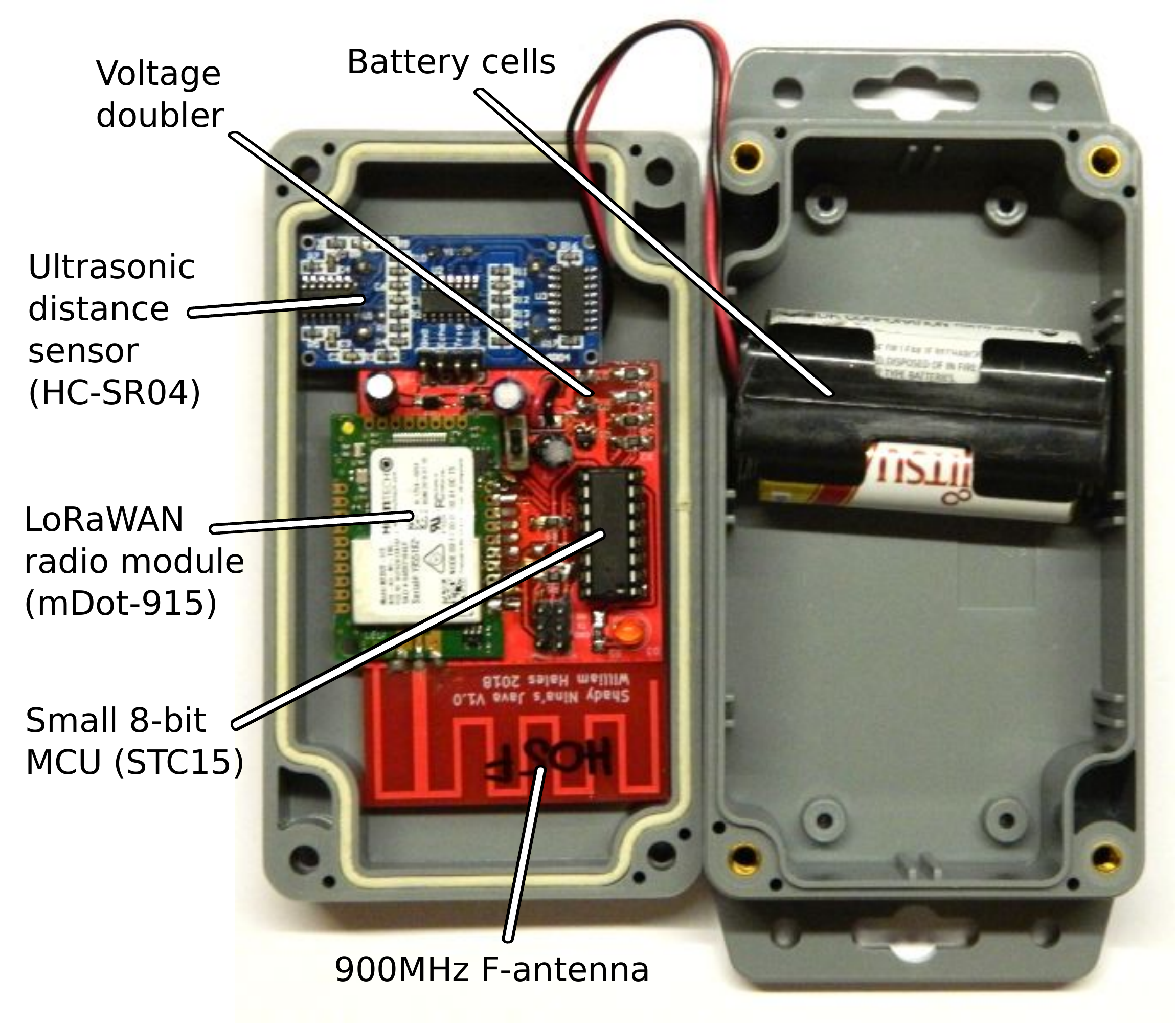}
	\caption{Interior of a PDU.}
	\label{fig:PDU_interior}
\end{figure}

\subsection{Implementing People Detector Units}\label{sec:pdu}


We designed and manufactured ten people detector units (PDUs) using Ultrasonic distance sensors and LoRaWAN radio modules. Fig.~\ref{fig:PDU_interior} shows the components of a PDU located in a weatherproof case.
Sensor units were designed to measure the presence of a bus queue as shown in Fig.~\ref{fig:busstop_topview}.  An array of small single-spatial sensors was chosen over a single multi-spacial sensor (such as a camera) due to portability and battery life requirements, as discussed in  \S\ref{spaciality}. Each component of the PDU and important design decisions taken during the implementation are described below:

\subsubsection{Control board}

A red circuit board containing a small 8-bit micro-controller and miscellaneous periphery components was used to coordinate each PDU. We adopted an ``wake-and-return-to-sleep'' design to maximise battery life. Each PDU spends as much time as possible sleeping, only waking to take brief measurements or transmit measured data.
PDUs are also configured to stop transmitting if they are unable to take any successful measurements. This greatly reduces the amount of transmissions when there is low occupancy at the stop, both increasing battery life and reducing interference to external devices using the same radio spectrum.
Furthermore, the message payload is optimised to reduce message length.  Distances are stored as 8-bit values with the resolution of distance measurement of 2 cm (permitting a theoretical measurement range of 0-510 cm). No checksum or other metadata are included in the payload, instead we rely upon the LoRaWAN protocols for reliability and error detection.


\subsubsection{Ultrasonic Sensor}\label{sect:built-ultrasonic}

The HC-SR04 ultrasonic distance sensor is a low-cost off-the-shelf greymarket part\footnote{``Greymarket'' parts are manufactured by a variety of vendors with no central authority and often no relevant/accurate data-sheet.  Use of these parts in common in most manufactured electronics today, however the increased variance in quality \& performance demands more attention be paid by users to their testing \& implementation.}, most variants of which can provide practical distance measurements from 0.01 m to 3 m with better than 0.02 cm of resolution depending on the target object being observed.
Our testing reveals most HC-SR04 sensors are unreliable for detecting soft and uneven objects such as humans beyond approximately 1 meter, where successful measurement rates fall below 50\%.  A failed measurement (``infinite distance'') indicates that either a queue member in front of a sensor has been missed (false negative) or that there is no detection within 3.5 meters (true negative).  
False positives (incorrectly detecting people in empty queue) are rare events for our sensors. 
In order to address the issue of unreliable detection, we configure the ultrasonic sensor to repeatedly make up to 5 measurement attempts (with a millisecond gap) until a valid (non-infinite) measurement is obtained. This configuration allows us to afford a high accuracy while maintaining a low false positive and negative rate. In operation, the sensors are configured to measure and record the distance every 10 seconds, and transmit (via LoRaWAN interface) a batch of 6 measurement record every minute in order to reduce the packet transmission rate.

Multiple PDUs operating in the same location could potentially pick up each other's ultrasonic pulses, and hence provide incorrect measurements. To statistically minimise this effect, we design the sensor units to be as non-deterministic in their timing as possible; where small changes in measurement success (number of cycles), transmit time and clock speed would prevent the units from synchronising their behaviour. Units are RC-oscillator controlled (rather than crystal controlled) and do not contain any form of real-time clock.  Even if all units are turned on simultaneously they naturally spread out their timings.
Note that the directionality of ultrasound sensors is important for the accuracy of the detection. Therefore, the devices should be firmly affixed to avoid any positional change.

\subsubsection{LoRaWAN Radio}

Multitech mDot radios\footnote{\url{https://www.multitech.com/brands/multiconnect-mdot}} are chosen as they were the only market-available option at the time of implementation that was compatible with the local (legal) LoRa frequency plan, available off-the-shelf (in-stock) and provided a well-documented UART interface for simple operation.
We choose to utilise a public LoRaWAN base station already provisioned by our campus Estate Management. 
The use of a public tower necessitated the use of an intermediary organisation, The Things Network (TTN)\footnote{TTN operates a large network of public-use LoRaWAN base stations, see \url{https://www.thethingsnetwork.org/}}, in our data collection pathway.




\subsubsection{Power supply}
Two 1.5V AA alkaline cells are used due to the concern about the longevity of rechargeable solutions in enclosures placed in direct sunlight. The 3V supplied is enough to operate the micro-controller and radio, however it is not sufficient for the ultrasonic distance sensor, which requires 5V. We use a discrete voltage doubler (operating at a fixed frequency by the micro-controller) as a simple solution to this problem. It has near-zero quiescent power draw and was later modified (see \S\ref{sect:energy}) to provide extended battery life operation.

\subsection{Performance Evaluation and Tuning}
In this subsection, we evaluate the performance of our PDUs in terms of detection range of the ultrasonic sensor, communication reliability of the LoRaWAN radio, and energy consumption of the entire unit.

\subsubsection{Detection Range}\label{sect:ultrasonic}


\begin{figure}[t!]
\centering
	\subfloat[Unmodified ultrasonic sensor.]				{\includegraphics[width=0.45\columnwidth]{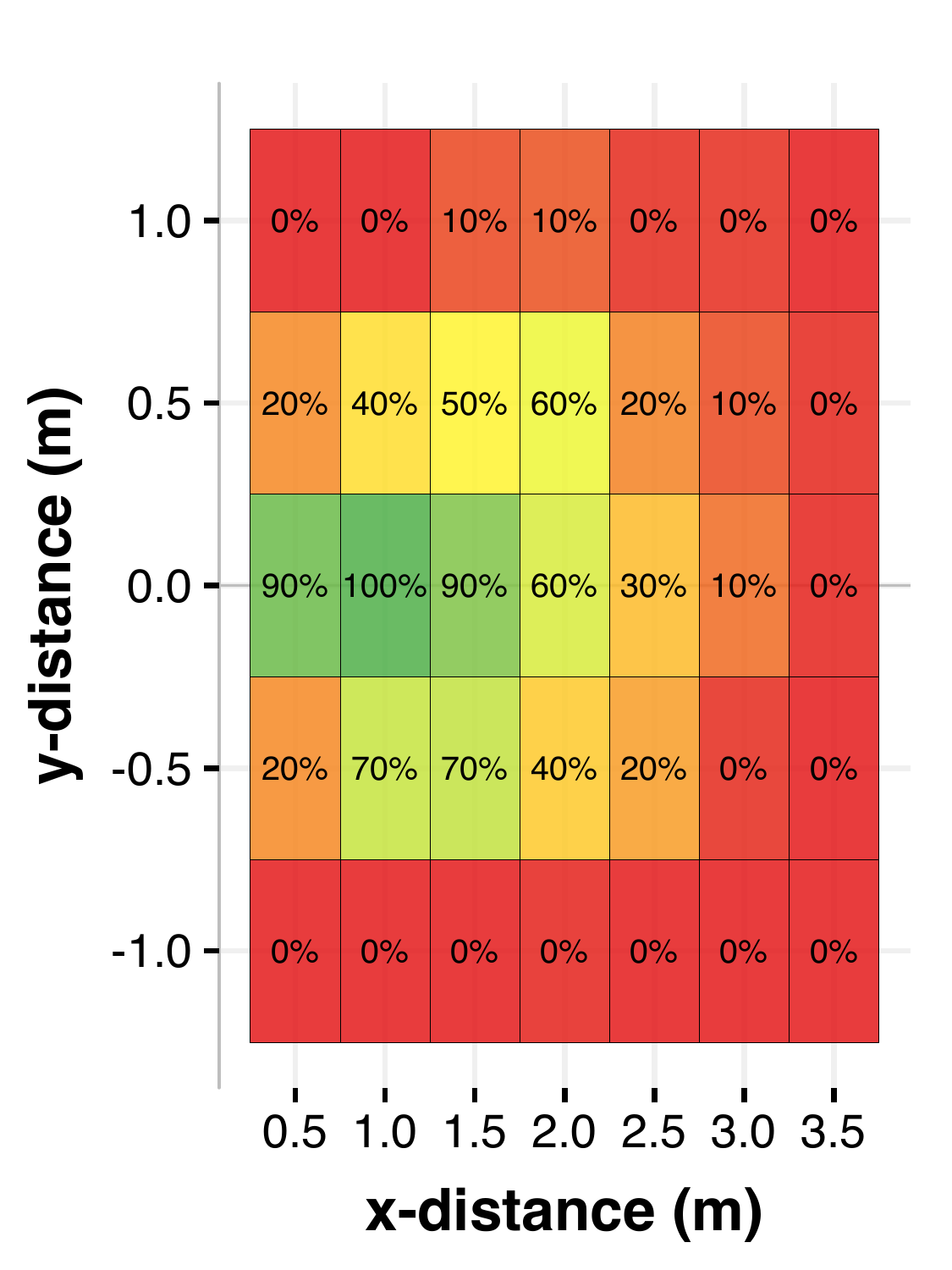}\label{fig:unmodified}}
	\subfloat[Modified ultrasonic sensor.]	{\includegraphics[width=0.45\columnwidth]{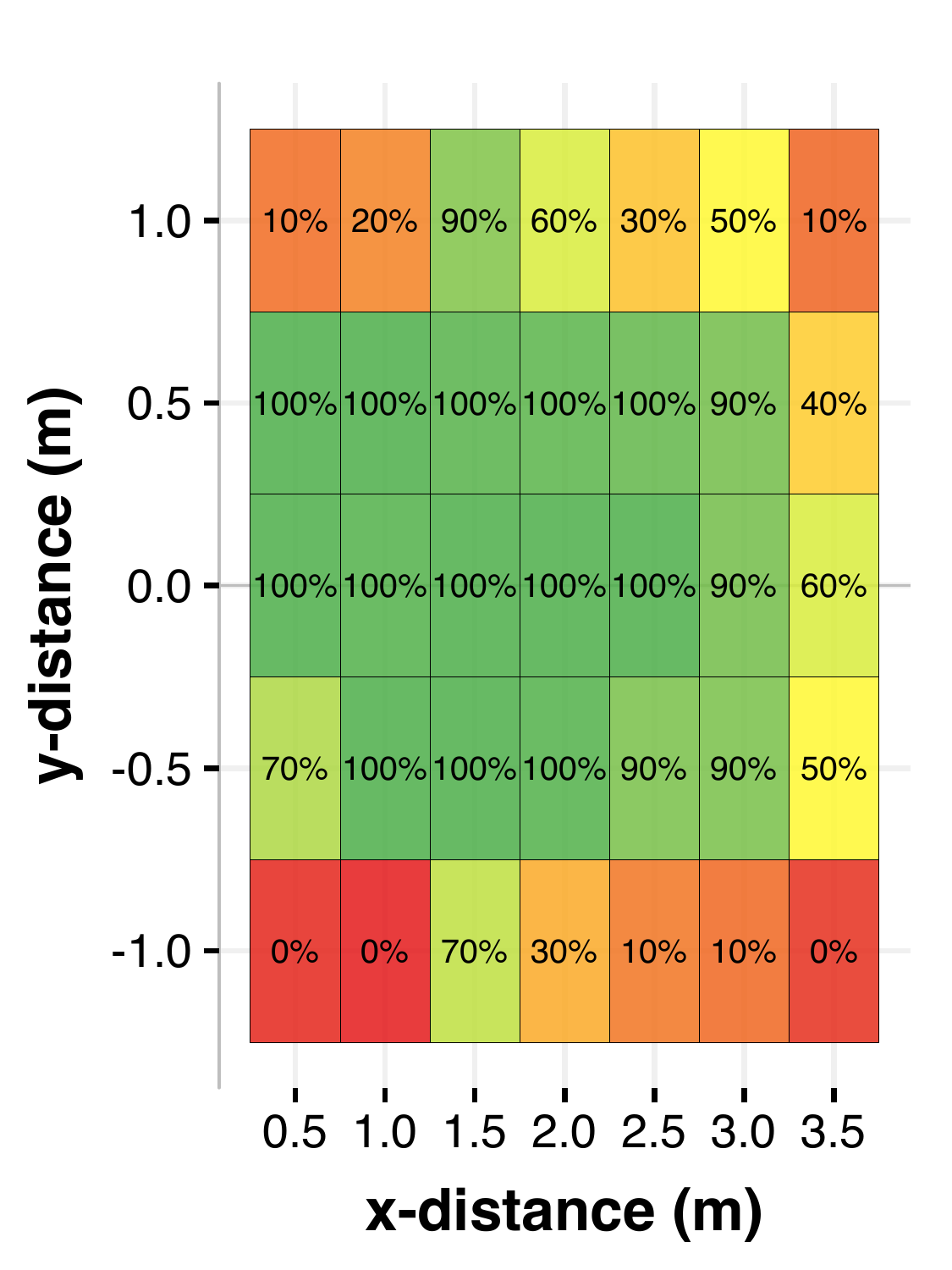}\label{fig:modified}}
	\caption{Probability of detecting a human at various proximities of the sensor (the sensor is located at origin).}
\label{fig:ultrasonic_range}
\end{figure}

Our PDU utilises a HC-SR04 ultrasonic sensor to return the distance to the closest object from the sensor. We first evaluate the detection ability in our laboratory. Our initial testing verified that the sensor performed well in returning the distance to large solid objects, however detection of people was erratic. We found that the sensor underperformed when the human subject moved more than 1.5 meters away from it. Also, the detection ability varied depending on the type of clothing (e.g., soft long sleeved top) worn by the subject. The cause of such poor detection ability with people is due to ultrasonic waves not being reflected strongly off people to be detected at the sensor. Soft clothing is also not a good reflector of ultrasonic waves when compared to bare human skin.

Emil \cite{emil2014:hcsr04} has done an in depth analysis of the HC-SR04 performance and has reverse engineered it. The author found that the band-pass filter used on the receiver circuit for detecting the reflected ultrasonic wave is centred at 18kHz. As the wave emitted by the sensor is 40kHz, the reflected wave back to the sensor is also expected to be 40kHz and having the received wave filtered by a band-pass centred at 18kHz reduces the receivers sensitivity. Therefore, as suggested by Emil, we modified two resistors used in the filter circuit of the HC-SR04, and hence enhanced the sensitivity of the receiver.

We then evaluate two groups of sensors namely three units of ``unmodified'' and one unit of ``modified'' HC-SR04's. For each sensor, the human subject stood at cells on a clearly marked grid map.  We collected 100 readings at each position, then the average was used for each group of sensors. The results of this experiment are shown in Fig.~\ref{fig:ultrasonic_range},
where the sensor is located at coordinate (0,0) facing right. Each cell value indicates the probability of human detection (green cells show high detection and red cells show low detection).
It is clearly seen that the modified sensor (Fig.~\ref{fig:modified}) has a better field of view, detecting the test subject in most situations even if they are not standing directly in front of the sensor. The detection range is also greatly increased. We also observe that the detection ability of the unmodified sensors tapers off after 1.5 meters whereas the detection ability of the modified sensors starts to taper off at a further distance of 3 meters.
The results of the experiment is conclusive in demonstrating that the modified sensor has significantly better ability in detecting people than the original version. 



\subsubsection{LoRaWAN Communications Reliability}\label{sect:commsrelia}

\begin{table}[t]
	\begin{center}
		\caption{LoRaWAN data rate settings and associated theoretical time-on-air for 6-byte payload. }
		\label{table:datarateLORA}
		\begin{tabular}{l c c c c c c}
			\toprule 
			\textbf{Data rate} & DR0 & DR1 & DR2 & DR3 & DR4 &DR5  \\
			\midrule
			
			\textbf{{Spreading Factor}} & 12 & 11 & 10 & 9 & 8 & 7   \\
			\textbf{{BW (kHz)}} & 125 & 125 & 125 & 125 & 125 & 125    \\
			\makecell{\textbf{Time-on-air (ms)} \\ (for 6-byte payload)} & 1319 & 741 & 330 & 185 & 103 & 51   \\
			\bottomrule
		\end{tabular}
	\end{center}
\end{table}
\label{table:DR}

We quantify the reliability of LoRaWAN by using packet delivery ratio (PDR) as a performance metric. PDR is defined as the fraction of packets received by the gateway out of the total number of packets sent.
Reliability of LoRaWAN notably depends on the deployment environment as well as settings on LoRaWAN physical layer.
There are three main parameters that can be tuned including spreading factor, SP (number of bits used to represent a symbol, which directly impacts bit rate and thus time-on-air of packets); bandwidth, BW (defines the range of frequencies over which LoRa signal spreads, higher bandwidth allows for faster data transmission rate but reduces receiver sensitivity and communication range); and transmission power, TXP (amount of energy used to transmit a packet).
LoRaWAN introduces a new variable called data rate (DR), which is an arbitrary configuration parameter used to define different combinations of SP and BW settings. The respective configurations (available in our country) that we evaluated are shown in Table~\ref{table:datarateLORA}. 
In theory, lower data rate decreases the signal to noise ratio (SNR) limit at the receiver gateway (causing the receiver gateway to become more sensitive to the received signal), thus creating a more reliable LoRaWAN link. However, this will also raise time-on-air of the transmitted packet, which in turn increases the chance of collision.

To evaluate the performance of various configurations of DR and TXP, we wrote a script to randomly cycle the setting combinations for all of our 10 PDUs. We consider six data rates (DR0-DR5) and six transmission power settings (TXP values of  2, 6, 10, 14, 18, and 36 dBm), thus a total of 36  combinations. The PDUs were set to transmit a 6-byte data packet every minute, and a total of 30 samples were collected for each setting combination per PDU. This yields a total sample size of 10,800 transmissions (from ten units).
The random cycling of settings helps reduce sampling bias caused by environmental factors such as temperature, humidity, and line-of-sight that can impact the reliability of LoRa links. Furthermore, repeated measurement attempts (explained in \S\ref{sect:built-ultrasonic}) were made during this phase of experiment to introduce random delays between successive transmissions, and thus preventing PDUs from getting synchronised, thus minimising collisions.
We installed our 10 PDUs on the fence (at the bus stop) to collect data for reliability analysis. 

\begin{figure}[t!]
	\centering
	\includegraphics[width=0.7\textwidth]{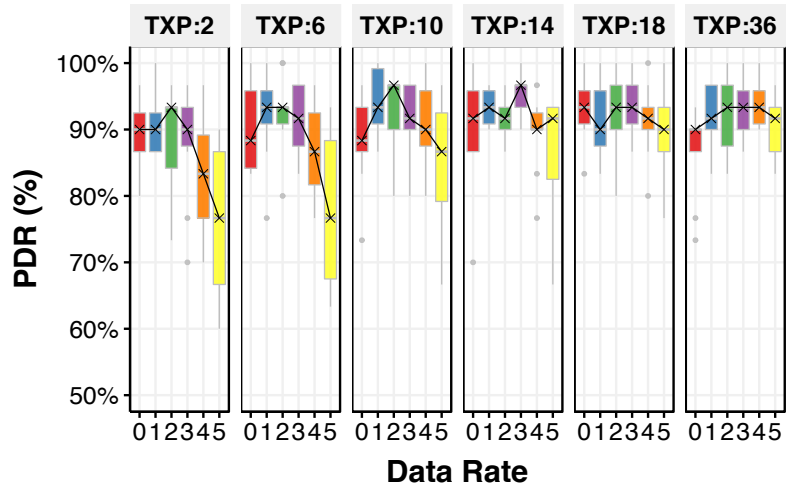}
	\caption{Packet delivery rate (PDR) varies by data rates (DR) and transmission power (TXP). }\label{fig:pdr_dr_txp}
\end{figure}

Fig.~\ref{fig:pdr_dr_txp} shows PDR (y-axis) for various combinations of DR settings (x-axis) and transmission power settings (facets). We observe that for transmission power up to 10 dBm, PDR has a general declining pattern in data rate (from DR0 to DR5) and is widely spread in values between 70\% and 90\% -- this behavior complies with other experimental studies on LoRaWAN performance \cite{cattani2017:LoRa_experimental,sanchez2018:performance}, where a low data rate was reported to result in more reliable link and thus gives a better PDR. 
By increasing the transmission power (TXP of 14 and 18 dBm), no obvious pattern is observed across various DR values, suggesting that the delivery rate is less susceptible to the changes in LoRaWAN data rate setting. This is due to the fact that higher transmission power yields stronger signals that are less prone to attenuation caused by the environment such as humidity, rain and traffic which are likely to have biggest effect on PDR.

\subsubsection{Energy Consumption}\label{sect:energy}

\begin{figure}[t!]
\centering
	\subfloat[Worst case.]
	{%
	  \includegraphics[width=0.45\textwidth]{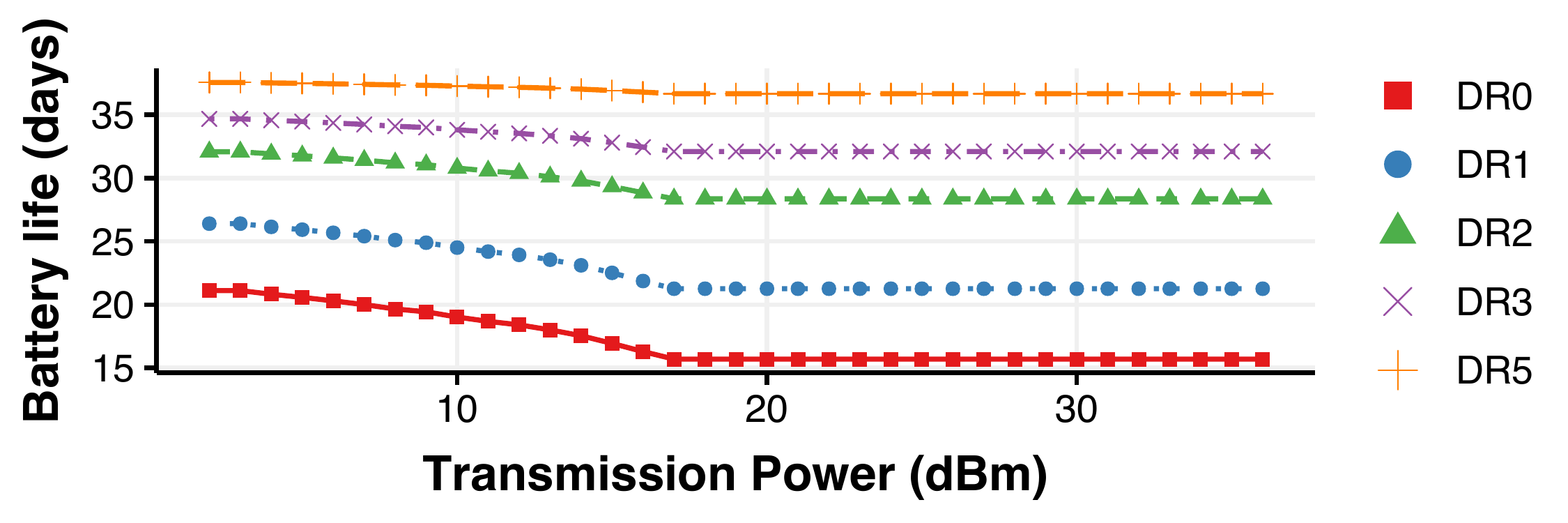}%
	}
	\vspace{-0.5mm}
	\subfloat[Best case.]{%
	  \includegraphics[width=0.45\textwidth]{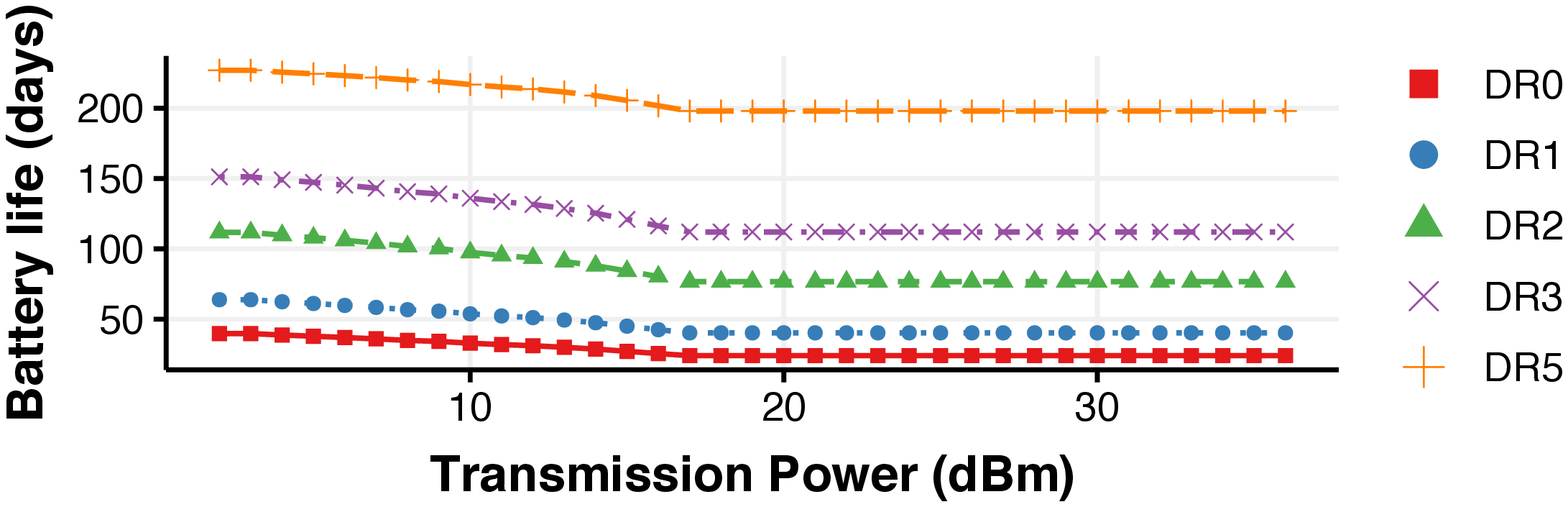}%
	}
	\vspace{-1mm}
	\caption{Estimate of battery life with different LoRaWAN parameter settings.}
	\label{fig:battery_life_worst}
	\vspace{-4mm}
\end{figure}

We evaluate the energy consumption of our PDUs by computing the estimated battery life. Battery lifetime (in days) can be computed by dividing the usable energy available in the batteries (found from manufacturers data sheet \cite{fdk:battery}) by the total daily energy consumed.
We start by computing the energy consumed for one packet. In our PDUs, the LoRaWAN radio and the ultrasonic sensor contribute to the majority of energy consumption -- the total energy consumed per packet can be calculated by adding energy consumed by LoRaWAN radio and by ultrasonic sensor. We experimentally measure the energy usage of both components using a small resistive shunt and an oscilloscope.

Energy consumed by the ultrasonic sensor varies depending on the number of retries required to make a valid measurement. Best case scenario is when the first measurement in a session successfully detects the queuer and worst case is when all retry measurements are executed. With 5 retries available, we found the  energy consumed by the ultrasonic sensor under best and worst cases to be $0.04$J and $0.29$J respectively.
The LoRaWAN radios energy consumption varied depending on the configured data rate and transmit power. Lower data rates consume more energy due to longer time on air, and high transmit power settings (from 17dBm to 36dBm) consume the maximum amount of energy for a given data rate -- it saturates due to limitations of our radio. The best case for our radio's energy consumption is with the highest data rate and lowest transmit power ($0.01$J), whereas worst case is lowest data rate and highest transmit power ($0.43$J).

Once we calculate the energy consumed per packet, the total daily energy consumption of a PDU can be obtained by multiplying this value by the maximum number of packets sent per day. As a result, battery lifetime (in days) can be computed by dividing the usable energy available in the battery by the total daily energy consumed.
Fig.~\ref{fig:battery_life_worst} represents the estimated battery life for each combination of LoRaWAN parameters with the best and worst case scenarios (for ultrasonic detection) respectively. The trends show that the data rate has a greater impact on the battery life compared to the transmit power. Ultrasonic energy dominates for the worst case scenario leading to a compressed range (\ie 15 to 38 days in Fig.~\ref{fig:battery_life_worst}) of battery life. Our PDU's energy consumption is dominated by the ultrasonic sensor due to its retry property when no valid measurement is returned, suggesting high energy consumption arises from an empty queue.

\textbf{LoRaWAN Parameters Decision:}
For our deployment scenario, we have found that: an acceptable packet delivery ratio (more than 85\% on average) can be achieved across all data rates when transmission power is configured to a value above 14 dBm (\S\ref{sect:commsrelia}); higher data rate setting can notably enhance battery life, for instance 208-day battery life can be achieved for DR5 setting while only 28-day life for DR0 setting when the transmission power is set at 14 dBm (\S\ref{sect:energy}).   
Furthermore, based on the theoretical time-on-air calculation per data rate setting shown in Table~\ref{table:datarateLORA}, faster data rate (\eg DR5) reduces air time for our sensors, which is limited to 30 seconds per-day according to TNN guidelines. In summary, data rate DR5 and transmission power of 14dBm are chosen because of acceptable packet delivery (reliability), longer battery life, and shorter time-on-air.




\begin{figure}[t!]
	\centering
	\vspace{1mm}
	\includegraphics[width=0.7\textwidth]{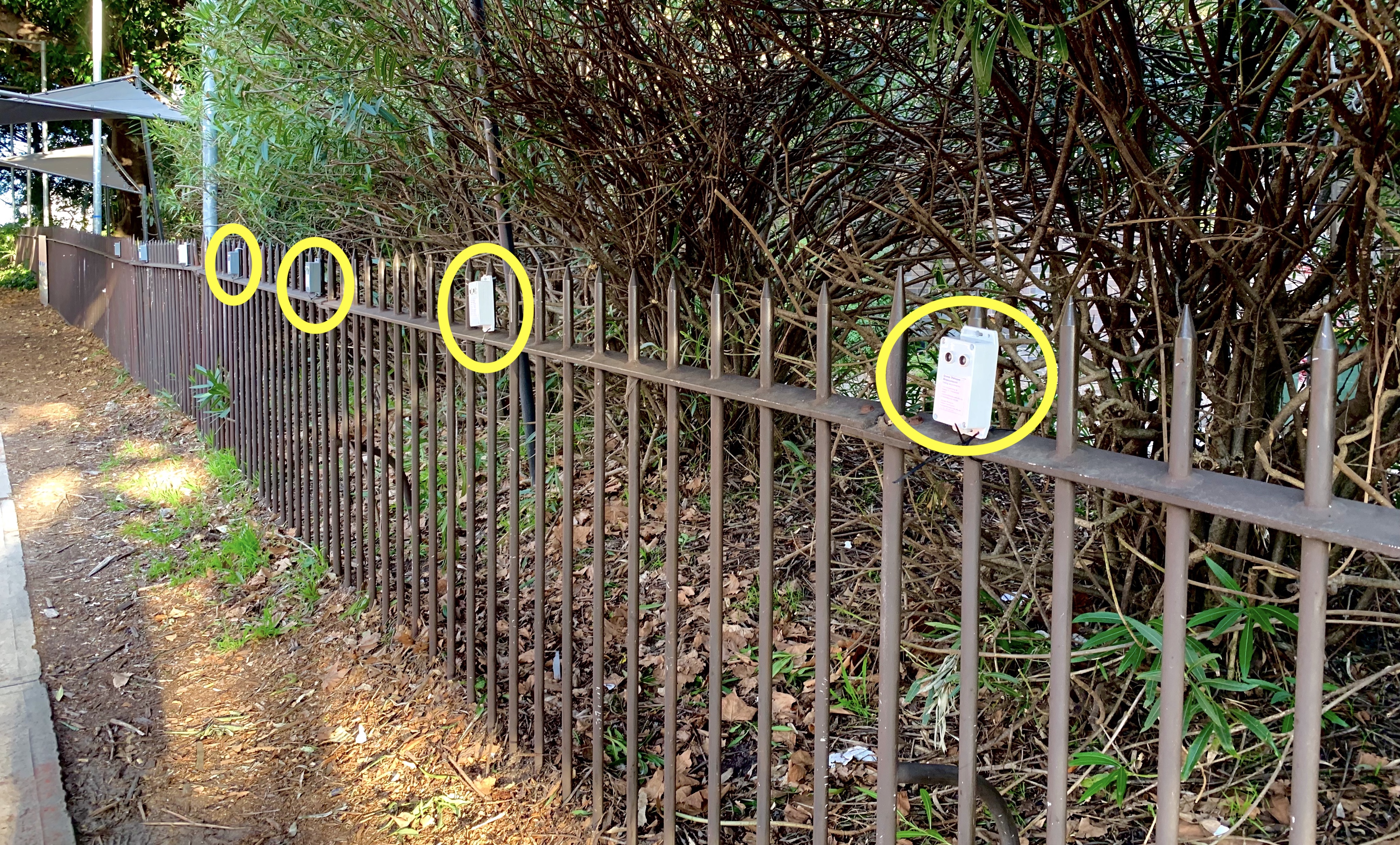}
	\caption{Sensors mounted on the campus fence bordering the bus stop.}\label{fig:sensor_mounting}
\end{figure}

\section{Inferring Queue Length from Sensed Data}\label{sect:Qlength}

In this section, we first illustrate the deployment of our people detector units (PDUs) at an on-campus bus stop for queue length detection. Next, we develop and evaluate an algorithm to infer queue length from the sensed data. Lastly, we evaluate the efficacy of our algorithm against ground-truth data obtained from the field.

\subsection{Experimental Setup}\label{sect:exp-setup}

We deployed ten PDUs at a main campus bus stop that serves as the first stop for express bus service to the city center. 
The sensors were affixed to the campus fence adjacent to the pedestrian footpath (about 3 meters in width) at a regular spacing of 6-meter, where each unit was installed 1.2-meter above ground. Our deployment of PDUs is shown in Fig.~\ref{fig:sensor_mounting}. The queue length measurement range was limited to the length of the fence since beyond the fence is an open area with no object for sensors to be affixed to. We observed that typically passengers would form a queue along the footpath at a distance of about 2-3 meters from the PDUs. The arrangement of our PDUs in relative to the queue formation is shown earlier in Fig.~\ref{fig:busstop_topview}.

In order to evaluate the accuracy of our queue inference algorithm, we manually collect ground truth data by recording timestamps when each passenger arrives at the queue and boards the bus. 
This data is used to compute the actual queue length over the measurement period. This data was collected during peak hours (3-6 pm) on a weekday.

\begin{figure}
	\centering
	\includegraphics[width=0.7\textwidth]{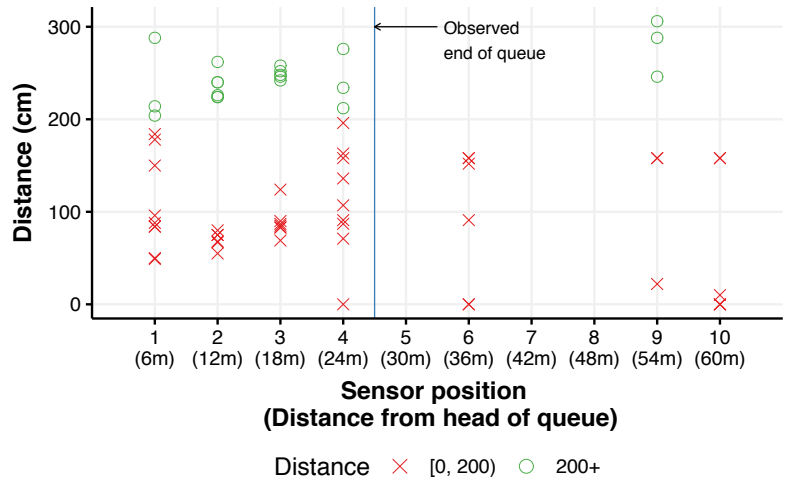}
	\caption{Raw sensed distance measurements at each PDU position during a 2-minute interval snapshot (4.10pm-4.12pm on 24 Sep 2019). Green circles indicate distance where queue is expected  and red crosses indicate detection that is not part of the queue.}\label{fig:instant_sensor_data}
\end{figure}

\subsection{Sensed Data}
Data collected from each PDU is a time-series of distance measurements between the sensor and the closest object. Since PDUs are deployed along the footpath which also serves as a pedestrian walkway, there is a chance that the sensors detect people who are walking by or standing within close proximity but are not part of the passenger queue. Therefore, further data processing is needed to filter such noisy observations in order to infer the correct queue length.

Fig.~\ref{fig:instant_sensor_data} shows a 2 minute snapshot of collected raw measurements (infinite values as mentioned in \S\ref{sect:built-ultrasonic} are not shown here), where y-axis shows the distance (in cm) measured by each PDU and the x-axis denotes the position of each PDU with 1 and 10 denoting the head and tail of the queue, respectively. 
Measurements over 200cm are denoted as green while those that are lower are marked red. This is because the former are highly likely to be passengers in the queue, while the latter are likely passers-by.
During the time period of this snapshot, between 40-58 queuers are observed, this is shown by a vertical blue line which indicates the approximate end of queue corresponded to the actual queue length. In the other word, the first 4 (or possibly 5) PDUs are expected to detect the queue while the rest are expected to report no detection.
Examining the sensed data, we see that all PDUs at positions 1 to 4 have successfully detected the existence of the queue, though some of the measured distances are not within the expected range (possibly from people walking pass by on the footpath). A false positive can also be seen at the position 9, where an object is detected within the range of the queue.

\begin{figure}
	\centering
	\includegraphics[width=0.7\textwidth]{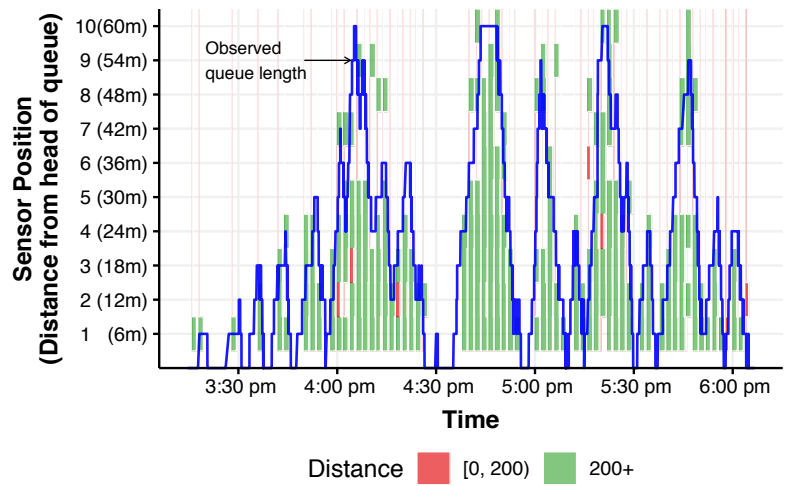}
	\caption{Time-trace of measurements from people detector units overlaid by ground-truth queue information. }\label{fig:raw_data_vs_gt}
\end{figure}

To further visualize the sensed data, we plot a time trace of PDU measurements and ground truth queue length in Fig.~\ref{fig:raw_data_vs_gt}, where x-axis denotes time at 2 minute resolution and y-axis represents the position of each deployed PDU. Green cells indicate detection of the queue by the corresponding PDU at the corresponding time.
We can see that PDUs are able to fairly track the dynamics of the queue, evidenced by the true position detections (green fills that lies under the blue ground truth trajectory). Nonetheless, several false detections can be observed including false positives where a queue is detected when there is no queue formation (\eg position 8 at 4.14pm) and false negatives where queue detection is missed (\eg positions 5 and 6 between 5:44pm and 5:48pm).

\subsection{Queue Inference Algorithm}\label{sect:q-length-algo}

Herein, we present a robust queue inference algorithm that attempts to eliminate the false positives and negatives reported above. The algorithm consists of three steps. The first step aims to address time synchronization of sensors since data from each PDU is recorded at different timestamps as explained in \S\ref{sect:built-ultrasonic}. In order to synchronize the records, we map the received data into its corresponding time bin of \textit{t} minute interval.

For each time interval, the second step decides whether a PDU detects a queue or not. If the fraction of positive measurements (detection within 2-3 m) over the time interval is greater than a pre-defined threshold percentage (\textit{th}), we deem that the PDU detects a queue. Executing this step yields a vector of 10 binary values (1: queue detected and 0: not detected), corresponding to the ten PDUs. 
Since there is typically no substantial gap in the middle of the queue, the derived binary vector may contain false negative values which causes a disjoint sequence of 1's in the vector. The last step therefore aims to fix such errors if present by employing the concept of hamming distance to correct the measured vector (code) to one of the following valid codes:

\[
valid \  codes 
\in
\scriptsize \Set{
  \begin{bmatrix}
    1 \\ 0 \\ 0 \\ 0 \\ \vdots \\ 0 
  \end{bmatrix} ,
    \begin{bmatrix}
    1 \\ 1 \\ 0 \\ 0 \\ \vdots \\ 0 
  \end{bmatrix} ,
    \begin{bmatrix}
     1 \\ 1 \\ 1 \\ 0 \\ \vdots \\ 0 
  \end{bmatrix} ,
  \dots\,
  \begin{bmatrix}
     1 \\ 1 \\ 1 \\ 1 \\ \vdots \\ 1 
  \end{bmatrix} 
  }
\]

The selected valid code is then used to derive passenger queue length. From our spot measurement, we found that there are on average 10 passengers standing between two adjacent PDUs. Hence the queue length can be deduced by multiplying the number of detected bits in the vector code by 10.

\begin{figure}[t!]
	\centering
	\includegraphics[width=0.7\textwidth]{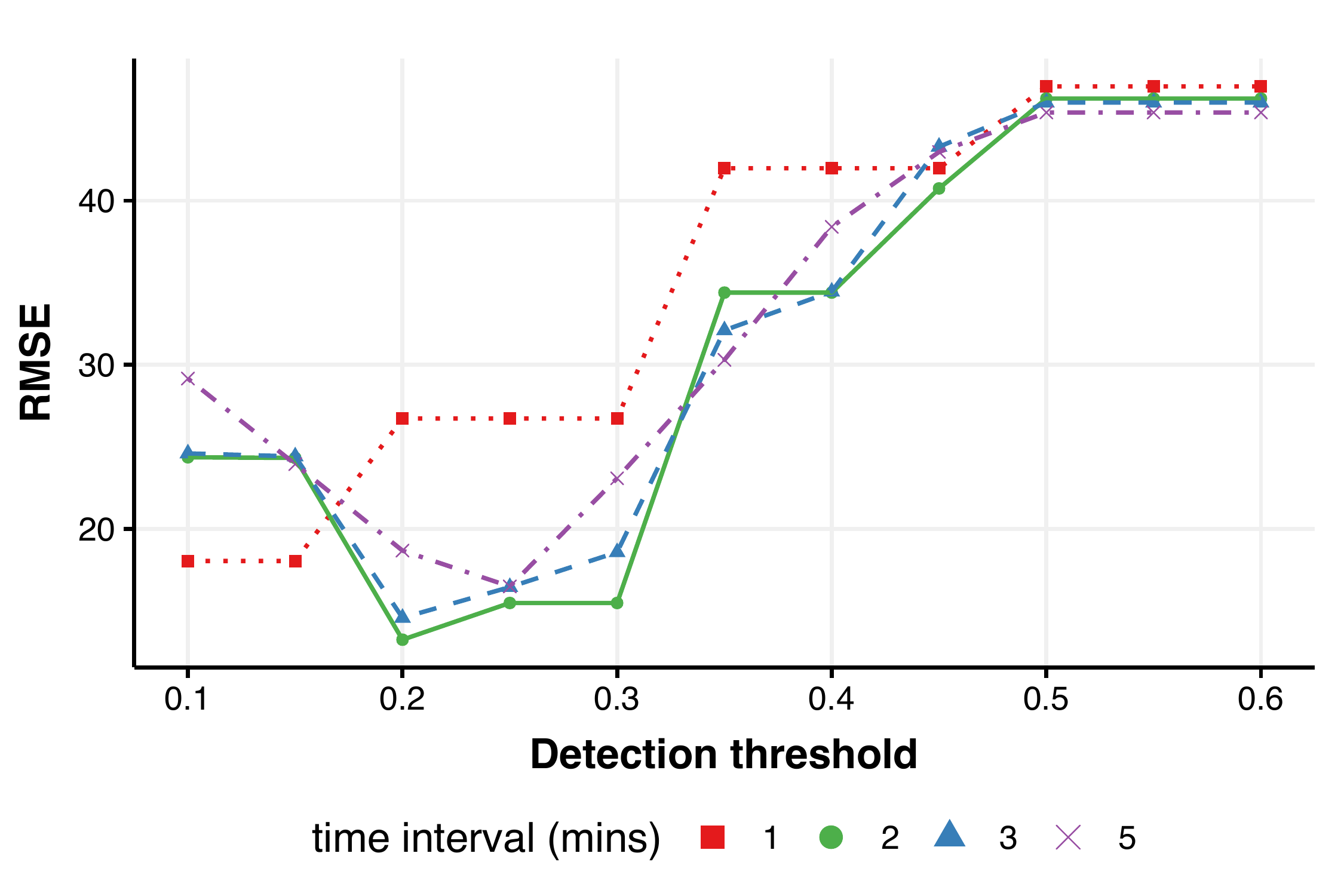}

	\caption{Comparison of RMSE across different tuning parmeters.}\label{fig:algo_mae_tuning}
\end{figure}

\begin{figure}[t!]
	\centering
	\includegraphics[width=0.7\textwidth]{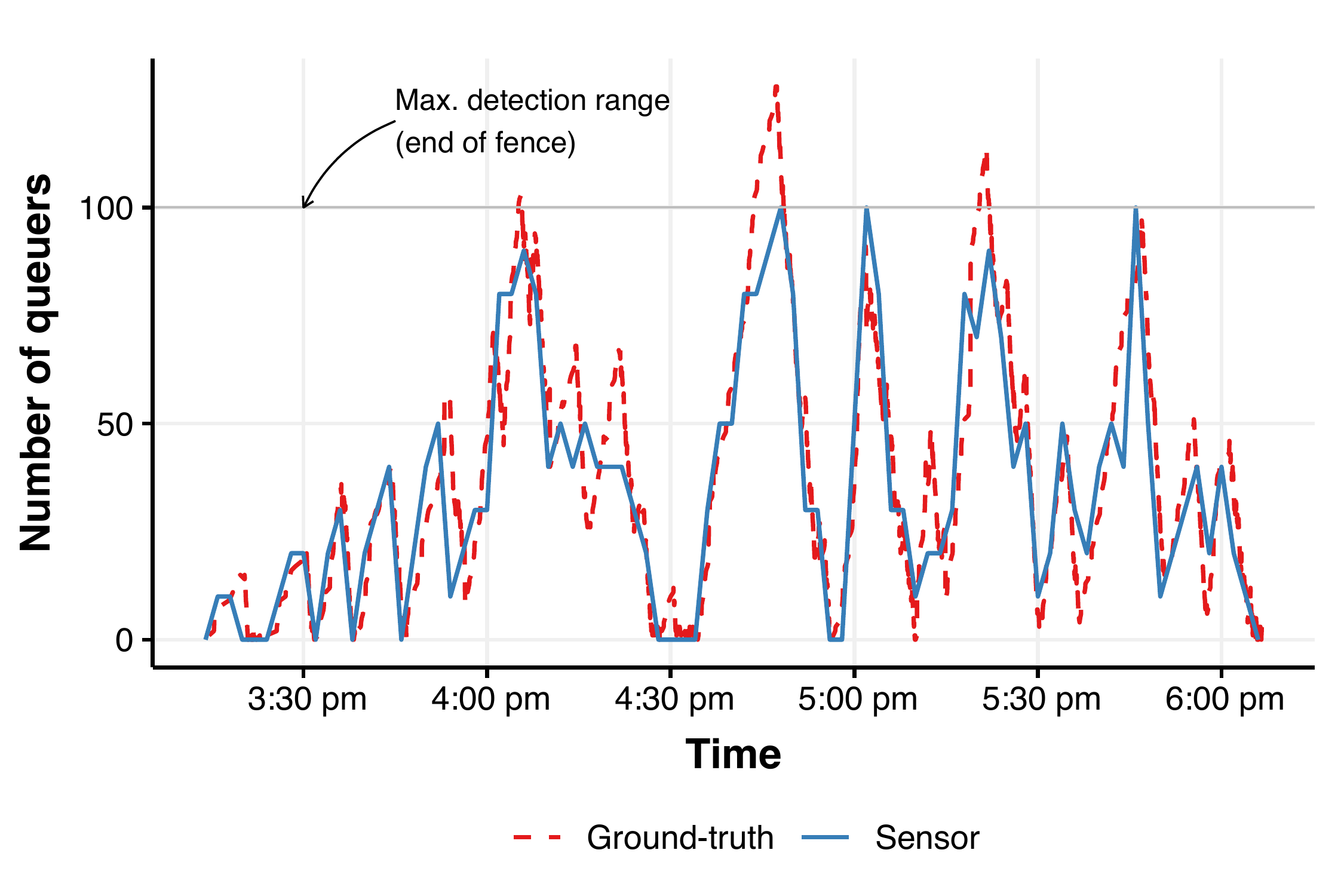}
	\caption{Real-time queue length inferred from sensor data vs. ground-truth.}\label{fig:algo_result}
\end{figure}

\textbf{Algorithm Evaluation and Tuning:} 
From the aforementioned description of the queue inference algorithm, there are two tunable parameters, namely time bin \textit{t} (in minutes) and detection threshold \textit{th} (in percentage), that are needed to be decided upon. In order to select the optimum parameter values, we apply our algorithm to the sensor data collected during the 3 hour field experiment (where ground truth occupancy was obtained)  and observe the impact of varying \textit{t} and \textit{th}. Root Mean Square Error (RMSE) is used as the performance metric.  
Fig.~\ref{fig:algo_mae_tuning} shows RMSE (y-axis) as a function of detection threshold (x-axis). Different line graphs shown the impact of varying the time bin duration. 
Note that a time interval of 2 min (solid green line) achieves the lowest RMSE. Also, choosing the detection threshold at 0.2 gives an acceptable RMSE of $13.25$ and MAE of $10.75$. 
The output queue length calculated by our algorithm, using the optimal values of \textit{t} and \textit{th}, is visualized in Fig~\ref{fig:algo_result}, where ground-truth is shown by dotted red lines and the calculated queue occupancy is shown by solid blue lines. We can see that our algorithm is able to closely track the real queue with low error. We note that our solution can only measure the queue up to the end of fence area (as noted earlier), hence the calculated queue length is capped at 100 people. Note that this is an artifact of our deployment and does not impact the algorithm accuracy.

\subsection{Deployment of uniBusQ}\label{sect:5daydeployment}
We deployed our PDUs at the bus stop for a period of five days, starting from 3pm on Monday (23 Sep 2019) until the end of Friday (27 Sep 2019). The tuned algorithm described in \S\ref{sect:q-length-algo} is applied to the collected data (from our deployment) to infer the temporal profile of queue length. We compute the arrival rate (every minute) of passengers by integrating the queue length (deduced from our sensors data) with a publicly available (and real-time) data of bus schedules in the form of General Transit Feed Specification (GTFS) . This GTFS data reports the actual arrival/departure time of buses to/from transit stops as well as buses type and model, allowing us to estimate the capacity of each dispatched bus. The computed rate of passengers arrival is highly fluctuating. Therefore, we apply moving average (15-minute window) to smooth the arrival rate data. Fig.~\ref{fig:static-schedule} illustrates the smoothed rate of passengers arrival per minute (black line) across the five days of our deployment. The bus arrival times are depicted in the same figure as red lines.

It is clearly seen in Fig.~\ref{fig:static-schedule} that bus arrivals are consistent across all weekdays despite variations in the arrival pattern of passengers. We observe that the passenger arrival pattern varies across different days especially on Friday where the peak occurs at 4:30pm as opposed to 5pm for other weekdays. However, buses are statically scheduled to be dispatched more frequently around the 5pm mark, suggesting a mismatch between supply and demand and thus inefficient use of resources. 
This observation suggests for a more optimal bus scheduling to be considered, so that the passengers experience is improved by reducing their waiting time at the bus stop but without adding extra bus services.

\section{Demand-Based Optimisation of Bus Scheduling}\label{sect:queue_opt}

We saw in \S\ref{sect:5daydeployment} that existing static bus schedules are not particularly efficient. In this section, we develop an optimization model that schedules the arrival of buses based on the actual passenger demand in order to minimize the total wait time of passengers at the stop. We formulate the problem with an assumption that the passengers demand is given and known. Note that the future demand may be predicted by way of modeling the pattern of historical data, but predicting the demand is beyond the scope of this paper given the limited data collected from our deployment.   


\subsection{Optimisation Formulation}\label{sect:opt-model}

Several research works\cite{huang2019novel, feng2018design} have developed dynamic bus scheduling schemes to improve the transit experience of passengers, with the common objective being minimizing passenger wait times.
However, many existing models estimate passenger wait time without considering the limited capacity of buses \cite{chen2020multiobjective,gkiotsalitis2018bus}. 
During peak hours particularly, certain passengers (depending on the queue length and their location in the queue) may need to wait for more than two buses before they get on board.
Our formulation, therefore, aims to incorporate the additional wait time of those passengers who may get left behind due to overloaded buses.

\begin{figure}[t!]
	\centering
	\includegraphics[width=0.8\textwidth]{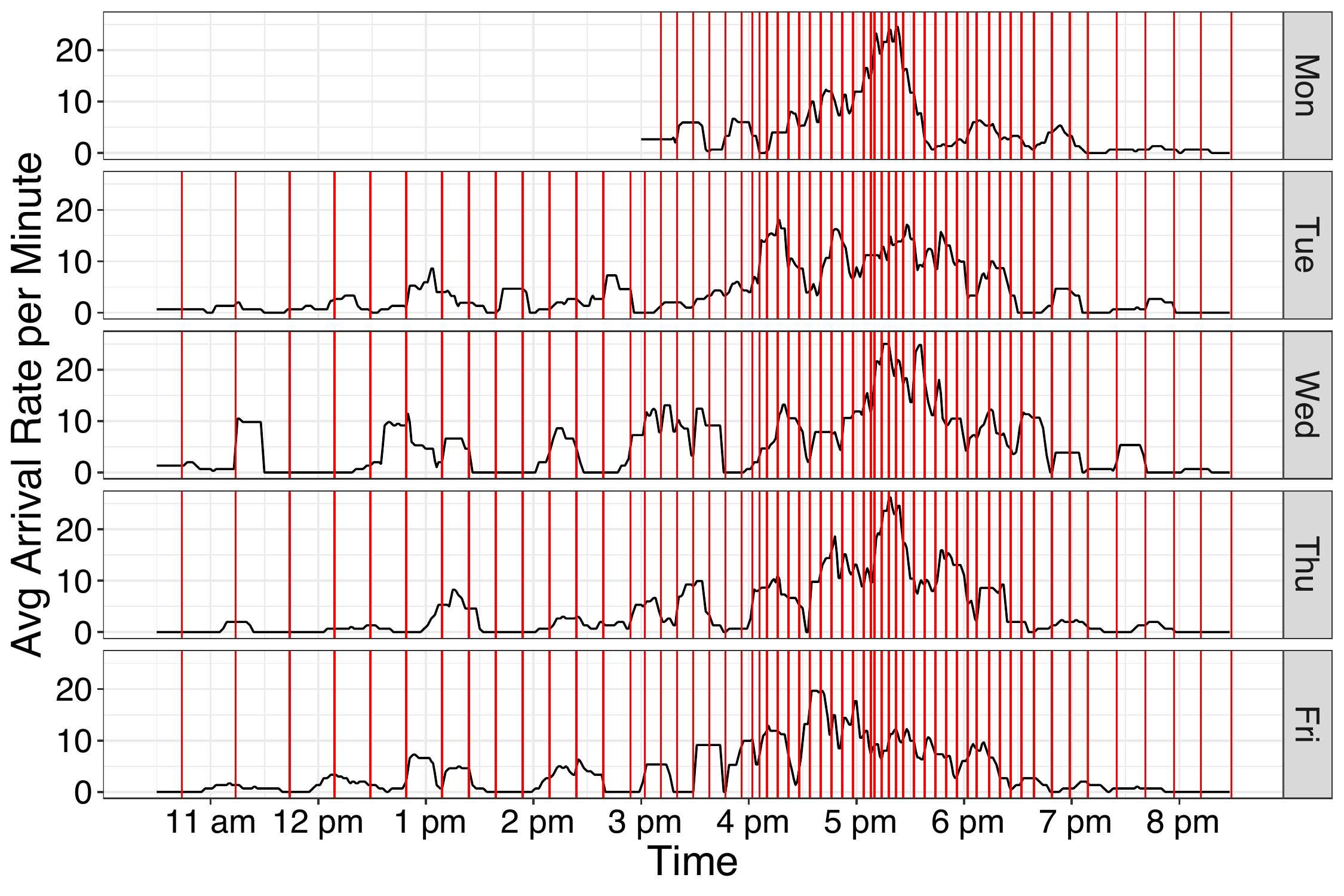}
	\caption{Static scheduling of bus dispatch time.}\label{fig:static-schedule}
\end{figure}

For our optimization problem, let there be $B$ buses available for use in a day. The arrival time and seating capacity of bus $i$ are respectively denoted by  $d_i$ and $C_i$ where $1 \leq i \leq B$.
Passengers demand, captured by their rate of arrival to the stop as shown in Fig.~\ref{fig:static-schedule}, is denoted by $\lambda(t)$.
The total daily wait time of passengers is obtained by adding 
the following components: 

(a) Time spent by passengers waiting for their first arriving bus --  the time between the arrival of individual passengers and the arrival of their first bus to the stop. 
We borrow from \cite{luo2018dynamic} the model of this factor of wait time for $B$ buses during a day, which is given by:


\begin{equation}
W^{first} = \sum_{i=1}^{B} \int_{d_{i-1}}^{d_i} (d_i - t) \lambda(t)~ dt
\end{equation}



(b) Time spent by waiting passengers who missed to board previous bus(es) due to capacity limit. For each arriving bus $i$, $N_i^{left}$ denotes the number of leftover passengers that arrived before and missed the previous bus $i-1$.  The total wait time of leftover passengers can be formulated as:

\begin{equation}
W^{left} = \sum_{i=1}^{B} N_i^{left} (d_i - d_{i-1})
\end{equation}

Given $N_1^{left}=0$, the number of leftover passengers can be calculated recursively by:

\begin{equation}
N_i^{left}= N_{i-1}^{left} + \int_{d_{i-2}}^{d_{i-1}}  \lambda(t)~ dt - C_{i-1}
\end{equation}


The objective of our optimization formulation is to minimize the total daily wait time of passengers, which can be written as:

\begin{equation}\label{eq:obj_function}
\begin{aligned}
\min \quad & W^{first}  + W^{left} \\
\end{aligned}
\end{equation}


In our optimization model, we assume an ordered set of buses, each with certain capacity. Our decision variable is the arrival time of these buses. There are a couple of constraints to be considered: 

(a) Time interval between consecutive buses (headway) need to be less than an hour ($H_{max}$) in order to maintain a minimum frequency of bus services. We also note that the headway cannot be less than a minute ($H_{min}$) as buses are not to be scheduled simultaneously. Therefore, our first constraint is given by:

\begin{equation}\label{eq:minmax}
H_{min} \leqslant d_{i} - d_{i-1} \leqslant H_{max}
\end{equation}

(b) The last bus (of a given day) needs to be scheduled at the end of the daily optimization period ($t_e$), consistent with the current bus schedules (say, 8:30pm).

\begin{equation}\label{eq:constraint_lastbus}
{\color{black}d_{B} = t_e}
\end{equation}



\subsection{Genetic Algorithm for Solving Optimisation Problem}

Our optimization model is a non-linear combinatorial optimization problem \cite{korte2011combinatorial}. Hence, a meta-heuristic approach is considered to solve this model. Genetic algorithm (GA) is one of the most well-studied algorithm used in the field of mass transit optimization \cite{huang2019novel, luo2018dynamic} and has proven to be effective in solving combinatorial optimization problems \cite{khuri1994evolutionary}. We thus use GA model to solve our optimization problem stated in \S\ref{sect:opt-model}.
GA is inspired by the process of natural evolution. The algorithm selects optimal solutions that yield the best fitness (based on the objective function) for reproduction in the next generation, hoping to produce better offspring (solutions with better fitness scores than the parents). In what follows, we describe each component of our GA model.

For our chromosome design, we let genes within a chromosome represent headways of a fixed ordered set of buses.
Therefore, each chromosome is expressed as $[H_1, H_2,...,H_{B-1}]$ where $H_i$ {\color{black}(a gene)} denotes the headway between bus $i-1$ and bus $i$. Note that the last available bus on a day needs to be scheduled at the end of daily window according to the second constraint in Eq.~\ref{eq:constraint_lastbus}, Hence there are total number of $B-1$ genes for the chromosome.
Note that we use bus headways instead of dispatching time in order to narrow down the search space. 


We use the optimization objective in Eq.~\ref{eq:obj_function} (minimizing the total daily wait time) as the fitness function of our genetic algorithm.
For cross-over operation where two parent solutions exchange genes, we adopt a local arithmetic cross-over \cite{dumitrescu2000evolutionary}. This operator takes weighted sum of the two parents to create a new gene where weights are randomly selected for each gene location. This method is chosen to increase the diversity in generated offspring. In other words, the objective is to create new genes which are not present in parents \cite{pavai2017crossover}.
Lastly, for mutation operation, we adopt a uniform random mutation (for simplicity) where genes are selected randomly from a uniform distribution over a range $[H_{min}, H_{max}]$, the lower and upper bounds in Eq.~\ref{eq:minmax}.



We use the GA library in R \cite{scrucca2013ga} to implement our genetic algorithm for solving the optimization problem. The default probability values of 0.8 and 0.2 are used respectively for cross-over and mutation operations. 
We run the algorithm for 1000 generations, each consisting of a population of 50 individuals. The search halts when the best solution score remains unchanged for 100 generations. 
 
In our optimisation problem statement, the number of available buses is fixed to the current bus schedule and therefore single objective optimisation algorithm can be sufficiently used to solve the problem aiming to minimise passenger wait-time. We note that for cases where additional buses are allowed to be added into the system, multi-objective optimisation should be considered where cost of extra buses can be used as the second objective.

\begin{figure}[t!]
	\centering
	\includegraphics[width=0.8\textwidth]{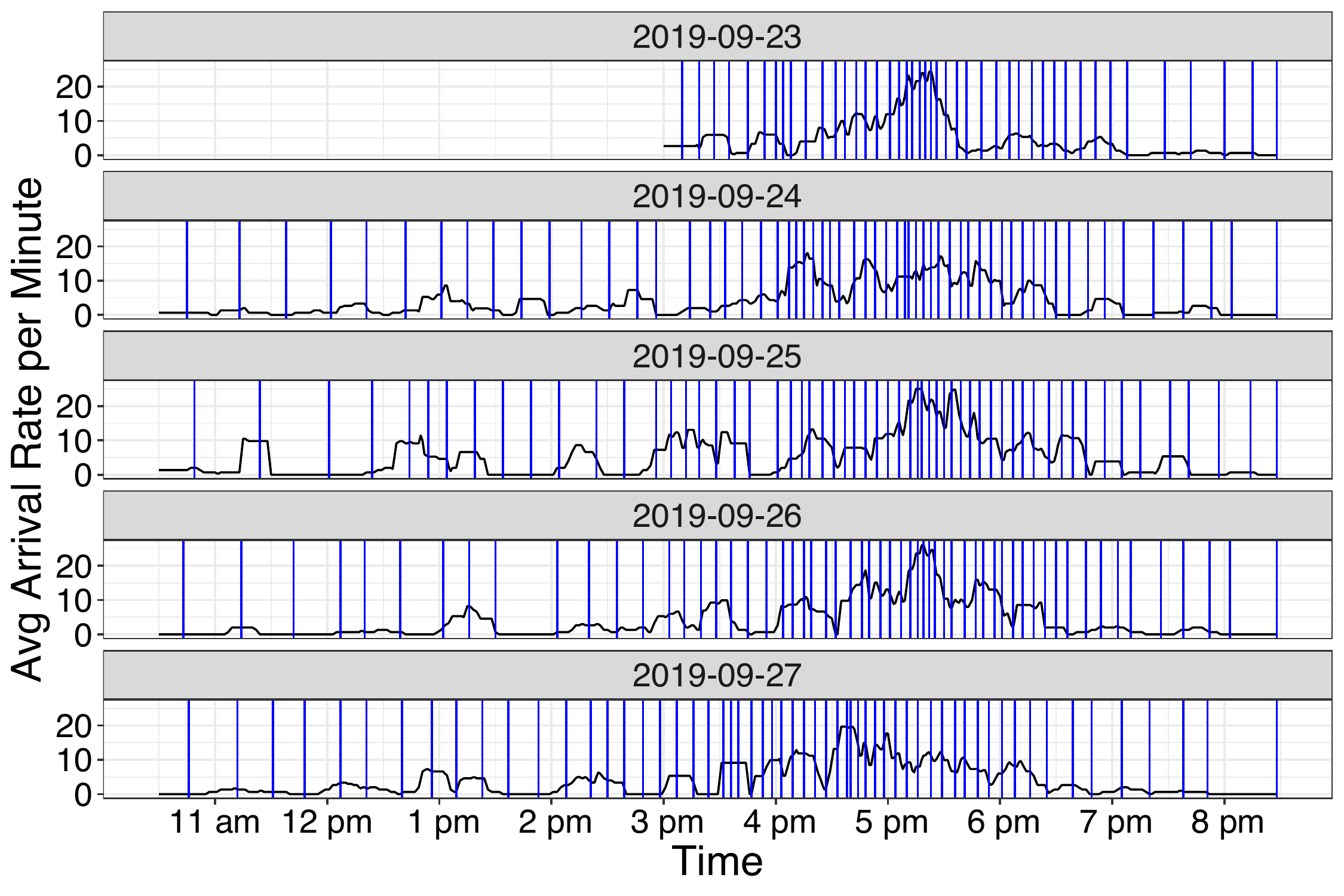}
	\caption{Optimised scheduling of bus dispatch time.}\label{fig:optimized-schedule}
\end{figure}

\subsection{Optimisation results}

We use the passenger demand data from our deployment (described in \S\ref{sect:5daydeployment}) as input to our optimization model. 
The optimal bus dispatching times (for each day) are illustrated in Fig.~\ref{fig:optimized-schedule}.

\begin{figure}[t!]
	\centering
	\includegraphics[width=0.7\textwidth]{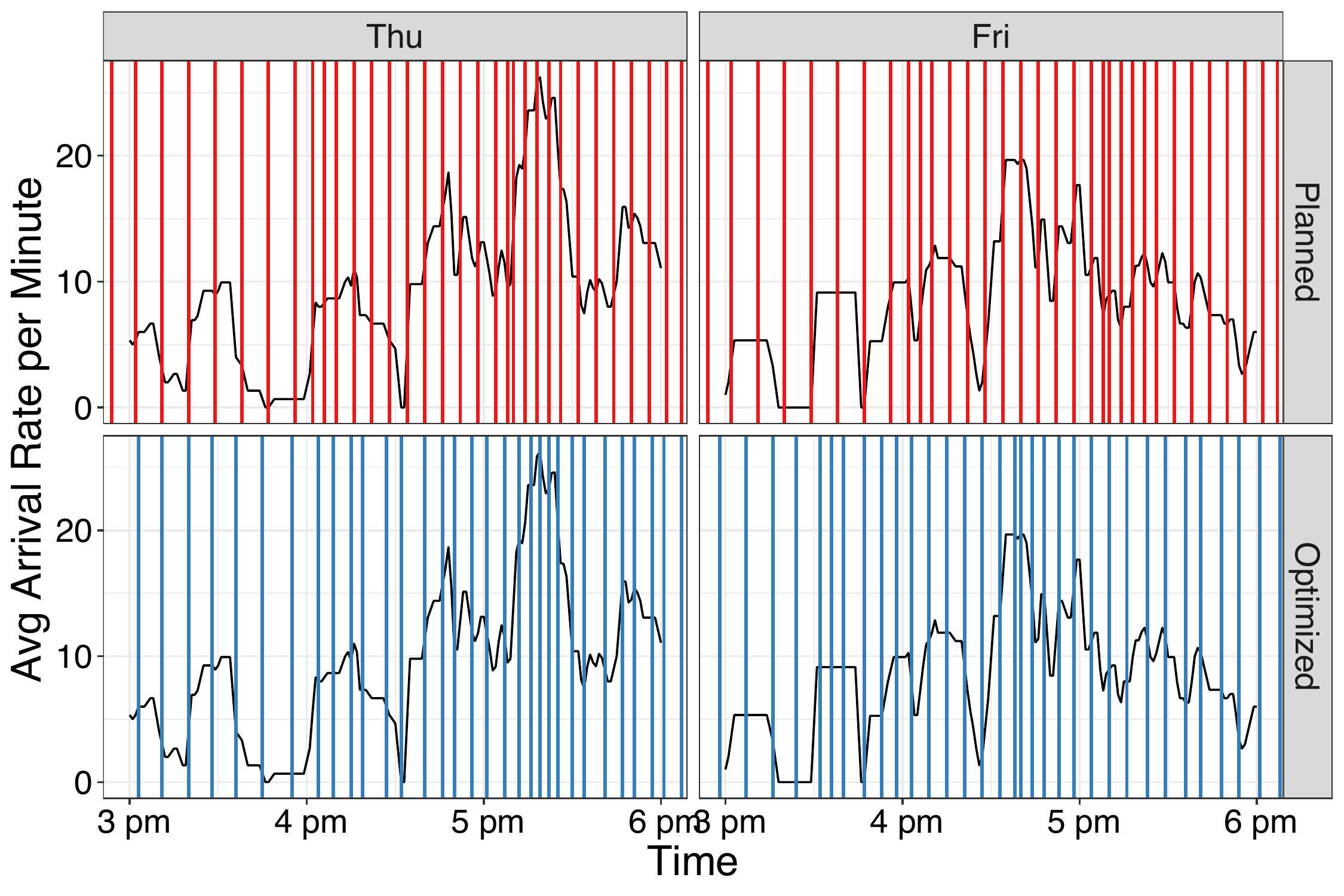}
	\caption{Comparison of existing bus dispatching time and optimised bus dispatching time for Monday and Friday.}\label{fig:compare-schedule}
\end{figure}

Comparing the two schedules, \ie static (in Fig.~\ref{fig:static-schedule}) with optimal (in Fig.~\ref{fig:optimized-schedule}), we observe that they match to a great extent from Monday to Thursday, displaying a higher frequency pattern during the peak time (5:00pm-5:30pm). Also, some minor variations can be observed, specially during times closer to the peak.
For example, an additional peak period is seen between 4pm and 4:30pm on Tuesday, prompting $40$\% more buses to be scheduled for that time compared to other weekdays.
The most notable difference between the two schedules is observed on Friday, where buses are scheduled more frequently by our optimal policy between 4.30pm and 5pm (as shown in Fig.~\ref{fig:optimized-schedule}) to fulfill the actual peak demand. To better visualize the shift in the peak demand, we illustrate in Fig.~\ref{fig:compare-schedule} a comparison of static schedule (top) with the optimal schedule (bottom) for Thursday (left) and Friday (right) during 3-6pm. 
It is clearly seen that static and optimal schedules highly correlate on Thursday with 7 buses scheduled during the peak (5pm-5:30pm). On the other hand, on Friday, the optimal schedule has two additional services during the revised peak period (4:30pm - 5pm) but two fewer buses during 5pm - 5:30pm.

We compute the average wait-time per passenger to compare the performance of static versus optimal bus schedules. 
Table~\ref{table:opt_waittime_tb} summarizes the average wait-time across each day of our 5-day trial, highlighting the reduction in wait-time that can be achieved by employing the optimal (dynamic) scheduling, compared to static scheduling.
As expected, the optimal approach yields lower wait-times on average across all five days -- the largest improvement ($42.93$\% reduction in wait-time) is observed on Friday, while the smallest improvement is achieved ($10.98$\% reduction in wait-time) on Thursday. 
Focusing on Friday, the average wait-time from the static bus schedule is $7.69$ minutes which can be reduced to $4.39$ minutes with the optimal bus schedule. This relatively large gap highlights to the inefficiency of existing static bus schedule which falls short in accommodating the shift in peak demand. 

\begin{table}[t]
	\begin{center}
		\caption{Average wait time (min) of passengers: static versus dynamic scheduling.}
		\label{table:optimisation}
		\begin{tabular}{l c c c c c c }
			\toprule
\multicolumn{1}{c}{} & \multicolumn{3}{c}{All day} & \multicolumn{3}{c}{Peak hour} \\
\cmidrule(l{3pt}r{3pt}){2-4} \cmidrule(l{3pt}r{3pt}){5-7}
date & actual & optimised & reduction &actual & optimised & reduction\\
\midrule
Mon & 4.43 & 3.22 & 27.21\% & 4.52 & 3.16 & 30.12\%\\
Tue & 6.55 & 5.07 & 22.72\% & 4.75 & 3.90 & 17.90\%\\
Wed & 10.22 & 6.78 & 33.66\% & 6.57 & 5.39 & 17.91\%\\
Thu & 6.79 & 6.04 & 10.98\% & 6.84 & 5.92 & 13.43\%\\
Fri & 7.69 & 4.39 & 42.93\% & 6.80 & 3.38 & 50.23\%\\
\bottomrule
		\end{tabular}
	\end{center}
\end{table}
\label{table:opt_waittime_tb}

\begin{figure}[t!]
	\centering
	\includegraphics[width=0.7\textwidth]{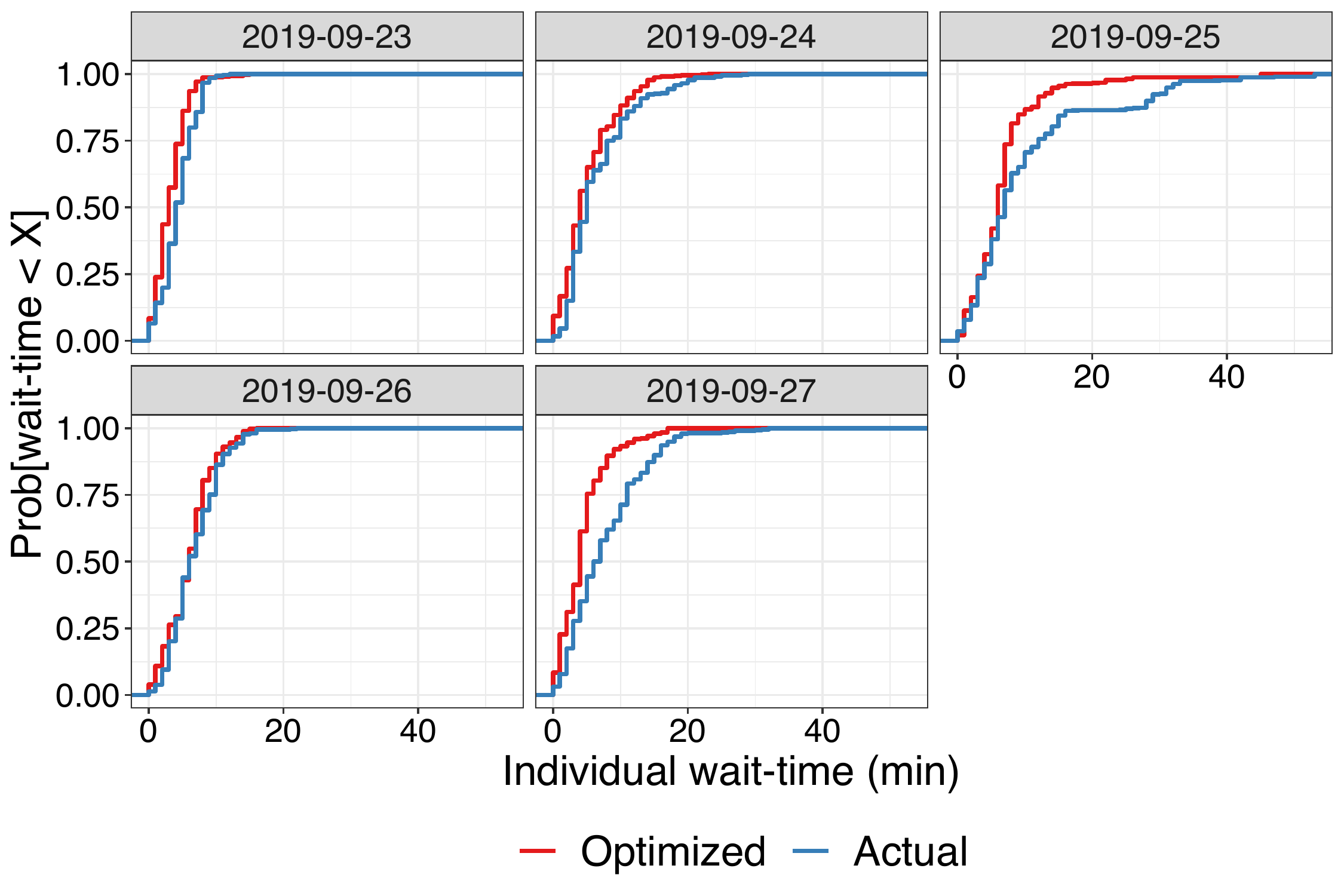}
	\caption{Cumulative Frequency Function (CDF) graph of individual wait-time from optimised and actual bus schedule for five deployment days.}\label{fig:wait_time_CDF}
\end{figure}

Lastly, we plot in Fig.~\ref{fig:wait_time_CDF} the cumulative distribution function (CDF) of  the wait-time of individual passengers. It can be seen that optimal and static curves exhibit very minor differences on Monday, Tuesday and Thursday, with the former slightly shifted to the left (slightly lower wait times overall). In contrast, we observe a more noticeable difference on Wednesday and Friday. On Friday, the static schedule requires more than 80\% of the passengers to wait up to 12 minutes. In contrast, the optimal schedule reduces the wait time to 7 minutes. On Friday, the longest wait time is 32 minutes with the static schedule, while this reduces to 17 minutes for the optimal scheme. The difference between the two curves on Wednesday can be attributed to the unexpected higher demand during the non-peak time period 11:30am - 1:00 pm (see Fig.~\ref{fig:static-schedule}). The optimal scheme adapts to this variation by scheduling additional buses during this time period (see Fig.~\ref{fig:optimized-schedule}).

\section{Conclusion}\label{sect:bus_conclusion}

In this chapter, we proposed an end-to-end system for queue length measurement using ultrasonic sensors and LoRaWAN for data communications. Our proposed solution can be used in a variety of outdoor settings (e.g. stadiums, airports) where an orderly queue is formed.
We first described the implementation of our people sensing devices that are battery-operated, able to communicate wirelessly, and have water-proof exterior, allowing the solution to be easily deployed in an outdoor environment. Next, we developed an algorithm to infer queue length information from the collected data. Our algorithm achieved a decent accuracy with MAE of 10.7 people for a queue size of up to 100 passengers. Finally, we showed that our queue detection system and algorithm can be used to optimize bus scheduling, revealing a potential wait time reduction of 42.93\% over a day by adopting a demand-driven bus scheduling.

\chapter{Conclusions and Future Work}
	\label{chap:conclusion}
	\vspace{-5mm}
	\minitoc
	
	\section{Conclusions}

Higher educational institutions are experiencing a steady growth in enrolment, which puts pressure on the demand for campus resources such as classrooms, parking spaces, and public transportation. Meanwhile, these resources are generally confronted with the issue of low efficiency, stemming from the lack of usage monitoring system that would allow campus managers to make informed decision in regards to asset management. Together, this can obstruct the university from fulfilling its anticipated demand due to inefficient use of limited resources. The emerging technology of Internet of Thing (IoT) has presented opportunities to transform campuses in new ways, not only to increase operational efficiency, but also to improve the campus community experiences. Despite myriad of promises IoT can bring, there's still a limited number of research exploring benefits of such technology in enabling a smarter management of resources in the context of smart campuses. 

In this thesis, we proposed a framework that exploits IoT technology to revolutionise the way campuses are managing their resources. Our framework involves the deployment of IoT devices for monitoring usage of campus resources, building of predictive models to forecast future demand patterns, and formulation of optimisation models that allow campus managers to make informed decisions about the allocations and management of their assets. By complying to this framework and using UNSW Sydney as an experimental platform, we demonstrated tangible benefits IoT can bring in three application domains, namely classroom utilisation, parking space management, and bus scheduling.

We summarise the important contributions of this thesis towards realising a smarter campus in the following:

		\begin{itemize}
			\item We undertook a comprehensive evaluation of IoT technologies for occupancy sensing and selected the most suitable solution for the deployment at 9 lecture theatres on UNSW campus. We then used the collected occupancy data to develop machine learning models that can predict future class attendance, where our models achieved a root mean square error (RMSE) of less than 0.16. Lastly, we developed an optimisation model for classroom allocations aiming to minimise usage of room costs. We showed that by allocating classrooms based on the predicted demand rather than the enrolment number, our university can reduce the usage of classroom resources by over 10\%.
			
			\item We instrumented an on-campus parking infrastructure with License Plate Recognition (LPR) cameras in order to monitor the parking behaviour of users. The data, collected over a period of 15 months, was then used to forecast future parking demand at multiple time horizons in the future, ranging from one day to up 10 weeks. Our model achieved a mean absolute error (MAE) of 4.58 cars per hour for 5-day prediction horizon. Finally, we proposed an optimisation framework employing a non-homogeneous time-continuous Markov Model and Mixed Integer Linear Programming to assist campus managers in partitioning the parking spaces to accommodate new paradigms of car use. 
			
			\item We designed and implemented queue sensing solution using ultrasonic sensor and LoRaWAN communications. The solution was deployed at our main campus bus stop and was shown to yield a reasonable detection accuracy with MAE of 10.7 people for a queue length of up to 100 people. Based on the collected public transport demand, we developed a demand-driven optimisation model to reschedule bus arrival times in order to minimise total wait time of passengers. As a result, we demonstrated that public transport experience of campus community can significantly enhance by 42.93\%. 
			
\end{itemize}

The aforementioned applications that were explored in this thesis have demonstrated some tangible use cases of a Smart Campus in the area of resource management. Even though the usability and accuracy of the results reported in this work was only applied to our University, the methodologies developed therein can set the ground for far-reaching applications in the Smart city domain.

\section{Future Work}
	
Our work is a significant milestone in realising tangible benefits of a Smart Campus, powered by IoT technology. Some of the key enhancements and refinements that can be extended in the future are outlined below:
		
		\begin{itemize}
			\item In this thesis, we deployed a single sensor type solution to monitor each campus resource usage, such as beam counter for occupancy detection and LPR cameras for parking detection. Fusion of sensing data from multiple sensors can be considered in order to improve the accuracy of occupancy detection. Some examples of inexpensive sensors that can be incorporated into the existing occupancy monitoring system are $CO_2$, temperature, and sound sensors. Exploiting data from existing WiFi-infrastructure is another interesting domain to be explored.
			
			\item We developed prediction models to forecast future usage of campus resources using historical data. The prediction accuracy can be improved by integrating external data into the models in order to capture any non-typical events that may impact demand behaviour. For instance, data on precipitation and special events occurring on campus are likely to impact student attendance rate and overall number of people coming to the campus, leading to higher demand on parking spaces and public transport.
			
			\item In our classroom optimisation model, the allocation only consider the timing of the lecture events without contemplating the convenience of students and lecturers. A more comprehensive models may consider the location of classes that an individual need to attend, so that the allocation can also minimise the travel distance of students and lecturers between consecutive classes. This factor is crucial, particularly for campuses with vast area of land, as the travel time between classes can be notoriously high.
			
			\item Our formulation for optimising bus scheduling assumes prior knowledge of passenger number, which is of course not practical, but is meant to establish a benchmark upper bound on the potential wait time reduction. A practical implementation can take advantage of a long-term data collection to forecast future pattern of bus demand. Furthermore, real-time prediction of passenger arrival rate will allow scheduling of buses to be more dynamic, allowing the decision to be made in a more spontaneous manner.

			\item Other criteria can be incorporated into the optimisation problems in addition to resource efficiency. For instance, student experience and costs (extra buses).
			
		\end{itemize}
	
We hope other researchers will explore the future directions identified above.

\backmatter
\renewcommand{\refname}{References} 

\appto{\bibsetup}{\sloppy}
\apptocmd{\sloppy}{\hbadness 10000\relax}{}{}
\begin{singlespace}
	\setlength\bibitemsep{10pt}   
	\printbibliography[heading=bibintoc,title={References}]

@article{caron2016internet2,
  title={The Internet of Things (IoT) and its impact on individual privacy: An Australian perspective},
  author={Caron, Xavier and Bosua, Rachelle and Maynard, Sean B and Ahmad, Atif},
  journal={Computer Law \& Security Review},
  volume={32},
  number={1},
  pages={4--15},
  year={2016},
  publisher={Elsevier}
}

@inproceedings{beltran2013thermosense,
  title={Thermosense: Occupancy thermal based sensing for hvac control},
  author={Beltran, Alex and Erickson, Varick L and Cerpa, Alberto E},
  booktitle={Proceedings of the 5th ACM Workshop on Embedded Systems For Energy-Efficient Buildings},
  pages={1--8},
  year={2013}
}

@article{sutjarittham2019experiences,
  title={Experiences with IoT and AI in a smart campus for optimizing classroom usage},
  author={Sutjarittham, Thanchanok and Gharakheili, Hassan Habibi and Kanhere, Salil S and Sivaraman, Vijay},
  journal={IEEE Internet of Things Journal},
  volume={6},
  number={5},
  pages={7595--7607},
  year={2019},
  publisher={IEEE}
}

@INPROCEEDINGS{sutjarittham:carparkusage,
	author={T. {Sutjarittham} and G. {Chen} and H. {Habibi Gharakheili} and V. {Sivaraman} and S. S. {Kanhere}},
	booktitle={Proc. IEEE WoWMoM}, 
	pages={1--10},
	title={{Measuring and Modeling Car Park Usage: Lessons Learned from a Campus Field-Trial}}, 
	year={2019},
	month={June},
	address={Washington DC, USA}
}

@article{sutjarittham2020monetizing,
  title={Monetizing Parking IoT Data via Demand Prediction and Optimal Space Sharing},
  author={Sutjarittham, Thanchanok and Gharakheili, Hassan Habibi and Kanhere, Salil S and Sivaraman, Vijay},
  journal={IEEE Internet of Things Journal},
  year={2020},
  publisher={IEEE}
}

@inproceedings{sutjarittham:ANTS, 
	author={Sutjarittham, Thanchanok and Habibi Gharakheili, Hassan and Kanhere, Salil S. and Sivaraman, Vijay}, 
	title={{Realizing a Smart University Campus: Vision, Architecture, and Implementation}}, 
	booktitle={Proc. IEEE ANTS}, 
	year={2018},
	month={Dec},
	address = {Indore, India}
}

@inproceedings{sutjarittham:classroomusage, 
	title={{Data-Driven Monitoring and Optimization of Classroom Usage in a Smart Campus}}, 
	booktitle={Proc. ACM/IEEE IPSN}, 
	author={T. Sutjarittham and others}, 
	year={2018},
	address = {Porto, Portugal}
}

@article{lewis2008survey,
  title={A survey of metaheuristic-based techniques for university timetabling problems},
  author={Lewis, Rhydian},
  journal={OR spectrum},
  volume={30},
  number={1},
  pages={167--190},
  year={2008},
  publisher={Springer}
}

@inproceedings{ferris2010onebusaway,
  title={OneBusAway: results from providing real-time arrival information for public transit},
  author={Ferris, Brian and Watkins, Kari and Borning, Alan},
  booktitle={Proceedings of the SIGCHI Conference on Human Factors in Computing Systems},
  pages={1807--1816},
  year={2010}
}

@article{valks2019smart,
  title={Smart campus tools 2.0 exploring the use of real-time space use measurement at universities and organizations},
  author={Valks, Bart and Arkesteijn, Monique and Den Heijer, Alexandra},
  journal={Facilities},
  year={2019},
  publisher={Emerald Publishing Limited}
}

@article{webb2018campus,
  title={Campus IoT collaboration and governance using the NIST cybersecurity framework},
  author={Webb, James and Hume, Dustin},
  year={2018},
  publisher={IET}
}

@inproceedings{hengliang2016construction,
  title={The Construction of Intelligent Transportation System Based on the Construction of Wisdom Campus-Take Soochow University as an example},
  author={Hengliang, Tang and Chuanrong, Chen},
  booktitle={2016 Eighth International Conference on Measuring Technology and Mechatronics Automation (ICMTMA)},
  pages={711--714},
  year={2016},
  organization={IEEE}
}

@inproceedings{bandara2016smart,
  title={Smart campus phase one: Smart parking sensor network},
  author={Bandara, HMAPK and Jayalath, JDC and Rodrigo, ARSP and Bandaranayake, AU and Maraikar, Z and Ragel, RG},
  booktitle={2016 Manufacturing \& Industrial Engineering Symposium (MIES)},
  pages={1--6},
  year={2016},
  organization={IEEE}
}

@article{buckman2014smart,
  title={What is a smart building?},
  author={Buckman, Alex H and Mayfield, Martin and Beck, Stephen BM},
  journal={Smart and Sustainable Built Environment},
  year={2014},
  publisher={Emerald Group Publishing Limited}
}

@article{kolokotsa2016development,
  title={Development of a web based energy management system for University Campuses: The CAMP-IT platform},
  author={Kolokotsa, Dionysia and Gobakis, Kostas and Papantoniou, S and Georgatou, C and Kampelis, N and Kalaitzakis, Kostas and Vasilakopoulou, K and Santamouris, Mat},
  journal={Energy and Buildings},
  volume={123},
  pages={119--135},
  year={2016},
  publisher={Elsevier}
}

@article{habibi2016smart,
  title={Smart innovation systems for indoor environmental quality (IEQ)},
  author={Habibi, Shahryar},
  journal={Journal of Building Engineering},
  volume={8},
  pages={1--13},
  year={2016},
  publisher={Elsevier}
}

@techreport{gartner:iotforecast,
	title = {{Forecast: Internet of Things — Endpoints and Associated Services, Worldwide, 2017}},
	author       = { Gartner}, 
	type = {Market Report},
  	url={https://www.gartner.com/en/documents/3840665/forecast-internet-of-things-endpoints-and-associated-ser},
 	month        = {Dec.},
 	 year         = {2017}
}

@article{de2015brescia,
  title={The Brescia Smart Campus Demonstrator. Renovation toward a zero energy classroom building},
  author={De Angelis, Enrico and Ciribini, Angelo Luigi Camillo and Tagliabue, Lavinia Chiara and Paneroni, Michela},
  journal={Procedia engineering},
  volume={118},
  pages={735--743},
  year={2015},
  publisher={Elsevier}
}

@inproceedings{hipwell2014developing,
  title={Developing smart campuses—A working model},
  author={Hipwell, Steven},
  booktitle={2014 International Conference on Intelligent Green Building and Smart Grid (IGBSG)},
  pages={1--6},
  year={2014},
  organization={IEEE}
}

@electronic{curtin:smartcampus,
	Year = {2017},
	Title = {{Curtin Uni, Hitachi team up for smart campus initiative}},
	url = {https://www.iothub.com.au/news/curtin-uni-hitachi-team-up-for-smart-campus-initiative-460501}
}

@electronic{melborne:smartcampus,
	Title = {{Case Study: University of Melbourne}},
	year = {2017},
	url = {https://www.cisco.com/c/en/us/about/case-studies-customer-success-stories/univ-melbourne.html}
}

@electronic{deakin:smartcampus,
	Title = {Building a Smarter Campus: Deakin University},
	Year = {2016},
	url = {https://blogs.cisco.com/education/building-a-smarter-campus-deakin-university}
}

@electronic{gasglow:smartcampus,
	Title = {Smart Campus - University of Glasgow},
	Year = {2015},
	url = {http://futurecities.catapult.org.uk/project/smart-campus-university-of-glasgow/}
}

@article{wang2018understanding,
  title={Understanding occupancy and user behaviour through Wi-Fi-based indoor positioning},
  author={Wang, Yan and Shao, Li},
  journal={Building Research \& Information},
  volume={46},
  number={7},
  pages={725--737},
  year={2018},
  publisher={Taylor \& Francis}
}

@inproceedings{hentschel2016supersensors,
  title={Supersensors: Raspberry Pi devices for smart campus infrastructure},
  author={Hentschel, Kristian and Jacob, Dejice and Singer, Jeremy and Chalmers, Matthew},
  booktitle={2016 IEEE 4th International Conference on Future Internet of Things and Cloud (FiCloud)},
  pages={58--62},
  year={2016},
  organization={IEEE}
}

@article{lopez2017overcrowding,
  title={Overcrowding detection in indoor events using scalable technologies},
  author={Lopez-Novoa, Unai and Aguilera, Unai and Emaldi, Mikel and L{\'o}pez-De-Ipi{\~n}a, Diego and P{\'e}rez-De-Albeniz, Iker and Valerdi, David and Iturricha, Ibai and Arza, Eneko},
  journal={Personal and Ubiquitous Computing},
  volume={21},
  number={3},
  pages={507--519},
  year={2017},
  publisher={Springer}
}

@inproceedings{nyarko2013cloud,
  title={Cloud based passive building occupancy characterization for attack and disaster response},
  author={Nyarko, Kofi and Wright-Brown, Cecelia},
  booktitle={2013 IEEE International Conference on Technologies for Homeland Security (HST)},
  pages={748--753},
  year={2013},
  organization={IEEE}
}

@inproceedings{mohottige2020evaluating,
  title={Evaluating Emergency Evacuation Events Using Building WiFi Data.},
  author={Mohottige, Iresha Pasquel and Gharakheili, Hassan Habibi and Vishwanath, Arun and Kanhere, Salil S and Sivaraman, Vijay},
  booktitle={IoTDI},
  pages={116--127},
  year={2020}
}

@inproceedings{anagnostopoulos2015robust,
  title={Robust waste collection exploiting cost efficiency of IoT potentiality in Smart Cities},
  author={Anagnostopoulos, Theodoros and Zaslavsky, Arkady and Medvedev, Alexey},
  booktitle={2015 International conference on recent advances in internet of things (RIoT)},
  pages={1--6},
  year={2015},
  organization={IEEE}
}

@inproceedings{folianto2015smartbin,
  title={Smartbin: Smart waste management system},
  author={Folianto, Fachmin and Low, Yong Sheng and Yeow, Wai Leong},
  booktitle={2015 IEEE Tenth International Conference on Intelligent Sensors, Sensor Networks and Information Processing (ISSNIP)},
  pages={1--2},
  year={2015},
  organization={IEEE}
}

@inproceedings{anagnostopoulos2015top,
  title={Top--k query based dynamic scheduling for IoT-enabled smart city waste collection},
  author={Anagnostopoulos, Theodoros and Zaslavsy, Arkady and Medvedev, Alexey and Khoruzhnicov, Sergei},
  booktitle={2015 16th IEEE International Conference on Mobile Data Management},
  volume={2},
  pages={50--55},
  year={2015},
  organization={IEEE}
}

@inproceedings{khan2020iot,
  title={IoT Based University Garbage Monitoring System for Healthy Environment for Students},
  author={Khan, Muhammad Nasir and Naseer, Fawad},
  booktitle={2020 IEEE 14th International Conference on Semantic Computing (ICSC)},
  pages={354--358},
  year={2020},
  organization={IEEE}
}

@inproceedings{robles2014internet,
  title={An internet of things-based model for smart water management},
  author={Robles, Tom{\'a}s and Alcarria, Ram{\'o}n and Mart{\'i}n, Diego and Morales, Augusto and Navarro, Mariano and Calero, Rodrigo and Iglesias, Sofia and L{\'o}pez, Manuel},
  booktitle={2014 28th International Conference on Advanced Information Networking and Applications Workshops},
  pages={821--826},
  year={2014},
  organization={IEEE}
}

@inproceedings{mudumbe2015smart,
  title={Smart water meter system for user-centric consumption measurement},
  author={Mudumbe, Mduduzi John and Abu-Mahfouz, Adnan M},
  booktitle={2015 IEEE 13th International Conference on Industrial Informatics (INDIN)},
  pages={993--998},
  year={2015},
  organization={IEEE}
}

@inproceedings{gabrielli2014smart,
  title={Smart water grids for smart cities: A sustainable prototype demonstrator},
  author={Gabrielli, Leonardo and Pizzichini, Mirco and Spinsante, Susanna and Squartini, Stefano and Gavazzi, Roberto},
  booktitle={2014 European Conference on Networks and Communications (EuCNC)},
  pages={1--5},
  year={2014},
  organization={IEEE}
}

@inproceedings{verma2015towards,
  title={Towards an IoT based water management system for a campus},
  author={Verma, Prachet and Kumar, Akshay and Rathod, Nihesh and Jain, Pratik and Mallikarjun, S and Subramanian, Renu and Amrutur, Bharadwaj and Kumar, MS Mohan and Sundaresan, Rajesh},
  booktitle={2015 IEEE First International Smart Cities Conference (ISC2)},
  pages={1--6},
  year={2015},
  organization={IEEE}
}

@article{nainan2013rfid,
  title={RFID technology based attendance management system},
  author={Nainan, Sumita and Parekh, Romin and Shah, Tanvi},
  journal={arXiv preprint arXiv:1306.5381},
  year={2013}
}

@article{chew2015sensors,
  title={Sensors-enabled smart attendance systems using NFC and RFID technologies},
  author={Chew, Cheah Boon and Mahinderjit-Singh, Manmeet and Wei, Kam Chiang and Sheng, Tan Wei and Husin, Mohd Heikal and Malim, NHAH},
  journal={Int. J. New Comput. Archit. Appl},
  volume={5},
  pages={19--29},
  year={2015}
}

@article{wang2013toward,
  title={Toward a Green Campus with the Internet of Things--the Application of Lab Management},
  author={Wang, H},
  journal={world},
  volume={7},
  pages={8},
  year={2013}
}

@electronic{evolveplus:overhead,
	Year = {2017},
	Title = {{EvolvePlus: Overhead Thermal Counter}},
	url = {https://goo.gl/ugtW1L}
}

@electronic{evolveplus:beamcounter,
	Year = {2017},
	Title = {{EvolvePlus: Wireless People Counters with Remote Data Viewing}},
	url = {https://goo.gl/dAUe8G}
}

@electronic{steinel:PIR,
	Year = {2017},
	Title = {{Steinel Australia: Presence Control PRO IR Quattro HD}},
	url = {https://goo.gl/T8zvNF}
}

@electronic{data2018occupancy,
	Year = {2018},
	Title = {{UNSW Smart Campus - Classroom Usage Monitoring and Optimization}},
	url = {https://smartcampus.unsw.edu.au/roomoccupancy/data}
}

@ELECTRONIC{tara:data2018,
	Year = {2018},
	Title = {{UNSW Smart Campus - Classroom Usage Monitoring and Optimization}},
	url = {https://smartcampus.unsw.edu.au/roomoccupancy/data/IoT-J/}
}

@inproceedings{wifiLCN2018,
	author = {I. P. Mohottige and others},
	title = {{Estimating Room Occupancy in a Smart Campus Using WiFi Soft Sensors}},
	booktitle = {Proc. IEEE LCN},
	year = {2018},
	month = {October},
	address = {Chicago, USA}
}

@article{devadoss:attendance,
  title={Evaluation of factors influencing student class attendance and performance},
  author={Devadoss, Stephen and Foltz, John},
  journal={American Journal of Agricultural Economics},
  volume={78},
  number={3},
  pages={499--507},
  year={1996},
  publisher={Wiley Online Library}
}

@article{Breiman:rf,
  author= {L. Breiman and others},
  title={Random Forests},
  journal={Machine Learning},
  volume={45},
  number={1},
  year={2001},
  pages={5-32}
}

@Manual{R,
    title = {R: A Language and Environment for Statistical Computing},
    author = {{R Core Team}},
    organization = {R Foundation for Statistical Computing},
    address = {Vienna, Austria},
    year = {2013},
    url = {http://www.R-project.org/}
}

@article{kuhn:caret,
  title={Caret package},
  author={M. Kuhn and others},
  journal={Journal of statistical software},
  volume={28},
  number={5},
  pages={1--26},
  year={2008}
}

@inproceedings{drucker:SVR,
  title={Support vector regression machines},
  author={H. Drucker and others},
  booktitle={Advances in neural information processing systems},
  pages={155--161},
  year={1997}
}

@online{google:ORTOOL,
  title = {Google Optimization Tools},
  year = 2018,
  url = {https://developers.google.com/optimization/},
  urldate = {2018-07-20}
}

@book{rossi:handbook,
  title={Handbook of constraint programming},
  author={F. Rossi and others},
  year={2006},
  publisher={Elsevier}
}

@inproceedings{melfi2011measuring,
  title={Measuring building occupancy using existing network infrastructure},
  author={Melfi, Ryan and Rosenblum, Ben and Nordman, Bruce and Christensen, Ken},
  booktitle={2011 International Green Computing Conference and Workshops},
  pages={1--8},
  year={2011},
  organization={IEEE}
}

@article{liu2015occupancy,
  title={Occupancy inference using pyroelectric infrared sensors through hidden Markov models},
  author={Liu, Pengcheng and Nguang, Sing-Kiong and Partridge, Ashton},
  journal={IEEE Sensors Journal},
  volume={16},
  number={4},
  pages={1062--1068},
  year={2015},
  publisher={IEEE}
}

@article{dodier2006building,
  title={Building occupancy detection through sensor belief networks},
  author={Dodier, Robert H and Henze, Gregor P and Tiller, Dale K and Guo, Xin},
  journal={Energy and buildings},
  volume={38},
  number={9},
  pages={1033--1043},
  year={2006},
  publisher={Elsevier}
}

@article{duarte2013revealing,
  title={Revealing occupancy patterns in an office building through the use of occupancy sensor data},
  author={Duarte, Carlos and Van Den Wymelenberg, Kevin and Rieger, Craig},
  journal={Energy and buildings},
  volume={67},
  pages={587--595},
  year={2013},
  publisher={Elsevier}
}

@inproceedings{wahl2012distributed,
  title={A distributed PIR-based approach for estimating people count in office environments},
  author={Wahl, Florian and Milenkovic, Marija and Amft, Oliver},
  booktitle={2012 IEEE 15th International Conference on Computational Science and Engineering},
  pages={640--647},
  year={2012},
  organization={IEEE}
}

@inproceedings{raykov2016predicting,
  title={Predicting room occupancy with a single passive infrared (PIR) sensor through behavior extraction},
  author={Raykov, Yordan P and Ozer, Emre and Dasika, Ganesh and Boukouvalas, Alexis and Little, Max A},
  booktitle={Proceedings of the 2016 ACM International Joint Conference on Pervasive and Ubiquitous Computing},
  pages={1016--1027},
  year={2016}
}

@electronic{movingpir:dynamicpricing,
	author={Swagatam},
	Year = {2019},
	Title = {{Detecting Static Human with PIR}},
	url = {https://www.homemade-circuits.com/pir-circuit-for-detecting-static-or/},
	note = {Accessed on: 2020-11-30}
}

@article{wang1999experimental,
  title={Experimental validation of CO2-based occupancy detection for demand-controlled ventilation},
  author={Wang, Shengwei and Burnett, John and Chong, Hoishing},
  journal={Indoor and built environment},
  volume={8},
  number={6},
  pages={377--391},
  year={1999},
  publisher={Karger Publishers}
}

@article{jiang2016indoor,
  title={Indoor occupancy estimation from carbon dioxide concentration},
  author={Jiang, Chaoyang and Masood, Mustafa K and Soh, Yeng Chai and Li, Hua},
  journal={Energy and Buildings},
  volume={131},
  pages={132--141},
  year={2016},
  publisher={Elsevier}
}

@article{zuraimi2017predicting,
  title={Predicting occupancy counts using physical and statistical Co2-based modeling methodologies},
  author={Zuraimi, MS and Pantazaras, A and Chaturvedi, KA and Yang, JJ and Tham, KW and Lee, SE},
  journal={Building and Environment},
  volume={123},
  pages={517--528},
  year={2017},
  publisher={Elsevier}
}

@article{rahman2017occupancy,
  title={Occupancy estimation based on indoor CO2 concentration: Comparison of neural network and bayesian methods},
  author={Rahman, Haolia and Han, Hwataik},
  journal={International Journal of Air-Conditioning and Refrigeration},
  volume={25},
  number={03},
  pages={1750021},
  year={2017},
  publisher={World Scientific}
}

@article{dong10:ITEST, 
	title={An information technology enabled sustainability test-bed (ITEST) for occupancy detection through an environmental sensing network}, 
	volume={42}, 
	number={7}, 
	journal={Energy and Buildings}, 
	author = {B. Dong and B. Andrews and K. P. Lam and M. {H{\"{o}} Ynck} and R. Zhang and Y. Chiou and D. Benitez},
	year={2010}, 
	pages={1038-1046},
	month={July}
}

@inproceedings{deyr:Namatad,
  title={{Namatad: Inferring occupancy from building sensors using machine learning}},
  author={A. Dey and X. Ling and A. Syed and Y. Zheng and B. Landowski and D. Anderson and K. Stuart and M. E. Tolentino},
  booktitle={Proc. IEEE WF-IoT},
  year={2016},
  address = {Reston, VA, USA}
}

@article{zimmermann:2017fusion,
  title={{Fusion of Non-Intrusive Environmental Sensors for Occupancy Detection in Smart Homes}},
  author={L. Zimmermann and R. Weigel and G. Fischer},
  journal={IEEE Internet of Things Journal},
  year={2017},
  publisher={IEEE},
  volume={5}, 
  pages={2343-2352},
  number={4}
  
}

@inproceedings{yang2012:multi,
  title={A multi-sensor based occupancy estimation model for supporting demand driven HVAC operations},
  author={Z. Yang and N. Li and B. Becerik-Gerber and M. Orosz},
  booktitle={Proc. Symposium on Simulation for Architecture and Urban Design},
  year={2012},
  adress={Orlando, Florida}
}

@article{chen2016fusion,
  title={A fusion framework for occupancy estimation in office buildings based on environmental sensor data},
  author={Chen, Zhenghua and Masood, Mustafa K and Soh, Yeng Chai},
  journal={Energy and Buildings},
  volume={133},
  pages={790--798},
  year={2016},
  publisher={Elsevier}
}

@inproceedings{masood2015real,
  title={Real-time occupancy estimation using environmental parameters},
  author={Masood, Mustafa K and Soh, Yeng Chai and Chang, Victor W-C},
  booktitle={2015 international joint conference on neural networks (IJCNN)},
  pages={1--8},
  year={2015},
  organization={IEEE}
}

@article{stancil2008active,
  title={Active multicamera networks: From rendering to surveillance},
  author={Stancil, Brian A and Zhang, Cha and Chen, Tsuhan},
  journal={IEEE Journal of selected topics in signal processing},
  volume={2},
  number={4},
  pages={597--605},
  year={2008},
  publisher={IEEE}
}

@inproceedings{trivedi2000intelligent,
  title={Intelligent environments and active camera networks},
  author={Trivedi, Mohan and Kohsia, Huang and Mikic, Ivana},
  booktitle={Smc 2000 conference proceedings. 2000 ieee international conference on systems, man and cybernetics.'cybernetics evolving to systems, humans, organizations, and their complex interactions'(cat. no. 0},
  volume={2},
  pages={804--809},
  year={2000},
  organization={IEEE}
}

@article{hou_pang_2011, 
		title={{People Counting and Human Detection in a Challenging Situation}}, 
		volume={41}, 
		number={1}, 
		journal={IEEE Transactions on Systems, Man, and Cybernetics -- Part A: Systems and Humans}, 
		author = {Y. Hou and G. K. H. Pang},
		year={2011}, 
		month={January},
		pages={24-33}
}

@inproceedings{li2011:robust,
  title={Robust people counting in video surveillance: Dataset and system},
  author={J. Li and L. Huang and C. Liu},
  booktitle={Proc. IEEE AVSS},
  year={2011},
  address={Klagenfurt, Austria}
}

@inproceedings{paci2014:classroomimage, 
	title={0, 1, 2, many -- A classroom occupancy monitoring system for smart public buildings},
	booktitle={Proc. Conference on Design and Architectures for Signal and Image Processing}, 
	author={F. Paci and D. Brunelli and L. Benini},
	year={2014},
	address={Madrid, Spain}
}

@article{sangogboye2017:performance,
  title={Performance comparison of occupancy count estimation and prediction with common versus dedicated sensors for building model predictive control},
  author={F. C. Sangogboye and K. Arendt and A. Singh and C. T. Veje and M. B. Kj{\ae}rgaard and B. N. J{\o}rgensen},
  booktitle={Building Simulation},
  volume={10},
  number={6},
  pages={829--843},
  year={2017},
  organization={Springer}
}

@inproceedings{liu2017:kinect,
	author = {K. S. Liu and J. Francis and C. Shelton and S. Lin},
	bookTitle={ACM/IEEE IPSN},
	title = {{Long Term Occupancy Estimation in a Commercial Space : An Empirical Study}},
	year = {2017},
	month={April},
	address={Pittsburgh, USA}
}

@inproceedings{wren2007merl,
  title={The MERL motion detector dataset},
  author={Wren, Christopher R and Ivanov, Yuri A and Leigh, Darren and Westhues, Jonathan},
  booktitle={Proceedings of the 2007 workshop on Massive datasets},
  pages={10--14},
  year={2007}
}

@inproceedings{yuan2020leveraging,
  title={Leveraging Fine-Grained Occupancy Estimation Patterns for Effective HVAC Control},
  author={Yuan, Yukun and Liu, Kin Sum and Munir, Sirajum and Francis, Jonathan and Shelton, Charles and Lin, Shan},
  booktitle={2020 IEEE/ACM Fifth International Conference on Internet-of-Things Design and Implementation (IoTDI)},
  pages={92--103},
  year={2020},
  organization={IEEE}
}

@inproceedings{fierro2012:wifi, 
	title={Demo Abstract: Zone-level occupancy counting with existing infrastructure}, 
	booktitle={Proc. ACM BuildSys}, 
	author = {G. Fierro and O. Rehmane and A. Krioukov and D. Culler}, 
	year={2012},
	address={Toronto, Ontario, Canada}
}

@inproceedings{balaji2013:wifi, 
	title={{Sentinel: Occupancy Based HVAC Actuation using Existing WiFi Infrastructure within Commercial Buildings}}, 
	booktitle={Proc. ACM SenSys}, 
	author = {B. Balaji and J. Xu and A. Nwokafor and R. Gupta and Y. Agarwal},
	year={2013},
	address = {Roma, Italy}
}

@inproceedings{ghai2012:wifi, 
	title={Occupancy detection in commercial buildings using opportunistic context sources}, 
	booktitle={Proc. IEEE PerCom},
	author = {S. K. Ghai and L. V. Thanayankizil and D. P. Seetharam and D. Chakraborty},
	year={2012},
	month={March},
	address={Lugano, Switzerland}
}

@inproceedings{conte2014bluesentinel,
  title={BlueSentinel: a first approach using iBeacon for an energy efficient occupancy detection system.},
  author={Conte, Giorgio and De Marchi, Massimo and Nacci, Alessandro Antonio and Rana, Vincenzo and Sciuto, Donatella and others},
  booktitle={BuildSys@ SenSys},
  pages={11--19},
  year={2014},
  organization={Citeseer}
}

@inproceedings{yang2016using,
  title={Using iBeacon for intelligent in-room presence detection},
  author={Yang, Yang and Li, Zhouchi and Pahlavan, Kaveh},
  booktitle={2016 IEEE International Multi-Disciplinary Conference on Cognitive Methods in Situation Awareness and Decision Support (CogSIMA)},
  pages={187--191},
  year={2016},
  organization={IEEE}
}

@inproceedings{ni2003landmarc,
  title={LANDMARC: indoor location sensing using active RFID},
  author={Ni, Lionel M and Liu, Yunhao and Lau, Yiu Cho and Patil, Abhishek P},
  booktitle={Proceedings of the First IEEE International Conference on Pervasive Computing and Communications, 2003.(PerCom 2003).},
  pages={407--415},
  year={2003},
  organization={IEEE}
}

@article{li2012measuring,
  title={Measuring and monitoring occupancy with an RFID based system for demand-driven HVAC operations},
  author={Li, Nan and Calis, Gulben and Becerik-Gerber, Burcin},
  journal={Automation in construction},
  volume={24},
  pages={89--99},
  year={2012},
  publisher={Elsevier}
}

@article{tesoriero2010improving,
  title={Improving location awareness in indoor spaces using RFID technology},
  author={Tesoriero, Ricardo and Tebar, R and Gallud, Jos{\'e} A and Lozano, Mar{\'\i}a Dolores and Penichet, Victor M Ruiz},
  journal={Expert Systems with Applications},
  volume={37},
  number={1},
  pages={894--898},
  year={2010},
  publisher={Elsevier}
}

@inproceedings{zhen2008indoor,
  title={An indoor localization algorithm for lighting control using RFID},
  author={Zhen, Zi-Ning and Jia, Qing-Shan and Song, Chen and Guan, Xiaohong},
  booktitle={2008 IEEE Energy 2030 Conference},
  pages={1--6},
  year={2008},
  organization={IEEE}
}

@article{want1992active,
  title={The active badge location system},
  author={Want, Roy and Hopper, Andy and Falcao, Veronica and Gibbons, Jonathan},
  journal={ACM Transactions on Information Systems (TOIS)},
  volume={10},
  number={1},
  pages={91--102},
  year={1992},
  publisher={ACM New York, NY, USA}
}

@article{ahmad2020occupancy,
  title={Occupancy detection in non-residential buildings--A survey and novel privacy preserved occupancy monitoring solution},
  author={Ahmad, Jawad and Larijani, Hadi and Emmanuel, R and Mannion, M and Javed, A},
  journal={Applied Computing and Informatics},
  year={2020},
  publisher={Emerald Publishing Limited}
}

@inproceedings{burke1997space,
  title={Space allocation: An analysis of higher education requirements},
  author={Burke, Edmund K and Varley, DB},
  booktitle={International Conference on the Practice and Theory of Automated Timetabling},
  pages={20--33},
  year={1997},
  organization={Springer}
}

@techreport{mccollum2004cornerstone,
  title={The cornerstone of effective management and planning of space},
  author={McCollum, B and McMullan, P},
  year={2004},
  institution={Technical report, Realtime Solutions Ltd}
}

@inproceedings{reis2000language,
  title={A language for specifying complete timetabling problems},
  author={Reis, Lu{\'\i}s Paulo and Oliveira, Eug{\'e}nio},
  booktitle={International Conference on the Practice and Theory of Automated Timetabling},
  pages={322--341},
  year={2000},
  organization={Springer}
}

@article{beyrouthy2009towards,
  title={Towards improving the utilization of university teaching space},
  author={Beyrouthy, Camille and Burke, Edmund K and Landa-Silva, Dario and McCollum, Barry and McMullan, Paul and Parkes, Andrew J},
  journal={Journal of the Operational Research Society},
  volume={60},
  number={1},
  pages={130--143},
  year={2009},
  publisher={Taylor \& Francis}
}

@article{abdelhalim2016:utilization,
  title={A utilization-based genetic algorithm for solving the university timetabling problem (uga)},
  author={E. A. Abdelhalim and G. A. El Khayat},
  journal={Alexandria Engineering Journal},
  volume={55},
  number={2},
  pages={1395-1409},
  year={2016},
  publisher={Elsevier}
}

@inproceedings{kostuch2004:university,
  title={The university course timetabling problem with a three-phase approach},
  author={P. Kostuch},
  booktitle={International Conference on the Practice and Theory of Automated Timetabling},
  pages={109--125},
  year={2004},
  organization={Springer}
}

@article{burke2003:time,
  title={A time-predefined approach to course timetabling},
  author={E. Burke and Y. Bykov and J. Newall and S. Petrovi{\'c}},
  journal={Yugoslav Journal of Operations Research},
  volume={13},
  number={2},
  pages={139--151},
  year={2003}
}

@article{badoni2015hybrid,
  title={A hybrid algorithm for university course timetabling problem},
  author={Badoni, Rakesh P and Gupta, DK},
  journal={Innovative Systems Design and Engineering},
  volume={6},
  number={2},
  pages={60--66},
  year={2015}
}

@article{el2015genetic,
  title={Genetic algorithm for solving course timetable problems},
  author={El-Sherbiny, Mahmoud M and Zeineldin, Ramadan A and El-Dhshan, Abdallah M},
  journal={International Journal of Computer Applications},
  volume={124},
  number={10},
  year={2015},
  publisher={Citeseer}
}

@inproceedings{socha2003ant,
  title={Ant algorithms for the university course timetabling problem with regard to the state-of-the-art},
  author={Socha, Krzysztof and Sampels, Michael and Manfrin, Max},
  booktitle={Workshops on Applications of Evolutionary Computation},
  pages={334--345},
  year={2003},
  organization={Springer}
}

@article{sabar2012honey,
  title={A honey-bee mating optimization algorithm for educational timetabling problems},
  author={Sabar, Nasser R and Ayob, Masri and Kendall, Graham and Qu, Rong},
  journal={European Journal of Operational Research},
  volume={216},
  number={3},
  pages={533--543},
  year={2012},
  publisher={Elsevier}
}

@electronic{webUI2018,
	Year = {2018},
	Title = {{Visualization of classroom occupancy}},
	url = {https://smartcampus.unsw.edu.au/roomoccupancy/}
}

@article{koenker2001:quantile,
  title={Quantile regression},
  author={Koenker, Roger and Hallock, Kevin F},
  journal={Journal of economic perspectives},
  volume={15},
  number={4},
  pages={143--156},
  year={2001}
}

@article{haupt2011:cross,
  title={Cross-validating fit and predictive accuracy of nonlinear quantile regressions},
  author={Haupt, Harry and Kagerer, Kathrin and Schnurbus, Joachim},
  journal={Journal of Applied Statistics},
  volume={38},
  number={12},
  pages={2939--2954},
  year={2011},
  publisher={Taylor \& Francis}
}

@article{filipovitch2016:excessparking,
  title={A systems model for achieving optimum parking efficiency on campus: The case of Minnesota State University},
  author={Filipovitch, Anthony and Boamah, Emmanuel Frimpong},
  journal={Transport Policy},
  volume={45},
  pages={86--98},
  year={2016},
 month={Jan.}
}

@article{dell2018:methodology,
  title={A methodology based on parking policy to promote sustainable mobility in college campuses},
  author={dell'Olio, Luigi and Cordera, Ruben and Ibeas, Angel and Barreda, Rosa and Alonso, Borja and Moura, Jose Luis},
  journal={Transport Policy},
  volume={80},
  pages={148--156},
  month={Aug.},
  year={2019},
  publisher={Elsevier}
}

@techreport{BritishCouncil,
	title = {{The shape of	things to come: higher education global trends and emerging opportunities to 2020}},
	author = {British Council},
	group = {csg},
	year = {2012},
	institution = {British Council},
	month = {Jan.}
}

@techreport{gminsight:carsharing,
	title = {{Car Sharing Market Size By Model (P2P, Station-Based, Free-Floating), By Business Model (Round Trip, One Way), By Application (Business, Private), Industry Analysis Report, Regional Outlook, Application Potential, Price Trend, Competitive Market Share \& Forecast, 2020 – 2026}},
	author       = { Wadhwani, Preeti Wadhwani and Saha, Prasenjit}, 
	institution = {Global Market Insights},
	type = {Market Report},
  	url={https://www.gminsights.com/industry-analysis/carsharing-market},
 	month        = {Apr.},
 	 year         = {2020}
}

@article{mimbela2007summary,
  title={Summary of vehicle detection and surveillance technologies used in intelligent transportation systems},
  author={Mimbela, Luz-Elena Y and Klein, Lawrence A and others},
  year={2007},
  publisher={United States. Joint Program Office for Intelligent Transportation Systems}
}

@techreport{mouskos2007technical,
  title={Technical solutions to overcrowded park and ride facilities},
  author={Mouskos, Kyriacos C and Boile, Maria and Parker, Neville and others},
  year={2007},
  institution={New Jersey. Dept. of Transportation}
}

@electronic{SFpark:dynamicpricing,
	Year = {2017},
	Title = {{SFpark}},
	url = {\url{http://sfpark.org/}},
	note = {Accessed on: 2020-09-17}
}

@inproceedings{rajabioun2013:intelligent,
  title={Intelligent parking assist},
  author={Rajabioun, Tooraj and Foster, Brandon and Ioannou, Petros},
  booktitle={Proc. IEEE Mediterranean Conference on Control and Automation},
  year={2013},
  month={June},
  address={Chania, Greece}
  
}

@article{gongjun2011smartparking,
  title={SmartParking: A secure and intelligent parking system},
  author={Yan, Gongjun and Yang, Weiming and Rawat, Danda B and Olariu, Stephan},
  journal={IEEE Intelligent Transportation Systems Magazine},
  volume={3},
  number={1},
  pages={18--30},
  year={2011},
  month={Apr}
}

@article{lu2010:MGCC,
  title={An intelligent secure and privacy-preserving parking scheme through vehicular communications},
  author={Lu, Rongxing and Lin, Xiaodong and Zhu, Haojin and Shen, Xuemin},
  journal={IEEE Trans. Veh. Technol.},
  volume={59},
  number={6},
  pages={2772--2785},
  month={July},
  year={2010}
}

@inproceedings{caliskan2007:MMCC,
  title={Predicting parking lot occupancy in vehicular ad hoc networks},
  author={Caliskan, Murat and others},
  booktitle={Proc. IEEE Vehicular Technology Conference},
  month={Apr.},
  year={2007},
  pages={277--281},
  address={Dublin, Ireland}

}

@inproceedings{pala2007:RFID,
  title={Smart parking applications using RFID technology},
  author={Pala, Zeydin and Inanc, Nihat},
  booktitle={Proc. IEEE Annual RFID Eurasia},
  month={Sep},
  year={2007},
  pages={1--3},
  address={Istanbul, Turkey}
}

@article{Bong2008:cctv,
author = {Bong, DBL and Ting, KC and Lai, KC},
journal = {IAENG International Journal of Computer Science},
title = {{Integrated approach in the design of car park occupancy information system (COINS)}},
year = {2008},
volume={35},
number={1},
month={Feb.}
}

@inproceedings{bin2009design,
  title={A design of parking space detector based on video image},
  author={Bin, Zhang and Dalin, Jiang and Fang, Wang and Tingting, Wan},
  booktitle={2009 9th International Conference on Electronic Measurement \& Instruments},
  pages={2--253},
  year={2009},
  organization={IEEE}
}

@article{albiol2011detection,
  title={Detection of parked vehicles using spatiotemporal maps},
  author={Albiol, Antonio and Sanchis, Laura and Albiol, Alberto and Mossi, Jos{\'e} M},
  journal={IEEE Transactions on Intelligent Transportation Systems},
  volume={12},
  number={4},
  pages={1277--1291},
  year={2011},
  publisher={IEEE}
}

@inproceedings{sevillano2014towards,
  title={Towards smart traffic management systems: Vacant on-street parking spot detection based on video analytics},
  author={Sevillano, Xavier and M{\`a}rmol, Elena and Fernandez-Arguedas, Virginia},
  booktitle={17th International Conference on Information Fusion (FUSION)},
  pages={1--8},
  year={2014},
  organization={IEEE}
}

@inproceedings{larisis2012u,
  title={U-Park: Parking management system based on wireless sensor network technology},
  author={Larisis, N and Perlepes, L and Kikiras, P and Stamoulis, G},
  booktitle={International conference on sensor technologies and applications (SENSORCOMM)},
  pages={170--177},
  year={2012}
}

@inproceedings{manni2010smart,
  title={Smart sensing and time of arrival based location detection in parking management services},
  author={M{\"a}nni, U},
  booktitle={2010 12th Biennial Baltic Electronics Conference},
  pages={213--214},
  year={2010},
  organization={IEEE}
}

@article{kianpisheh2012smart,
  title={Smart parking system (SPS) architecture using ultrasonic detector},
  author={Kianpisheh, Amin and Mustaffa, Norlia and Limtrairut, Pakapan and Keikhosrokiani, Pantea},
  journal={International Journal of Software Engineering and Its Applications},
  volume={6},
  number={3},
  pages={55--58},
  year={2012},
  publisher={Citeseer}
}

@inproceedings{lee2008intelligent,
  title={Intelligent parking lot application using wireless sensor networks},
  author={Lee, Sangwon and Yoon, Dukhee and Ghosh, Amitabha},
  booktitle={2008 International Symposium on Collaborative Technologies and Systems},
  pages={48--57},
  year={2008},
  organization={IEEE}
}

@inproceedings{vishnubhotla2010zigbee,
  title={ZigBee based multi-level parking vacancy monitoring system},
  author={Vishnubhotla, R and Rao, P Sudhakara and Ladha, A and Kadiyala, S and Narmada, A and Ronanki, B and Illapakurthi, S},
  booktitle={2010 IEEE International Conference on Electro/Information Technology},
  pages={1--4},
  year={2010},
  organization={IEEE}
}

@electronic{laexpresspark,
	title         = {{LA Express Park}},
	url           = {http://www.laexpresspark.org},
	year = {2018},
	note = {Accessed: 2020-12-12}
}

@inproceedings{quinones2015design,
  title={Design of a smart parking system using wireless sensor network},
  author={Qui{\~n}ones, Manuel and Gonazalez, Victor and Quinones, Luis and Valdivieso, Carlos and Yaguana, Willian},
  booktitle={2015 10th Iberian Conference on Information Systems and Technologies (CISTI)},
  pages={1--6},
  year={2015},
  organization={IEEE}
}

@electronic{sfparkoverview2014,
	title         = {{SFpark: Putting Theory Into Practice}},
	url           = {https://www.sfmta.com/reports/sfpark-pilot-overview},
	year = {2014},
	note = {Accessed: 2020-12-12}
}

@electronic{Chinaparking,
	title={China Parking},
	year = {2014},
	url           = {https://www.sfmta.com/reports/sfpark-pilot-overview},
	note = {Accessed: 2020-12-19}
}

@electronic{melbournedata,
	title={City of Melbourne - Open Data},
	year = {2020},
	url           = {http://data.melbourne.vic.gov.au},
	note = {Accessed: 2020-12-12}
}

@electronic{AREASpain,
	title={Barcelona's AREA},
	year = {},
	url           = {https://www.areaverda.cat/en/information/barcelona-area},
	note = {Accessed: 2020-12-19}
}

@inproceedings{yang2012smart,
  title={Smart parking service based on wireless sensor networks},
  author={Yang, Jihoon and Portilla, Jorge and Riesgo, Teresa},
  booktitle={IECON 2012-38th Annual Conference on IEEE Industrial Electronics Society},
  pages={6029--6034},
  year={2012},
  organization={IEEE}
}

@inproceedings{tang2006intelligent,
  title={An intelligent car park management system based on wireless sensor networks},
  author={Tang, Vanessa WS and Zheng, Yuan and Cao, Jiannong},
  booktitle={2006 First International Symposium on Pervasive Computing and Applications},
  pages={65--70},
  year={2006},
  organization={IEEE}
}

@inproceedings{rinne2014mobile,
  title={Mobile crowdsensing of parking space using geofencing and activity recognition},
  author={Rinne, Mikko and T{\"o}rm{\"a}, Seppo and Kratinov, D},
  booktitle={10th ITS European Congress, Helsinki, Finland},
  pages={16--19},
  year={2014}
}

@inproceedings{yang2013ipark,
  title={iPark: Identifying parking spaces from trajectories},
  author={Yang, Bin and Fantini, Nicolas and Jensen, Christian S},
  booktitle={Proceedings of the 16th International Conference on Extending Database Technology},
  pages={705--708},
  year={2013}
}

@inproceedings{nandugudi2014pocketparker,
  title={Pocketparker: Pocketsourcing parking lot availability},
  author={Nandugudi, Anandatirtha and Ki, Taeyeon and Nuessle, Carl and Challen, Geoffrey},
  booktitle={Proceedings of the 2014 ACM International Joint Conference on Pervasive and Ubiquitous Computing},
  pages={963--973},
  year={2014}
}

@inproceedings{koster2014recognition,
  title={Recognition and recommendation of parking places},
  author={Koster, Andrew and Oliveira, Allysson and Volpato, Orlando and Delvequio, Viviane and Koch, Fernando},
  booktitle={Ibero-American Conference on Artificial Intelligence},
  pages={675--685},
  year={2014},
  organization={Springer}
}

@inproceedings{stenneth2012phonepark,
  title={PhonePark: Street parking using mobile phones},
  author={Stenneth, Leon and Wolfson, Ouri and Xu, Bo and Philip, S Yu},
  booktitle={2012 IEEE 13th international conference on mobile data management},
  pages={278--279},
  year={2012},
  organization={IEEE}
}

@inproceedings{nawaz2013parksense,
  title={Parksense: A smartphone based sensing system for on-street parking},
  author={Nawaz, Sarfraz and Efstratiou, Christos and Mascolo, Cecilia},
  booktitle={Proceedings of the 19th annual international conference on Mobile computing \& networking},
  pages={75--86},
  year={2013}
}

@inproceedings{ma2014updetector,
  title={Updetector: Sensing parking/unparking activities using smartphones},
  author={Ma, Shuo and Wolfson, Ouri and Xu, Bo},
  booktitle={Proceedings of the 7th ACM SIGSPATIAL international workshop on computational transportation science},
  pages={76--85},
  year={2014}
}

@article{bagula2015design,
  title={On the design of smart parking networks in the smart cities: An optimal sensor placement model},
  author={Bagula, Antoine and Castelli, Lorenzo and Zennaro, Marco},
  journal={Sensors},
  volume={15},
  number={7},
  pages={15443--15467},
  year={2015},
  publisher={Multidisciplinary Digital Publishing Institute}
}

@inproceedings{abdullah2013integrating,
  title={Integrating Zigbee-based mesh network with embedded passive and active RFID for production management automation},
  author={Abdullah, Samihah and Ismail, Widad and Halim, Zaini Abdul and Zulkifli, Che Zalina},
  booktitle={2013 IEEE International Conference on RFID-Technologies and Applications (RFID-TA)},
  pages={1--6},
  year={2013},
  organization={IEEE}
}

@article{rashid2012automatic,
  title={Automatic parking management system and parking fee collection based on number plate recognition},
  author={Rashid, Muhammad Mahbubur and Musa, A and Rahman, M Ataur and Farahana, N and Farhana, A},
  journal={International Journal of Machine Learning and Computing},
  volume={2},
  number={2},
  pages={94},
  year={2012},
  publisher={IACSIT Press}
}

@article{tian2014design,
  title={Design of intelligent parking management system based on license plate recognition},
  author={Tian, Qing and Guo, Teng and Qiao, Shuai and Wei, Yun and Fei, Wei-wei},
  journal={Journal of Multimedia},
  volume={9},
  number={6},
  pages={774},
  year={2014},
  publisher={Citeseer}
}

@article{hartigan1979:kmean,
  title={Algorithm AS 136: A k-means clustering algorithm},
  author={Hartigan, John A and Wong, Manchek A},
  journal={Journal of the Royal Statistical Society. Series C (Applied Statistics)},
  volume={28},
  number={1},
  pages={100--108},
  year={1979},
  publisher={JSTOR}
}

@article{chang2004:LPR,
  title={Automatic license plate recognition},
  author={Chang, Shyang-Lih and Chen, Li-Shien and Chung, Yun-Chung and Chen, Sei-Wan},
  journal={IEEE Trans. Intell. Transp. Syst},
  volume={5},
  number={1},
  pages={42--53},
  year={2004},
  month={Mar.}
}

@electronic{nedap:camera,
	year = {2019},
	title = {{ANPR Access HD - HD license plate camera for vehicle access control}},
	url = {\url{https://bit.ly/2HixxU5}}
}

@conference{LevenshteinDist,
	Author = {Satoshi Takahashi and Takashi Izumi},
	Title = {{Travel Time Measurement by Vehicle Sequence Matching Method - Evaluation Method of Vehicle Sequence using Levenshtein Distance}},
	Booktitle = {Proc. SICE-ICASE},
	address = {Busan, Korea},
	Year = {2006},
	month={Oct.},
}

@manual{AnprGuidance,
	Address = {Centre for Applied Science and Technology, Sandridge, St Albans, AL49HQ United Kingdom},
	Author = {Vivienne Lyons},
	Edition = {1},
	Month = {Mar.},
	Organization = {Home Office of UK government},
	Title = {Guidance on ANPR Performance Assessment and Optimisation},
	Year = {2014}
}

@electronic{fairwork:AUworkhour,
	year = {2019},
	title = {{Maximum weekly hours}},
	organization={Australian Government Fair Work Ombudsman},
	url = {\url{https://bit.ly/1I14M9L}},
	note = {Accessed on: 2020-07-08}
}

@article{sidje1998:expokit,
  title={Expokit: A software package for computing matrix exponentials},
  author={Sidje, Roger B},
  journal={ACM Transactions on Mathematical Software (TOMS)},
  volume={24},
  number={1},
  pages={130--156},
  year={1998},
  publisher={ACM New York, NY, USA}
}

@article{sorjamaa2007methodology,
  title={Methodology for long-term prediction of time series},
  author={Sorjamaa, Antti and Hao, Jin and Reyhani, Nima and Ji, Yongnan and Lendasse, Amaury},
  journal={Neurocomputing},
  volume={70},
  number={16-18},
  pages={2861--2869},
  year={2007},
  publisher={Elsevier}
}

@article{chevillon2007direct,
  title={Direct multi-step estimation and forecasting},
  author={Chevillon, Guillaume},
  journal={Journal of Economic Surveys},
  volume={21},
  number={4},
  pages={746--785},
  year={2007},
  publisher={Wiley Online Library}
}

@electronic{UNSW:parkingfee,
year = {2020},
  title = {UNSW Estate Management Parking Rates 2020},
  url = {https://www.estate.unsw.edu.au/getting-here/parking/parking-rates-2020},
  note = {Accessed on: 2020-06-16}
}

@techreport{PBA:carsharingreport,
  title={The impact of car share services in Australia},
  year={2016},
  institution={Phillip Boyle \& Associates}
}

@inproceedings{bontempi2012machine,
  title={Machine learning strategies for time series forecasting},
  author={Bontempi, Gianluca and Taieb, Souhaib Ben and Le Borgne, Yann-A{\"e}l},
  booktitle={European business intelligence summer school},
  pages={62--77},
  year={2012},
  organization={Springer}
}

@inproceedings{zheng2015parking,
  title={Parking availability prediction for sensor-enabled car parks in smart cities},
  author={Zheng, Yanxu and Rajasegarar, Sutharshan and Leckie, Christopher},
  booktitle={Proc. IEEE ISSNIP},
  year={2015},
  month={May},
  address={Singapore, Singapore}
}

@article{rajabioun2015street,
  title={On-street and off-street parking availability prediction using multivariate spatiotemporal models},
  author={Rajabioun, Tooraj and Ioannou, Petros A},
  journal={IEEE Trans. Intell. Transp. Syst.},
  volume={16},
  number={5},
  pages={2913--2924},
  year={2015},
  month={May}
}

@article{vlahogianni2016real,
  title={A real-time parking prediction system for smart cities},
  author={Vlahogianni, Eleni I and Kepaptsoglou, Konstantinos and Tsetsos, Vassileios and Karlaftis, Matthew G},
  journal={Journal of Intelligent Transportation Systems},
  volume={20},
  number={2},
  pages={192--204},
  year={2016},
  month={June},
  publisher={Taylor \& Francis}
}

@inproceedings{pack2011using,
  title={Using metaheuristics and queueing models to optimize schedules in the academic enterprise},
  author={Pack, Charles D and Christensen, Edward W and Potter, Ronda M and Forys, Leonard and Erramilli, Ashok},
  booktitle={2011 IEEE Symposium on Computational Intelligence in Scheduling (SCIS)},
  pages={1--8},
  year={2011},
  organization={IEEE}
}

@inproceedings{shao2018LSTM,
  title={Parking availability prediction with long short term memory model},
  author={Shao, Wei and Zhang, Yu and Guo, Bin and Qin, Kai and Chan, Jeffrey and Salim, Flora D},
  booktitle={Int. Conf. Green, Pervasive, and Cloud Computing},
  pages={124--137},
  year={2018},
  month={May},
  organization={Springer}
}

@inproceedings{camero2018evolutionary,
  title={Evolutionary deep learning for car park occupancy prediction in smart cities},
  author={Camero, Andr{\'e}s and Toutouh, Jamal and Stolfi, Daniel H and Alba, Enrique},
  booktitle={Int. Conf. Learning and Intelligent Optimization},
  pages={386--401},
  year={2018},
  organization={Springer}
}

@article{klappenecker2014finding,
  title={Finding available parking spaces made easy},
  author={Klappenecker, Andreas and Lee, Hyunyoung and Welch, Jennifer L},
  journal={Ad Hoc Networks},
  volume={12},
  pages={243--249},
  year={2014},
  month={Jan.},
  publisher={Elsevier}
}

@article{burns1992econometric,
  title={An econometric forecasting model of revenues from urban parking facilities},
  author={Burns, Malcolm R and Faurot, David J},
  journal={Journal of Economics and Business},
  volume={44},
  number={2},
  pages={143--150},
  year={1992},
  month={May},
  publisher={Elsevier}
}

@article{wilkins2018lag,
  title={To lag or not to lag?: Re-evaluating the use of lagged dependent variables in regression analysis},
  author={Wilkins, Arjun S},
  journal={Political Science Research and Methods},
  volume={6},
  number={2},
  pages={393--411},
  year={2018},
  publisher={Cambridge University Press}
}

@article{zhang2013iterated,
  title={Iterated time series prediction with multiple support vector regression models},
  author={Zhang, Li and Zhou, Wei-Da and Chang, Pei-Chann and Yang, Ji-Wen and Li, Fan-Zhang},
  journal={Neurocomputing},
  volume={99},
  pages={411--422},
  year={2013},
  publisher={Elsevier}
}

@article{becker2017comparing,
  title={Comparing car-sharing schemes in Switzerland: User groups and usage patterns},
  author={Becker, Henrik and Ciari, Francesco and Axhausen, Kay W},
  journal={Transportation Research Part A: Policy and Practice},
  volume={97},
  pages={17--29},
  year={2017},
  publisher={Elsevier}
}

@book{anderson2012continuous,
  title={Continuous-time Markov chains: An applications-oriented approach},
  author={Anderson, William J},
  year={2012},
  publisher={Springer Science \& Business Media}
}

@article{shaheen2015one,
  title={One-way carsharing's evolution and operator perspectives from the Americas},
  author={Shaheen, Susan A and Chan, Nelson D and Micheaux, Helen},
  journal={Transportation},
  volume={42},
  number={3},
  pages={519--536},
  year={2015},
  publisher={Springer}
}

@article{stasko2013carsharing,
  title={Carsharing in a university setting: Impacts on vehicle ownership, parking demand, and mobility in Ithaca, NY},
  author={Stasko, Timon H and Buck, Andrew B and Gao, H Oliver},
  journal={Transport Policy},
  volume={30},
  pages={262--268},
  year={2013},
  publisher={Elsevier}
}

@article{pierce2015optimizing,
  title={Optimizing the use of public garages: Pricing parking by demand},
  author={Pierce, Gregory and Willson, Hank and Shoup, Donald},
  journal={Transport Policy},
  volume={44},
  pages={89--95},
  year={2015},
  publisher={Elsevier}
}

@book{litman2016parking,
  title={Parking management: strategies, evaluation and planning},
  author={Litman, Todd},
  year={2016},
  publisher={Victoria Transport Policy Institute Victoria, BC}
}

@article{shoup1997high,
  title={The high cost of free parking},
  author={Shoup, Donald C},
  journal={Journal of planning education and research},
  volume={17},
  number={1},
  pages={3--20},
  year={1997},
  publisher={Sage Publications Sage CA: Thousand Oaks, CA}
}

@inproceedings{pala2007smart,
  title={Smart parking applications using RFID technology},
  author={Pala, Zeydin and Inanc, Nihat},
  booktitle={2007 1st Annual RFID Eurasia},
  pages={1--3},
  year={2007},
  organization={IEEE}
}

@inproceedings{wang2011reservation,
  title={A reservation-based smart parking system},
  author={Wang, Hongwei and He, Wenbo},
  booktitle={2011 IEEE Conference on Computer Communications Workshops (INFOCOM WKSHPS)},
  pages={690--695},
  year={2011},
  organization={IEEE}
}

@inproceedings{wang2014:tracking,
  title={Tracking human queues using single-point signal monitoring},
  author={Y. Wang and others},
  booktitle={Proc. ACM MobiSys},
  pages={42--54},
  year={2014},
  organization={ACM},
  address={New Hampshire, USA}
}

@article{tang2018:indoor,
  title={Indoor crowd density estimation through mobile smartphone wi-fi probes},
  author={X. Tang and others},
  journal={IEEE transactions on systems, man, and cybernetics: systems},
  year={2018},
  publisher={IEEE},
  volume={},
  issue={},
  pages = {}
 }

@inproceedings{schauer2014:estimating,
  title={Estimating crowd densities and pedestrian flows using wi-fi and bluetooth},
  author={L. Schauer and others},
  booktitle={Proc. ACM Mobiquitous},
  pages={171--177},
  year={2014},
  address={London, UK},
  organization={ACM}
}

@article{shu2016:queuing,
  title={Queuing time prediction using WiFi positioning data in an indoor scenario},
  author={H. Shu and others},
  journal={Sensors},
  volume={16},
  number={11},
  pages={1958},
  year={2016},
  publisher={Multidisciplinary Digital Publishing Institute}
}

@inproceedings{okoshi2015:queuevadis,
  title={Queuevadis: Queuing analytics using smartphones},
  author={T. Okoshi and others},
  booktitle={Proc. ACM IPSN},
  pages={214--225},
  year={2015},
  organization={ACM},
  address={Seattle, USA}
}

@inproceedings{li2014:queuesense,
  title={Queuesense: Collaborative recognition of queuing on mobile phones},
  author={Q. Li and others},
  booktitle={Proc. IEEE SECON},
  pages={230--238},
  year={2014},
  organization={IEEE},
  address={Singapore, Singapore}
}

@inproceedings{elhamshary2018:crowdmeter,
  title={Crowdmeter: Congestion level estimation in railway stations using smartphones},
  author={M. Elhamshary and others},
  booktitle={Proc. IEEE PerCom},
  pages={1--12},
  year={2018},
  organization={IEEE},
  address={Athens, Greece}
}

@inproceedings{goncalves2016:crowdsourcing,
  title={Crowdsourcing queue estimations in situ},
  author={J. Goncalves and others},
  booktitle={Proc. ACM CSCW},
  pages={1040--1051},
  year={2016},
  organization={ACM},
  address={San Francisco, California, USA}
}

@inproceedings{segen1996:camera,
  title={A camera-based system for tracking people in real time},
  author={Segen, Jakub},
  booktitle={Proc. IEEE ICPR},
  pages={63--67},
  year={1996},
  organization={IEEE},
  address={Vienna, Austria}
}

@article{masoud2001:novel,
  title={A novel method for tracking and counting pedestrians in real-time using a single camera},
  author={O. Masoud and others},
  journal={IEEE transactions on vehicular technology},
  volume={50},
  number={5},
  pages={1267--1278},
  year={2001},
  publisher={IEEE}
}

@inproceedings{aubert1999:passengers,
  title={Passengers queue length measurement},
  author={Aubert, Didier},
  booktitle={Proc. IEEE ICIAP},
  pages={1132--1135},
  year={1999},
  organization={IEEE},
  address={Venice, Italy}
}

@inproceedings{leibe2005:pedestrian,
  title={Pedestrian detection in crowded scenes},
  author={B. Leibe and others},
  booktitle={Proc. IEEE CVPR},
  volume={1},
  pages={878--885},
  year={2005},
  organization={IEEE},
  address={San Diego, CA, USA}
}

@inproceedings{weppner2013:bluetooth,
  title={Bluetooth based collaborative crowd density estimation with mobile phones},
  author={Weppner, Jens and others},
  booktitle={Proc. IEEE PerCom},
  pages={193--200},
  year={2013},
  organization={IEEE},
  address={San Diego, CA, USA}
}

@inproceedings{musa2012:tracking,
  title={Tracking unmodified smartphones using wi-fi monitors},
  author={Musa, ABM and others},
  booktitle={Proc. ACM SenSys},
  pages={281--294},
  year={2012},
  organization={ACM},
  address={Toronto, Canada}
}

@MISC{emil2014:hcsr04,
   author={Emil},
   title={Making a better HC-SR04 echo locator},
   howpublished={\url{https://uglyduck.vajn.icu/ep/archive/2014/01/Making_a_better_HC_SR04_Echo_Locator.html}},
   note={Accessed: 19/05/2019},
   year={2014}
}

@article{cattani2017:LoRa_experimental,
  title={An experimental evaluation of the reliability of lora long-range low-power wireless communication},
  author={M. Cattani and others},
  journal={Journal of Sensor and Actuator Networks},
  volume={6},
  number={2},
  pages={7},
  year={2017},
  publisher={Multidisciplinary Digital Publishing Institute}
}

@article{sanchez2018:performance,
  title={Performance evaluation of LoRa considering scenario conditions},
  author={R. Sanchez-Iborra and others},
  journal={Sensors},
  volume={18},
  number={3},
  pages={772},
  year={2018},
  publisher={Multidisciplinary Digital Publishing Institute}
}

@MISC{arun_pressuremat,
	title={Product page for Arun Pressure Pads},
	howpublished={\url{http://www.arun-electronics.co.uk/pressure_mat.htm}},
	year={2014},
	note={Accessed: 20/10/2019}
}

@MISC{lorawan_spec,
	title={LoRaWAN 1.0.3 Specification},
	author={Sornin Et Al},
	year={2018},
	publisher={LoRa Alliance Technical Committee},
	howpublished={\url{https://lora-alliance.org/sites/default/files/2018-07/lorawan1.0.3.pdf}}
}

@MISC{lorawan_security,
	title={LoRaWAN Security, FULL  END–TO–END  ENCRYPTION  FOR IoT APPLICATION PROVIDERS},
	publisher={Gemalto, Actility AND Semtech},
	year={2017},
	howpublished={\url{https://lora-alliance.org/sites/default/files/2019-05/lorawan_security_whitepaper.pdf}}
}

@MISC{ttn_freqplans,
	title={LoRaWAN Frequencies Overview},
	publisher={The Things Network},
	howpublished={\url{https://www.thethingsnetwork.org/docs/lorawan/frequency-plans.html}},
	note = {Accessed: 20/10/2019},
	year={2019}
}

@article{daniels2013:paradox,
  title={The paradox of public transport peak spreading: Universities and travel demand management},
  author={R. Daniels and others},
  journal={International Journal of Sustainable Transportation},
  volume={7},
  number={2},
  pages={143--165},
  year={2013},
  publisher={Taylor \& Francis}
}

@MISC{fdk:battery,
   author={FDK Corporation},
   title={Technical information: alkaline manganese batteries AA size / LR6G07 Premium},
   howpublished={\url{https://media.digikey.com/pdf/Data\%20Sheets/FDK/LR6G07.pdf}},
   note={Accessed: 15/10/2019}
}

@article{huang2019novel,
  title={A novel bus-dispatching model based on passenger flow and arrival time prediction},
  author={Huang, Zhao and Li, Qingquan and Li, Fan and Xia, Jizhe},
  journal={IEEE Access},
  volume={7},
  pages={106453--106465},
  year={2019},
  publisher={IEEE}
}

@article{luo2018dynamic,
  title={Dynamic bus dispatching using multiple types of real-time information},
  author={Luo, Xinggang and Liu, Yingxin and Yu, Yang and Tang, Jiafu and Li, Wei},
  journal={Transportmetrica B: Transport Dynamics},
  year={2018},
  publisher={Taylor \& Francis}
}

@article{scrucca2013ga,
  title={{GA}: a package for genetic algorithms in {R}},
  author={Scrucca, Luca and others},
  journal={Journal of Statistical Software},
  volume={53},
  number={4},
  pages={1--37},
  year={2013},
  publisher={Citeseer}
}

@book{dumitrescu2000evolutionary,
  title={Evolutionary computation},
  author={Dumitrescu, Dumitru and Lazzerini, Beatrice and Jain, Lakhmi C and Dumitrescu, Alexandra},
  year={2000},
  publisher={CRC press}
}

@article{sun2016optimization,
  title={Optimization Model and Algorithm Design of Bus Lines Non-Fixed Headways Problem Based on Passenger Arrival Rates},
  author={Sun, Qimeng and Zhang, Xiaoning},
  journal={Open Journal of Transportation Technologies5},
  volume={1},
  pages={7--16},
  year={2016}
}

@article{mokhtari2018integration,
  title={Integration of efficient multi-objective ant-colony and a heuristic method to solve a novel multi-objective mixed load school bus routing model},
  author={Mokhtari, Naz-afarin and Ghezavati, Vahidreza},
  journal={Applied Soft Computing},
  volume={68},
  pages={92--109},
  year={2018},
  publisher={Elsevier}
}

@article{tirachini2013crowding,
  title={Crowding in public transport systems: effects on users, operation and implications for the estimation of demand},
  author={Tirachini, Alejandro and Hensher, David A and Rose, John M},
  journal={Transportation research part A: policy and practice},
  volume={53},
  pages={36--52},
  year={2013},
  publisher={Elsevier}
}

@article{smith1997traffic,
  title={Traffic flow forecasting: comparison of modeling approaches},
  author={Smith, Brian L and Demetsky, Michael J},
  journal={Journal of transportation engineering},
  volume={123},
  number={4},
  pages={261--266},
  year={1997},
  publisher={American Society of Civil Engineers}
}

@article{faraway1998time,
  title={Time series forecasting with neural networks: a comparative study using the air line data},
  author={Faraway, Julian and Chatfield, Chris},
  journal={Journal of the Royal Statistical Society: Series C (Applied Statistics)},
  volume={47},
  number={2},
  pages={231--250},
  year={1998},
  publisher={Wiley Online Library}
}

@article{chen2016prediction,
  title={Prediction of shanghai metro line 16 passenger flow based on time series analysis-with Lingang avenue station as a study case},
  author={Chen, YL and Sha, YW and Zhu, XL and Zhang, XH},
  journal={Oper. Res. Fuzzy},
  volume={6},
  number={1},
  pages={15--26},
  year={2016}
}

@article{suwardo2010arima,
  title={ARIMA models for bus travel time prediction},
  author={Suwardo, W and Napiah, Madzlan and Kamaruddin, Ibrahim},
  journal={Journal of the institute of engineers Malaysia},
  pages={49--58},
  year={2010}
}

@article{zhou2014direct,
  title={Direct ridership forecast model of urban rail transit stations based on spatial weighted LS-SVM},
  author={Zhou, JZ and Zhang, DY},
  journal={Journal of the China Railway Society},
  volume={36},
  number={1},
  pages={1--7},
  year={2014}
}

@article{tsai2009neural,
  title={Neural network based temporal feature models for short-term railway passenger demand forecasting},
  author={Tsai, Tsung-Hsien and Lee, Chi-Kang and Wei, Chien-Hung},
  journal={Expert Systems with Applications},
  volume={36},
  number={2},
  pages={3728--3736},
  year={2009},
  publisher={Elsevier}
}

@article{wang2007real,
  title={Real-time freeway traffic state estimation based on extended Kalman filter: A case study},
  author={Wang, Yibing and Papageorgiou, Markos and Messmer, Albert},
  journal={Transportation science},
  volume={41},
  number={2},
  pages={167--181},
  year={2007},
  publisher={INFORMS}
}

@article{tang2003comparison,
  title={Comparison of four modeling techniques for short-term AADT forecasting in Hong Kong},
  author={Tang, YF and Lam, William HK and Ng, Pan LP},
  journal={Journal of Transportation Engineering},
  volume={129},
  number={3},
  pages={271--277},
  year={2003},
  publisher={American Society of Civil Engineers}
}

@inproceedings{zhang2017short,
  title={Short-term passenger flow forecasting based on phase space reconstruction and LSTM},
  author={Zhang, Yong and Zhu, Jiansheng and Zhang, Junfeng},
  booktitle={International Conference on Electrical and Information Technologies for Rail Transportation},
  pages={679--688},
  year={2017},
  organization={Springer}
}

@article{luo2019new,
  title={A new framework of intelligent public transportation system based on the Internet of Things},
  author={Luo, Xing-Gang and Zhang, Hong-Bo and Zhang, Zhong-Liang and Yu, Yang and Li, Ke},
  journal={IEEE Access},
  volume={7},
  pages={55290--55304},
  year={2019},
  publisher={IEEE}
}

@article{pang2017scheduling,
  title={Scheduling optimization of intelligent public transport system based on MAST},
  author={Pang, Ming-bao and Chen, Mao-lin and Zhang, Ning},
  journal={J. Transp. Syst. Eng. Inf. Technol.},
  volume={17},
  number={1},
  pages={1009--6744},
  year={2017}
}

@inproceedings{sun2017unsupervised,
  title={Unsupervised mechanisms for optimizing on-time performance of fixed schedule transit vehicles},
  author={Sun, Fangzhou and Samal, Chinmaya and White, Jules and Dubey, Abhishek},
  booktitle={2017 IEEE International Conference on Smart Computing (SMARTCOMP)},
  pages={1--8},
  year={2017},
  organization={IEEE}
}

@article{chen2020multiobjective,
  title={A multiobjective single bus corridor scheduling using machine learning-based predictive models},
  author={Chen, Bing and Bai, Ruibin and Li, Jiawei and Liu, Yueni and Xue, Ning and Ren, Jianfeng},
  journal={International Journal of Production Research},
  pages={1--16},
  year={2020},
  publisher={Taylor \& Francis}
}

@article{feng2018design,
  title={Design of intelligent bus positioning based on Internet of Things for smart campus},
  author={Feng, Xiaojian and Zhang, Jiuling and Chen, Jinhong and Wang, Guoqing and Zhang, Liuye and Li, Runze},
  journal={IEEE Access},
  volume={6},
  pages={60005--60015},
  year={2018},
  publisher={IEEE}
}

@article{eberlein1998real,
  title={The real-time deadheading problem in transit operations control},
  author={Eberlein, Xu Jun and Wilson, Nigel HM and Barnhart, Cynthia and Bernstein, David},
  journal={Transportation Research Part B: Methodological},
  volume={32},
  number={2},
  pages={77--100},
  year={1998},
  publisher={Elsevier}
}

@article{zhang2018two,
  title={Two-way-looking self-equalizing headway control for bus operations},
  author={Zhang, Shuyang and Lo, Hong K},
  journal={Transportation research part B: methodological},
  volume={110},
  pages={280--301},
  year={2018},
  publisher={Elsevier}
}

@article{daganzo2009headway,
  title={A headway-based approach to eliminate bus bunching: Systematic analysis and comparisons},
  author={Daganzo, Carlos F},
  journal={Transportation Research Part B: Methodological},
  volume={43},
  number={10},
  pages={913--921},
  year={2009},
  publisher={Elsevier}
}

@article{pavai2017crossover, 
	author = {Pavai, G. and Geetha, T. V.}, 
	title = {A Survey on Crossover Operators}, 
	year = {2016}, 
	publisher = {Association for Computing Machinery}, 
	address = {New York, NY, USA}, 
	volume = {49}, number = {4}, 
	journal = {ACM Comput. Surv.},
	month = dec, articleno = {72},
 	numpages = {43}
  }

@inproceedings{khuri1994evolutionary,
  title={An Evolutionary Approach to Combinatorial Optimization Problems.},
  author={Khuri, Sami and B{\"a}ck, Thomas and Heitk{\"o}tter, J{\"o}rg},
  booktitle={ACM Conference on Computer Science},
  pages={66--73},
  year={1994}
}

@book{korte2011combinatorial,
  title={Combinatorial optimization},
  author={Korte, Bernhard H and Vygen, Jens and Korte, B and Vygen, J},
  volume={1},
  year={2011},
  publisher={Springer}
}

@inproceedings{gkiotsalitis2018bus,
  title={Bus operations scheduling subject to resource constraints using evolutionary optimization},
  author={Gkiotsalitis, Konstantinos and Kumar, Rahul},
  booktitle={Informatics},
  volume={5},
  number={1},
  pages={9},
  year={2018},
  organization={Multidisciplinary Digital Publishing Institute}
}
\end{singlespace}



\end{document}